\documentstyle[12pt,doublespace]{report}
\input epsf
\begin{document}
\pagenumbering{roman}
%\include{frontend}

%%TiTLe Page

\thispagestyle{empty}
\vspace*{\fill}
\begin{center}
{\Huge The Phenomenology of Neutral} \\[.2in]
{\Huge Heavy Leptons} \\[.2in]
by \\[.2in]
Ivan Melo, RNDr. \\[.4in]
A thesis submitted to\\
the Faculty of Graduate Studies and Research \\
in partial fulfilment of\\
the requirements for the degree of\\[.1in]Doctor of Philosophy\\[.5in]
Department of Physics \\[.2in]
Ottawa-Carleton Institute for Physics\\
Ottawa, Ontario\\
February, 1996\\
%\today\\
\copyright\ copyright 1996, Ivan Melo \\
\end{center}
\vspace*{\fill}

\thispagestyle{empty}
\vspace*{\fill}
\begin{center}
The undersigned hereby recommend to\\
 the Faculty of Graduate Studies and Research\\
acceptance of the thesis,\\[.5in]
{\Large The Phenomenology of Neutral Heavy Leptons}\\[.5in]
submitted by Ivan Melo, RNDr. \\
in partial fulfilment of the requirements for \\
the degree of Doctor of Philosophy \\[.5in]
\rule{3in}{.5pt}\\
Chair, Department of Physics\\[.5in]
\rule{3in}{.5pt}\\
Thesis Supervisor\\[.5in]
\rule{3in}{.5pt}\\
External Examiner\\[.5in]
Carleton University\\[.5in]
Date: \rule{1.5in}{.5pt}
%\today
\end{center}
\vspace*{\fill}

\setcounter{page}{0}
%% Abstract
\newpage
\vspace*{\fill}
%\addcontentsline{toc}{chapter}{Abstract}
{\Huge Abstract} \vspace{.5in}

Naturally small neutrino masses can arise in some grand unified models. The mechanism of neutrino mass generation in these models typically requires the existence of neutral heavy leptons. We study the low-energy phenomenology of these new fermions. Concentrating on loop corrections due to neutral heavy leptons, we examine how the flavour-conserving leptonic decays of the Z boson, universality breaking in these decays, and the W boson mass depend on the mass and mixings of the neutral heavy leptons. Working within the framework of a superstring-inspired $SU(2)_{L} \times U(1)_{Y}$ model,
we show that these flavour-conserving processes have some virtues over the traditionally considered flavour-violating decays.
\vspace*{\fill}

%% Acknowledgements
\newpage
\vspace*{\fill}
%\addcontentsline{toc}{chapter}{Acknowledgements}
{\Huge Acknowledgements} \vspace{.5in}

Having come to Carleton from overseas, I was lucky to find here excellent conditions for my work, from the top research carried out at the department, to even more importantly, a very friendly, stimulating and inspiring atmosphere created by the Physics faculty, staff and students. 
I am indebted to everyone who helped me to feel at home. Especially one person is dear to my family. In late Roselyn Tighe
we found an exceptional friend who stood by us in both good and bad times. She became our Canadian mom. Thank you, Roz, for everything.

I was priviledged to work closely with two fine physicists who helped me to achieve my best. My thanks go to Pat Kalyniak who helped my dream of doing theoretical physics to come true, for her guidance and support; and to Peter Watson for the many discussions his true physics spirit made so exciting.

To my parents, for their support and prayers, I am grateful.

To Jakub, Matej and Katka, for giving me the strength to pursue my goal, my love.
\vspace*{\fill}

\tableofcontents
\newpage
         
\listoftables
\addcontentsline{toc}{chapter}{List of Tables}
\newpage
\addcontentsline{toc}{chapter}{List of Figures}
\listoffigures
\newpage

\pagenumbering{arabic}
\setcounter{page}{1}
\markright{}

\pagestyle{myheadings}
\thispagestyle{plain}

%\newpage

\chapter{Introduction}

\section{The problem of small neutrino masses}
\label{tposnm}

Why are macroscopic things around us massive ?
We know they, as composite objects, get their mass from the mass of their 
constituents and from
the interactions among the constituents. But if the constituents are
elementary particles, where is {\it their} mass coming from ?
For theoretical physicists mass represents a rather unwelcome pollution of their
elegant massless theories. They believe the world is essentially symmetric and
it is the mass that makes this symmetry hard to see (symmetry is hidden or 
spontaneously broken). Therefore they build
massless theories possessing beautiful symmetries and then break these
symmetries to give particles their masses. The exact mechanism of symmetry
breaking is not known and may not be known for a long time. 
It is one of the most attractive
problems of particle physics. Some argue we have
to wait for the ultimate solution until physics at the Planck scale 
($10^{19}$ GeV) is understood, others
suggest optimistically that $10^{5}$ GeV might do. The ultimate theory should 
predict the masses of all elementary particles in agreement with experiment.

It is believed the origin of the weak boson masses
(associated with the electroweak symmetry breaking) 
is better understood than that of the fermion masses (flavour
symmetry breaking).
Concentrating on the fermions, we list the quark
and lepton content of the so-called standard model of electroweak
interactions in Table \ref{tparticles}, along with their masses \cite{griff}. 
\begin{table}[htb]
\begin{center}
\begin{tabular}{|l|c|} \hline
   1.       & Mass [MeV]           \\
  \hline
$u$         &   4.2   \\
$d$         &   7.5   \\
$\nu_{e}$ &     0   \\
$e$         &   0.5110   \\
\hline
\end{tabular} $\;\;\;\;\;\;$
\begin{tabular}{|l|c|} \hline
   2.       & Mass [MeV]           \\
  \hline
$c$           &   1 100   \\
$s$           &     150   \\
$\nu_{\mu}$ &       0   \\
$\mu$       &   105.6   \\
\hline
\end{tabular} $\;\;\;\;\;\;$
\begin{tabular}{|l|c|} \hline
   3.       & Mass [GeV]           \\
  \hline
$t$            &   176   \\
$b$            &   4.2   \\
$\nu_{\tau}$ &   0   \\
$\tau$       &   1.784   \\
\hline
\end{tabular}
\end{center}
\caption{The three families of quarks ($u,c,t,d,s,b$) and leptons (neutrinos
$\nu_{e}, \nu_{\mu}, \nu_{\tau}$; charged leptons $e, \mu, \tau$) 
and their masses according to
the standard model of electroweak interactions}
\label{tparticles}
\end{table}
These masses are not predicted by the standard model.
The quark and charged lepton masses represent an experimental input, $9$ 
parameters out of the total $17$ present in the standard model
\footnote{Of the remaining 8 parameters, the four CKM mixing matrix parameters 
are also closely related to the origin of the quark masses.}. 
The neutrinos are postulated as massless. This postulate, however,
is based on the assumption that 
the neutrino is the only fermion without a right-handed field - an asymmetry
going against the spirit of symmetric theories which work with both left-handed
and right-handed fields. Therefore the massless neutrinos are not natural in
the standard model and none of the twelve fermion masses is actually predicted. 

While the ultimate solution for mass prediction may be
far from us, it is worthwhile to think of partial steps which could bring us
closer to it.
A popular strategy aims at the reduction of the twelve independent mass
parameters. From grand unified theories (GUT's) which describe strong and
electroweak interactions as different manifestations of a single 
force, there are hints the fermion masses are related to each other by simple
formulae at very high energies (GUT scale $\sim 10^{16}$ GeV) 
\footnote{For example in the simplest version of $SO(10)$ GUT, all masses in a
given family are equal at the GUT scale.} 
and it is in the 
transition to our low-energy world  
\footnote{By low energies we mean here energies up to a few hundred GeV and by
very low energies those up to a few GeV.}
where masses pick up 
different corrections and end up in the array
of seemingly unrelated numbers shown in Table~\ref{tparticles}.

In the past the efforts to partially explain fermion masses in grand unified 
models
experienced great difficulties facing the problem of small  
neutrino masses. In the discussion above, we dismissed the way the standard
model postulates the zero mass neutrinos. What is then the experimental basis 
for the claim the neutrino masses are small, possibly zero ? Ironically, 
even though neutrinos
are possibly the second most abundant elementary species in the Universe, we
do not know exactly what their mass is.
This is partly due to the smallness of this mass
and partly because it is so hard
to detect neutrinos (a typical cross section is $10^{-43} \mbox{ cm}^{2}$ for a
$10$ MeV
neutrino, more than twenty orders of magnitude below the cross section for an
electron of the same energy). Experiments set
the following upper limits on the masses of the electron, muon and tau
type neutrino \cite{pdb}:
\begin{equation}
\begin{array}{lcr}
m_{\nu_{e}}    & < & 7.2  \;{\rm eV}     \\
m_{\nu_{\mu}}  & < & 270  \;{\rm keV}    \\
m_{\nu_{\tau}} & < & 31   \;{\rm MeV}
\end{array}
\end{equation}
The neutrino masses are, even at their allowed maxima, strikingly small when 
compared with the masses of the
charged leptons and quarks within their families.
For example,
within the first family of the standard model, $u$ and $d$ quarks
have masses of a few MeV, and the electron has a mass of $0.511$ MeV (see Table
\ref{tparticles}).
The simple formulae relating the fermion masses we mentioned 
earlier,
put different fermion masses on the same scale: it is thus conceivable in GUT's
that masses of the first family are scattered around a few MeV, but the mass of
the electron type neutrino is a deep mystery, being at least five orders of
magnitude below this natural scale for the first family. The difference of five
orders cannot be explained by corrections  masses pick up running from
the GUT scale to low energies. Something else must be
involved. Similar behaviour, although perhaps less pronounced, can be observed 
among the second and third family members.
         
\section{Solutions to the problem}
\label{sttp}

%There are two approaches to the problem of the small neutrino masses. One
%assumes neutrinos are massless particles; the other suggests they are massive.
%The more popular class of theories is the one with massive neutrinos. We will
%therefore discuss it first.

Before we start to discuss possible solutions to our problem, let us be more 
specific about the relation between GUT's and
low-energy theories, such as the standard model. While the standard model 
\cite{key1,key2,key3,key4} has a
natural scale of $10^{2}$ GeV, GUT models \cite{gut} describe physics actively 
operating at the GUT scale of $10^{16}$ GeV; at the same time they should 
explain low-energy data at least as well as the standard model.
GUT models, although more elegant than the standard model, are also
more complicated: they have a richer gauge structure and symmetry breaking
sector and, especially in the case of $E_{6}$, a fermion sector with more
particles.
A thorough discussion of neutrino masses in these models is beyond the scope of
this work. Here we are mainly interested in the phenomenology of neutral
heavy leptons at low energies. Fortunately, at low energies
most of the extraneous baggage associated with the complicated structure
of GUT models has very little impact - it is integrated out and the remaining
effective theory often
represents just a minimal extension of the standard model of electroweak
interactions.
This is in fact no big surprise. It simply reflects a very good
agreement of experimental data with the standard model and the fact the GUT's
extend the standard model rather than replace it.
The physics active at high energies thus decouples at the scales currently
accessible to us.
For example, in the model studied in this thesis, 
it is just two extra neutrino fields
per family which enlarge the standard model.
Therefore we will focus our discussion on these minimal extensions,
referring to unification models
as motivation for a particular simple extension of the standard model. 
Interestingly, basic directions in the theoretical treatment of neutrino masses
can be followed with just minor extensions of the fermion sector of the standard
model, leaving the gauge structure $SU(2)_{L}\times U(1)_{Y}$
and symmetry breaking sector intact.

The simplest model with massive neutrinos one can immediately think of is a 
straightforward extension of the standard model. One can introduce the
right-handed neutrino field missing in the standard model and treat the
neutrino in the same way as all other fermions - as a massive Dirac particle.
This means the neutrino mass is still not predicted and the simplest model
thus fails to address the problem of the small neutrino masses.

A possible solution was found by Yanagida and Gell-Mann, Ramond and Slansky
in the famous see-saw mechanism \cite{guts,seesaw}.
A simple low-energy see-saw model has the same fermion content as the simplest
model just described, however, mass terms violating the
total lepton number $L$ are allowed.
This leads to the description of neutrinos as Majorana fermions rather than 
Dirac; neutral heavy leptons (NHL's) are introduced into this theory as a 
necessary ingredient.
In Chapter 2 we describe the
difference between Majorana and Dirac
fermions more formally. Here it suffices to say that a Dirac
neutrino is a particle like all other fermions with left-handed and
right-handed particle and antiparticle states, while a Majorana neutrino is a
particle which is its own antiparticle and therefore comes in just two states
described by a left-handed and a right-handed field.
The neutrino is the only particle of
the fermion content of the standard model that can possibly be described as 
a Majorana particle because it is neutral. 
                                                               
The see-saw mechanism comes with the following relation for
the mass of a neutrino $m_{\nu}$:
\begin{eqnarray}
\label{ss1}
m_{\nu} = D^{2}/M_{N}.
\end{eqnarray}
The mass $D$ is a typical family mass (say $1-2$ MeV for the first family)
and $M_{N}$ is the NHL mass. This relation tells us the neutrinos become very
light with respect to $D$ due to a very large mass $M_{N}$. Their mass is
not predicted ($M_{N}$ is unknown);
nevertheless its smallness is understood since a very
large mass scale required for the $M_{N}$ is naturally expected in GUT's.
                                                  
Whether neutrinos are Dirac or Majorana particles is an issue by itself.
Experiments which try to address it rely on the fact that Majorana
neutrinos break lepton number conservation, whereas Dirac neutrinos respect it.
A clear answer would be provided by the observation of a neutrinoless double 
beta
decay, a so far unobserved process mediated only by Majorana
neutrinos~\footnote{Another possibility of proving the Majorana nature of the 
neutrino was
suggested as a result of theoretical studies of the
origin of the lepton symmetry breaking. This symmetry can be broken explicitly
(it is not a symmetry of the Lagrangian at any energy or temperature); it can
also be broken spontaneously. If the lepton number $L$ is a global symmetry 
which is
broken spontaneously, a massless, pseudoscalar Majoron arises \cite{majoron}.
The discovery of this particle would prove the Majorana nature of the
neutrino.} \cite{betabeta}.

Theories with massive neutrinos are popular since some motivation for nonzero
neutrino masses comes also from outside particle physics.
Massive neutrinos could explain the mystery of missing solar neutrinos
\cite{solarnu} through 
matter enhanced time dependent
neutrino oscillations, the so-called MSW effect \cite{msw}.
%A similar problem is with the missing atmospheric muon neutrinos \cite{atmo}.
In cosmology, massive neutrinos could explain at least part of the dark matter
puzzle. The possibility of this increased after COBE data on the anisotropy
of the cosmic microwave background radiation were analysed: there are hints
some $10 - 30$ \% of the dark matter is hot \cite{hot} and neutrinos with
masses between 2 - 7 eV make a good candidate.  Cosmology further constrains
(subject to plausible assumptions) the mass of each of the three neutrinos 
to $m_{\nu} < 25$ eV (see Ref.
\cite{mohapatra}, Sec. 15.3.1). 
%Heavier neutrinos would make the density of the
%Universe larger than the critical density which is believed to be the
%actual density of the Universe in inflationary scenarios.

%Indeed, if the neutrinos (assuming the three neutrino
%species are degenerate in mass) were
%heavier than $2.5$ eV (for Hubble constant $50\mbox{ km s}^{-1}\mbox{Mps}^{-1}$
%), they would
%contribute more than $10$ \% of the critical density (see Ref.
%\cite{mohapatra}, Sec. 15.3.1) which is
%believed to be the
%actual density of the Universe in inflationary scenarios. This argument gives
%also the upper limit on the neutrino mass of $25$ eV, with the same assumptions
%applied.

These astrophysical and cosmological indications are quite appealing and
some authors argue on their basis that an $SO(10)$ GUT
with three nearly degenerate 
Majorana neutrino masses of about $2$ eV could be the correct unified theory
\cite{hints}.

The see-saw mechanism is not the only one which addresses the issue of the
smallness of the neutrino mass. 
An alternative was worked out in
the class of models wherein neutrinos remain massless while other neutral
leptons can acquire a large mass.
Here it is argued there can be some global symmetry
present, such as lepton number $L$, which prevents neutrinos from becoming 
massive.
%One such model could arise as a low-energy limit of a supersymmetry-inspired 
%$SO(10)$ GUT \cite{wolfe} 
%or a superstring-inspired $E_{6}$ GUT \cite{vallemo}.
%The supersymmetry inspiration of the former, and the superstring inspiration of
%the latter consists of yet another (left-handed)
%neutrino field added to the fermion content of the standard
%model. 
One such model could arise as a low-energy limit of a superstring-inspired
$E_{6}$ GUT \cite{vallemo}. The superstring inspiration consists of yet another
(left-handed) neutrino field added to the fermion content of the standard
model.
Therefore we have three neutrino fields per family: the standard 
left-handed one, the right-handed one required by GUT's and the one needed
by superstrings.
This field content and the lepton number symmetry 
\footnote{We note the conserved lepton number 
is also present in the standard model and in the simplest model with massive
neutrinos. However, in the former the conserved $L$
is a consequence of the missing right-handed neutrino field and in the latter,
the two neutrino fields per family are not enough to keep neutrinos massless.}
give rise to 
three massless neutrinos and to
three massive Dirac NHL's, as described in Chapter 3. In this thesis I study 
phenomenological implications of this model (henceforth called 'our' model), 
with emphasis put on signatures of NHL's
in precision data from LEP collider at CERN (leptonic widths of the Z boson)
and Tevatron collider at Fermilab (the mass of the W boson, $M_{W}$).

Both theories with massive and massless neutrinos (our model) 
naturally exhibit some of the properties familiar
from the quark sector of the standard model. For instance, neutrino mixing 
may arise with an analogue of
the Cabibbo-Kobayashi-Maskawa (CKM) mixing matrix for the leptonic sector.
Individual lepton family
numbers are expected to be violated, as is lepton universality and CP
symmetry. The total lepton number $L$ has already been discussed.
What is not predicted by our model, in contrast with see-saw models, are
time dependent neutrino oscillations and neutrinoless double beta decay.
The physics of light neutrinos is thus richer in see-saw models.
On the other hand the physics of neutral heavy leptons looks more promising
in our model. The NHL's contribution to low-energy observables is proportional 
to their mixings and these depend on the ratio $\frac{D}{M_{N}}$. In see-saw
models this ratio is normally very small since the ratio 
$\frac{D^{2}}{M_{N}} = m_{\nu}$ is very small. In our model with 
massless neutrinos
there is no such restriction and, consequently, mixings can be relatively large.

To show that the differences between theories with massive neutrinos and 
our model should not be taken too seriously,
we note that there is a variant of our model where the lepton 
number symmetry is
slightly broken and the neutrinos are given a small mass \cite{vallemo,garval}
and there are variants
of see-saw models where the restriction from the ratio 
$\frac{D^{2}}{M_{N}}$ is avoided (assuming certain symmetries in the 
neutrino mass
matrix) and possibly large mixings of NHL's arise as a result
\cite{pilaftsis2,Ng1}.

\section{The phenomenology of NHL's}
                
To investigate the phenomenology of NHL's more closely, let us be more specific
about how they enter low-energy observables via their mixings. For simplicity
we neglect interfamily mixings in this section. The key point, formally derived
and discussed in Chapters 2 and 3, is
that the neutrinos taking part in weak interactions, $\nu_{l}\; (l = e, \mu,
\tau)$ are no longer states of definite mass but are combinations of light
(massless in our model) neutrinos $\nu^{'}$ and NHL's $N$:
\begin{equation}
\nu_{l} = K_{L}\:\nu^{'} \; + \; K_{H} N.
\end{equation}
Here $K_{L}$ and $K_{H}$ are mixing parameters 
related through $K_{L}^{2} + K_{H}^{2} = 1$. 
The chance of finding the signature of an NHL is proportional to the size of 
$K_{H}$,
which is equal to the ratio we discussed in the previous section, $K_{H} =
\frac{D}{M_{N}}$. 
This picture is model independent, whether NHL's come from our model, 
or a see-saw one. In the standard model 
we have $K_{H} = 0,
K_{L} = 1$ and the weak eigenstates $\nu_{l}$ become identical with the mass
eigenstates $\nu^{'}$.

The eigenstates $\nu_{l}$ interact weakly by coupling to the mediators of 
electroweak
interactions, W and Z bosons. Through the mixing $K_{H}$, so do NHL's.
For example, the simplest way to discover NHL's would be to directly
produce them at the Z boson factories (LEP collider at CERN and SLC at
SLAC). However, to be
produced directly, their mass would have to be smaller than the Z boson mass -
this is the energy available at the Z factories.
Since there has been so far no evidence for NHL production, we conclude that 
either they are more massive than the Z, or
their mixing is small enough to suppress their production to such a degree that
they escape detection.

Another method to probe mixings of NHL's is an indirect one, through the
measurement of the mixing parameter $K_{L}$ of the light neutrinos in, for
example, pion and beta decays. NHL's are not produced in the decays, 
nevertheless their
existence could be revealed
if $K_{L}$ is far enough from 1 to reduce the decay rates beyond the
experimental uncertainties. These so-called universality constraints give us
the best limits on mixings.
 
While the direct production is sensitive to NHL masses only up to the 
mass of the Z boson $M_{Z}$, 
indirect methods are sensitive only to mixings, not masses. 
What if the NHL is heavier than the Z boson ?
Can we obtain some information on its mass ?
In cases when particle physicists face a problem of 
probing masses of hypothetical particles which are larger than the energy
currently available, they study contributions of these particles
in radiative corrections (loops) to some
observables. Since loops are higher order terms of perturbation theory,
they represent just a small correction to the lowest
order (tree-level) calculation and often their size 
is not greater than 
experimental uncertainties.  For example, only recently have precision 
experiments 
seen some evidence for the genuine electroweak loop corrections of the standard
model. 
Despite their smallness, loop corrections in combination with precision data
can impose important restrictions on the parameter
space of various models. 

The major part of my thesis studies how NHL's
contribute via loops to the regular leptonic decays of the Z boson, a lepton
universality breaking parameter (both studied in precision tests at LEP) 
and the mass of the W boson. This is a novel approach to the study of NHL's in
loops. The previous studies concentrated mainly on flavour-violating
decays, such as $\mu \rightarrow e \gamma$ at very low energies or 
$Z \rightarrow
e^{+}\mu^{-}$ at the Z factories' energy. We argue that our calculations probe 
NHL masses and mixings more
efficiently than these traditional studies of flavour-violating processes.
The limits on NHL mass we obtain from the leptonic decays of the Z boson and
the
lepton universality breaking parameter are comparable to the limit derived from
the considerations of perturbative theory breakdown, discussed in Sec.
\ref{breakdown}.
%Although carried out in the context of the superstring-inspired model, our
%analysis is qualitatively valid also for see-saw models.

This work is organized as follows:
In Chapter 2 we treat models of neutrino mass formally.  
After specifying the classical Lagrangian of the standard model,
we investigate how the peculiar nonzero energy density of the vacuum
associated with the existence of a fundamental scalar Higgs field accommodates
fermion masses in the standard model. We then move on to describe 
neutrino masses in
$SU(2)_{L}\times U(1)_{Y}$ models beyond the standard model; we consider
the simplest extension
of the standard model leading to massive Dirac neutrinos and a
see-saw model with Majorana neutrinos as examples.  
We also briefly describe neutrino mass in grand unified models and show where
motivation for our model comes from.

In Chapter 3 we discuss our model, a superstring-inspired
$SU(2)_{L}\times U(1)_{Y}$ model of neutrino mass, in detail.
We define the fermion
content and the neutrino mass matrix and show how massless neutrinos and 
NHL's arise through the diagonalization of the mass matrix in the case of one 
family
and also in the general case of three families. The mixing matrix is described
and phenomenologically relevant mixing parameters are defined. 
In the second part of that chapter we review existing
constraints on NHL's.

As a prerequisite for one-loop calculations, the standard model at the one-loop
level is discussed formally in Chapter 4. A key ingredient of the 
calculations is the
renormalization of the standard model and we spend some time dealing with its
salient features.

In Chapter 5 we revisit the flavour-violating leptonic decays of the Z boson
in our model. Our results are followed by a discussion of flavour-violating
processes in general.

In Chapter 6 we study the impact NHL's have through loops on flavour-conserving
leptonic decays of the Z boson, lepton universality breaking in these decays,
and the W boson mass. One-loop corrections are classified and
calculated and the most important diagrams are identified. Violation of the
decoupling theorem by the NHL's and its relevance is discussed. 
Implicit dependence of our results on muon decay is clarified.

Chapter 7 completes calculations from the previous chapter by considering
the full set of diagrams contributing to the muon decay.

In the last chapter we conclude by summarizing our main results.

\newpage

\chapter{The models of neutrino mass}

In this chapter we treat the models of neutrino mass formally.
We start with the standard model, then we proceed with models beyond the
SM. As discussed in the
Introduction, we concentrate mainly on minimal extensions of the standard
model
- models with $SU(2)_{L} \times U(1)_{Y}$ gauge structure.

In Sec. \ref{classical1} we present the classical electroweak Lagrangian 
defining the standard model. We rather state than discuss the
principles upon which it is based. The purpose is to specify our notation and
to give a practical reference point for the discussions of the models of 
neutrino mass (Chapters 2 and 3) and for one-loop calculations (Chapters 5,6,7)
\footnote{One-loop calculations actually require the quantization and the 
subsequent extension of the classical Lagrangian by additional terms; this is 
discussed in Chapter 4, which directly precedes one-loop 
calculations in Chapters 5,6 and 7.}.

In Sec. \ref{fermion2} we examine the fermion 
masses in the standard model,
in Sec. \ref{neutrino3} we describe the generation of neutrino masses 
in simple
extensions of the standard model 
(a simple model with Dirac neutrinos and a see-saw model
with Majorana neutrinos) and finally, in Sec. \ref{grand4} we briefly deal 
with neutrino masses in grand unified models. 
The superstring-inspired minimal extension
of the standard model, motivated in this last section, 
is discussed in detail in Chapter 3.
                
\section{Classical electroweak Lagrangian}
\label{classical1}

The standard model of electroweak interactions (SM) \cite{key1,key2,key3,key4}
is a gauge theory based on a local
$SU(2)_{L} \times U(1)_{Y}$ symmetry group, which describes 
electromagnetic and weak
interactions as manifestations of a single electroweak force.
The $SU(2)_{L}$ part is the group of the weak isospin $I$ and the $U(1)_{Y}$ 
part is the group
of the weak hypercharge $Y$. The quantum numbers $I_{3}$ (the third component
of $I$) and $Y$ are related to the electric charge $Q$ via
\begin{eqnarray}
Q & = & I_{3} + \frac{Y}{2}.
\end{eqnarray}
The $SU(2)_{L}
\times U(1)_{Y}$ symmetry is spontaneously broken to $U(1)_{em}$ (the group of
the electric charge $Q$) and particle
masses are generated by the nonsymmetric vacuum.             

We now list the fermion content of the SM
\footnote{In fact, we already did it in Table
\ref{tparticles}; here we use more formal description.}.
 The particles, represented by chiral (left-handed and right-handed)
fields, form three families and therefore can be represented as three-component
vectors in family space.
With the theoretical treatment of fermion masses in mind we differentiate
between the field content of an unbroken electroweak theory and that of a
broken one.
The left-handed lepton fields of the unbroken theory,
\begin{eqnarray}
\psi_{L} & \equiv & (\psi_{1_{L}},\psi_{2_{L}},\psi_{3_{L}}) \; = \; \left[
\left( \begin{array}{c}
                       \nu_{e} \\
                          e
               \end{array} \right)_{L},\;
\left( \begin{array}{c}
                       \nu_{\mu} \\
                       \mu
               \end{array} \right)_{L},\;
\left( \begin{array}{c}
                       \nu_{\tau} \\
                       \tau
               \end{array} \right)_{L}\right] , 
\end{eqnarray}
transform as doublets
($I = \frac{1}{2}$) under $SU(2)_{L}$ with the hypercharge $Y = -1$. In short,
their quantum numbers are ($\frac{1}{2}, -1$).
                 
The right-handed lepton fields (0,-2) are
\begin{eqnarray}
\psi_{R} & \equiv & (\psi_{1_{R}},\psi_{2_{R}},\psi_{3_{R}}) \; = \;
(e_{R},\mu_{R},\tau_{R}). 
\end{eqnarray}
Note there are no right-handed neutrino fields in the SM.
        
The left-handed ($\frac{1}{2}, \frac{1}{3}$) and the right-handed up
($0, \frac{4}{3}$) and down ($0, -\frac{2}{3}$) quark fields of the unbroken
theory are respectively
\begin{eqnarray}
q_{L}^{'} & \equiv & (q_{1_{L}}^{'},q_{2_{L}}^{'},q_{3_{L}}^{'}) \; = \; \left[
\left( \begin{array}{c}
                       u_{1}^{'} \\
                       d_{1}^{'}
               \end{array} \right)_{L},\;
\left( \begin{array}{c}
                       u_{2}^{'} \\
                       d_{2}^{'}
               \end{array} \right)_{L},\;
\left( \begin{array}{c}
                       u_{3}^{'} \\
                       d_{3}^{'}
               \end{array} \right)_{L}\right], \nonumber \\
\nonumber \\
u_{R}^{'} & \equiv & (u_{1_{R}}^{'},u_{2_{R}}^{'},u_{3_{R}}^{'}), \nonumber \\
d_{R}^{'} & \equiv  & (d_{1_{R}}^{'},d_{2_{R}}^{'},d_{3_{R}}^{'}). 
\end{eqnarray}
The quark fields (weak eigenstates) of the broken theory are different from
the fields of the
unbroken theory (and also from the quark mass eigenstates) :
\begin{eqnarray}
\label{weakstate}
q_{L} & \equiv  & (q_{1_{L}},q_{2_{L}},q_{3_{L}}) \; = \; \left[
\left( \begin{array}{c}
                       u \\
                       \tilde{d}
               \end{array} \right)_{L},\;
\left( \begin{array}{c}
                       c \\
                       \tilde{s}
               \end{array} \right)_{L},\;
\left( \begin{array}{c}
                       t \\
                       \tilde{b}
               \end{array} \right)_{L}\right], \nonumber \\
\nonumber \\
u_{R} & \equiv &  (u_{1_{R}},u_{2_{R}},u_{3_{R}}) \; = \; (u_{R},c_{R},t_{R}),
\nonumber \\
\tilde{d_{R}} & \equiv & 
(\tilde{d_{1_{R}}},\tilde{d_{2_{R}}},\tilde{d_{3_{R}}}) \; = \;
                (\tilde{d_{R}},\tilde{s_{R}},\tilde{b_{R}}),
\end{eqnarray}
where $\;\;\tilde{d} = V_{CKM} d\;\;$ are weak eigenstates of the broken
theory obtained
from mass eigenstates $d$ through the Cabibbo-Kobayashi-Maskawa (CKM) matrix 
$V_{CKM}$ (see also Eq. \ref{charcur}).

The classical electroweak Lagrangian is the sum of the fermion
part, the gauge part and
the Higgs part:
\begin{eqnarray}
{\cal L}_{EW} = {\cal L}_{G} + {\cal L}_{F} + {\cal L}_{H}.
\end{eqnarray}
                    
The fermion part, which describes the fermions and their interactions, 
is given by
\begin{eqnarray}
\label{fermionl}
{\cal L}_{F} & = & i \sum_{j=1}^{3}\left\{\overline{\psi_{j_{L}}}\;
\gamma^{\mu}\;{\cal D}_{\mu}\;
\psi_{j_{L}} + \overline{\psi_{j_{R}}}\;\gamma^{\mu}\;{\cal D}_{\mu}
\;\psi_{j_{R}}
+ \overline{q_{j_{L}}}\;
\gamma^{\mu}\;{\cal D}_{\mu}\;
q_{j_{L}}  \right. \nonumber \\
& + & \left. \overline{u_{j_{R}}}\;\gamma^{\mu}\;{\cal D}_{\mu}\;u_{j_{R}}
+ \overline{d_{j_{R}}}\;\gamma^{\mu}\;{\cal D}_{\mu}\;d_{j_{R}}\right\},
\end{eqnarray}
where
\begin{eqnarray}
{\cal D}_{\mu} \psi_{L} & = & \left[ \left(\partial_{\mu}
+ i\frac{g_{1}}{2} Y B_{\mu}\right){\bf I}
- i\frac{g_{2}}{2} \mbox{{\boldmath
$\vec{\tau}$}} \cdot \vec{W_{\mu}} \right] \psi_{L}, \\
{\cal D}_{\mu} \psi_{R} & = & \left(\partial_{\mu}
+ i\frac{g_{1}}{2} Y B_{\mu}\right) \psi_{R},
\end{eqnarray}
are covariant derivatives for left and right-handed fields respectively.
These derivatives ensure the gauge invariance of the ${\cal L}_{F}$
by introducing a weak isospin triplet of gauge fields $\vec{W_{\mu}} \equiv
(W_{\mu}^{1},W_{\mu}^{2},W_{\mu}^{3})$ and a weak isospin singlet gauge
field $B_{\mu}$. The gauge fields interact with the fermions with the strength
$g_{2}$, the $SU(2)_{L}$ coupling constant and the strength $g_{1}$, the
$U(1)_{Y}$ coupling constant. ${\bf I}$ is a $2 \times 2$ unit matrix in
isospin space and $\frac{1}{2}\mbox{{\boldmath $\vec{\tau}$}}$ are generators
of $SU(2)_{L}$ transformations in two-dimensional representation;
$\mbox{{\boldmath $\vec{\tau}$}}$ are Pauli matrices (see Appendix \ref{Acko}).
Weak isospin $I$ is the eigenvalue of the operator $\left( \frac{1}{2}
\mbox{{\boldmath $\vec{\tau}$}}\right)^{2}$ and $I_{3}$ is the eigenvalue of
the operator $\frac{1}{2} \mbox{{\boldmath $\tau_{3}$}}$.
                
The gauge part of the Lagrangian describes the gauge fields and their
self-interac-{\linebreak}tions; it is given by
\begin{eqnarray}
{\cal L}_{G} = - \frac{1}{4} W^{a}_{\mu \nu} W^{\mu \nu ,a}
-\frac{1}{4} B_{\mu \nu} B^{\mu \nu},
\end{eqnarray}
where  $ a=1,2,3 $ is the $SU(2)$ index and
\begin{eqnarray}
W^{a}_{\mu \nu} & = & \partial_{\mu} W_{\nu}^{a} -\partial_{\nu} W_{\mu}^{a}
+ g_{2} \: \epsilon_{abc} \: W_{\mu}^{b} W_{\mu}^{c}, \nonumber \\
B_{\mu \nu} & = & \partial_{\mu} B_{\nu} - \partial_{\nu} B_{\mu},
\end{eqnarray}
are field strength tensors for the isotriplet $W_{\mu}^{a}$ and the isosinglet
$B_{\mu}$ fields respectively.
           
The Higgs part of the Lagrangian, responsible for spontaneous electroweak
symmetry breaking, is the sum of two terms:
\begin{eqnarray}
{\cal L}_{H} = {\cal L}_{HG} + {\cal L}_{HF}.
\end{eqnarray}
Here
${\cal L}_{HG}$ describes the Higgs-gauge interactions and ${\cal L}_{HF}$
the Higgs-fermion or so-called 
Yukawa interactions. ${\cal L}_{HG}$ has the form
\begin{eqnarray}
{\cal L}_{HG} = ({\cal D}_{\mu}\Phi)^{\dagger} ({\cal D}^{\mu}\Phi) - V(\Phi),
\end{eqnarray}
where
\begin{eqnarray}
 \Phi & = & \left( \begin{array}{c}
                       \phi^{+} \\
                       \phi^{0}
               \end{array} \right),  \nonumber \\   \nonumber \\   
{\cal D}_{\mu}\Phi & = & \left[(\partial_{\mu} +
i\frac{g_{1}}{2} Y B_{\mu}){\bf I} -
i\frac{g_{2}}{2} \mbox{{\boldmath $\vec{\tau}$}}
\cdot \vec{W_{\mu}}\right]\Phi,            \nonumber \\ \nonumber \\
V(\Phi) & = & -\mu^{2}\Phi^{\dagger}\Phi + \lambda(\Phi^{\dagger}\Phi)^{2},
\;\;\;\; \lambda > 0.
\end{eqnarray}
${\cal D}_{\mu}\Phi$ is the covariant derivative for the $Y = 1$ Higgs doublet
$\Phi$ $(\frac{1}{2},1)$ with the charged component $\phi^{+}$ and the neutral
component $\phi^{0}$;
$V(\Phi)$
is the Higgs potential constructed so it can lead to the vacuum in which the
average value (vacuum expectation value) of the Higgs doublet, denoted
$\langle \Phi \rangle$, is nonzero. To keep $U(1)_{em}$ unbroken, it is the
neutral component $\phi^{0}$ which develops the vacuum expectation value:
\begin{eqnarray}
\label{vev}
\langle \Phi \rangle & = & \frac{1}{\sqrt{2}}\left( \begin{array}{c}
                       0 \\
                       v
               \end{array} \right), \;\;\;\; v = \frac{\mu}{\sqrt{\lambda}}.
\end{eqnarray}
The symmetry is broken spontaneously because the electroweak Lagrangian 
is symmetric under $SU(2)_{L} \times U(1)_{Y}$
transformations but the lowest energy state, the vacuum, is not 
(here $\langle \Phi \rangle$ is not symmetric).
                 
The Higgs doublet can be written now as
\begin{eqnarray}
\label{higgsplus}
 \Phi & = & \left( \begin{array}{c}
                       \phi^{+} \\
                       \phi^{0}
               \end{array} \right) =
\left( \begin{array}{c}
                       \phi^{+} \\
     \frac{1}{\sqrt{2}}(v+H+i\chi)
               \end{array} \right),
\end{eqnarray}
where $\phi^{\pm}$ and $\chi$ are unphysical Higgs fields and $H$ is 
the physical Higgs field. 

The 
%Higgs mechanism of the 
spontaneous symmetry breaking 
%\footnote{There are three aspects of the electroweak symmetry breaking:
%the spontaneous breaking of the global symmetry, the Higgs mechanism which
%applies in case of the the spontaneous breaking of the local symmetry and the
%set of fields and interactions represented by ${\cal L}_{HG}$. We described only
%the first one. The Higgs mechanism is relevant for gauge boson masses, not
%fermion masses so we do not discuss it here. The third aspect is currently
%a very attractive issue nowadays since there can be many different fields and
%interactions (not just simple $\Phi$ and $V(\Phi)$) which can reproduce gauge 
%boson and fermion masses.} 
gives rise to massive gauge fields $W_{\mu}^{\pm}$ and
$Z_{\mu}$, mediators of weak charged and neutral interactions leaving the
massless photon field $A_{\mu}$, the mediator of electromagnetic interactions:
\begin{eqnarray}
W_{\mu}^{\pm} & = & \frac{1}{\sqrt{2}}\left(W_{\mu}^{1} \mp
i W_{\mu}^{2}\right), \nonumber \\
Z_{\mu}       & = &  + \cos \theta_{W} W_{\mu}^{3} + \sin \theta_{W} B_{\mu},
\nonumber \\
A_{\mu}       & = &  - \sin \theta_{W} W_{\mu}^{3} + \cos \theta_{W} B_{\mu}.
\end{eqnarray}
The $W^{\pm}$ mass $M_{W}$ and the $Z$ mass $M_{Z}$ are given by
\begin{eqnarray}
M_{W} & = & \frac{v}{2} g_{2}, \;\;\;\;\;\; 
M_{Z} \; = \; \frac{v}{2} \sqrt{g_{1}^{2}+g_{2}^{2}}.
\end{eqnarray}
The Weinberg angle $\theta_{W}$ is defined as
\begin{eqnarray}
\cos \theta_{W} & = & \frac{M_{W}}{M_{Z}} \; = \; \frac{g_{2}}{\sqrt{g_{1}^{2} +
g_{2}^{2}}}.
\end{eqnarray}
The electric charge $e=\sqrt{4\pi \alpha}$ can be expressed as
\begin{eqnarray}
e = \frac{g_{1}g_{2}}{\sqrt{g_{1}^{2} + g_{2}^{2}}},
\end{eqnarray}
or
\begin{eqnarray}
g_{2} = \frac{e}{\sin \theta_{W}},\;\;\;\;\;\;g_{1} = \frac{e}{\cos
\theta_{W}}.
\end{eqnarray}

The second term of the Higgs part of the Lagrangian, ${\cal L}_{HF}$, is
discussed in the next section.

\section{Fermion masses in the SM}
\label{fermion2}

The spontaneous symmetry breaking is responsible also for fermion masses.
The starting point is ${\cal L}_{HF}$ which describes the Yukawa interactions
between fermions and the Higgs doublet:
\begin{eqnarray}
\label{yukawa}
{\cal L}_{HF} & = & - \sum_{i=1}^{3} \sum_{j=1}^{3}\left[\tilde{G}_{ij}
\overline{u_{i_{R}}^{'}}
(\tilde{\Phi}^{\dagger}q_{j_{L}}^{'}) +
G_{ij}\overline{d_{i_{R}}^{'}}(\Phi^{\dagger}q_{j_{L}}^{'})\right]
 + h.c.  \nonumber \\
& - & \;\;\;\;\sum_{i=1}^{3}\left[\makebox[1.03in][c]{ } \;\;\;\: +
h_{i} \overline{\psi_{i_{R}}} (\Phi^{\dagger}\psi_{i_{L}})\right] +h.c.,
\end{eqnarray}
where
\begin{eqnarray}
\label{higgsminus}
\tilde{\Phi} & = & i \tau_{2} \Phi^{*} = \left( \begin{array}{c}
                                  {\phi^{0}}^{*} \\
                                   -\phi^{-}
             \end{array} \right)
\end{eqnarray}
is the $Y = -1$ Higgs doublet $(\frac{1}{2},-1)$
and $\tilde{G}_{ij},G_{ij}, h_{i}$ are arbitrary Yukawa couplings which are
free
parameters in the SM. The purpose of the empty space in the second line of
Eq. \ref{yukawa} will be clarified below.
                                           
To generate masses, one substitutes in Eq. \ref{yukawa} the vacuum expectation
value $\langle \Phi \rangle$ (see Eq. \ref{vev})
for $\Phi$. Thus the first term (plus its h.c.) in the first line of 
Eq.~\ref{yukawa}
gives mass to $u,c,t$ quarks; the second term gives mass to $d,s,b$ quarks and
the term in the second line gives mass to charged leptons.

Let us study the charged lepton case first. 
We get the electron mass $m_{e}$ from the
second line of Eq. \ref{yukawa} for $i = 1$. The
substitution of $\langle \Phi \rangle$ yields
\begin{eqnarray}
& - & \frac{1}{\sqrt{2}}h_{1}\overline{e_{R}}
{\left( \begin{array}{c}
                       0 \\
                       v
               \end{array} \right)}^{\dagger}
\left( \begin{array}{c}
                       \nu_{e_{L}} \\
                       e_{L}
               \end{array} \right)
-\frac{1}{\sqrt{2}}h_{1}
\overline{\left( \begin{array}{c}
                       \nu_{e_{L}} \\
                       e_{L}
               \end{array} \right)}
\left( \begin{array}{c}
                       0 \\
                       v
               \end{array} \right)e_{R} \;   \nonumber \\  \nonumber \\
& = & \; -\frac{1}{\sqrt{2}}h_{1}\overline{e_{R}}\:v\:e_{L} 
  -\frac{1}{\sqrt{2}}h_{1}\overline{e_{L}}\:v\:e_{R} \; = \;
  -\frac{1}{\sqrt{2}}h_{1} v \left( \overline{e_{R}}e_{L}+
   \overline{e_{L}}e_{R}\right) \;  \nonumber \\  \nonumber \\
& \equiv & \; - m_{e}\left( \overline{e_{R}}e_{L}+
   \overline{e_{L}}e_{R}\right) \; = \; - m_{e}\overline{e}e,
\end{eqnarray}
which is the familiar form of the Dirac mass term.
Without any inter-generation couplings in the lepton part
of Eq. \ref{yukawa} (Yukawa couplings $h_{i}$ are simple  numbers as opposed 
to matrices $\tilde{G}_{ij},G_{ij}$), there are no mixings among leptons in the
SM
\footnote{When discussing fermion masses, one cannot avoid the question
of possible mixings among fermions. It is because we look for
mass effects in various weak processes where the states of definite weak 
quantum numbers
(weak interaction eigenstates) participate rather than the states of
definite mass (mass eigenstates). Mixings then relate mass
eigenstates to weak eigenstates.
Further, mixings and masses are connected through their common origin derived
from Yukawa couplings and the vacuum expectation value of the Higgs doublet. 
For these reasons we will study mixings along with masses.}.
As a result, lepton family numbers (flavours) are separately
conserved and there are no
lepton flavour-violating processes. The total lepton number $L$ is also
conserved since it is the sum of lepton family numbers.
% charged or neutral currents involving leptons.

Quark masses are more involved, because inter-generation couplings are allowed
($ \tilde{G}_{ij},G_{ij}$ are nondiagonal matrices in flavour space). As a
consequence, $q^{'}_{j_{L}}, u^{'}_{i_{R}}$ and $d^{'}_{i_{R}}$, the weak 
eigenstates of the unbroken theory, are different from the mass eigenstates 
$u,c,t$ and $d,c,b$. They are related 
through the unitary matrices $A_{L}, A_{R}, B_{L}$, $B_{R}$,~\cite{barger}:
\begin{eqnarray}
\left( \begin{array}{c}
                       u^{'}_{1_{L,R}} \\
                       u^{'}_{2_{L,R}} \\
                       u^{'}_{3_{L,R}}
               \end{array} \right) = A_{L,R}
\left( \begin{array}{c}
                       u_{L,R} \\
                       c_{L,R} \\
                       t_{L,R}
               \end{array} \right) ,\;\;\;\;\;\;\;\;\;\;
\left( \begin{array}{c}
                       d^{'}_{1_{L,R}} \\
                       d^{'}_{2_{L,R}} \\
                       d^{'}_{3_{L,R}}
               \end{array} \right) = B_{L,R}
\left( \begin{array}{c}
                       d_{L,R} \\
                       s_{L,R} \\
                       b_{L,R}
               \end{array} \right). 
\end{eqnarray}
To generate quark masses we again substitute $\langle \Phi \rangle$ for $\Phi$,
now in the first line of Eq. \ref{yukawa}.
We obtain mass matrices $\frac{v}{\sqrt{2}}\tilde{G}_{ij}$ and
$\frac{v}{\sqrt{2}}G_{ij}$ which are diagonalized by the matrices $A$ and
$B$ to yield masses $m_{u}, ..., m_{b}$ of $u, ..., b$ quarks:
\begin{eqnarray}
\frac{v}{\sqrt{2}}A_{R}^{-1}\tilde{G}A_{L} = 
\left( \begin{array}{lll}
                       m_{u} & 0 & 0 \\
                       0 & m_{c} & 0 \\
                       0 & 0 & m_{t}  
               \end{array} \right) ,\;\;\;\;\;
\frac{v}{\sqrt{2}}B_{R}^{-1}GB_{L} =
\left( \begin{array}{lll}
                       m_{d} & 0 & 0 \\
                       0 & m_{s} & 0 \\
                       0 & 0 & m_{b}
               \end{array} \right). 
\end{eqnarray}
Mixings arise in the charged current interactions of quarks:
the quark charged current Lagrangian (part of ${\cal L}_{F}$, Eq.
\ref{fermionl}) is given as
\begin{eqnarray}
\label{charcur}
{\cal L}_{cc} & = & 
\frac{g_{2}}{\sqrt{2}}W^{\mu}
\overline{(u^{'}_{1_{L}},u^{'}_{2_{L}},u^{'}_{3_{L}})}
\gamma_{\mu}
\left( \begin{array}{c}
                       d^{'}_{1_{L}} \\
                       d^{'}_{2_{L}} \\
                       d^{'}_{3_{L}}
               \end{array} \right) 
\; = \; \frac{g_{2}}{\sqrt{2}}W^{\mu}\overline{(u_{L},c_{L},t_{L})}
\:A_{L}^{\dagger}B_{L}\:\gamma_{\mu}
\left( \begin{array}{c}
                       d_{L} \\
                       s_{L} \\
                       b_{L}
               \end{array} \right),  
\nonumber \\
\end{eqnarray}
where $V_{CKM} \equiv A_{L}^{\dagger}B_{L}$ is 
the $3\times 3$ unitary CKM mixing matrix \cite{ckm}. 
It is a nondiagonal matrix inducing transitions between families in charged
current interactions. Acting on mass eigenstates $d, s, b$, it gives us weak
eigenstates $\tilde{d}, \tilde{s}, \tilde{b}$ (see Eq. \ref{weakstate}).
There is no mixing in the neutral current Lagrangian,
hence no flavour-changing neutral currents at the lowest order of 
perturbation theory, although they can arise at the one-loop level.

For neutrino masses, there is an empty space in the second line of Eq. 
\ref{yukawa} because no right-handed neutrino fields $\nu_{R}$ are included. 
Thus, there cannot be nonzero neutrino masses in the SM 
\footnote{There actually could be nonzero neutrino masses without right-handed
neutrino fields if the Higgs sector of the SM was appropriately extended
\cite{mohapatra}.}. 

The generation of fermion masses in the SM, as we have just described it, 
is considered
to be the least satisfactory part of the SM. Each mass enters as an unknown
parameter (Yukawa couplings are not predicted) which has to be supplied by 
experiment. The SM  rather accommodates
fermion masses than predicts them. The problem of fermion masses, and neutrino
masses in particular, has been a top priority for particle physicists for
some time now.

\section{Neutrino masses in $SU(2)_{L}\times U(1)_{Y}$ models beyond the SM}
\label{neutrino3}

As noted in the Introduction, basic directions in the theoretical treatment
of neutrino masses can be followed in the class of models based on the same
symmetry group as the SM, on $SU(2)_{L} \times U(1)_{Y}$. This fixes
the gauge sector; the fermion content and the Higgs (symmetry breaking) sector
offer some
freedom which is used by different models within the class. 
We keep here also the symmetry breaking sector of the SM untouched and extend 
the fermion sector only. We examine two such models in this section and the
third one, our model, in Chapter 3.

\subsection{A simple model of neutrino mass}
\label{asimple31}
In this straightforward extension of the SM one postulates one right-handed
neutrino field $\nu_{R}$ per family with the $SU(2)_{L}\times U(1)_{Y}$ quantum 
numbers (0,0). 
Neutrinos are then treated in the same manner as all other fermions in the SM.
The presence  of right-handed
neutrino fields allows new Yukawa interactions,
\begin{eqnarray}
{\cal L}_{new} & = & - \sum_{i=1}^{3} \sum_{j=1}^{3}\tilde{h}_{ij}
\overline{\nu_{i_{R}}}
(\tilde{\Phi}^{\dagger}\psi_{j_{L}}) + h.c.\;. 
\end{eqnarray}
This is the term missing from Eq. \ref{yukawa}.
 Neutrinos acquire Dirac mass by analogy
with up type quarks in the SM (see Sec. \ref{fermion2}); the only minor 
difference is that here
we do not introduce mixings among the charged leptons. Neutrino mass
eigenstates are then different from weak eigenstates
$\nu_{e},\nu_{\mu},\nu_{\tau}$, leading to neutrino mixing and the violation of
family lepton numbers.

The shortcoming of this model is that it provides no answer to the problem of
smallness of neutrino masses. We can make masses small by tuning Yukawa
couplings $\tilde{h}_{ij}$ but this is not satisfactory, as there is no good
reason why the $\tilde{h}_{ij}$ should themselves be small.

\subsection{See-saw mechanism in an  $SU(2)_{L}\times U(1)_{Y}$ model,
Majorana neutrinos and Majorana NHL's}
\label{see-saw32}
  
Charged fermions are formally described by Dirac spinors. Neutrinos are
described in the same way in the simple
extension discussed above. However, because neutrinos are neutral, another
possibility opens up. They could be Majorana particles. To illustrate the
difference between Dirac and Majorana neutrinos, let us decompose the Dirac mass
term into  its components, which form the so-called Majorana basis of a matrix
representation of mass terms (see Ref. \cite{mohapatra}, Sec. 4.5),
\begin{eqnarray}
\label{decomp}
 m \;\overline{\nu}\nu & = &
 m \:\left(\overline{\nu_{L}}\nu_{R}
 + \overline{\nu_{R}}\nu_{L}
\right) = m \;\overline{\nu_{L}} \nu_{R} + h.c. =
 \frac{1}{2} m \:\left(\overline{\nu_{L}}\nu_{R}+\overline{\nu_{L}^{c}}
\nu_{R}^{c}\right)
 + h.c. \; = \nonumber \\ \nonumber \\
& = & \frac{1}{2}\left(\overline{\nu^{c}_{L}}\;\overline{\nu_{L}}\right)
\left( \begin{array}{ll}
                       0 & m  \\
                       m & 0   
               \end{array} \right)
\left( \begin{array}{c}
                       \nu_{R}     \\
                       \nu_{R}^{c} 
               \end{array} \right) + h.c.     \nonumber \\
& = &
\frac{1}{2}\left(\overline{\nu_{L}}\;\overline{\nu_{L}^{c}}\right)
\left( \begin{array}{ll}
                       0 & m  \\
                       m & 0
               \end{array} \right)
\left( \begin{array}{c}
                       \nu_{R}^{c} \\
                       \nu_{R}
               \end{array} \right) + h.c.,  
\end{eqnarray}
where $\nu^{c} = C\gamma_{0}\nu^{*}$ ($C= i \gamma^{2} \gamma^{0}$)
is the charge conjugate field of $\nu$,
$\nu_{R}^{c} \equiv \frac{1}{2}(1+\gamma_{5})\nu^{c}$
 is the charge conjugate of the field
$\nu_{L}$, and $\nu_{L}^{c}$ is the charge conjugate of the field
$\nu_{R}$.
In the above, we used  the identity
$\overline{\nu_{L}^{c}}\nu_{R}^{c} = \overline{\nu_{L}}\nu_{R}$,
proven in Appendix \ref{proof}.
From Eq. \ref{decomp} it is obvious that the Dirac mass term has a
very special mass matrix in the Majorana basis, namely, 
the two diagonal terms
are zero. Can we make these two matrix elements nonzero ? The answer is yes, if
we are willing to accept the violation of the total lepton number $L$, or 
equivalently baryon minus lepton ($B-L$) number 
\footnote{The relevance of $B-L$, rather than $L$, is discussed in Ref.
\cite{mohapatra}, Sec. 2.4.}.
We already broke individual lepton family numbers and there
is nothing sacred about $B-L$ symmetry either.

In an $SU(2)_{L}\times U(1)_{Y}$ see-saw model we introduce the following mass
matrix (written for the case of one family),
\begin{eqnarray}
\label{majorana}
-{\cal L}_{mass} & = &
 \frac{1}{2}\left(\overline{\nu_{L}}\;\overline{n_{L}^{c}}\right)
\left( \begin{array}{ll}
                       0 & D  \\
                       D & M   
               \end{array} \right)
\left( \begin{array}{c}
                       \nu_{R}^{c} \\
                       n_{R} 
               \end{array} \right) + h.c.,
\end{eqnarray}   
so the fermion content is the same as that of the simple model of Sec.
\ref{asimple31}
\footnote{Note that in Eq. \ref{decomp} we use notation
 $\nu_{L}, \nu_{R}$ for left-handed
and right-handed chiral fields respectively; in contrast, here we use $n_{R}$
rather than $\nu_{R}$ for the right-handed field. The reason lies in the fact
that for a Dirac neutrino two independent fields $\nu_{L}, \nu_{R}$ combine to
form a single particle, while in this case $\nu_{L}$ with its partner
$\nu_{R}^{c}$ form (in the limit $M \gg D$) a light Majorana neutrino 
and $n_{R}$ with its
partner $n_{L}^{c}$ form a Majorana NHL; hence we use a different
notation for the field describing a different particle.}
, but here we
allow Majorana mass terms breaking $B-L$ number conservation
\footnote{For $B-L$ to be conserved, ${\cal L}_{mass}$ must be
invariant under the following transformations:
$\nu_{L} \rightarrow e^{- i (B - L) \alpha} \nu_{L} = e^{- i \alpha} \nu_{L};\;
n_{R} \rightarrow e^{- i \alpha} n_{R};\; n^{c}_{L} \rightarrow e^{+ i \alpha}
n^{c}_{L};\; \overline{n^{c}_{L}} \rightarrow e^{- i \alpha} \overline{n^{c}_{L}}
$. 
The term in Eq.~\ref{blbreak} transforms as $\overline{n^{c}_{L}} n_{R}
\rightarrow e^{- i \alpha} e^{- i \alpha} \: \overline{n^{c}_{L}} n_{R} \neq
\overline{n^{c}_{L}} n_{R}$, i.e., it breaks $B-L$ conservation.},
\begin{eqnarray}
\label{blbreak}
 \frac{1}{2}M \:\overline{n_{L}^{c}}n_{R} + h.c.\;.
\end{eqnarray}
The matrix ${\cal M} = \left( \begin{array}{ll}
                       0 & D  \\
                       D & M   
               \end{array} \right)$ now describes two massive Majorana
neutrinos rather than a single Dirac one. To see that, we have to diagonalize
 ${\cal M}$ (e.g. Ref. \cite{mohapatra}, Sec.~5.1.4),
\begin{eqnarray}
{\cal M} & = & O^{T} \left( \begin{array}{ll}
                       m_{1} & 0  \\
                       0 & m_{2}   
               \end{array} \right)
 \left( \begin{array}{rl}
                       -1 & 0  \\
                        0 & 1   
               \end{array} \right) O,
\end{eqnarray}
where $m_{1,2} = \frac{1}{2}\big(\sqrt{M^{2}+4 D^{2}}\mp M \big)$ are the 
masses of the two Majorana neutrinos. In the above,
\begin{eqnarray}
O & = & \left( \begin{array}{lr}
                       \cos \theta & -\sin \theta  \\
                        \sin \theta & \cos \theta   
               \end{array} \right), \;\;\;\;\; \tan 2\theta = \frac{2D}{M},
\end{eqnarray}
is an orthogonal rotation matrix defining massive Majorana neutrinos 
$\nu^{'}, N$ as
\begin{eqnarray}
\left( \begin{array}{c}
                       \nu^{'}_{L} \\
                       N_{L}
               \end{array} \right) \equiv
O \left( \begin{array}{c}
                       \nu_{L} \\
                       n_{L}^{c} 
               \end{array} \right),
\;\;\;\;\; \left( \begin{array}{c}
                       \nu^{'}_{R} \\
                       N_{R} 
               \end{array} \right) \equiv
 \left( \begin{array}{rl}
                       -1 & 0  \\
                        0 & 1   
               \end{array} \right) O
\left( \begin{array}{c}
                       \nu_{R}^{c} \\
                       n_{R}
               \end{array} \right). 
\end{eqnarray}
From here we can show 
\begin{eqnarray}
\nu^{'} & = & \nu^{'}_{L}+ \nu^{'}_{R} \; = \; \cos \theta (\nu_{L}-\nu_{R}^{c})
         - \sin \theta ( n_{L}^{c}-  n_{R}) \; = \; -\nu^{'c}, \nonumber \\
N & = & N_{L}+ N_{R} \; = \; \sin \theta (\nu_{L}+\nu_{R}^{c})
         + \cos \theta ( n_{L}^{c}+  n_{R}) \; = \; N^{c},  
\end{eqnarray}
that is, $\nu^{'}$ and $N$  are their own charge conjugates, their own 
antiparticles;
therefore they are Majorana neutrinos.

We see how this model explains the small neutrino masses when we assume
that $M \gg D$. In this limit, the masses $m_{1}$ of $\nu^{'}$ and $m_{2}$ of
 $N$ become
\begin{eqnarray}
\label{see}
m_{1} \doteq \frac{D^{2}}{M}, \;\;\;\;\; m_{2} \doteq M;
\end{eqnarray}
and using $\sin 2\theta \doteq \tan 2\theta \doteq 2\theta = \frac{2 D}{M}$,
we find for the weak eigenstate $\nu_{L}$
\begin{eqnarray}
\label{ss2}
\nu_{L} & \doteq & \nu^{'}_{L} + \frac{D}{M} N_{L} \; \doteq \; \nu^{'}_{L}.
\end{eqnarray}
Eq. \ref{see} is the famous see-saw mass relation (see also Eq. \ref{ss1}),
whereby a weakly interacting neutrino, $\nu_{L} \doteq \nu^{'}_{L}$, gets 
very light compared to the typical family fermion mass $D$ thanks to the very 
large Majorana mass $M$. Assuming $D \sim m_{\tau}, \;M$ has to be greater 
than about $10^{8} \; {\rm GeV}$ in order to meet the cosmological bound 
(see Sec.~\ref{sttp}) $m_{\nu} \; < 25$ eV. 
%\begin{equation}
%\label{eigen2}
%\begin{array}{rcl}
%n_{1} & \doteq & \nu_{L}-\nu_{R}^{c} - \frac{D}{M}( n_{L}^{c}-  n_{R})
%\nonumber \\
%n_{2} & \doteq & \frac{D}{M}(\nu_{L}+\nu_{R}^{c}) +  n_{L}^{c}+  n_{R} 
% \nonumber
%\end{array}  
%\end{equation}                                                              

It looks as though we have replaced the problem of the smallness of the 
neutrino mass
with another one, the problem of the big mass $M$. Indeed, in the context of
an $SU(2)_{L} \times U(1)_{Y}$ model, the origin of the big mass $M$ is a 
mystery.
%This shows the limitation of our restriction to study  
%$SU(2)_{L} \times U(1)_{Y}$
%models of neutrino masses. 
At this point  we invoke our motivational grounds, 
the unification models (see the next section). 
There are in fact large scales in these models
associated with the unification energies. The $SU(2)_{L} \times U(1)_{Y}$
 see-saw model could be a low-energy limit of some GUT theory.

In this thesis we are specifically interested in NHL's,
described in this section by the field $N$ with the mass $m_{2} \equiv M$.
From
Eqs. \ref{see}, \ref{ss2} it is obvious that NHL's in this model are,
first, too
heavy to be observed directly in the near future, and second, their contribution
to left-handed weak eigenstates is so small that there is little hope to see 
even their indirect effects. See-saw models tend to be phenomenologically
uninteresting.
There are however models with special forms of the Dirac and Majorana mass 
matrices (in the general
case of $n$ families, masses $D$ and $M$ in Eq. \ref{majorana} become Dirac 
and Majorana
$n \times n$ mass matrices) that avoid this suppression 
\cite{pilaftsis2,Ng1}. 
For instance, Pilaftsis
\cite{pilaftsis2} finds a relation among the elements of $D$ and $M$ matrices
that leads
to massless neutrinos at the tree-level and small Majorana masses are generated
radiatively. The cosmological constraint on the scale $M$ is much weaker in
this model and consequently, the mixing of NHL's ($K_{H} \sim D/M$) is not
suppressed.
We shall refer to such models as see-saw models with enhanced
mixings.               
%In that case, however, it is not clear why these special forms should be
%preferred over the general ones.
Although calculations in this work were carried out in the context of a
superstring-inspired model, our
analysis is qualitatively valid also for this class of see-saw models.

\section{Neutrino mass in grand unified models}
\label{grand4}

Here we briefly touch the question of neutrino mass in grand unified models
(GUT's).
A nice short review of the subject can be found in Ref. \cite{mohapatra}; the
case of $SO(10)$ is discussed also in Ref. \cite{seesaw}, and
that of $E_{6}$ in Ref. \cite{strings}.

There are 15 chiral fermion fields per generation currently known, $e_{L},
e_{R}, \nu_{e}$ and twelve $u$ and $d$ quark fields. In the simplest GUT model,
$SU(5)$, these $15$
fields are assigned to $\{10\}$ and $\overline{\{5\}}$-dimensional 
representations.
There is
no right-handed neutrino postulated; therefore, one cannot generate a Dirac mass
for the neutrino and also it is not possible to generate Majorana mass as
described in Sec. \ref{see-saw32}. One can still generate Majorana mass 
without a
right-handed neutrino if an appropriate Higgs field is introduced. The problem
is that this Higgs field is introduced {\it ad hoc} and, as a result, neutrino
masses do not arise in $SU(5)$ naturally. Moreover, $SU(5)$ is ruled out by
the proton decay measurements \cite{proton}.

The next popular group is $SO(10)$. This group contains left-right symmetric
$SU(2)_{L} \times SU(2)_{R} \times SU(4)_{C}$ as its subgroup, which implies
automatically the right-handed neutrino. The number of chiral fermion fields
per generation is thus $16$, filling the fundamental $\{16\}$ representation.
With a right-handed neutrino in the  fundamental representation, neutrino
masses in $SO(10)$ can arise naturally via the see-saw mechanism (see Sec.
\ref{see-saw32}). The actual values of neutrino masses  are sensitive to the 
Majorana
mass matrix $M$ (see Eq. \ref{majorana}), which in turn can tell us about the
particular branch of the $SO(10)$ breaking down to low-energy $SU(3)_{C}
\times SU(2)_{L} \times U(1)_{Y}$. $SO(10)$ predicts a lower rate for 
proton decay than does $SU(5)$.

Finally, a lot of attention is paid to $E_{6}$ based GUT's \cite{strings}.
This is thanks
to their superstring connections. Green and Schwarz \cite{green} showed
that string theory in ten dimensions is anomaly free for the gauge group
$E_{8} \times E^{'}_{8}$ and that the compactification of the additional six
dimensions can result in the breaking of $E_{8}$ down to $E_{6}$, which
becomes an effective GUT group. 

The fundamental representation of $E_{6}$ is $\{27\}$-dimensional, implying $27$
chiral fermion fields per generation, $11$ more than we had in $SO(10)$. These 
eleven fields must be new particles, often referred to as exotics. They include
a colour triplet weak isosinglet quark and its antiparticle and five new
leptons. Of the new leptons, four (two charged and two neutral) form two weak
isodoublets and the fifth one is a weak isosinglet.

Curiously, the superstring-inspired $E_{6}$ model experiences certain
difficulties in understanding the small neutrino masses
\cite{vallemo,bernabeu1} :
there are no appropriate Higgs fields to 
provide the large Majorana mass $M$ for the see-saw mass matrix
(see Eq.~\ref{majorana}) and therefore the see-saw mechanism does not operate
here.  Interesting solutions to this
problem suggest that besides the fundamental $\{27\}$-plet there exists
an additional, $E_{6}$ singlet neutral fermion field $S_{L}$.
At low energies, $S_{L}$, along with the right-handed neutrino $n_{R}$, can
enrich the neutral lepton spectrum of the SM. 
The other three neutral exotic leptons decouple from the low-energy spectrum. 
The mass matrix formed by $\nu_{L}, n_{R}$ and $S_{L}$
offers an alternative to the see-saw mechanism in generating naturally light
(in fact massless) neutrinos.

The phenomenological implications of such a superstring-inspired
low-energy model, which is just a minimal extension of the SM, are studied in 
this thesis. 
The model itself is described in detail
in the next chapter.

\newpage

\chapter{A superstring-inspired $SU(2)_{L} \times U(1)_{Y}$ model of neutrino
 mass}

In this thesis we study phenomenological aspects of an $SU(2)_{L} \times 
U(1)_{Y}$
model, which 
extends the neutral fermion spectrum of the SM by two new fields,
the right-handed neutrino $n_{R}$ and a left-handed field $S_{L}$.
The model could arise as a low-energy limit of a superstring-inspired 
$E_{6}$ GUT 
\cite{vallemo,bernabeu1}; it was also suggested as a low-energy limit of a 
supersymmetry-inspired $SO(10)$ GUT \cite{wolfe}.
%The model was motivated by a supersymmetric grand unified
%$E_{6}$ model inspired by the heterotic superstring 
%\cite{vallemo,valle,bernabeu1}.
Superstring-inspired 
%$E_{6}$ 
GUT's have an
interesting problem with neutrino masses (see the discussion in 
Sec.~\ref{grand4}):  the see-saw mechanism does not apply here
and unacceptably large neutrino masses arise as
a consequence \cite{vallemo,bernabeu1}.
The existence of the field $S_{L}$ was suggested
as a potential solution to this problem. $S_{L}$ is an $E_{6}$ singlet
which may be present in superstring models. At low energies it can remain
in the neutral fermion spectrum along with the right-handed neutrino $n_{R}$
and the usual left-handed neutrino $\nu_{L}$. These three fields together with
imposed $B - L$ conservation form a mass matrix leading to
an alternative to the see-saw mechanism in
addressing the problem of the smallness of neutrino masses.

In this chapter we define the model and give a detailed
treatment of
neutrino masses and mixing matrix, and the neutrino interaction Lagrangian.

\section{Fermion content and mass matrix}
\label{content1}

In this superstring-inspired model we keep, in line with introductory arguments 
in Sec. \ref{neutrino3}, the gauge
sector and the Higgs sector of the SM untouched. The fermion content is 
enlarged by two neutrino fields, $n_{R}$ and $S_{L}$, per family. Their
$SU(2)_{L}\times U(1)_{Y}$ quantum numbers are $(0,0)$. 
The field $n_{R}$ is a right-handed neutrino, while $S_{L}$ is an $E_{6}$ 
singlet
neutrino field. In a single family, we thus have the following leptons (given
with their quantum numbers):
\begin{equation}
\begin{array}{cccc}
\left( \begin{array}{c}
                       \nu_{e} \\
                         e    
               \end{array} \right)_{L} &
e_{R} &  n_{R} & S_{L}  \\ \\
\Big(\frac{1}{2},-1\Big) & (0,-2) & (0,0) & (0,0)
\end{array}
\end{equation}
The definition of the
model is completed by specifying the mass matrix ${\cal M}$. 
In the Majorana basis it is given by
\begin{eqnarray}
\label{ourmatrix}
-{\cal L}_{mass} & = & \frac{1}{2}{\cal M} \; = \;
 \frac{1}{2}\left(\overline{\nu_{L}}\;\overline{n_{L}^{c}}\;\overline{S_{L}}
\right)
\left( \begin{array}{lll}
                       0     & D & 0      \\
                       D^{T} & 0 & M^{T}  \\
                       0     & M & 0
               \end{array} \right)
\left( \begin{array}{c}
                       \nu_{R}^{c} \\
                       n_{R}      \\
                       S^{c}_{R}
               \end{array} \right) + h.c..
\end{eqnarray}
Each $\nu_{L},n_{R},S_{L}$ represents now a collection of three fields,
one for each family, e.g. $\nu_{L} = (\nu_{e},\nu_{\mu},\nu_{\tau})$ is the
vector of the three SM weak eigenstate neutrinos. $D$ and $M$ are $3 \times 3$
Dirac mass matrices.
The top diagonal element must vanish unless we extend
the symmetry breaking sector of our model. A weak isotriplet Higgs field could
allow this term. However, we retain the symmetry breaking sector of the SM.
%affects the $\rho$ parameter of the weak interactions
%\footnote{The $\rho$ parameter is the ratio of the strength of the neutrino
%low-energy
%effective charged current interaction to that of the neutral current
%interaction. The SM prediction, $\rho = 1$, is confirmed by experiment.}; it
%has to vanish if we take $\rho = 1$ as a phenomenological constraint
%\cite{kayser}.
The middle element is zero due to the absence of the
appropriate Higgs fields that would provide the Majorana mass. This is enforced
by imposed $B-L$ number conservation, which is also responsible for all other
zeros in the mass matrix ${\cal M}$. Only terms preserving the $B-L$ number,
$\overline{\nu_{L}}\;D\;n_{R}+ h.c.$ and
$\overline{S_{L}}\;M\;n_{R} + h.c.$, remain (see footnote 7, p. 23 on $B-L$ 
conservation). 
%We will see the consequences this has on neutrino masses when we diagonalize 
%the mass matrix.

%The bottom
%diagonal element could actually be nonzero as will be discussed below (see Eq. 
%\ref{muelement}).

%what may
%look imposed from the point of view of $SU(2)_{L}\times U(1)_{Y}$ model, but 
%the
%origin of this symmetry, as well as the origin of the matrix $M$,
%can be traced back to $SO(10)$ GUT model \cite{Wolfe} or $E(6)$ supergravity
%\cite{Vallemo}. In contrast, the $B-L$ conservation in the SM is just an
%automatic consequence of the missing right-handed neutrino fields.

To find the physical neutrino states of the model, we have to diagonalize the
mass matrix ${\cal M}$. We will do it within a single family first and then
we will generalize the procedure for the three families.

\subsection{Diagonalization of ${\cal M}$ for a single family}
\label{diagon2}

In the case of a single family, $\nu_{L},n_{R},S_{L}$ 
represent each only one field and matrices $D,M$ become simple numbers.
We perform the following rotation,
\begin{eqnarray}
\label{rotation}
\left( \begin{array}{c}
                       \nu_{L} \\
                       n_{L}^{c}      \\
                       S_{L}
               \end{array} \right) \equiv O 
\left( \begin{array}{c}
                       \nu_{L}^{'} \\
                       n_{L}^{c}      \\
                       S_{L}^{'}
               \end{array} \right),\;\;\;\;\;\;\;\;\;\;
\left( \begin{array}{c}
                       \nu_{R}^{c} \\
                       n_{R}      \\
                       S^{c}_{R}
               \end{array} \right)  \equiv O
\left( \begin{array}{c}
                       \nu_{R}^{'c} \\
                       n_{R}      \\
                       S^{'c}_{R}
               \end{array} \right), 
\end{eqnarray}
where
\begin{eqnarray}
O & = & \left( \begin{array}{ccc}
                       c_{\theta}  & 0 & s_{\theta}   \\
                             0      & 1 &      0  \\
                       - s_{\theta}  & 0 & c_{\theta}
               \end{array} \right), \;\;\;\;\; 
c_{\theta} = \cos \theta, \; s_{\theta} = \sin \theta, \; 
\tan \theta =\frac{D}{M}.
\end{eqnarray}
The mass matrix ${\cal M}$ becomes
\begin{eqnarray}
{\cal M} & = & \left(\overline{\nu_{L}^{'}}\; \overline{n_{L}^{c}}\; 
\overline{S_{L}^{'}} \right) O^{T} 
\left( \begin{array}{lll}
                       0     & D & 0      \\
                       D     & 0 & M      \\
                       0     & M & 0
               \end{array} \right) O
\left( \begin{array}{c}
                       \nu_{R}^{'c} \\
                       n_{R}      \\
                       S^{'c}_{R}
               \end{array} \right) + h.c.  \nonumber   \\
\nonumber \\
& = & \left(\overline{\nu_{L}^{'}}\; \overline{n_{L}^{c}}\;
\overline{S_{L}^{'}} \right)
\left( \begin{array}{ccc}
  0                  & D c_{\theta} - M s_{\theta} &               0      \\
  D c_{\theta} - M s_{\theta}     &   0 & D s_{\theta} + M c_{\theta}   \\
  0                  & D s_{\theta} + M c_{\theta} &               0
               \end{array} \right)
\left( \begin{array}{c}
                       \nu_{R}^{'c} \\
                       n_{R}      \\
                       S^{'c}_{R}
               \end{array} \right) + h.c.  \nonumber \\
\nonumber \\
& = & \left(\overline{\nu_{L}^{'}}\; \overline{n_{L}^{c}}\;
\overline{S_{L}^{'}} \right)
\left( \begin{array}{ccc}
  0                  &        0          &            0                 \\
  0                  &        0          &   \sqrt{D^{2} + M^{2}}       \\
  0                  &    \sqrt{D^{2} + M^{2}}   &               0
               \end{array} \right) 
\left( \begin{array}{c}
                       \nu_{R}^{'c} \\
                       n_{R}      \\
                       S^{'c}_{R}
               \end{array} \right) + h.c. ,
\end{eqnarray}
yielding a massless neutrino $\nu^{'}$.
Moreover, we recognize the submatrix
\begin{eqnarray}
\left(\overline{n_{L}^{c}}\;
\overline{S_{L}^{'}} \right)
\left( \begin{array}{cc}
      0          &   \sqrt{D^{2} + M^{2}}       \\
     \sqrt{D^{2} + M^{2}}   &               0
               \end{array} \right)
\left( \begin{array}{c}
                       n_{R}      \\
                       S^{'c}_{R}
               \end{array} \right)
\end{eqnarray}
as the matrix representation of a Dirac mass term (see Eq. \ref{decomp})
 in the
Majorana basis. Indeed, putting
\begin{eqnarray}
\label{identif}
n_{L}^{c} \equiv N^{c}_{L},\; \; \; S^{'c}_{R} \equiv N^{c}_{R},\; \; \;  n_{R}
\equiv N_{R},\; \; \; S_{L}^{'} \equiv N_{L},
\end{eqnarray}
we reproduce Eq. \ref{decomp} and therefore, besides the massless neutrino
$\nu^{'}$,
we generate a Dirac neutral heavy lepton $N$ with the mass $M^{'} =
\sqrt{D^{2} + M^{2}}$.
%Inverting Eq. \ref{rotation} we get
%\begin{eqnarray}
%\left( \begin{array}{c}
%                       \nu_{L}^{'} \\
%                       n_{L}^{c}      \\
%                       S_{L}^{'}
%               \end{array} \right) = O^{T}
%\left( \begin{array}{c}
%                       \nu_{L} \\
%                       n_{L}^{c}      \\
%                       S_{L}
%               \end{array} \right).
%\end{eqnarray}
The weak eigenstate $\nu_{L}$ is given  by
\begin{eqnarray}
\nu_{L} & = & \cos \theta \; \nu^{'}_{L} + \sin \theta \; S_{L}^{'} \nonumber
\\
        & = & \frac{M}{\sqrt{D^{2} + M^{2}}} \; \nu^{'}_{L} +      
              \frac{D}{\sqrt{D^{2} + M^{2}}} \;  N_{L}     \nonumber \\
        & \equiv & K_{L} \nu^{'}_{L} + K_{H} N_{L},     
\end{eqnarray}
where $K_{L}, K_{H}$ are mixing factors (matrices in the case of three families)
for massless neutrinos and NHL's, respectively.

For $ M \gg D $ the mass of the NHL, $M^{'}$, and the weak eigenstate $\nu_{L}$
are approximately
\begin{eqnarray}
\label{eigenmix}
M^{'} & \doteq & M,   \nonumber    \\
\nu_{L} & \doteq & \nu^{'}_{L} + \frac{D}{M} N_{L}, 
\end{eqnarray}
that is, the mixing of NHL's is $K_{H} \doteq \frac{D}{M}$.

\subsection{Mass matrix diagonalization in case of three families}
\label{diagon3}

We leave the matrix representation of the mass matrix observing that
\begin{eqnarray}
\label{fa3}
-{\cal L}_{mass} & = & \frac{1}{2}{\cal M} \; = \;
 \frac{1}{2}\left(\overline{\nu_{L}}\;\overline{n_{L}^{c}}\;\overline{S_{L}}
\right)
\left( \begin{array}{lll}
                       0     & D & 0      \\
                       D^{T} & 0 & M^{T}  \\
                       0     & M & 0
               \end{array} \right)
\left( \begin{array}{c}
                       \nu_{R}^{c} \\
                       n_{R}      \\
                       S^{c}_{R}
               \end{array} \right) + h.c. \nonumber \\
%& = &
%\frac{1}{2}\left(\overline{\nu_{L}}\;\overline{n_{L}^{c}}\;\overline{S_{L}}
%\right)
%\left( \begin{array}{c}
%                       D \; n_{R}      \\
%                       D^{T} \; \nu_{R}^{c} + M^{T} \; S^{c}_{R} \\
%                       M \; n_{R}
%               \end{array} \right) + h.c. \nonumber \\
& = & \frac{1}{2}\left(\overline{\nu_{L}}\;D\;n_{R} + 
\overline{n_{L}^{c}}\; D^{T} \; \nu_{R}^{c} +
\overline{n_{L}^{c}}\;M^{T}\;S^{c}_{R} +
\overline{S_{L}}\;M\;n_{R} \right) + h.c. \nonumber \\
& = & \frac{1}{2}\left(\overline{\nu_{L}}\;D\;n_{R} +
\overline{\nu_{L}}\;D\;n_{R} +
\overline{S_{L}}\;M\;n_{R}  +
\overline{S_{L}}\;M\;n_{R} \right) + h.c. \nonumber \\
& = & 
\overline{\nu_{L}}\;D\;n_{R} +
\overline{S_{L}}\;M\;n_{R} + h.c. \;.
\end{eqnarray}
In the above we used the identity (for which the proof is almost identical with 
that for
$\overline{\nu_{L}^{c}}\nu_{R}^{c} = \overline{\nu_{L}}\nu_{R}$, see Appendix
\ref{proof})
\begin{equation}
\overline{n_{L}^{c}}D^{T}\nu_{R}^{c}  =  \overline{\nu_{L}} D n_{R}\;.
\end{equation}
Performing the following rotation
\begin{eqnarray}
\label{matrixg}
\left( \begin{array}{c}
                       \nu_{L}^{'} \\
                       S_{L}^{'}
               \end{array} \right) = G
\left( \begin{array}{c}
                       \nu_{L} \\
                       S_{L}
               \end{array} \right) =
\left( \begin{array}{cc}
      U_{1}          &   U_{2}   \\
      U_{3}          &   U_{4}
               \end{array} \right)
\left( \begin{array}{c}
                       \nu_{L} \\
                       S_{L}
               \end{array} \right),
\end{eqnarray}
where $G$ is a unitary matrix, we get
\begin{eqnarray}
-{\cal L}_{mass} & = & \left( \overline{\nu_{L}^{'}} U_{1} +
\overline{S_{L}^{'}}U_{3}\right)D n_{R}   
+ \left( \overline{\nu_{L}^{'}} U_{2} +
\overline{S_{L}^{'}}U_{4}\right) M n_{R}+ h.c. \nonumber \\
& = & \overline{\nu_{L}^{'}}\left(U_{1} D + U_{2} M\right) n_{R}+ h.c.
+\overline{S_{L}^{'}}\left(U_{3}D + U_{4} M\right)n_{R} + h.c.\;. 
\end{eqnarray}
We put (compare with $ D \cos \theta - M \sin \theta = 0$ for a single
generation case)
\begin{equation}
U_{1} D + U_{2} M = 0,
\end{equation}
hence
\begin{eqnarray}
-{\cal L}_{mass} & = & \overline{S_{L}^{'}}\left(U_{3}D + U_{4} M\right)n_{R} +
h.c. = \overline{S_{L}^{'}}M^{'}n_{R} + h.c.\;.
\end{eqnarray}
There is no mass term for $\nu^{'}_{L}$, therefore $\nu^{'}_{L}$ is a massless
neutrino. $M^{'}$, unlike in the one family case, has to be further diagonalized
with rotations ($Z, T$ unitary matrices) in the NHL basis:
\begin{eqnarray}
S_{L}^{''} = T S_{L}^{'},\;\;\;\; n_{R}^{''} = Z n_{R},
\end{eqnarray}
yielding
\begin{eqnarray}
\label{mprimed}
-{\cal L}_{mass} & = & \overline{S_{L}^{''}}\left(TU_{3}DZ^{\dagger} +
TU_{4}MZ^{\dagger}\right)n_{R}^{''} + h.c. \nonumber \\
& = & \overline{ S_{L}^{''}}M^{''} n_{R}^{''} + h.c.,
\end{eqnarray} 
such that $M^{''}$ is diagonal.
Identifying (see also Eq. \ref{identif})
\begin{eqnarray}
S_{L}^{''} & \equiv & N_{L},  \nonumber \\
n_{R}^{''} & \equiv & N_{R},
\end{eqnarray}
we arrive at
\begin{eqnarray}
-{\cal L}_{mass} & = &
\overline{N} M^{''} N                   
\; = \;
M_{N_{4}} \overline{N_{4}}N_{4} +
M_{N_{5}} \overline{N_{5}}N_{5} +  M_{N_{6}} \overline{N_{6}}N_{6}.
\end{eqnarray}
Here $M^{''}$ is diagonal with elements $M_{N_{4}}, M_{N_{5}}, M_{N_{6}}$
being masses of  three Dirac NHL's $N_{4}, N_{5}, N_{6}$.
The weak eigenstate vector, $\nu_{L}$, is given by
\begin{eqnarray}
\label{weakeig}
\nu_{L} \equiv \nu_{l} \equiv \left(\nu_{e},\nu_{\mu},\nu_{\tau}\right)
& = & U_{1}^{\dagger}\nu_{L}^{'} + U_{3}^{\dagger}S_{L}^{'}
= U_{1}^{\dagger}\nu_{L}^{'} + U_{3}^{\dagger}T^{\dagger}S_{L}^{''}  \nonumber
\\
& \equiv & K_{L}\nu_{L}^{'} + K_{H}S_{L}^{''} = K_{L} \nu_{L}^{'} + K_{H}N_{L}.
\end{eqnarray}

\subsection{Discussion of mass eigenstates}
\label{discus3}

The diagonalization of the mass matrix yielded
three
massless neutrinos $\nu_{L}^{'} \equiv
\left(\nu_{1_{L}}^{'},\nu_{2_{L}}^{'},\nu_{3_{L}}^{'}\right)$ along with three
Dirac NHL's $N \equiv \left(N_{4},N_{5},N_{6}\right)$ with mass
$M_{N} \sim M$. 
The masslessness of the neutrinos is the consequence of the assumed
$B-L$ symmetry.
This symmetry also prevents neutrinos from acquiring small masses in radiative
corrections. The neutrinos are massless due to $B-L$ symmetry also in the SM,
but the difference is that in the SM the $B-L$ symmetry is an automatic
consequence of the missing right-handed neutrino fields, while here this
symmetry is imposed with right-handed neutrinos present.
Note
that massless neutrinos
imply there are no time dependent neutrino oscillations and no neutrinoless
double beta decays.

In the light of arguments for massive neutrinos (see Sec. \ref{sttp})
it may seem
surprising that this model yields massless neutrinos. However, small neutrino
masses can be generated in a variant of our model by introducing a 
small Majorana mass  term $\mu$ 
\cite{vallemo,garval} in the mass matrix ${\cal M}$
in Eq. \ref{ourmatrix} :
\begin{eqnarray}
\label{muelement}
-{\cal L}_{mass} & = & \frac{1}{2}{\cal M} \; = \;
 \frac{1}{2}\left(\overline{\nu_{L}}\;\overline{n_{L}^{c}}\;\overline{S_{L}}
\right)
\left( \begin{array}{lll}
                       0     & D & 0      \\
                       D^{T} & 0 & M^{T}  \\
                       0     & M & \mu
               \end{array} \right)
\left( \begin{array}{c}
                       \nu_{R}^{c} \\
                       n_{R}      \\
                       S^{c}_{R}
               \end{array} \right) + h.c.\;.\;\;\;\;
\end{eqnarray}
Besides, now that we have a superstring motivation for the field content of 
our model, there is nothing unusual about massless neutrinos either. 
The unnaturalness of the SM treatment of neutrino masses is 
removed, dark matter has other candidates and the solar neutrino puzzle
has alternative explanations. 
Whether neutrinos have any mass at all may actually be of secondary interest
for a theorist who is trying to come up with some plausible explanation of the
low experimental limits on this mass.
%Perhaps the key problem associated with neutrinos 
%is the smallness of their masses rather than they massive or not.

The weak eigenstates in our model ($\nu_{L}$) are dominated by massless
neutrinos $\nu_{L}^{'}$ with a small admixture
 ($K_{H} \sim D/M$, see Eq. \ref{eigenmix}) of NHL's $N$. 
In see-saw models, the NHL mixing is generally suppressed 
due to the scale $M$ 
(see Eq. \ref{see} and the discussion afterwards)
\footnote{An exception are the see-saw models with enhanced mixings discussed at
the end of Sec. \ref{see-saw32}.}, which
has to be very large to explain small neutrino masses, $m_{\nu} \sim
\frac{D^{2}}{M}$, dictated by experiments and cosmological arguments.
The NHL mixing in our
model is not, however, restricted by the dependence of neutrino masses 
$m_{\nu^{'}}$ on scales $D$ and
$M$ (there is no such dependence as $m_{\nu^{'}} = 0$). Therefore, the scale 
$M$ can be much lower than in the case of see-saw models
and hence rates for many interesting phenomena
can be large. This means that signatures of NHL's might be found
even at current accelerator energies and luminosities. Our model is thus
attractive not only conceptually, but also practically.

%We note that see-saw models with enhanced mixings (discussed at the end 
%of Sec. \ref{see-saw32}) are equally attractive in this respect.
%, there are special forms 
%of the
%see-saw matrices that avoid the suppression of the NHL mixing. 
%For instance, Pilaftsis
%\cite{pilaftsis2} finds a relation among the elements of $D$ and $M$ matrices
%that leads
%to massless neutrinos at the tree-level and small Majorana masses are generated
%radiatively. The cosmological constraint on the scale $M$ is much weaker in
%this model and consequently, the mixing of NHL's ($K_{H} \sim D/M$) is not
%suppressed. 

\section{Properties of the mixing matrix}
\label{properties3}

The weak interaction eigenstates $\nu_{l}$ are related to six mass eigenstates
$\nu^{'}, N$ via a $3 \times 6$ mixing matrix $K$ with 
components
$K_{l \alpha}$; $l = e, \mu, \tau$ and $\alpha = \nu^{'}_{1}, \nu^{'}_{2},
\nu^{'}_{3}, N_{4}, N_{5}, N_{6}$ (see Eq. \ref{weakeig})
\begin{eqnarray}
  \left( \begin{array}{c}
                       \nu_{e} \\
                       \nu_{\mu}      \\
                       \nu_{\tau}
               \end{array} \right)
& = & {\left( \begin{array}{llllll}
  K_{e\nu^{'}_{1}} & K_{e\nu^{'}_{2}} & K_{e\nu^{'}_{3}} &
  K_{eN_{4}}   & K_{eN_{5}}   & K_{eN_{6}} \\
  K_{\mu\nu^{'}_{1}} & K_{\mu\nu^{'}_{2}} & K_{\mu\nu^{'}_{3}} &
  K_{\mu N_{4}}   & K_{\mu N_{5}}   & K_{\mu N_{6}} \\
  K_{\tau\nu^{'}_{1}} & K_{\tau\nu^{'}_{2}} & K_{\tau\nu^{'}_{3}} &
  K_{\tau N_{4}}   & K_{\tau N_{5}}   & K_{\tau N_{6}} \\
\end{array}    \right)}
\left( \begin{array}{c}
                       \nu^{'}_{L} \\
                       N_{L}    
               \end{array} \right)   \nonumber \\
 & \equiv & (K_{L} \; K_{H})
\left( \begin{array}{c}
                       \nu^{'}_{L} \\
                       N_{L}
               \end{array} \right)  ; \;\;\;\;
\nu^{'} \; = \; \left( \begin{array}{c}
                       \nu^{'}_{1} \\
                       \nu^{'}_{2}      \\
                       \nu^{'}_{3}
               \end{array} \right), \;\;
N \; = \; \left( \begin{array}{c}
                       N_{4}      \\
                       N_{5}      \\
                       N_{6}
               \end{array} \right).
  \end{eqnarray}
Alternatively, we can write
\footnote{Where not indicated in this work, 
indices $i,j,k$ run through $1,2,3$ and $a,b,c$ 
through $4,5,6$.}
\begin{eqnarray}
\label{alter}
\nu_{l} & = & \sum_{i=1,2,3}\big(K_{L}\big)_{li} \nu^{'}_{i_{L}} 
+ \sum_{a=4,5,6} \big(K_{H}\big)_{la} N_{a_{L}} 
\; = \; \big(K_{L}\big)_{li} \nu^{'}_{i_{L}} + \big(K_{H}\big)_{la} N_{a_{L}}.
\end{eqnarray}
A quick inspection tells us the matrix $K$ has $3 \times 6$ complex parameters
 = $36$ degrees of freedom.
Unitarity implies an important property often used throughout this work,
\begin{eqnarray}
\label{KLKH}
K_{L}K^{\dagger}_{L} +K_{H}K^{\dagger}_{H} = 1.
\end{eqnarray}
This property reduces the number of degrees of freedom by $9$ to $27$.
Further elimination of unphysical parameters via redefinition (rephasing)
\footnote{This operation is also done in the SM when one parametrizes the CKM 
matrix.}
of physical mass eigenstates leaves us
$3^{2}$ angles and ${(3-1)}^{2}$ phases \cite{branco}.
This allows for
possible lepton flavour violation, universality
violation and CP violation.
                                       
The mixing factor which typically governs flavour-conserving
processes, $ll_{mix}$,
is given by
\begin{eqnarray}
ll_{mix} & = & \sum_{a=4,5,6} \big(K_{H}\big)_{la} 
\big(K_{H}^{\dagger}\big)_{al}\;\; ;
\makebox[.5in] [c] { } l= e, \mu, \tau
\end{eqnarray}
and the flavour-violating mixing factor   $l{l^{'}}_{mix}$
is defined as 
\begin{eqnarray}
l{l^{'}}_{mix} & = & \sum_{a=4,5,6} \big(K_{H}\big)_{la}
\big(K_{H}^{\dagger}\big)_{al^{'}}\;\; ;
\makebox[.5in] [c] { } l,l^{'} = e, \mu, \tau,
\makebox[.2in] [c] { }l \neq l^{'}.
\end{eqnarray}
Further, the following important inequality holds
\begin{eqnarray}
\label{ineq}
|{l{l^{'}}_{mix}}|^2 & \leq & {ll}_{mix}\:\:{l^{'}l^{'}}_{mix},
\makebox[.5in] [c] { } l \neq l^{'}.
\end{eqnarray}
This implies  that one might observe nonstandard effects in flavour-conserving
processes even if they are absent in flavour-violating processes.

\section{Interaction Lagrangians}
\label{inter3}

The charged and neutral current Lagrangians are obtained from the corresponding
terms in the SM Lagrangian substituting for $\nu_{l}$ from Eq. \ref{alter}.
The charged current Lagrangian is given by
\begin{eqnarray}
\label{ccur}
{\cal L}_{cc} & = & \frac{1}{2 \sqrt{2}} g_{2} W^{\mu} \sum_{l=e, \mu ,\tau}
 \Big\{ \; \sum_{i} \bar
{l} \gamma_{\mu} (1-\gamma_{5}) \big(K_{L}\big)_{li} \nu^{'}_{i} 
+ \sum_{a} \bar{l} \gamma_{\mu} (1-\gamma_{5}) \nonumber \\
& \times & \big(K_{H}\big)_{la} N_{a} \Big\} + h.c.
\end{eqnarray}
and the neutral current Lagrangian as 
\begin{eqnarray}
{\cal L}_{nc} &  = & \frac{g_{2}}{4c_{W}} Z^{\mu} \sum_{i,a}
 \bar{\nu_{i}}^{'}
{(K_{L}^{\dagger}K_{H})}_{ia} \gamma_{\mu} (1-\gamma_{5})N_{a} + h.c.
 \nonumber \\
& + &  \frac{g_{2}}{4c_{W}} Z^{\mu} \sum_{a,b}
 \bar{N_{a}}
{(K_{H}^{\dagger}K_{H})}_{ab} \gamma_{\mu} (1-\gamma_{5})N_{b} \nonumber   \\
& + & \frac{g_{2}}{4c_{W}} Z^{\mu} \sum_{i,j}
 \bar{\nu_{i}}^{'}
{(K_{L}^{\dagger}K_{L})}_{ij} \gamma_{\mu} (1-\gamma_{5})\nu_{j}^{'}.  
\end{eqnarray}

We will also need Lagrangians with neutrinos and NHL's interacting with the 
Higgs
$H$ and with unphysical Higgs $\phi^{+},\phi^{-}$ and $\chi$.
The starting point is the Yukawa Lagrangian describing the 
interactions of neutrino
fields $\nu_{l}, n_{R}$ with the Higgs doublet~$\tilde{\Phi}$,
\begin{eqnarray}
\label{yukawa1}
{\cal L} & = & - (\overline{\nu_{l}}\:\:\overline{l_{L}})\:\tilde h \:
\tilde{\Phi}\:n_{R} + h.c.,
\end{eqnarray}
where $\tilde h$ is a matrix (in family space) of Yukawa couplings.
The $S_{L}$ field does not couple to $\tilde{\Phi}$, but rather might couple
to a new $SU(2)_{L} \times U(1)_{Y}$ singlet Higgs field 
(responsible also for mass $M$) 
present in some superstring models \cite{bernabeu1}. We do not introduce such a
field.

The physical Higgs part, derived in Appendix \ref{Becko}, is given by
\begin{eqnarray}
{\cal L}_{H} &  = & - \frac{g_{2}}{2 M_{W}}\overline{N}(K_{H}^{\dagger}K_{H})
M_{N} N H \nonumber \\
& - & \frac{g_{2}}{2 M_{W}}\overline{\nu^{'}}(K_{L}^{\dagger}K_{H})M_{N}
\frac{1+\gamma_{5}}{2}N H \nonumber \\
& - & \frac{g_{2}}{2 M_{W}}\overline{N}(K_{H}^{\dagger}K_{L})M_{N}
\frac{1-\gamma_{5}}{2}\nu^{'} H, 
\end{eqnarray}
the unphysical neutral Higgs $\chi$ part,
\begin{eqnarray}
{\cal L}_{\chi} &  = & + i \frac{g_{2}}{2 M_{W}}
\overline{N}(K_{H}^{\dagger}K_{H})
M_{N} \gamma_{5} N \chi \nonumber \\
& + & i \frac{g_{2}}{2 M_{W}}\overline{\nu^{'}}(K_{L}^{\dagger}K_{H})M_{N}
\frac{1+\gamma_{5}}{2}N \chi \nonumber \\
& - & i \frac{g_{2}}{2 M_{W}}\overline{N}(K_{H}^{\dagger}K_{L})M_{N}
\frac{1-\gamma_{5}}{2}\nu^{'} \chi, 
\end{eqnarray}
and the unphysical charged Higgs $\phi^{+},\phi^{-}$ parts
\begin{eqnarray}
{\cal L}_{\phi^{-}} &  = &  + \frac{g_{2}}{\sqrt{2} M_{W}}\overline{e_{L}}K_{H}
M_{N}N_{R}\phi^{-} + O\big(\frac{m_{l}}{M_{W}}\big) + h.c. \;.
\end{eqnarray}
Feynman rules corresponding to these Lagrangians are listed in Appendix
\ref{Cecko}.

\section{Review of existing constraints on NHL's}
\label{review3}

Constraints on neutral heavy lepton masses and mixings come from three
different sources.

{\bf i)} First, there is the possibility of direct production of NHL's.
 At $e^{+}e^{-}$
colliders such as LEP I or SLC, they could be produced in Z decays:
\footnote{In the rest of this work we drop the prime from $\nu^{'}$.}
\begin{eqnarray}
Z \rightarrow N_{a} + \nu
\end{eqnarray}
and subsequently decay via neutral or charged currents:
\begin{center}
\setlength{\unitlength}{1in}
\begin{picture}(6,1.5)
\put(0.1,+0.125){\mbox{\epsfxsize=5.5in\epsffile{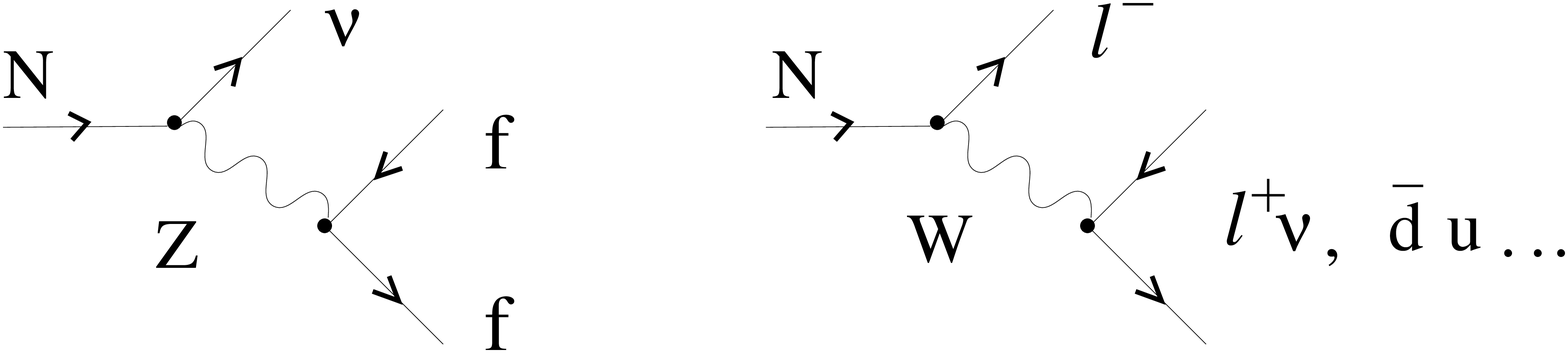}}}
\end{picture}
\end{center}
The rate for $Z$ decays into an NHL and a light neutrino has been given
previously \cite{Dittmar} as
\begin{eqnarray}
\Gamma(Z \rightarrow N_{a} + \nu) & = & a_{mix}
(1-\frac{{M_{N_{a}}}^{2}}{{M_{Z}}^{2}})(1+\frac{{M_{N_{a}}}^{2}}{{2
M_{Z}}^{2}})\Gamma(Z \rightarrow \nu + \nu),
\end{eqnarray}
where
\begin{eqnarray}
a_{mix} & = & \sum_{l=e,\mu,\tau} {|\left(K_{H}\right)_{la}|}^{2}.
\end{eqnarray}
The subsequent NHL decay rate (for $M_{N} \leq M_{W}$) is then
given by
\begin{eqnarray}
\Gamma_{N} & = & a_{mix} (\frac{M_{N}}{m_{\mu}})^{5}
\Phi_{l}\Gamma_{\mu},
\end{eqnarray}
where $\Gamma_{\mu}$ is the muon decay rate and $\Phi_{l}$ is the
effective number of decay channels available to the NHL \cite{Gronau}.
LEP data effectively (better than indirect constraints, see below) probe
NHL mixings for NHL mass up to $80$ GeV \cite{Dittmar} .

NHL production at $pp$ supercolliders was studied in Ref. \cite{Gour}. It was
concluded that the CERN Large Hadron Collider (LHC) has the potential to push
the limits on $a_{mix}$ below the LEP constraints for NHL mass up to $110$ GeV.

%Given the absence of experimental evidence for such direct production,
%we will consider only NHL's with mass greater than
%the $Z$ mass.

{\bf ii)} Second, there are constraints on NHL mixing parameters from a 
variety of low
energy experiments and from experiments at LEP I where neutral heavy leptons
are not directly present.
NHL's however, do affect observables indirectly: due to unitarity properties of
the mixing
matrix $K$, a nonzero NHL mixing slightly reduces the couplings of light
neutrinos from their SM values
\footnote{For example in case of $We\nu$ vertex, the mixing is changed from
SM value $= 1$ to $K_{L}$, see Eq.~\ref{ccur}.}, thus affecting rates
for nuclear $\beta$ decays, $\tau$ and $\pi$ decays, and for $Z$ decays.
The following upper limits are consistent with experiment \cite{Nardi}
\begin{eqnarray}
\label{limits1}
      ee_{mix} & \leq & 0.0071 \nonumber \\
      \mu\mu_{mix} & \leq  & 0.0014 \nonumber \\
      \tau\tau_{mix} & \leq & 0.033 
\end{eqnarray}
The limit on $\tau\tau_{mix}$ is improved to $\leq 0.024$ if the 
invisible width of
the Z boson is included in the analysis \cite{Nardi}.
The limits in Eq. \ref{limits1} are model independent and hold for any value 
of the NHL mass. They
arise from a global analysis of results including lepton universality
experiments, CKM matrix unitarity tests, $W$ mass measurements and neutral
current data from LEP I  experiments. Note
that the LEP I neutral current data analysis did not include NHL loop effects 
but, rather, only coupling constant modifications due to mixing. We consider NHL
loop effects in this work.

Since the limit on the parameter $\tau\tau_{mix}$ plays
(as the least stringent one) the most important role in our analysis, 
we will pay further attention to its source. It comes from
the $\mu - \tau$ universality test based on the $\tau$ leptonic decays compared
to the $\mu$ leptonic decays. The result of the test is given as the ratio of
the couplings of $\tau$ and $\mu$ to the W boson, $g_{\tau}/g_{\mu}$
(in the SM we have $g_{\tau} = g_{\mu} = g_{2}$). The ratio is found from
\begin{eqnarray}
\frac{\Gamma(\tau \rightarrow e \nu \nu) / 
\Gamma^{SM}(\tau \rightarrow e \nu \nu)}
{\Gamma(\mu \rightarrow e \nu \nu) /
\Gamma^{SM}(\mu \rightarrow e \nu \nu)}
& = & \Big(\frac{g_{\tau}}{g_{\mu}}\Big)^{2} \; = \;
\frac{1-\tau\tau_{mix}}{1-\mu\mu_{mix}}.
\end{eqnarray}
Setting $\mu\mu_{mix}=0$, we get
\begin{eqnarray}
\tau\tau_{mix} & = & 1 - \Big(\frac{g_{\tau}}{g_{\mu}}\Big)^{2},
\end{eqnarray}
with \cite{Nardi}
\begin{eqnarray}
\Big(\frac{g_{\tau}}{g_{\mu}}\Big)^{2} & = & 0.989 \pm 0.016 .
\end{eqnarray}

{\bf iii)} Finally, the NHL masses and mixings can be constrained via their
contribution in loops to various processes. The calculation 
to the one-loop level of the perturbation theory is naturally more involved 
than the mostly tree-level considerations required for direct and indirect
constraints. In return we can probe regions in the mixings vs NHL mass
parameter space currently inaccessible to the direct and indirect methods.
For example, as we will see, we can place upper limits on the NHL mass. We
caution though that these limits depend on the mixings and they will be relaxed
should tighter bounds on mixings be achieved. Still this is an improvement
over the direct and indirect methods which are blind to NHL masses larger than
$M_{Z}$.
%Such constraints are NHL mass
%dependent 

\begin{figure}
\begin{center}
\setlength{\unitlength}{1in}
\begin{picture}(6,1.8)
\put(1.6,+0.125){\mbox{\epsfxsize=2.5in\epsffile{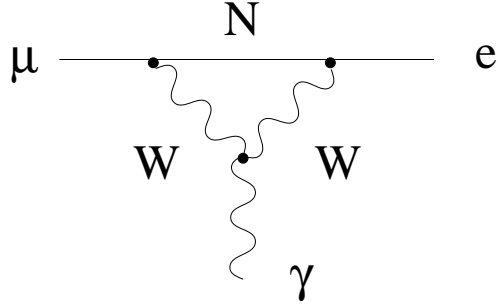}}}
\end{picture}
\end{center}
\caption{A one-loop diagram  leading to $\mu \rightarrow e \gamma$ decay}
\label{muegama}
\end{figure}

There are two classes of these processes, lepton flavour-violating and lepton
flavour-conserving. 
Lepton flavour-violating decays have been considered a hot candidate for a new 
physics manifestation in general for many decades. 
They include so far unobserved, so-called rare decays of $\mu$ and 
$\tau$ leptons and $\mu\:-\:e$ conversion in nuclei (A,Z):
\begin{equation}
\label{fvdecay3}
\begin{array}{c}
\mu \rightarrow e \gamma,\;\;\;\;\;\tau \rightarrow e \gamma,\;\;\;\;\;\tau
\rightarrow \mu \gamma, \;\;\;\;\;\;\; \\
\mu, \tau \rightarrow e e^{+} e^{-},\;\;\;\;\;\tau \rightarrow \mu e^{+} e^{-},
\;\;\;\;\;\tau \rightarrow e \mu^{+} \mu^{-}, \;\;\;\;\;\tau \rightarrow \mu
\mu^{+} \mu^{-}, \;\;\;\;\;\;\;\;\;\;\;\;\;\;\;\;\;\; \nonumber \\
\mu^{-}(A,Z) \rightarrow e^{-}(A,Z), 
 \nonumber 
 \end{array}
\end{equation}
and the Z boson decays
\begin{eqnarray}
\label{fvz3}
Z \rightarrow e^{\pm}\mu^{\mp},\;\;\;\;\;Z \rightarrow
e^{\pm}\tau^{\mp},\;\;\;\;\;Z \rightarrow \mu^{\pm}\tau^{\mp}.
 \end{eqnarray}
We will discuss these decays in Chapter 5; here we at least
mention  the decay $\mu \rightarrow e \gamma$, which underwent an 
intensive
experimental scrutiny and its stringent upper limit places a tough constraint 
on the mixing parameter $e\mu_{mix}$.
For illustration, 
one of the diagrams contributing to $\mu \rightarrow e \gamma$ is shown in Fig.
\ref{muegama}.
In Sec. \ref{fvple} we will show that $\mu \rightarrow e \gamma$ 
gives the following
upper limit on the mixing parameter $e \mu_{mix}$:
\begin{eqnarray}
\label{limits3}
|e\mu_{mix}| & \leq & 0.00024.
\end{eqnarray}
By combining the indirect constraints obtained from the global analysis
(see Eq.~\ref{limits1}) with the inequality relations of Eq. \ref{ineq} one
obtains the following upper limits on the mixing factors
\begin{eqnarray}
\label{limits2}
|e\mu_{mix}| & \leq & 0.0032 \nonumber \\
|\mu\tau_{mix}| & \leq & 0.0068 \nonumber \\
|e\tau_{mix}| & \leq & 0.015. 
\end{eqnarray}
For the mixings $\mu\tau_{mix}$ and $e\tau_{mix}$, these are the
strongest available constraints.
                   
The second class consists of lepton flavour-conserving processes with NHL's in 
loops. The main part of this thesis (Chapters 6 and 7)
is devoted to two of these processes, 
$Z \rightarrow l^{+}l^{-}$ with partial leptonic
width $\Gamma_{ll}$ and universality breaking parameter $U_{br}$ as observables
; and $\mu \rightarrow
e \nu_{e} \nu_{\mu}$ with the W mass $M_{W}$ as observable. 
We will argue that the flavour-conserving processes can be competitive
with and even have some advantages over the flavour-violating ones.

\newpage

\chapter{Standard model at the one-loop level}

As a prerequisite for one-loop calculations in Chapters 5,6 and 7, we discuss 
here the standard model of electroweak interactions at the one-loop level.
The classical electroweak Lagrangian was specified in Sec. \ref{classical1}.
One-loop corrections require treatment within the framework of the quantum
field theory: the classical Lagrangian has to be quantized and extended to 
include some new terms. We present these quantum field
theoretical 'amendments' to the classical Lagrangian
in Sec. \ref{quant1}. 

One-loop corrections calculated from the full Lagrangian typically suffer
from divergences. A systematic way of removing these divergences, the
renormalization of the SM, is discussed in Sec. \ref{renor}.
There are a lot of different schemes used to renormalize the SM. We opted for
the on-shell renormalization scheme of W. Hollik \cite{key6,key9}, 
introduced in Sec. \ref{onshell1}.

\section{Quantization}
\label{quant1}

The quantum field theory requires two more terms to be added to the classical
Lagrangian, ${\cal L}_{{\it gfix}}$ and ${\cal L}_{{\it ghost}}$:
\begin{eqnarray}
{\cal L}_{EW} = {\cal L}_{G} + {\cal L}_{F} + {\cal L}_{H} +
{\cal L}_{{\it gfix}} + {\cal L}_{{\it ghost}}.
\end{eqnarray}
The gauge fixing term ${\cal L}_{{\it gfix}}$ is required in order to define
meaningful propagators of the gauge fields which are otherwise singular
\cite{key5}.
 The linear gauge fixing of the 't~Hooft type is given by \cite{key6}
\begin{eqnarray}
{\cal L}_{{\it gfix}} = -\frac{1}{2}\left(F_{\gamma}^{2} + F_{Z}^{2} +
2F_{+}F_{-}\right),
\end{eqnarray}
where
\begin{eqnarray}
F_{\pm} & = & \frac{1}{\sqrt{\xi^{W}}}\left(\partial^{\mu}W_{\mu}^{\pm}
\mp i M_{W}\xi^{W}\phi^{\pm}\right), \nonumber \\
F_{Z}   & = & \frac{1}{\sqrt{\xi^{Z}}}\left(\partial^{\mu}Z_{\mu}
- M_{Z}\xi^{Z} \chi\right), \nonumber  \\
F_{\gamma} & = & \frac{1}{\sqrt{\xi^{\gamma}}}\partial^{\mu}A_{\mu},
\end{eqnarray}
and $\xi^{W}, \xi^{Z}, \xi^{\gamma}$ are gauge fixing parameters.
In the 't Hooft type gauge the vector boson propagators have the form 
($V = W, Z$)
\begin{eqnarray}
\label{bosonprop}
\frac{i}{k^{2}-M_{V}^{2}+i\epsilon}\left(-g^{\mu\nu} + 
\frac{\left(1-\xi^{V}\right)
k^{\mu}k^{\nu}}{k^{2}-\xi^{V}M_{V}^{2} + i\epsilon}\right),
\end{eqnarray}
and propagators of unphysical Higgs particles $\phi^{\pm},\chi$ are given by
\begin{eqnarray}
\label{unphysprop}
\frac{i}{k^{2}-\xi^{V}M_{V}^{2}+i\epsilon}.
\end{eqnarray}
The unitary gauge is defined by $\xi^{V}\rightarrow \infty$. We can see
that in this gauge unphysical Higgs freeze out (their propagators vanish)
and only physical particles appear in Feynman diagrams.

In this work we use the Feynman gauge defined by
 $\xi^{V} = \xi^{\gamma} = 1$. In this gauge there are
unphysical Higgs present, but the positive trade-off is the particularly 
simple form of the gauge boson propagators of Eq. \ref{bosonprop}.

The ${\cal L}_{{\it ghost}}$ term \cite{key5,key6,key7} is specific to 
nonabelian theories where the
%In electroweak (nonabelian) theory adding ${\cal L}_{{\it gfix}}$ to the 
%Lagrangian spoils the gauge invariance. 
% forget this !!! (not in abelian quantum electrodynamics, QED). 
one-loop
self-energies of the gauge bosons computed from ${\cal L}_{G} + {\cal L}_{F} +
{\cal L}_{H} + {\cal L}_{{\it gfix}}$ do not satisfy gauge invariance and
unitarity. ${\cal L}_{{\it ghost}}$ removes this difficulty with 
scalar anticommuting ghost fields 
$u^{\pm},u^{Z},u^{\gamma}$ (Fadeev-Popov ghosts) which appear
naturally in Fadeev-Popov quantization based on the path-integral method
\cite{key8}.
$u^{\pm},u^{Z}$ propagators are the same as the propagators of unphysical
Higgs, Eq. \ref{unphysprop}, while the $u^{\gamma}$ propagator is given by
\begin{eqnarray}
\frac{i}{k^{2}+i\epsilon}.
\end{eqnarray}

\section{Renormalization}
\label{renor}

In ${\cal L}_{EW}$ there are five independent parameters (showing only
lepton
Yukawa couplings $h_{i}$ and counting them as one):
\begin{eqnarray}
g_{2},\;g_{1},\; \lambda,\;v,\;h_{i}. \nonumber
\end{eqnarray}
After symmetry breaking we can replace them by an equivalent set (counting
$m_{f}$ as one)
\begin{eqnarray}
\label{parameters}
e,\;M_{W},\;M_{Z},\;M_{H},\;m_{f},
\end{eqnarray}
where $e^{2}/4\pi = \alpha$ and masses are those of $W,Z$ and Higgs
bosons and of a fermion, respectively.

  Originally these parameters were
identified with their physical values ($\alpha = 1/137, M_{Z} = 91.137$ GeV,
etc.). However, loop corrections calculated
in terms of the physical values of these parameters diverge and the parameters
themselves are modified - their physical values are changed by an infinite
amount. 
Renormalization takes care of these (so-called ultraviolet) infinities through 
the reexamination of 
the meaning of the Lagrangian parameters in Eq. \ref{parameters}.
We will illustrate the process on the electric charge $e$. In an effort to get
to the core of the one-loop renormalization of the SM, we present at times
simplified versions of the SM formulae. The reader is made aware of the
simplifications in a series of footnotes.

\subsection{Electric charge renormalization}
The
piece of ${\cal L}_{EW}$ defining the electric charge is the interaction term
of the QED Lagrangian
\begin{eqnarray}
{\cal L}_{em} & = & e\:\overline{l}\gamma_{\mu} l A^{\mu}.
\end{eqnarray}
We identify $\alpha = e^{2}/4\pi$ with its physical value $1/137$
\footnote{In this section we enforce $\alpha = 1/137.036$ using notation 
$\alpha_{137}, e_{137}$}
measured
in low-energy (Thomson limit $k^{2} = (p-q)^{2} \rightarrow 0$, see diagram
below) 
electron scattering.
%\begin{center}
%\setlength{\unitlength}{1in}
%\begin{picture}(6,1.2)
%\put(2.25,+0.125){\mbox{\epsfxsize=1.3in\epsffile{ren3.eps}}}
%\end{picture}
%\end{center}
The photon - charged lepton vertex corresponding to ${\cal L}_{em}$ is given by
\begin{center}
\setlength{\unitlength}{1in}
\begin{picture}(6,1.2)
\put(1.25,+0.125){\mbox{\epsfxsize=1.3in\epsffile{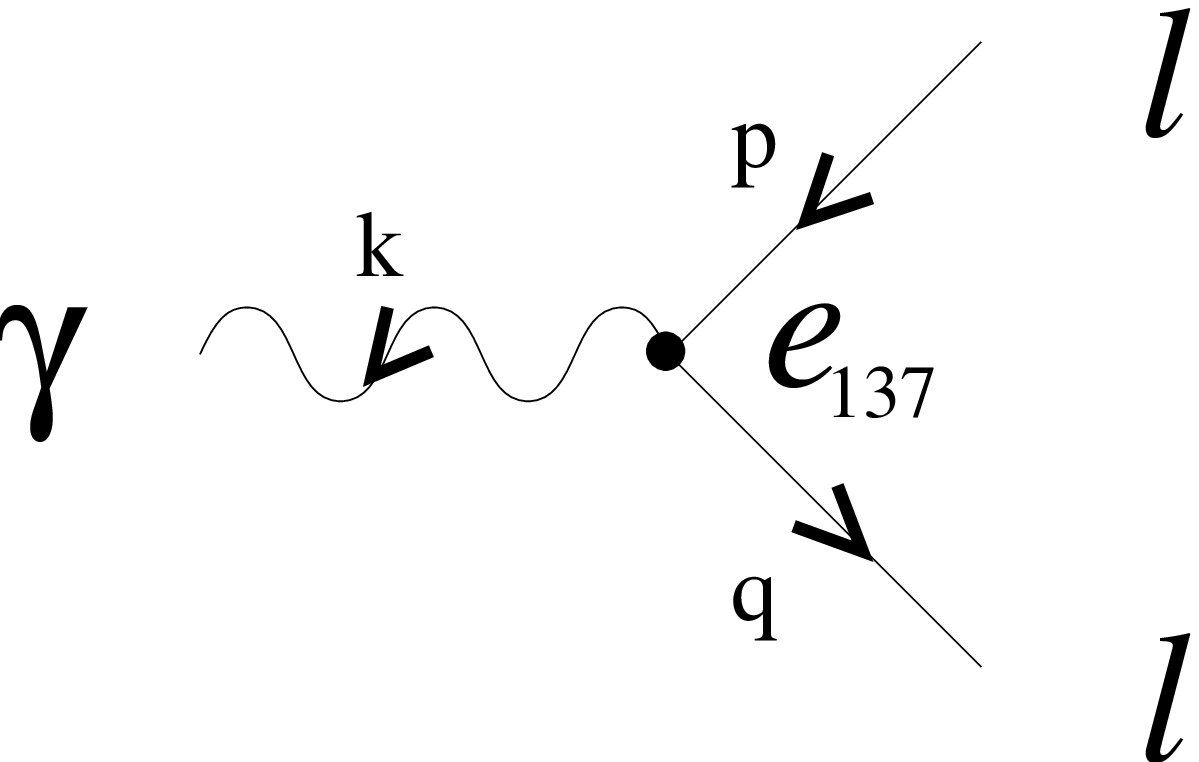}}}
\put(3.5,0.6){$\equiv \Gamma^{0}= i e_{137} \gamma_{\mu}.$}
\end{picture}
\end{center}
One-loop corrections to this vertex, calculated in terms 
of $e_{137}$, are
\footnote{Strictly speaking, lepton self-energies, $\gamma$ self-energy and
$\gamma - Z$ mixing also 
contribute to this vertex at the one-loop level. 
They are, however, each renormalized independently (in the on-shell scheme, for
example, they vanish in the Thomson limit) and they will not change the
essence of our arguments.}
\begin{center}
\setlength{\unitlength}{1in}
\begin{picture}(6,1.5)
\put(.1,+0.125){\mbox{\epsfxsize=3.0in\epsffile{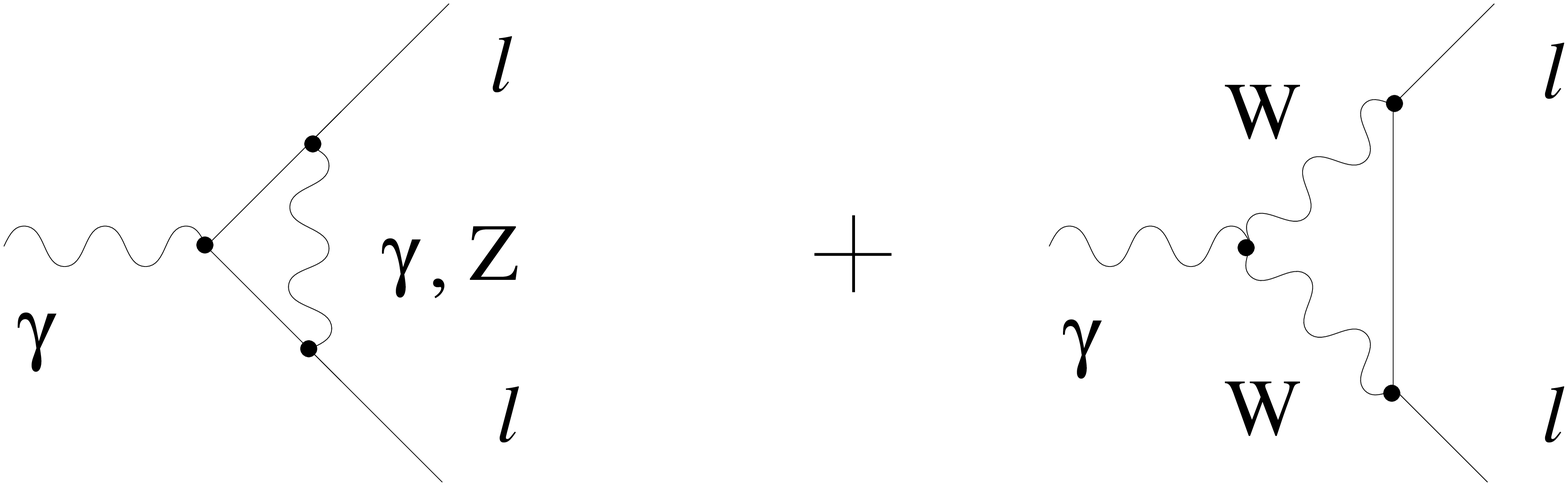}}}
\put(3.0,0.6){$\equiv \Gamma^{1}= - i e_{137} \gamma_{\mu}\left[\Lambda_{V}(0) +
F_{V}(k^{2})\right] + ...,$}
\end{picture}
\end{center}
where ellipses ... represent terms with  Lorentz structure 
different from $\gamma_{\mu}$
\footnote{$\gamma_{\mu}\gamma_{5}$ and $(p+q)_{\mu}$, renormalized
independently},
$F_{V}$ is the form factor split off so that $F_{V}(0) = 0$
and $\Lambda_{V}(0)$ is given by
\footnote{This form is exact only for the diagram with $\gamma$ in the loop,
the other two diagrams have the divergence multiplied by some combinations of
$s_{W}^{2}$. There is also an infrared divergence present in the former diagram
that we do not deal with here.}
\begin{eqnarray}
\label{lambda}
\Lambda_{V}(0) & = & -\frac{\alpha_{137}}{4\pi}\left(
\frac{2}{\epsilon} + finite \;\;constants\right),
\end{eqnarray}                                                 
where $2/\epsilon$ is the ultraviolet divergence ($\epsilon \rightarrow 0$)
regularized by dimensional regularization (see Appendix \ref{Decko} on 
dimensional regularization). 
The one-loop corrected vertex $\Gamma$ is thus given by
\begin{eqnarray}
\label{gamma}
\Gamma  =  \Gamma^{0} + \Gamma^{1} & = & i e_{137}
\gamma_{\mu}\left[1 - \Lambda_{V}(0) -
F_{V}(k^{2})\right] + ...,
\end{eqnarray}
At low energies (Thomson limit $k^{2}\rightarrow 0$)
\begin{eqnarray}
\label{thomson}
\Gamma(k^{2}\rightarrow 0) & = & i e_{137}
\gamma_{\mu}\left[1 - \Lambda_{V}(0)\right] + ...,
\end{eqnarray}
hence the charge is changed from its tree-level value in $\Gamma^{0}$ 
by an infinite amount:
\begin{eqnarray}
\label{charge}
e_{137} & \rightarrow & e_{137}\left[1 - \Lambda_{V}(0)\right].
\end{eqnarray}
                  
To explain this difficulty, we note the quantity measured by the experiment
as $e_{137}$ is a {\it loop corrected} charge.
But the quantity on
the right-hand side of Eq.~\ref{charge} {\it is} a loop (one-loop) corrected
charge, 
therefore the quantity multiplying the $1 - \Lambda_{V}(0)$ factor, denoted
$e_{137}$, cannot be
$e_{137}$. This implies the charge in the electromagnetic Lagrangian cannot be
identified with its physical value.
The correct approach is to admit the independent parameters 
appearing in
${\cal L}_{EW}$ are in fact 'bare', unrenormalized quantities
\begin{eqnarray}
e^{b},\;M_{W}^{b},\;M_{Z}^{b},\;M_{H}^{b},\;m_{f}^{b},
\end{eqnarray}
different from the physical values.
The vertex $\Gamma$ is then given by (compare with Eq.~\ref{gamma})
\begin{eqnarray}
\label{gamabare}
\Gamma & = & \Gamma^{0} + \Gamma^{1} \; = \; i e^{b}
\gamma_{\mu}\left[1 - \Lambda_{V}(0) -
F_{V}(k^{2})\right] + ...  \nonumber \\
& = & i e^{b}
\gamma_{\mu}\left[1  + \frac{\alpha^{b}}{4\pi}\left(
\frac{2}{\epsilon} + finite \;\;constants\right) -  
F_{V}(k^{2})\right] + ... \;.
\end{eqnarray}
Bare parameters are unambiguously fixed by the requirement that they
lead to correct physical values. For electric charge we demand that (in view of
the discussion above)
\begin{eqnarray}
\label{thomsonbare}
\Gamma(k^{2}\rightarrow 0)  = i e^{b}
\gamma_{\mu}\left[1 - \Lambda_{V}(0) \right] + ... & = & ie_{137}\gamma_{\mu},
\end{eqnarray}
or equivalently
\begin{eqnarray}
\label{chargebare}
e^{b}\left[1-\Lambda_{V}(0)\right] & = & e_{137}.
\end{eqnarray}
From here (after plugging in $\Lambda_{V}(0)$ calculated in terms of
$\alpha^{b}$) we can solve for $e^{b}$,
\footnote{We work to order $O(e_{137}\alpha_{137})$ in Sec. \ref{renor}. Higher
order terms are neglected.}
\begin{eqnarray}
\label{eb}
e^{b} & = & e_{137}\left[1-\frac{\alpha_{137}}{4\pi}\left(
\frac{2}{\epsilon} + finite\;\; constants\right)\right].
\end{eqnarray}
Using Eqs. \ref{chargebare}, \ref{eb}, the Eq. \ref{gamabare} can be written as
\begin{eqnarray}
\Gamma & = & i e_{137}\gamma_{\mu}\left[1 - F_{V}(k^{2})\right] + ... ,
\end{eqnarray}
where $F_{V}(k^{2})$ is of the order $O(\alpha_{137})$.
The infinity is thus removed from the vertex $\Gamma$
, i.e., the electromagnetic vertex (and the charge) is renormalized.

\subsection{Renormalization schemes}

The loop calculations are rarely carried out in terms of bare
parameters.
A widely used technique is to split the bare charge $e^{b}$, the bare fermion 
mass $m_{f}^{b}$ and the bare boson mass $M^{b}$ as
\begin{eqnarray}
\label{split}
e^{b} & = & \hat{e} + \delta e,  \nonumber \\
m_{f}^{b} & = & \hat{m_{f}} + \delta m_{f},  \nonumber          \\
M_{b}^{2} & = & \hat{M^{2}} + \delta M^{2},  
\end{eqnarray}
where $\hat{e}, \hat{m_{f}}, \hat{M}$ are renormalized (finite) charge, fermion,
and gauge boson masses
and $\delta e, \delta m_{f}, \delta M^{2}$ are infinite corrections, so-called 
counterterms.
This split introduces a
degree of freedom, as there is no unique way to perform it. Renormalized
charge and mass can take on different finite values including the physical ones.
This freedom leads in practice to many different ways of splitting the bare
parameters, i.e., to many different renormalization schemes (RS).

The difference between two renormalized charges coming from two different
RS is small, of the order $O(\alpha_{137})$, as every renormalized charge is
chosen to be equal to $e_{137}$ in the lowest order:
\begin{eqnarray}
\hat{e} & = & e_{137}\left[1 + O(\alpha_{137})\right].
\end{eqnarray}
With the substitution of Eq. \ref{split}, the ${\cal L}_{em}$ becomes
\begin{eqnarray}
{\cal L}_{em} = e^{b}\;\overline{l}\gamma_{\mu} l A^{\mu}  
 & = & {\hat e}\;\overline{l}\gamma_{\mu} l A^{\mu}
 +  \delta e\;\overline{l}\gamma_{\mu} l A^{\mu},
\end{eqnarray}
where the second term is called the counterterm Lagrangian.
The calculation of the vertex $\Gamma$ now leads to
\footnote{Note $\delta e/{\hat e}$ is of the order $O(\hat{\alpha}) =
O(\alpha_{137})$.}
\footnote{From now on we will use $\Gamma = i {\hat e}
\gamma_{\mu}\left[1 - \Lambda_{V}(0) -
F_{V}(k^{2})\right] $ for the unrenormalized vertex, and 
${\hat \Gamma}$ for the renormalized vertex,
${\hat \Gamma} = \Gamma +$ counterterm. The counterterm contains besides
$\delta e$ also wave function renormalization factors, not considered until
Sec. \ref{frenor}.}
\begin{eqnarray}
\label{gammacounter}
{\hat \Gamma}  =  \Gamma^{0} + \Gamma^{1} & = & i {\hat e}
\gamma_{\mu}\left[1 - \Lambda_{V}(0) + \delta e/{\hat e} -
F_{V}(k^{2})\right] + ...\;.
\end{eqnarray}
The conditions of Eqs. \ref{thomsonbare}, \ref{chargebare}  are now given by
\begin{eqnarray}
\label{thomsoncounter}
{\hat \Gamma}(k^{2}\rightarrow 0)  = i {\hat e}
\gamma_{\mu}\left[1 - \Lambda_{V}(0) + \delta e/{\hat e} \right] + ...
 & = & i e_{137} \gamma_{\mu},  \\
\label{chargecounter}
{\hat e}\left[1-\Lambda_{V}(0)+\delta e/{\hat e} \right] & = & e_{137}.
\end{eqnarray}
At this point we can illustrate  two different approaches to the
choice of renormalization scheme. 

{\bf i)} If we prefer to use some particular value of
${\hat e}$ in the calculation, say 
\begin{eqnarray}
{\hat e} = e_{137}\left[1 + b\;\alpha_{137}\right],
\end{eqnarray}
the counterterm
$\delta Z \equiv \delta e/{\hat e} $ is 
consequently fixed by
(see Eq. \ref{chargecounter})
\begin{eqnarray}
\delta Z \equiv \delta e/{\hat e} & = & \frac{e_{137}}{{\hat e}} + 
\Lambda_{V}(0)- 1  .
\end{eqnarray}
In fact, the most popular scheme in electroweak calculations is an on-shell
scheme (OS) where
\begin{eqnarray}
\label{OS}
 {\hat e} \equiv {\hat e}^{OS} & = & e_{137}\;\;\;\;\;(b=0), \nonumber \\
 \delta Z^{OS} \equiv \delta e/{\hat e}^{OS} & = & \Lambda_{V}(0)\; = \;
 -\frac{\alpha_{137}}{4\pi}\left(
\frac{2}{\epsilon} + finite \;\;constants\right),
\end{eqnarray}
and all masses 
assume their physical,
on-shell values  (${\hat M_{Z}}^{OS} = M_{Z} = 91.1884$ GeV,
 etc.).
The Eq. \ref{thomsoncounter} with ${\hat e} = e_{137}$,
\begin{eqnarray}
{\hat \Gamma}(k^{2}\rightarrow 0) & = & i e_{137} \gamma_{\mu},
\end{eqnarray}
is called an on-shell renormalization condition.

{\bf ii)} A different approach 
is to start with fixing the counterterm $\delta e$
instead of ${\hat
e}$. For instance we may require
that $\delta e$ ($\delta m_{f}, \delta M^{2}$ likewise) only contain infinities
(no finite terms)
\footnote{Compare this with OS where $\delta e/\hat{e}^{OS} = \Lambda_{V}(0)$ 
contains both infinite and finite terms.}.
This is the essence of minimal subtraction (MS) scheme. One chooses
\begin{eqnarray}
\label{MS}
\delta Z^{MS} \equiv \delta e/{\hat e}^{MS}
& = &
-\frac{\hat{\alpha}^{MS}}{4\pi}\frac{2}{\epsilon} \; \doteq \;
-\frac{\alpha_{137}}{4\pi}\frac{2}{\epsilon}, 
\end{eqnarray}
and the charge is consequently given by
\begin{eqnarray}
{\hat e}\equiv {\hat e}^{MS} & = & \frac{e_{137}}{[1-\Lambda_{V}(0)+
\delta Z^{MS}]} \;\doteq\; 
\frac{e_{137}}{1+\frac{\alpha_{137}}{4\pi}\times finite
\;\;\;terms},
\end{eqnarray}
hence the electric charge ${\hat e}^{MS}$ differs from $e_{137}$
and likewise masses $\hat{m_{f}}^{MS}, 
\hat{M}^{MS}$ do not assume their on-shell values. The MS scheme is
frequently used in quantum chromodynamics where on-shell  quark masses are
not well defined anyway.

%Another way to look at Eq. \ref{chargecounter} is to substitute for ${\hat e}$
% as a function
%of $e_{137}$, say ${\hat e} = b e_{137}$. After dividing both sides by $b$ we
%get
%\begin{eqnarray}
%\label{anothercharge}
%e_{137}\left[1 - \Lambda_{V}(0) + \delta e/{\hat e}\right]
% & = & \frac{e_{137}}{b}.
%\end{eqnarray}
%This can be interpreted in the following way. The loop calculation is done in
%terms of $e_{137}$ and the former degree of freedom (between ${\hat e}$ and
%$\delta e$) is now between $\delta e$ and the righthand side (RHS) of Eq.
%\ref{anothercharge}.
%Different schemes can be chosen by fixing the appropriate renormalization
%conditions, i.e. by fixing the RHS of the Eq. \ref{anothercharge}, and
%calculating
%$\delta e$ in turn. For OS we thus have $e_{137}$ on the RHS of the equation.

\subsection{Mass renormalization}

The analysis performed above for the electric charge is essentially valid also
for masses. The difference is in the form of renormalization condition.
Masses can be defined as the poles of the propagators. For instance, gauge
boson propagators $V = W,Z$ have poles at bare mass ${M_{V}^{b}}^{2}$:
% (factor $g_{\mu\nu}$ omitted):
\begin{center}
\setlength{\unitlength}{1in}
\begin{picture}(6,1.2)
\put(0.7,+0.6){\mbox{\epsfxsize=1.3in\epsffile{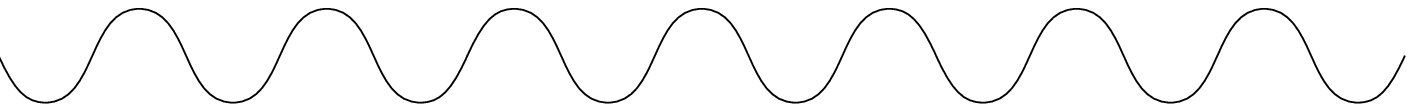}}}
\put(2.9,0.6){$\equiv P^{0}= \frac{-ig^{\mu\nu}}{k^{2}-{M_{V}^{b}}^{2}
+i\epsilon}.$}
\end{picture}
\end{center}
The one-loop correction to $P^{0}$ is given by
\begin{center}
\setlength{\unitlength}{1in}
\begin{picture}(6,1.2)
\put(0.7,+0.3){\mbox{\epsfxsize=1.3in\epsffile{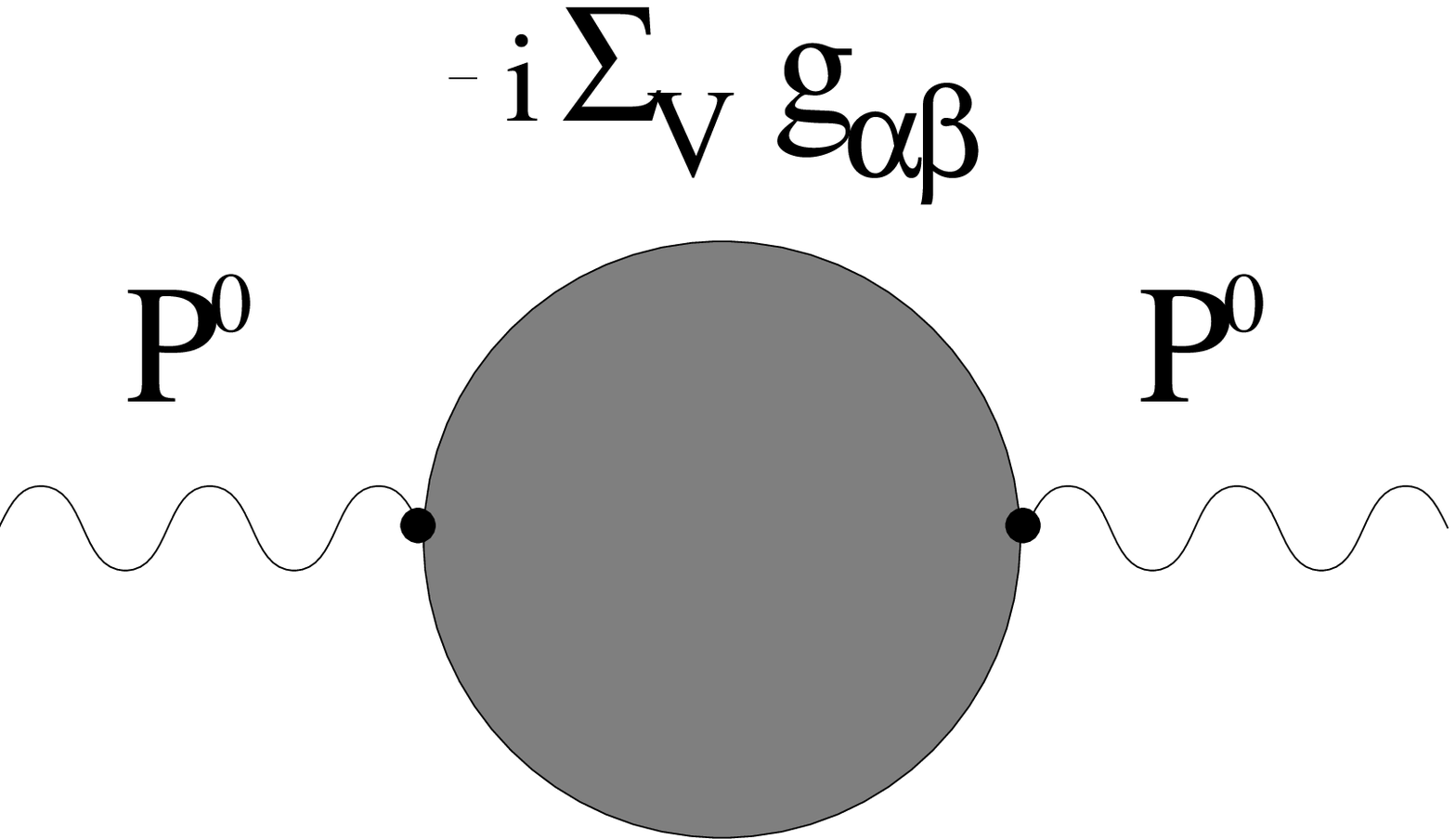}}}
\put(2.9,0.6){$\equiv P^{1}= \frac{-ig^{\mu\alpha}}{k^{2}-{M_{V}^{b}}^{2}
+i\epsilon}
(-i\:\Sigma_{V}\:g_{\alpha \beta})\frac{-ig^{\beta\nu}}
{k^{2}-{M_{V}^{b}}^{2}+i\epsilon}.$}
\end{picture}
\end{center}
For a close-up of the blob $-i\:\Sigma_{V}g_{\alpha \beta}$ 
(unrenormalized vector boson self-energy tensor), 
see Figs. \ref{zfd}, \ref{wfd} and the relevant discussion in Chapter 6.
The one-loop corrected renormalized
\footnote{Here, we mean renormalized as far as mass is concerned. 
There is still one
divergence remaining in $\Sigma_{V}$ which will be removed only after the field
renormalization, see Sec. \ref{frenor}. Therefore we will withhold the 
notation ${\hat P}$ until then.}
propagator is thus (compare with Eq. \ref{gammacounter})
\begin{eqnarray}
\label{propagator}
P = P^{0} + P^{1} & = & \frac{-ig^{\mu\nu}}{k^{2}-{M_{V}^{b}}^{2}+i\epsilon}
  \left[1 - \frac{\Sigma_{V}}{k^{2}-{M_{V}^{b}}^{2}+i\epsilon}\right]
\nonumber \\
& \cong & 
\frac{-ig^{\mu\nu}}{k^{2}-{M_{V}^{b}}^{2}+i\epsilon}
\frac{1}{\left(1 + \frac{\Sigma_{V}}{k^{2}-{M_{V}^{b}}^{2}+i\epsilon}\right)} 
\; = \;
\frac{-ig^{\mu\nu}}{k^{2}-{M_{V}^{b}}^{2}+\Sigma_{V}(k^{2})+i\epsilon}
\nonumber \\
& = & \frac{-ig^{\mu\nu}}{k^{2}-{\hat M}_{V}^{2}-\delta M_{V}^{2}
  + \Sigma_{V}(k^{2})+i\epsilon}.
\end{eqnarray}
We demand that the poles (masses) of renormalized propagators remain at
their physical values regardless of the choice of ${\hat M}_{V}$ or $\delta
M_{V}^{2}$ (compare with Eq. \ref{thomsoncounter}):
\begin{eqnarray}
P(k^{2} \rightarrow M_{V}^{2}) \;=\;
    \frac{-ig^{\mu\nu}}{k^{2}-{\hat M}_{V}^{2}-\delta M_{V}^{2}
  + \Sigma_{V}(M_{V}^{2})+i\epsilon}
& = & \frac{-ig^{\mu\nu}}{k^{2}-{\hat M}_{V}^{{OS}^{2}}+i\epsilon}.\;\;\;\;
\end{eqnarray}
To put the mass on shell we have to take
\begin{eqnarray}
\label{os}
{\hat M}_{V} = {\hat M}_{V}^{OS} \; \equiv \; M_{V},
\end{eqnarray}
or equivalently
\begin{eqnarray}
\label{orequiv}
\Sigma_{V}(M_{V}^{2}) - \delta M_{V}^{2} = 0.
\end{eqnarray}
              
\subsection{Field renormalization}
\label{frenor}

For physical S-matrix elements, the renormalization of the five parameters in 
Eq.~\ref{parameters}
is all that is required. However, if one wishes to also have finite Green
functions, then the renormalization of the fields is also required
(see, e.g., Ref. \cite{key6}).

While the particle masses are given by the poles of the propagators, the
normalization of the fields is given by the residues of the propagators. For
gauge boson fields, for example, we have:
\begin{eqnarray}
\frac{-i g^{\mu\nu}}{k^{2}-M_{V}^{2}+i\epsilon}
\end{eqnarray}
with the residue equal to one (ignoring $-i g^{\mu\nu}$).
The residue (field normalization) is changed by the loop corrections.
To show that, we expand $\Sigma_{V}(k^{2})$ about $k^{2} = M_{V}^{2}$:
\begin{eqnarray}
\Sigma_{V}(k^{2}) & = & \Sigma_{V}(M_{V}^{2}) +
(k^{2}-M_{V}^{2})\Sigma_{V}^{'}(M_{V}^{2}) + ...,  \nonumber \\
& & \Sigma_{V}^{'} \equiv \partial \Sigma_{V}/ \partial k^{2},
\end{eqnarray}
and substitute it into the propagator of Eq. \ref{propagator}:
\begin{eqnarray}
P & = & \frac{-i g^{\mu\nu}}{k^{2}-{\hat M}_{V}^{2}-\delta M_{V}^{2}
  + \Sigma_{V}(M_{V}^{2})+(k^{2}-M_{V}^{2})\Sigma_{V}^{'}(M_{V}^{2}) +
... + i\epsilon}.
\end{eqnarray}
Applying the on-shell condition Eqs. \ref{os}, \ref{orequiv} we get
\begin{eqnarray}
P & = & \frac{-i g^{\mu\nu}}{k^{2}-M_{V}^{2}
  +(k^{2}-M_{V}^{2})\Sigma_{V}^{'}(M_{V}^{2}) +
... + i\epsilon}    \nonumber \\
& \cong & 
\frac{-i g^{\mu\nu}}{(k^{2}-M_{V}^{2}+ i\epsilon)}
\frac{1}{[1+\Sigma_{V}^{'}(M_{V}^{2}) + ...]}.
\end{eqnarray}
with the (divergent) residue $1/[1+\Sigma_{V}^{'}(M_{V}^{2})]$ at $k^{2}
\rightarrow M_{V}^{2}$.
The problem is fixed by the field counterterms generated by the substitution
\begin{eqnarray}
V_{\mu} & \rightarrow & Z_{V}^{1/2}V_{\mu} \;=\; (1+\delta Z_{V})^{1/2}V_{\mu}
\; \doteq \;
(1+\frac{1}{2} \delta Z_{V})V_{\mu}.
\end{eqnarray}
The field counterterm $ \delta Z_{V}$ modifies the propagator as follows
\footnote{
\begin{eqnarray}
M_{V}^{2} V_{\mu} V^{\mu} & \rightarrow & (1 + \delta Z_{V})
M_{V}^{2} V_{\mu} V^{\mu} , \nonumber \\
(\Box + M_{V}^{2})V_{\mu} & \rightarrow & 
(1 + \delta Z_{V}) (\Box + M_{V}^{2})V_{\mu}. \nonumber 
\end{eqnarray}}
\begin{eqnarray}
\label{hatp1}
{\hat P} \;=\; \frac{1}{1+\delta Z_{V}}P \;=\;  \frac{1}{(1+\delta Z_{V})}
\frac{-i g^{\mu\nu}}{(k^{2}-M_{V}^{2}+ i\epsilon)}
\frac{1}{[1+\Sigma_{V}^{'}(M_{V}^{2})]}.
\end{eqnarray}
To renormalize the fields (enforce their normalization) we demand that the 
residues of the field propagators
be equal to one at the poles. This implies the following condition
(higher order term neglected):
\begin{eqnarray}
\label{os2}
\Sigma_{V}^{'}(M_{V}^{2})+\delta Z_{V} & = & 0.
\end{eqnarray}
It is easy to see from Eqs. \ref{propagator}, \ref{hatp1}    
that before being put on shell, ${\hat P}$ in terms of the renormalized self 
energy ${\hat \Sigma}_{V}$ is given by
\begin{eqnarray}
\label{hatp}
{\hat P} & = & \frac{-i g^{\mu\nu}}{k^{2}-\hat{M}_{V}^{2}
+{\hat \Sigma}_{V}(k^{2})+i\epsilon},
\\
\label{hatsigma}
{\hat \Sigma}_{V}(k^{2}) & = & \Sigma_{V}(k^{2})-\delta M_{V}^{2}+
    \delta Z_{V}(k^{2}-\hat{M}_{V}^{2}),
\end{eqnarray}
so that the on-shell renormalization conditions Eqs. \ref{orequiv}, \ref{os2} 
become
\begin{eqnarray}
\label{sigmam}
{\hat \Sigma}_{V}(M_{V}^{2}) & = & 0, \\
\label{parsigmam}
\frac{\partial {\hat \Sigma}_{V}(M_{V}^{2})}{\partial k^{2}} & = & 0.
\end{eqnarray}

Before we go further, one remark is in order.
So far we have been discussing a simplified renormalization of some parameters
at the one-loop
level. In the next section we will stay at one-loop level, however, we will
present the full set of counterterms and OS renormalization conditions required
for the processes studied in this thesis.

To prove the renormalizability of the SM, it has to be shown that the infinities
one encounters in 
loop calculations to any order can be removed by the finite number of 
counterterms. 
%generated
%by the split of the
%bare parameters (as described in this section).
% One can envisage a divergence
%with a Lorentz structure which
%has no counterpart in the counterterm Lagrangian and hence cannot be subtracted
%off.
% In that case we would be forced to introduce another parameter to the
%theory which would take care of the problem. This would however weaken the
%predictive power of the model. 
%To prove renormalizability of the electroweak theory one has to show that
%divergences in
%any order will be removed by the finite number of parameters (counterterms).
This was done by 't Hooft in Refs. \cite{thooftsmren,thooftdim} for a general
case of non-abelian theories with spontaneous symmetry breaking.

\section{The on-shell scheme of W. Hollik}
\label{onshell1}

There are many renormalization schemes used in the calculation of loop
corrections by different authors.
They are distinguished in the first place by the choice of independent input
parameters. The choice $e,\;M_{W},\;M_{Z},\;M_{H},\;m_{f}$ that we
are using is only one of several possible. 
Given the set of input parameters,
there are still infinitely many possibilities for choosing the renormalized
quantities ${\hat e}$, ${\hat m}$. The OS scheme is the most popular and
natural in the
standard model of electroweak interactions. 

Even then, within the OS itself, there are many
different approaches to renormalization. For instance, some opt for field
renormalization, others do not and those who do, do it with different numbers of
field renormalization constants.

  In this thesis we follow the OS scheme
($e,M_{W},M_{Z},M_{H},m_{f}$) of Wolfgang Hollik \cite{key6,key9}.
We introduce multiplicative renormalization constants for each free parameter
and each symmetry multiplet of fields 
\footnote{Multiplicative renormalization and only one constant per multiplet
guarantees the gauge invariance of the counterterm Lagrangian. To make
a connection with Eq. \ref{split}, note that
\begin{eqnarray}
{\hat e} & \rightarrow & Z\;{\hat e}\;=\;(1+\delta Z)\;{\hat e}\;=\;{\hat e}
+ \delta e. \nonumber
\end{eqnarray}}
at the level of the unbroken theory:
\begin{eqnarray}
\label{multi}
W_{\mu}^{a} & \rightarrow & \left(Z_{2}^{W}\right)^{\frac{1}{2}}W_{\mu}^{a}
\nonumber \\
B_{\mu}     & \rightarrow & \left(Z_{2}^{B}\right)^{\frac{1}{2}}B_{\mu} 
\nonumber \\
\psi_{j_{L}} & \rightarrow & \left(Z_{L}^{j}\right)^{\frac{1}{2}}\psi_{j_{L}}
\nonumber \\
\psi_{j_{R}} & \rightarrow & \left(Z_{R}^{j}\right)^{\frac{1}{2}}\psi_{j_{R}}
\nonumber \\
\Phi         & \rightarrow & \left(Z^{\Phi}\right)^{\frac{1}{2}}\Phi  
\nonumber \\
g_{2}        & \rightarrow & Z_{1}^{W}\left(Z_{2}^{W}\right)^{-\frac{3}{2}}g_{2}
\nonumber \\
g_{1}        & \rightarrow & Z_{1}^{B}\left(Z_{2}^{B}\right)^{-\frac{3}{2}}g_{1}
\nonumber \\
v            & \rightarrow & \left(Z^{\Phi}\right)^{\frac{1}{2}}(v-\delta v)
\nonumber \\
\lambda      & \rightarrow & Z^{\lambda} \left(Z^{\Phi}\right)^{-2}\lambda 
\nonumber \\
h_{j}        & \rightarrow & \left(Z^{\Phi}\right)^{-\frac{1}{2}}Z_{1}^{j}h_{j},
\end{eqnarray}
ten constants in all (counting Yukawa couplings as one). Five of them are 
associated with fields and five with coupling constants.
% It is the minimum set of renormalization
%constants required for the renormalization of the fermion processes studied in 
%this thesis.

To generate the counterterm Lagrangian $\delta {\cal L}$,
the renormalization constants are expanded as
\begin{eqnarray}
\label{deltaz}
Z = 1 + \delta Z,
\end{eqnarray}
and Eqs. \ref{multi} - \ref{deltaz} are applied to ${\cal L}_{EW}$.
The counterterms added to the unrenormalized quantities then yield the
renormalized self-energies given in Appendix~\ref{Ecko}, 
Eq. \ref{rselfe}; and
the renormalized
electromagnetic, weak neutral and charged current vertices given in Eq.
\ref{rvertexa}. These renormalized expressions can be compared with Eq.
\ref{gammacounter} and Eq. \ref{hatsigma}.
%The counterterms $\delta e$ and $\delta m$ 
%are given by the linear combinations of many $\delta Z$ parameters.

The ten independent counterterm constants are fixed by the nine 
on-shell renormalization conditions
\footnote{The condition
on $\hat{\Sigma}^{f}(k)$ in Eq. \ref{RC2} below fixes both $\delta Z_{L}$ and 
$\delta Z_{R}$ constants.}.
The first set of conditions puts the masses 
on-shell (compare with Eq. \ref{sigmam})
\footnote{Only real parts of self-energies enter these conditions. 
The imaginary parts are finite.}:
\begin{eqnarray}
\label{RC1}
\hat{\Sigma}^{W}(M_{W}^{2})\;\;\; =\;\;\; \hat{\Sigma}^{Z}(M_{Z}^{2})
\;\;\; = \;\;\; \hat{\Sigma}^{H}(M_{H}^{2})\;\;\; = \;\;\;
\hat{\Sigma}^{f}(m_{f}^{2})\;\;\; = \;\;\;0,
\end{eqnarray}
where $\hat{\Sigma}^{W}, \hat{\Sigma}^{Z}, \hat{\Sigma}^{H}$ and
$\hat{\Sigma}^{f}$ are the $W, Z,$ Higgs and fermion renormalized self-energies
respectively;
the second set of conditions is the generalization of the QED electric charge 
renormalization:
\begin{eqnarray}
\label{RC2}
\hat{\Gamma}^{\gamma ee}(k^{2} \rightarrow 0) \; \equiv \; 
\hat{\Gamma}(k^{2} \rightarrow 0) & = & i e \gamma_{\mu}
\nonumber \\
\nonumber \\
\hat{\Sigma}^{\gamma Z}(k^{2} \rightarrow 0) & = & 0
\nonumber \\
\nonumber \\
\displaystyle \left.
\left[\frac{\partial}{\partial
k^{2}}\hat{\Sigma}^{\gamma}\left(k^{2}\right)\right]
      \right|_{k^{2}=0} & = & 0
\nonumber \\
\nonumber \\
\displaystyle \left.
\frac{1}{{\not k}-m_{f}}\hat{\Sigma}^{f}(k) \right|_{{\not k}=m_{f}} & = & 0
\nonumber \\
\nonumber \\
\displaystyle \left.
\left[\frac{\partial}{\partial k^{2}}\hat{\Sigma}^{H}(k^{2})\right]
              \right|_{k^{2}=M_{H}^{2}} & = & 0,
\end{eqnarray}
where $\hat{\Sigma}^{\gamma}$ and $\hat{\Sigma}^{\gamma Z}$ are renormalized
photon self-energy and photon-Z mixing respectively.
The conditions involving $\hat{\Sigma}^{f},
\hat{\Gamma}^{\gamma ee}$ and $\hat{\Sigma}^{\gamma}$ come directly from QED.
Derivative conditions can be compared with Eq. \ref{parsigmam} derived for W and
Z bosons. 

When writing down renormalization conditions, one has to be careful not to
violate Ward (Slavnov - Taylor) identities \cite{wst}. 
These consequences of gauge symmetry 
also relate renormalization constants to one
another and can be used as a cross-check of the consistency of the
renormalization conditions. 
In the set above, for example, Ward identities make the axial part of 
$\hat{\Gamma}^{\gamma ee}$ vanish in the Thomson limit \cite{key6,key9}.

The renormalization constants calculated from Eqs. \ref{RC1}, \ref{RC2}
are given in Appendix \ref{Ecko}, Eq. \ref{rconstants}.

\newpage

\chapter{Lepton flavour-violating processes}

Among the processes with NHL's in the loops, the lepton flavour-violating
decays have so far received a lot more attention
\cite{Ng1,bernabeu1,Ng2,ggjv,Ilakovac,Jarlskog,Valle2,Korner,pilaftsis1}
than the flavour-conserving processes \cite{bernabeu2,melo}.
One of the reasons could be a certain preconception that the experimental
signature of the flavour violation is `much more dramatic'. It is our intention
to show in this and the next chapter that in many cases this expectation is
rather naive.

Another probable reason (this time justified) is that the calculation of the 
flavour-violating processes
is simpler, with the smaller number of contributing diagrams and without
having to actually renormalize. We will demonstrate this in Sec. \ref{fvzb1}
in case of the flavour-violating decays of the Z boson.
These rare processes were studied in the context of our model previously 
\cite{bernabeu1,Valle2};
however, the limit of large NHL mass was not fully investigated. This was
pointed out in Ref. \cite{Korner}, where the branching ratios for 
$Z \rightarrow l_{1}^{-} l_{2}^{+}\;
(e^{\pm}\mu^{\mp}, \mu^{\pm} \tau^{\mp}, e^{\pm} \tau^{\mp})$
were derived in the see-saw model of Ref. \cite{pilaftsis2}.
We therefore reexamine the flavour-violating leptonic decays of the Z boson
in our model, carefully treating the case of a large NHL mass. 
The diagrams are very similar to the flavour-conserving leptonic decays of the
Z boson discussed in Chapter 6 and we will borrow some results from there; our
intention here is to focus on the typical features of the flavour-violating
processes rather than on the calculational details.

In Sec. \ref{fvple} we continue with the discussion of the sensitivity of the
flavour-violating processes in general to the presence of NHL's.
We will refer to this discussion later, when in the main part of this work -
the calculation of the flavour-conserving processes in Chapters 6 and 7 -
we will confront our results with those for flavour-violating processes.

\section{Flavour-violating leptonic decays of the Z boson}
\label{fvzb1}

\begin{figure}
\begin{center}
\setlength{\unitlength}{1in}
\begin{picture}(6,3)
\put(.3,+0.125){\mbox{\epsfxsize=5.0in\epsffile{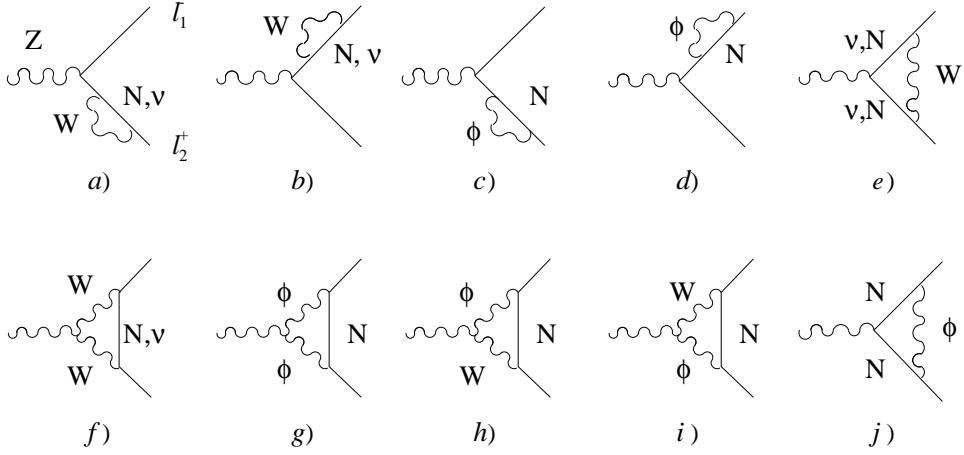}}}
\end{picture}
\end{center}
\caption{One-loop diagrams for flavour-violating leptonic decays of the Z 
boson.}
\label{odlisny}
\end{figure}

In the SM, the CKM matrix gives rise to the flavour-violating
hadronic decays of the Z boson at the one-loop level. In our model, 
by analogy, the mixing
matrix $K$ (see Sec. \ref{properties3}) induces the flavour-violating decays 
$Z \rightarrow l_{1}^{-} l_{2}^{+}$ at the one-loop level
\footnote{This feature is not exclusive to our model. Lepton flavour-violating
processes are typical for many other
nonstandard models with mixing in the lepton sector.}. 
One-loop Feynman diagrams generating these decays 
are given in Fig. \ref{odlisny}.
There is no tree-level contribution since there is no mixing between the
charged leptons in the neutral current Lagrangian.
We will be studying how these graphs contribute to the observable, the width
$\Gamma_{l_{1}^{-}l_{2}^{+}}$, in particular the dependence of the width on 
parameters from the neutrino
sector of our model - mixings and NHL masses. The analysis can be simplified
(without sacrificing the salient features) by assuming the three
NHL's are degenerate, with mass $M_{N}$.

%************** Vertices *************************************
\subsection{The amplitude and the width for $Z \rightarrow l_{1}^{-} l_{2}^{+}$}
The total amplitude ${\cal M}$ is given by the sum of partial amplitudes
corresponding to the graphs of Fig.
\ref{odlisny}a-j (subscripts in ${\cal M}_{W N} ...$ refer to  
particles in the loop)
\footnote{We work in the Feynman gauge, see Sec. \ref{quant1}}:
\begin{eqnarray}
{\cal M} & = & 
+ie\epsilon_{\mu}\gamma^{\mu}(1-\gamma_{5})\frac{\alpha}{4\pi}
\Big\{ k_{1}{\cal M}_{W N} - k_{1}{\cal M}_{W \nu} + k_{1}{\cal M}_{\phi N}
+ k_{4}{\cal M}_{\nu \nu W} \nonumber \\
& + &
k_{3}{\cal M}_{\nu NW} + k_{3}{\cal M}_{N\nu W} + k_{2}{\cal M}_{NNW}
+ k_{1}{\cal M}_{WWN} - k_{1}{\cal M}_{WW\nu} \nonumber \\
& + &
k_{1}{\cal M}_{\phi \phi N} + k_{1}{\cal M}_{\phi WN} + 
k_{2}{\cal M}_{NN \phi} \Big\},
\end{eqnarray}
where $k_{1}, k_{2}, k_{3}, k_{4}$ are mixing factors to be derived shortly
and $\epsilon_{\mu}$ is a polarization four-vector of the $Z$ boson.
Further, functions ${\cal M}_{W N}, ...$ depend on  masses and momenta 
of internal and external particles. 
${\cal M}_{WN}$ is the sum of diagrams \ref{odlisny}a, \ref{odlisny}b with $N$ 
in the loop, ${\cal M}_{W \nu}$ is the sum of \ref{odlisny}a, \ref{odlisny}b 
with $\nu$ in the loop, ${\cal M}_{\phi N}$ is the sum of diagrams 
\ref{odlisny}c, \ref{odlisny}d and
${\cal M}_{\phi W N}$ is the sum of equal contributions from 
Fig.~\ref{odlisny}h and \ref{odlisny}i.
Besides diagrams \ref{odlisny}a, \ref{odlisny}b also \ref{odlisny}f comes in 
both with massless $\nu$'s and NHL's. 
Diagram \ref{odlisny}e comes in
with four combinations of neutral lepton types.
A sample calculation of one function, ${\cal M}_{NN \phi}$, will be given in
Sec. \ref{irreduciblev}. Here we simply state results for all functions
($\frac{m_{l}^{2}}{M_{W}^{2}}$ terms neglected) :
\begin{eqnarray}
\label{amplitudes}
{\cal M}_{WN} & = & \frac{-\frac{1}{2}+s_{W}^{2}}{4 s_{W}^{3}c_{W}}
\left[\frac{1}{2}-\Delta_{\mu}
+\ln{M_{W}^{2}} +  f({\cal X})\right],
  \nonumber   \\
{\cal M}_{W \nu} & = & \frac{-\frac{1}{2}+s_{W}^{2}}{4 s_{W}^{3}c_{W}}
\left[\frac{1}{2}-\Delta_{\mu}
+\ln{M_{W}^{2}} \right], \nonumber  \\ 
{\cal M}_{\phi N} & = & \frac{-\frac{1}{2}+s_{W}^{2}}{4 s_{W}^{3}c_{W}}
\left[-\frac{1}{2}-\Delta_{\mu}
+\ln{M_{W}^{2}} +  f({\cal X})\right]
\frac{{\cal X}}{2}, \nonumber  \\
{\cal M}_{abW} & = & -\frac{1}
{8 s_{W}^{3} c_{W}}\Big\{2M_{Z}^{2}\big[(C_{23}(M_{a},M_{W},M_{b})
+ C_{11}(M_{a},M_{W},M_{b})\big] \nonumber  \\
& + & 
2-4C_{24}^{fin}(M_{a},M_{W},M_{b})-\Delta_{\mu} \Big\}, \;\;\; a,b =  N, \nu
\mbox{ ; } M_{\nu} = 0,
\nonumber \\
%&   & \mbox{where a,b run over }  N, \nu \mbox{ ; } M_{\nu} = 0.
%\nonumber  \\ 
{\cal M}_{WWa} & = & -  \frac{3 c_{W}}
{4 s_{W}^{3}}\Big\{\frac{2}{3}M_{Z}^{2}\Big[-C_{11}(M_{W},M_{a},M_{W})
-C_{23}(M_{W},M_{a},M_{W}) \nonumber \\
& - &  C_{0}(M_{W},M_{a},M_{W})\Big] 
+4C_{24}^{fin}(M_{W},M_{a},M_{W})-\frac{2}{3}+ \Delta_{\mu}\Big\}, \nonumber \\
{\cal M}_{\phi \phi N} & = & - \frac{1}
{2 s_{W}^{3}} \frac{1-2s_{W}^{2}}{2c_{W}}{\cal X}
\Big[C_{24}^{fin}(M_{W},M_{N},M_{W}) + \frac{1}{4}\Delta_{\mu}\Big],
 \nonumber \\
{\cal M}_{\phi WN} & = & +\frac{ M_{W}^{2}}
{2 s_{W}c_{W}}{\cal X}C_{0}(M_{W},M_{N},M_{W}), \nonumber \\
{\cal M}_{NN \phi} & = & +
\frac{ M_{W}^{2}}{8 s_{W}^{3} c_{W}}{\cal X}^{2}
C_{0}(M_{N},M_{W},M_{N}), 
\end{eqnarray}
where
\begin{eqnarray}
\label{calex}
{\cal X} & \equiv & \frac{M_{N}^{2}}{M_{W}^{2}}, 
\;\;\;\;\;\;\;\;s_{W}\;\equiv\;\sin \theta_{W},
\;\;\;\;\;\;\;\;c_{W}\;\equiv\;\cos \theta_{W}, \nonumber \\
\Delta_{\mu}  & = & \frac{2}{\epsilon}-\gamma+\ln 4\pi + \ln \mu^{2},
 \nonumber \\
%\Delta^{m} & = & \frac{2}{\epsilon }-\gamma +\ln 4\pi -
%\ln\frac{m^{2}}{\mu^{2}}, \nonumber \\
f({\cal X}) & = & \frac{{\cal X}^{2}\log {\cal X}}{({\cal X}-1)^{2}} + 
      \frac{{\cal X}}{1-{\cal X}}.
\end{eqnarray}
Our results are written in terms of the 't Hooft-Veltman integrals 
$C_{0}, C_{24}, C_{11}$, $C_{23}$~\cite{thooft}
defined  in the Appendix \ref{scalari}. 
They contain both finite and infinite parts
regularized by dimensional regularization \cite{thooftdim} (see
Appendix \ref{Decko}). Infinite parts are
parametrized by $\Delta_{\mu}$, where $\epsilon \rightarrow 0$ is not to be
confused with the polarization four-vector $\epsilon_{\mu}$.

We now evaluate the mixing parameters.
The parameter $k_{1}$ comes from diagrams with one NHL in the loop,
$k_{2}$ comes from diagrams with two NHL's, $k_{3}$ from diagrams with
one NHL and one massless neutrino and $k_{4}$ from the graph with two 
massless neutrinos. The case of one massless neutrino is trivial (-- $k_{1}$)
and is not shown below.
Starting with terms that come directly from the Feynman rules of Appendix
\ref{Cecko},
we work our way through to the final form using the properties
of the mixing matrix $K$ from Sec. \ref{properties3} (if not explicitly shown,
repeated indices are summed over):
\begin{eqnarray}
\label{fvmixings}
k_{1} & = & \big(K_{H}\big)_{la}\big(K_{H}^{\dagger}\big)_{al^{'}}
%\;=\; \sum_{a}\big(K_{H}\big)_{la}\big(K_{H}^{\dagger}\big)_{al^{'}} 
\;=\;
ll^{'}_{mix}, \nonumber  \\
k_{2} & = & \big(K_{H}\big)_{la}\big(K_{H}^{\dagger}K_{H}\big)_{ab}
            \big(K_{H}^{\dagger}\big)_{bl^{'}} \;=\;
\sum_{m=e,\mu,\tau}\big(K_{H}\big)_{la}
              \big(K_{H}^{\dagger}\big)_{am}
              \big(K_{H}\big)_{mb}
              \big(K_{H}^{\dagger}\big)_{bl^{'}}
\nonumber \\
      & = &   \sum_{m=e,\mu,\tau}lm_{mix}ml^{'}_{mix}, \nonumber \\
k_{3} & = & \big(K_{L}\big)_{li}\big(K_{L}^{\dagger}K_{H}\big)_{ia}
            \big(K_{H}^{\dagger}\big)_{al^{'}} \;=\;
\sum_{m=e,\mu,\tau}\big(K_{L}\big)_{li}
              \big(K_{L}^{\dagger}\big)_{im}
              \big(K_{H}\big)_{ma}
              \big(K_{H}^{\dagger}\big)_{al^{'}}
\nonumber  \\
      & = &   \sum_{m}\left[\delta_{lm} - \big(K_{H}\big)_{lb}
              \big(K_{H}^{\dagger}\big)_{bm}\right]\big(K_{H}\big)_{ma}
              \big(K_{H}^{\dagger}\big)_{al^{'}} \;=\;
              ll^{'}_{mix} - \sum_{m} lm_{mix} ml^{'}_{mix} \nonumber \\
      & = &   k_{1}-k_{2}, \nonumber \\
k_{4} & = &   \big(K_{L}\big)_{li}\big(K_{L}^{\dagger}K_{L}\big)_{ij}
            \big(K_{L}^{\dagger}\big)_{jl^{'}} \;=\;
             - 2k_{1} + k_{2}.
\end{eqnarray}
For $k_{4}$ we show only the initial and final step.

To address the question of infinities, we note that we do not have to actually
renormalize. Indeed, we easily observe
the mass independent divergences (in fact any terms independent of mass)
are cancelled in the sums
\begin{eqnarray}
{\cal M}_{W W \nu} + {\cal M}_{W W N},  \nonumber         \\
{\cal M}_{N \nu W} + {\cal M}_{\nu N W} + {\cal M}_{N N W} +
{\cal M}_{\nu \nu W}, \nonumber \\
{\cal M}_{W \nu} + {\cal M}_{W N}. 
\end{eqnarray}
The origin of this, so-called GIM cancellation can be
traced back to the unitarity of the mixing matrix $K$
\footnote{This cancellation is referred to as the GIM cancellation
since it has a similar origin as the cancellations due to the CKM matrix in
$K^{0} \rightarrow \mu^{+} \mu^{-}$ which lead to the postulation of the $c$
quark by Glashow, Iliopoulos and Maiani (GIM) \cite{key4}.}.
The remaining divergent amplitudes have their divergence multiplied by the mass
term ${\cal X}$ and therefore GIM cancellation does not apply here. However,
these divergences vanish in the sum of mass dependent diagrams, 
\begin{eqnarray}
{\cal M}_{\phi \phi N} + {\cal M}_{\phi N}.
\end{eqnarray}

Using Eq. \ref{fvmixings} it can be shown that the width for the 
flavour-violating decays of the 
Z boson to $l_{1}^{-}l_{2}^{+}$ is given in terms of $k_{1}$ and $k_{2}$ as
\begin{eqnarray}
\Gamma_{l_{1}^{-}l_{2}^{+}} 
& = &
 \frac{2}{3} \frac{\alpha^{3}}{(4\pi)^{2}} M_{Z}  
   |k_{1}{\cal M}_{1}+k_{2}{\cal M}_{2}|^{2},
\end{eqnarray}
where
\begin{eqnarray}
{\cal M}_{1} & = & {\cal M}_{\phi WN} + {\cal M}_{\phi \phi N}
 - {\cal M}_{WW\nu} + {\cal M}_{WWN} + {\cal M}_{N \nu W} + {\cal M}_{\nu N W}
 - 2 {\cal M}_{\nu \nu W}       \nonumber    \\
& - & {\cal M}_{W \nu} + {\cal M}_{\phi N} + {\cal M}_{WN}, \nonumber \\
{\cal M}_{2} & = & {\cal M}_{N N \phi} - {\cal M}_{N \nu W} 
 - {\cal M}_{\nu N W} + {\cal M}_{\nu \nu W} + {\cal M}_{N N W}.
\end{eqnarray}
The amplitude squared can be written as
\begin{eqnarray}
|k_{1}{\cal M}_{1}+k_{2}{\cal M}_{2}|^{2} & = & |k_{1}|^{2}|{\cal M}_{1}|^{2}
 + |k_{2}|^{2}|{\cal M}_{2}|^{2} + 2 Re\left(k_{1}k_{2}^{*}{\cal M}_{1}{\cal
M}_{2}^{*}\right).
\end{eqnarray}
The mixing factors $k_{1}, k_{2}$ are process dependent and the following
relations hold between CP conjugate final states:
\begin{eqnarray}
k_{1,2}\;\;\;\equiv\;\;\; k_{1,2}(l_{1}^{-}l_{2}^{+}) 
& = & k_{1,2}^{*}(l_{1}^{+}l_{2}^{-}),
\end{eqnarray}                     
implying
\footnote{Note the difference
\begin{eqnarray}
Re\left\{k_{1}(l_{1}^{-}l_{2}^{+}) \; k_{2}^{*}(l_{1}^{-}l_{2}^{+})
 \; {\cal M}_{1}{\cal M}_{2}^{*}\right\} & - &
Re\left\{k_{1}(l_{1}^{+}l_{2}^{-}) \; k_{2}^{*}(l_{1}^{+}l_{2}^{-})
 \; {\cal M}_{1}{\cal M}_{2}^{*}\right\} \nonumber \\
& = & 
-2 Im\left( k_{1}k_{2}^{*}\right)Im \left({\cal M}_{1}{\cal
M}_{2}^{*}\right) \nonumber
\end{eqnarray}
may lead to a CP violating asymmetry
\begin{eqnarray}
\eta \equiv \frac{\Gamma_{l_{1}^{-}l_{2}^{+}} -
\Gamma_{l_{1}^{+}l_{2}^{-}}}{\Gamma_{Z}} & = & - \frac{8}{3}
\frac{\alpha^{3}}{(4\pi)^{2}} \frac{M_{Z}}{\Gamma_{Z}}
 Im\left( k_{1}k_{2}^{*}\right)Im
\left({\cal M}_{1}{\cal
M}_{2}^{*}\right).  \nonumber
\end{eqnarray}
We found that the maximum value allowed, $\eta \leq 2.2 \times 10^{-14}$ for
$e\tau$ mode at $M_{N} = 5$ TeV, is very small (see experimental limits in Eq.
\ref{brra}).}
\begin{eqnarray}
Re\left\{k_{1}(l_{1}^{-}l_{2}^{+}) \; k_{2}^{*}(l_{1}^{-}l_{2}^{+})
 \; {\cal M}_{1}{\cal M}_{2}^{*}\right\} & + &
Re\left\{k_{1}(l_{1}^{+}l_{2}^{-}) \; k_{2}^{*}(l_{1}^{+}l_{2}^{-})
 \; {\cal M}_{1}{\cal M}_{2}^{*}\right\}  \nonumber \\
& = & 2 Re\left(k_{1}k_{2}^{*}\right)Re\left({\cal M}_{1}{\cal
M}_{2}^{*}\right),
\end{eqnarray}
giving the total rate for $Z \rightarrow l_{1}^{+}l_{2}^{-} + 
l_{1}^{-}l_{2}^{+}$
\begin{eqnarray}
\label{gfviol}
\Gamma_{l_{1}^{-}l_{2}^{+} + l_{1}^{+}l_{2}^{-}}
& = &
 \frac{4}{3} \frac{\alpha^{3}}{(4\pi)^{2}} M_{Z}
 \left\{|k_{1}|^{2}|{\cal M}_{1}|^{2}
 + |k_{2}|^{2}|{\cal M}_{2}|^{2} + 2 Re\left(k_{1}k_{2}^{*}\right) \right.
 \nonumber \\
& \times & \left. Re\left( {\cal M}_{1}{\cal M}_{2}^{*}\right)\right\}.
\end{eqnarray}

\subsection{Approximate relations in the limit of large NHL mass}
\label{arit}

While we can easily see how the width $\Gamma_{l_{1}^{-}l_{2}^{+} +
l_{1}^{+}l_{2}^{-}}$ depends on mixing
factors, the dependence on the NHL mass $M_{N}$ is obscured by the
algebraic complexity of the 't~Hooft - Veltman integrals. Fortunately, in 
the most
interesting case, which is that of a large $M_{N}$, the amplitudes become
particularly simple. It is the most interesting case since the signal is
the largest due to quadratic nondecoupling effects. This means that some of the
diagrams give rise to terms $O(M_{N}^{2})$. These effects, as well as
the question of how high we can go with the mass $M_{N}$ without disturbing
perturbation theory, will be discussed 
in Secs. \ref{appelc} and \ref{breakdown}. For now, we just state that by large
NHL mass we mean $M_{Z} < M_{N} < 5$~TeV.

In the limit of a large NHL mass $M_{N}$, the amplitudes exhibit the 
following behaviour:
\begin{eqnarray}
\label{aproxm}
{\cal M}_{WWN} & = & -  \frac{3 c_{W}}
{4 s_{W}^{3}}
\Big\{4C_{24}^{fin}(M_{W},M_{N},M_{W})-\frac{2}{3}+ \Delta_{\mu}\Big\},
\nonumber \\
{\cal M}_{abW} & = & -\frac{1}
{8 s_{W}^{3} c_{W}}\Big\{2-4C_{24}^{fin}(M_{a},M_{W},M_{b})-\Delta_{\mu}
\Big\},  \\
&   &  a,b \:=\:  N,\nu\;;\;\nu,N\;;\;N,N 
\mbox{ ; } \;\;M_{\nu} = 0. \nonumber 
\end{eqnarray}
These formulae differ by
less than one percent from the exact ones in Eq. \ref{amplitudes}, at 
$M_{N} = 500$ GeV and the
difference decreases with rising $M_{N}$ to less than $0.1$ percent at
$M_{N} = 5000$ GeV.
$C$ functions in the same limit behave as
\begin{eqnarray}
\label{aproxc}
C_{0}(M_{W},M_{N},M_{W}) & = & \frac{1}{M_{N}^{2}}\Big[\ln{{\cal X}}
+ 2 \sqrt{4c_{W}^{2} - 1}\left(\theta - \frac{\pi}{2}\right) + 1
\nonumber \\
& + & O\left({\cal X}^{-1}\right) \Big], \;\;\;\;\; 
\theta = \arctan{\sqrt{4c_{W}^{2} - 1}}, \nonumber \\
C_{0}(M_{N},M_{W},M_{N}) & = & \frac{1}{M_{N}^{2}}\Big[1 +
O\left({\cal X}^{-1}\right) \Big], \nonumber  \\
C_{24}^{fin}(M_{W},M_{N},M_{W}) & = & \frac{3}{8} - \frac{1}{4} \ln M_{N}^{2}
+ O\left({\cal X}^{-1}\right), 
\end{eqnarray}
and also $C_{24}^{fin}$ function of any other combination of arguments involving
 $M_{N}$ varies slowly as $\ln{M_{N}^{2}}$. 

With the help of Eqs. \ref{aproxm}, \ref{aproxc}, we can see there are three
amplitudes in Eq. \ref{amplitudes} with nondecoupling behaviour, namely
the
quadratic dependence on NHL mass.  They are ${\cal M}_{NN\phi}, {\cal
M}_{\phi\phi N}$ and ${\cal M}_{\phi N}$. However, as numerical calculations
 show,
${\cal M}_{\phi\phi N} \rightarrow - {\cal M}_{\phi N}$ for large $M_{N}$,
 leaving us ${\cal M}_{NN\phi}$ as the only amplitude with  the nondecoupling 
 behaviour.
${\cal M}_{NN\phi}$ gives the dominant contribution to ${\cal M}_{2}$  and,
moreover, it ensures that for large $M_{N}$
\begin{eqnarray}
|k_{2}|^{2}|{\cal M}_{2}|^{2} & > & |k_{1}|^{2}|{\cal M}_{1}|^{2},
\end{eqnarray}
despite the fact that the $|k_{2}|$ is quadratically small compared to the 
linear $|k_{1}|$. In Refs. \cite{bernabeu1,Valle2} the authors
neglected terms proportional to $|k_{2}|$, therefore their results do not apply
in the large $M_{N}$ limit.

\subsection{Numerical results}
\label{numeres}

In the numerical calculations, the term $Re\left(k_{1}k_{2}^{*}\right)$ from the
Eq. \ref{gfviol} was treated as follows.
We only have limits on $|k_{1}|, |k_{2}|$ (input parameters for our
calculations), not on
real and imaginary parts of $k_{1}, k_{2}$. 
Thus for given $|k_{1}|, |k_{2}|$, the real part of $k_{1}k_{2}^{*}$ can vary as
\begin{eqnarray}
- |k_{1}||k_{2}| & \leq &
Re\left(k_{1}k_{2}^{*}\right) \;\;\;\leq \;\;\; |k_{1}||k_{2}|.
\end{eqnarray}
In our calculations we set 
\begin{eqnarray}
Re\left(k_{1}k_{2}^{*}\right) & = & \delta |k_{1}||k_{2}|,
\end{eqnarray}
and $\delta$ is varied between $-1$ and $+1$ as an independent input 
parameter.
To find a numerical value of the parameter $|k_{2}|^{2}$, we express it
 in terms of $ll_{mix}$ and 
${l_{1}l_{2}}_{mix}$,
\begin{eqnarray}
|k_{2}|^{2} & \equiv & |k_{2}|^{2}(l_{1}^{-}l_{2}^{+})
\;\; = \;\; 
\left({l_{1}l_{1}}_{mix} + {l_{2}l_{2}}_{mix}\right)^{2}
|{l_{1}l_{2}}_{mix}|^{2}
+ |{l_{1}l_{3}}_{mix}|^{2}|{l_{3}l_{2}}_{mix}|^{2}  \nonumber \\
& + & 2\left({l_{1}l_{1}}_{mix} + {l_{2}l_{2}}_{mix}\right)
  Re\left\{{l_{1}l_{2}^{*}}_{mix} \; {l_{1}l_{3}}_{mix} \;{l_{3}l_{2}}_{mix}
  \right\}.
\end{eqnarray}
The smallness of $e\mu_{mix}$ effectively removes some of the terms.
For the $e\mu$ final state, the first and the third terms above are negligible,
leaving
\begin{eqnarray}
\label{quartica4}
|k_{2}|^{2}(e\mu) & \doteq & |{e\tau}_{mix}|^{2}|{\tau\mu}_{mix}|^{2},
\end{eqnarray}
while for $e\tau$ and $\mu\tau$ sector we have
\begin{eqnarray}
\label{quarticb4}
|k_{2}|^{2}(e\tau) & \doteq & \left({ee}_{mix} +
{\tau\tau}_{mix}\right)^{2}
|{e\tau}_{mix}|^{2} ,  \nonumber   \\
|k_{2}|^{2}(\mu\tau) & \doteq & \left({\mu\mu}_{mix} +
{\tau\tau}_{mix}\right)^{2}
|{\mu\tau}_{mix}|^{2}.   
\end{eqnarray}
The maximally allowed mixings (Eqs. \ref{limits1}, \ref{limits2}, \ref{limits3})
imply 
$|k_{1}| =$ $(0.00024,$ $0.015, 0.0068)$ and  
$|k_{2}| =$ $(0.0001, $ $0.0006, 0.00023)$
for $e\mu, e\tau, \mu\tau$ modes respectively.
As noted before, we assume degenerate NHL's with mass $M_{N}$.
Gauge boson masses used in the numerical calculations are
$M_{Z} = 91.1884$ GeV \cite{mt1} and $M_{W} = 80.410$~GeV \cite{mw1}. 
The total decay width of the Z
boson is taken as $\Gamma_{Z} = 2.4963$ GeV \cite{mt1}. 

\begin{figure}
\begin{center}
\setlength{\unitlength}{1in}
\begin{picture}(6,7)
\put(.1,+0.325){\mbox{\epsfxsize=5.0in\epsffile{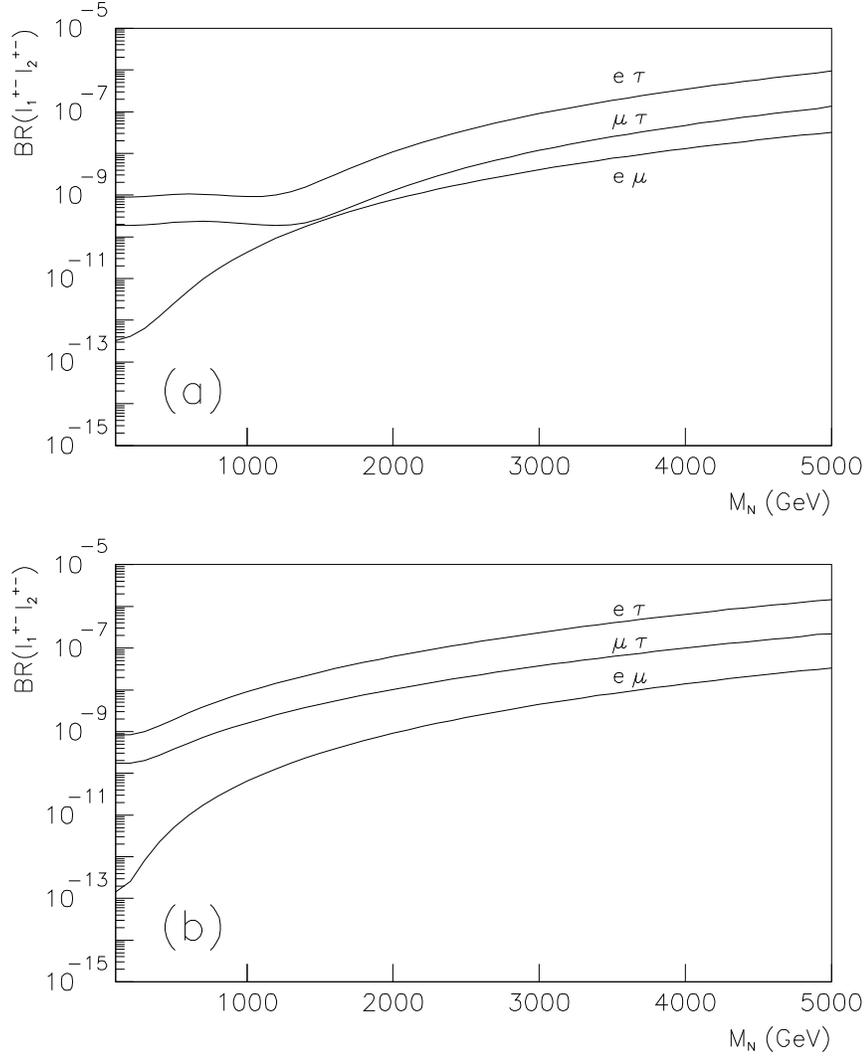}}}
\end{picture}
\end{center}
\caption{The branching ratio $Z \rightarrow l_{1}^{\pm}l_{2}^{\mp}$
as a function
of $M_{N}$
for (a) $\delta = -1$, (b) $\delta = +1$
.}
\label{fviolfig}
\end{figure}

The results are shown in Fig. \ref{fviolfig}a,b. They show how the branching 
ratio 
$ BR\left(l_{1}^{\pm}l_{2}^{\mp}\right) \equiv \Gamma_{l_{1}^{+}l_{2}^{-} +
l_{1}^{-}l_{2}^{+}}/\Gamma_{Z}$
 varies with the NHL mass. In Fig. \ref{fviolfig}a we set $\delta = -1$, 
in Fig. \ref{fviolfig}b
$\delta = +1$.
The graphs start at $M_{N} = 100$ GeV.
For NHL masses less than $M_{W}$, the rates are negligibly small.
A sudden rise in branching ratio just above $M_{N} = 1$ TeV for $e\tau$ and
$\mu\tau$ modes in Fig. \ref{fviolfig}a 
signals that at this point the $|k_{2}|^{2}|{\cal
M}_{2}|^{2}$ term overtakes the $|k_{1}|^{2}|{\cal M}_{1}|^{2}$ term and
the nondecoupling behaviour (generated by the Feynman graph 
Fig. \ref{odlisny}j) becomes dominant.
We predict the following branching ratio limits for $M_{N} = 5$ TeV and 
$\delta = +1$:
\begin{eqnarray}
BR_{th}(Z \rightarrow e^{\pm}\mu^{\mp}) & < & 3.3 \times 10^{-8},  \nonumber \\
BR_{th}(Z \rightarrow e^{\pm}\tau^{\mp}) & < & 1.4 \times 10^{-6}, \nonumber  \\
BR_{th}(Z \rightarrow \mu^{\pm}\tau^{\mp}) & < & 2.2 \times 10^{-7}. 
\end{eqnarray}
These results are similar to those of 
Ref. \cite{pilaftsis1}, where the
calculation was done in the context of a see-saw model with enhanced mixings. 

Current experimental upper limits on branching ratios are \cite{pdb}
\begin{eqnarray}
\label{brra}
BR_{exp}(Z \rightarrow e^{\pm}\mu^{\mp}) & < & 6 \times 10^{-6},
\;\;\;\;\;95 \%\;\;{\rm C.L.},  \nonumber \\
BR_{exp}(Z \rightarrow e^{\pm}\tau^{\mp}) & < & 1.3 \times 10^{-5},
\;\;\;\;\;95 \%\;\;{\rm C.L.}, \nonumber \\
BR_{exp}(Z \rightarrow \mu^{\pm}\tau^{\mp}) & < & 1.9 \times 10^{-5},
\;\;\;\;\;95 \%\;\;{\rm C.L.}\;.
\end{eqnarray}
Our prediction for $e\tau$ mode is thus at least one order of magnitude below 
the experimental limit.

\section{Flavour-violating processes at very low energies}
\label{fvple}

How does this compare with flavour-violating processes at very low energies ?
The rare decay $\mu \rightarrow e \gamma$ (see Fig. \ref{muegama})
is very well measured and supplies 
us with a stringent limit on $|e\mu_{mix}|$ (see Eq. \ref{limits3}), 
which we use as an input parameter for our calculations. We now derive this
limit.
The decay $\mu \rightarrow e \gamma$ was studied in the context of our model
and see-saw models 
with enhanced mixings 
%\footnote{The models with enhanced mixings: our model and the see-saw model of
%Pilaftsis \cite{pilaftsis2}.}
by several authors \cite{Ng1,Ng2,ggjv,Ilakovac,Jarlskog}. In our model,
with mass degenerate
NHL's, the $\mu \rightarrow e \gamma$ branching ratio is \cite{ggjv}
\begin{eqnarray}
\label{brmeg}
BR(\mu \rightarrow e \gamma) & = & \frac
{3\alpha}{32\pi}{|e\mu_{mix}|}^2 {|F_{\gamma}({\cal X})|}^{2},
\end{eqnarray} \\
where
\begin{eqnarray}
{|F_{\gamma}({\cal X})|}^{2} & = & - {\cal X}\frac{1- 5{\cal X}-2{\cal
X}^{2}}{(1-{\cal X})^{3}} + \frac{6{\cal X}^{3}}{(1-{\cal X})^{4}}\ln {\cal X},
\;\;{\cal X} = \frac{M_{N}^{2}}{M_{W}^{2}},
\end{eqnarray}
is an NHL mass dependent form factor. For NHL masses
$M_{N} > 500$ GeV, which we ultimately consider, the formfactor becomes
independent of mass,
\begin{eqnarray}
\label{formfactor}
F_{\gamma}({\cal X}) \rightarrow -2.
\end{eqnarray}
This is another example of the nondecoupling behaviour ($F_{\gamma}({\cal
X})$ does not vanish) of NHL's
\footnote{It is instructive to see how this result arises from the dimensional
analysis argument. The effective Lagrangian for the $\mu \rightarrow e \gamma$
is given by \cite{chengli}
\begin{eqnarray}
{\cal L}_{eff} & = & T({\cal X}){\overline e}_{L}\sigma_{\lambda \nu} \mu_{R}
F^{\lambda \nu}, \nonumber
\end{eqnarray}
where the field operators ${\overline e}_{L}, \mu_{R}, F^{\lambda \nu}$ have
mass dimensions 3/2, 3/2 and 2, respectively; hence $T({\cal X})$ has to have
dimension -1. For Fig. \ref{muegama}, the large mass $M_{N}$ dominance thus 
suggests
$T({\cal X}) \sim \frac{1}{M_{N}}$ on dimensional grounds. However, there is
also a possibility of $T({\cal X}) \sim \frac{m_{\mu}}{M_{N}^{2}}$; this
is indeed what happens since it is $m_{\mu}$ (or $m_{e}$) which gives the right
helicity flip to yield the required ${\cal L}_{eff}$ (these points can be
best understood after writing down the amplitude for the graph). 
The amplitude for
Fig. \ref{muegama} thus decouples quadratically.

There is another graph where two internal W's are replaced by the unphysical
Higgs $\phi$. The large $M_{N}$ behaviour is in this case boosted by $M_{N}$
dependent couplings of NHL's to $\phi$'s, so this dominant graph yields
$T({\cal X}) \sim F_{\gamma}({\cal X}) \sim \frac{m_{\mu}}{M_{N}^{2}} 
\frac{M_{N}^{2}}{M_{W}^{2}}
\sim const$, in agreement with Eq. \ref{formfactor}.}.
It is the mildest nondecoupling case; 
we encountered in the previous section 
amplitudes with quadratic dependence on NHL mass.
Given the current experimental limit on the
$\mu \rightarrow e \gamma$ branching ratio \cite{pdb},
\begin{eqnarray}
BR(\mu \rightarrow e \gamma) & \leq & 4.9 \times 10^{-11}\;\;\;\;\;90\%\;
{\rm C.L.},
\end{eqnarray}
Eqs. \ref{brmeg}, \ref{formfactor} yield an upper limit on the mixing of
$|e\mu_{mix}| \leq  0.00024$, given previously as Eq. \ref{limits3}.

On the other hand, experimental limits on $\tau \rightarrow e \gamma$ and
$\tau \rightarrow \mu \gamma$ \cite {pdb},
\begin{eqnarray}
BR_{exp}(\tau \rightarrow e \gamma) & < & 1.2 \times 10^{-4}, 
\;\;\;\;\;90 \%\;\;{\rm C.L.}\;, \nonumber \\
BR_{exp}(\tau \rightarrow \mu \gamma) & < & 4.2 \times 10^{-6},
\;\;\;\;\;90 \%\;\;{\rm C.L.}\;, 
\end{eqnarray}
are much weaker.
The predicted rate for both $e \gamma$ and $\mu \gamma$ modes is \cite{ggjv}
\begin{eqnarray}
BR_{th}(\tau \rightarrow e \gamma, \mu \gamma) & = & 
7 \times 10^{-7}, \;\;\;\;{\rm for}\;\;\;\;M_{N} > 500 \;{\rm GeV}.
\end{eqnarray}
However, the limits on mixing parameters used in Ref. \cite{ggjv}
 are out of date
now. With the current limits the predicted rate would be smaller by at least
one order of magnitude, implying that the theoretical result is
two orders of magnitude below the experimental upper limit for  
$\mu \gamma$ mode and about three orders for
$e \gamma$ mode.
This explains why we had to use indirect limits of Eq. \ref{limits2} for
$\mu\tau_{mix}$ and $e\tau_{mix}$.

Another well-measured muon decay mode is $\mu \rightarrow e^{-}e^{-}e^{+}$,
with \cite {pdb}
\begin{eqnarray}
\label{mueee4}
BR_{exp}(\mu \rightarrow e^{-}e^{-}e^{+}) & < & 1.0 \times
10^{-12},\;\;\;\;\;90 \% {\rm C.L.}\;.
\end{eqnarray}
This process was considered by Refs. \cite{Ng1,Ng2,Ilakovac,Jarlskog}.
The calculation shows the quadratic nondecoupling we encountered  in the
lepton flavour-violating decays of the Z boson. 
Ref. \cite{Jarlskog} gives (with an assumption
discussed therein) the following constraint for the parameters of the
superstring-inspired (our) model:
\begin{eqnarray}
\label{mnlimit4}
ee_{mix}|e\mu_{mix}| & \leq & 0.93 \times 10^{-5} \frac{1 {\rm
TeV}^{2}}{M_{N}^{2}({\rm TeV}^{2})},
\end{eqnarray}
which for $M_{N} \geq 3$ TeV is competitive with a constraint implied by Eqs.
\ref{limits1}, \ref{limits2}, \ref{limits3}:
\begin{eqnarray}
ee_{mix}|e\mu_{mix}| & \leq & 0.17 \times 10^{-5}.
\end{eqnarray}

Also considered in Refs. \cite{Ng1,Ng2,Jarlskog} is 
$\mu - e$ conversion in nuclei,
$\mu^{-}(A,Z) \rightarrow e^{-}(A,Z)$. The constraint on the product
$ee_{mix}|e\mu_{mix}|$ \cite{Jarlskog} is similar to the one above.

For the flavour-violating decays of the tau into three leptons ($\tau
\rightarrow$ $e^{-}e^{-}e^{+},$ $e^{-}\mu^{-}\mu^{+},$ etc) there is to my
knowledge no calculation studying the large (TeV) NHL mass limit 
in the context of our model. Within the see-saw model of
Ref. \cite{pilaftsis2}, Pilaftsis predicts with the current limits on mixings 
and for $M_{N} = 3$ TeV \cite{pilaftsis1}:
\begin{eqnarray}
BR_{th}(\tau \rightarrow e^{-}e^{-}e^{+}) & = & 5 \times 10^{-7}, \nonumber   \\
BR_{th}(\tau \rightarrow e^{-}\mu^{-}\mu^{+}) & = & 3 \times 10^{-7}. 
\end{eqnarray}
The current experimental limits are \cite{pdb}
\begin{eqnarray}
BR_{exp}(\tau \rightarrow e^{-}e^{-}e^{+}) & < & 1.4 \times 10^{-5},\;\;\;\;\;90
\%\;\;{\rm C.L.}\;,\nonumber \\
BR_{exp}(\tau \rightarrow e^{-}\mu^{-}\mu^{+})
& < & 1.4 \times 10^{-5},\;\;\;\;\;90 \%\;\;{\rm C.L.}\;.
\end{eqnarray}

Finally, hadronic decay modes of the $\tau$ lepton, $\tau \rightarrow l \eta,
l\pi^{0}$ \cite{ggjv} are disfavoured by loose limits, e.g. $BR(\tau
\rightarrow \mu^{-} \pi^{0}) < 4.4 \times 10^{-5}$ \cite{pdb}.

In conclusion, to probe large NHL masses, we would have to push experimental 
upper limits by at least one order of magnitude for flavour-violating leptonic
decays of the Z boson, and by one to two orders of magnitude for 
flavour-violating decays of the $\tau$ lepton. This most likely requires 
increased high luminosity running at LEP~I energy and $\tau$ factory
\cite{tfactory}. 
Note the dominant contribution
to the total rate for $Z \rightarrow l_{1}^{+}l_{2}^{-} + l_{1}^{-}l_{2}^{+}$,
$|k_{2}|^{2}|{\cal M}_{2}|^{2}$, depends quartically on small mixings (see Eqs.
\ref{quartica4}, \ref{quarticb4})  and also quartically on the NHL mass $M_{N}$.
Further mass independent limits on mixings will therefore suppress this
dominant contribution rather quickly, unless $M_{N}$ is very large.

\newpage

\chapter{Lepton flavour-conserving processes}
\label{chap6}
                                                            
In this and the following chapter we will examine two lepton flavour-conserving
processes: 
i) $Z \rightarrow l^{+}l^{-} \; (l = e, \mu, \tau)$ with observables
$\Gamma_{ll}$ (the width) and $U_{br}$ (universality breaking parameter); and
ii) $\mu \rightarrow e \nu \nu$ with observable $M_{W}$ (the W boson mass).
We will show that these observables probe the mixings vs NHL mass parameter
space of our model in many respects more efficiently than the flavour-violating
decays discussed in the previous chapter.
We work to the one-loop ($O(\alpha)$) level of perturbation theory. 

In Sec. \ref{sectree} we classify, closely following the SM case of 
Ref. \cite{key6}, one-loop corrections to $Z \rightarrow l^{+}l^{-}$ 
into three groups - oblique, vertex and QED corrections. Each group
is then individually studied in Secs. \ref{secqed} - \ref{secver}. 
We note that a large number of contributing diagrams comes directly from the SM
without being modified by NHL's.
In such cases we use the SM results of Ref. \cite{key6}.
As far as non-SM contributions are concerned, we present a detailed
calculation of two Feynman diagrams (one oblique and one vertex) and a
summary of results for the remaining ones. Divergent results are then
renormalized as discussed in Chapter 4.

Secs. \ref{secimp} - \ref{breakdown} are less technical and hopefully more 
intriguing. We show the
impact of loop corrections to a $\mu$-decay on the Z width, calculate the W
mass $M_{W}$, discuss the
violation of the decoupling theorem and the quadratic dependence of the loop
corrections on 
the NHL mass, and the limitations of the perturbative calculations.

Our numerical results are presented in Sec. \ref{seresu} and discussed in Sec.
\ref{conc}. We investigate here only a part of the mixings vs NHL mass 
parameter space by setting
$ee_{mix} = \mu\mu_{mix} = 0$. The full space is studied in Chapter 7.

\section{$\bf Z \rightarrow l^{+}l^{-}$: the tree-level and the 
corrections}
\label{sectree}

The tree-level leptonic width of the Z boson in the SM is given by 
\begin{eqnarray}
\label{treew}
\Gamma_{0} & = & \frac{\alpha}{3} M_{Z}(v_{l}^{2} + a_{l}^{2});
\end{eqnarray}
with 
$v_{l} = (-1+4s_{W}^{2})/(4s_{W}c_{W})$ and $a_{l} = -1/(4s_{W}c_{W})$ being,
respectively, the vector and axial vector couplings
of the charged leptons to $Z$.
We neglected terms proportional to
$m_{l}^{2}/M_{W}^{2}$. In this approximation, as a consequence of the lepton
universality of the
SM, the partial widths for all three modes ($ee, \mu\mu,
\tau\tau$) are equal.

One-loop corrected leptonic decays of the Z boson in the SM were
thoroughly discussed by W. Hollik in Ref. \cite{key6}. He parametrizes the 
leptonic width as
\begin{eqnarray}
\label{oneloop}
\Gamma_{ll} & = & \frac{\Gamma_{0}
 + \delta{\hat \Gamma_{ll}}}{1+{\hat \Pi}_{Z}
(M_{Z}^{2})}(1+\delta_{QED}).
\end{eqnarray}                                    
The one-loop electroweak corrections include 
Z boson propagator (so-called oblique) corrections ${\hat \Pi}_{Z}$;
vertex corrections $\delta{\hat \Gamma_{ll}}$ 
and QED corrections $\delta_{QED}$. 
To give the reader some feeling for the numbers involved, we note that the SM 
prediction
with $M_{Z} = 91.1884$ GeV, $m_{t} = 176$ GeV and $M_{H} = 200$ GeV is
\begin{eqnarray}
\Gamma_{0} & = & 81.45 \; {\rm MeV}, \nonumber  \\
\Gamma_{ll} & = & 84.03 \; {\rm MeV}, 
\end{eqnarray}
i.e., loops account for $\Gamma_{ll} - \Gamma_{0} \doteq 2.5$ MeV.
The current experimental value under the assumption of lepton universality is
\cite{mt1}
\begin{eqnarray}
\label{gamaexp}
\Gamma_{ll}^{exp} & = & 83.93 \pm 0.14 \; {\rm MeV}.
\end{eqnarray}
Without assuming universality \cite{mt1},
\begin{eqnarray}
\label{pwidths}
\Gamma_{ee}^{exp}      & = & 83.92 \pm 0.17 \; {\rm MeV}, \nonumber   \\
\Gamma_{\mu\mu}^{exp}  & = & 83.92 \pm 0.23 \; {\rm MeV}, \nonumber  \\
\Gamma_{\tau\tau}^{exp}& = & 83.85 \pm 0.29 \; {\rm MeV}. 
\end{eqnarray}

In our model, Eqs. \ref{treew}, \ref{oneloop} keep the same form. 
It is ${\hat \Pi}_{Z}$ and
$\delta{\hat \Gamma}_{ll}$ which are modified by the contribution of NHL's.
Also, $\Gamma_{0}$ is modified (via $s_{W}$) in an indirect way 
(see Sec. \ref{secimp}); the QED parameter $\delta_{QED}$
is not affected by NHL's.
We now address these corrections one by one, starting with $\delta_{QED}$.

\section{QED corrections}
\label{secqed}

\begin{figure}
\begin{center}
\setlength{\unitlength}{1in}
\begin{picture}(6,1.5)
\put(.1,+0.325){\mbox{\epsfxsize=5.0in\epsffile{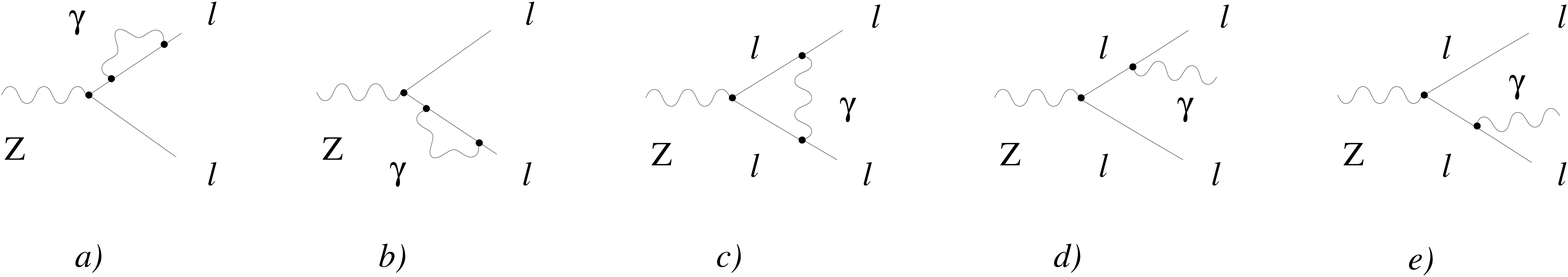}}}
\end{picture}
\end{center}
\caption{QED corrections.}
\label{qedfd}
\end{figure}

QED corrections (Fig. \ref{qedfd}) form a gauge invariant subset and 
therefore can be treated
independently of the genuine electroweak corrections \cite{key6}.  
The graphs of Fig. \ref{qedfd}
were calculated in Ref. \cite{qed} where the results were shown to modify
the Z~width by a factor $\delta_{QED}$ (see Eq. \ref{oneloop}),
\begin{eqnarray}
\delta_{QED} & = & \frac{3 \alpha}{4 \pi}.
\end{eqnarray}
Our inclusion of NHL's has no impact on this SM result.

\section{Z-propagator corrections ${\hat \Pi}_{Z}$}
\label{seczprop}

\begin{figure}
\begin{center}
\setlength{\unitlength}{1in}
\begin{picture}(6,3.0)
\put(.9,+0.4){\mbox{\epsfxsize=3.5in\epsffile{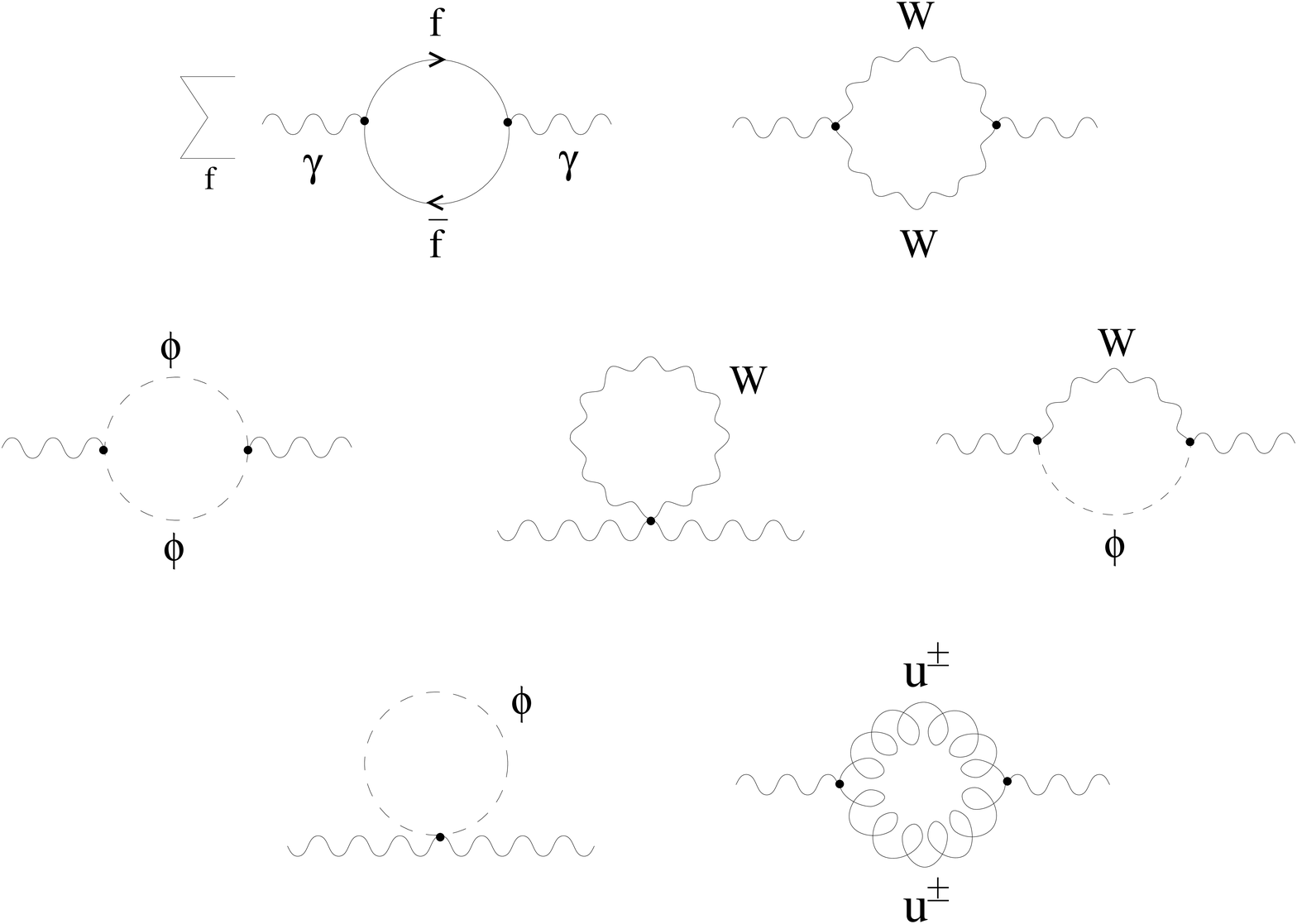}}}
%\put(1.65,-0.25){\footnotesize {\bf Figure \ref{pfd}:} Photon self-energy.}
\end{picture}
\end{center}
\caption{Photon self-energy.}
\label{pfd}

\begin{center}
\setlength{\unitlength}{1in}
\begin{picture}(6,3.0)
\put(.9,+0.125){\mbox{\epsfxsize=3.5in\epsffile{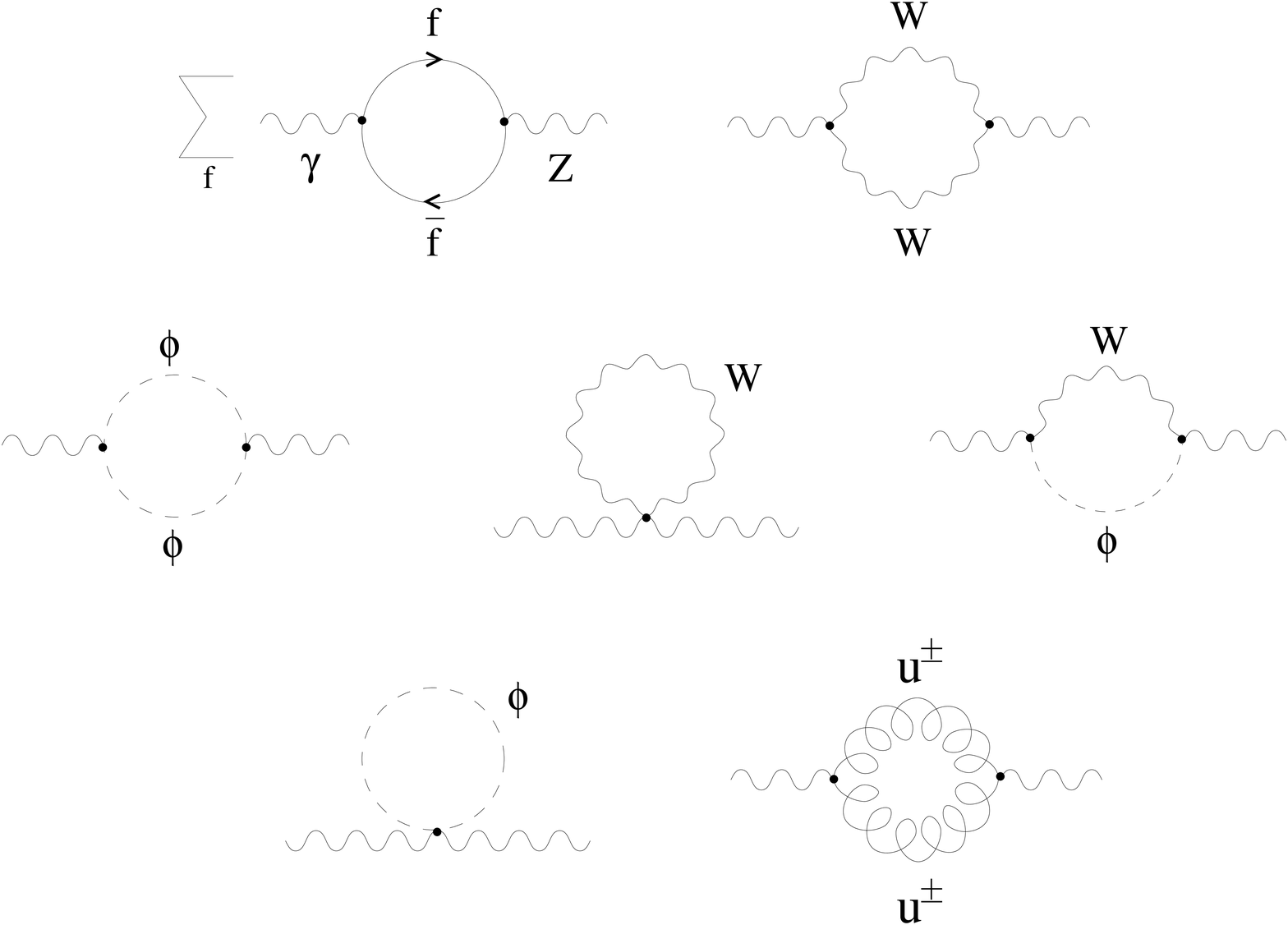}}}
%\put(1.65,-0.25){\footnotesize {\bf Figure \ref{pzfd}:}Photon-Z mixing energy.}
\end{picture}
\end{center}
\caption{Photon-Z mixing energy.}
\label{pzfd}
\end{figure}

\begin{figure}
\begin{center}
\setlength{\unitlength}{1in}
\begin{picture}(6,2.6)
\put(.6,+0.125){\mbox{\epsfxsize=4.5in\epsffile{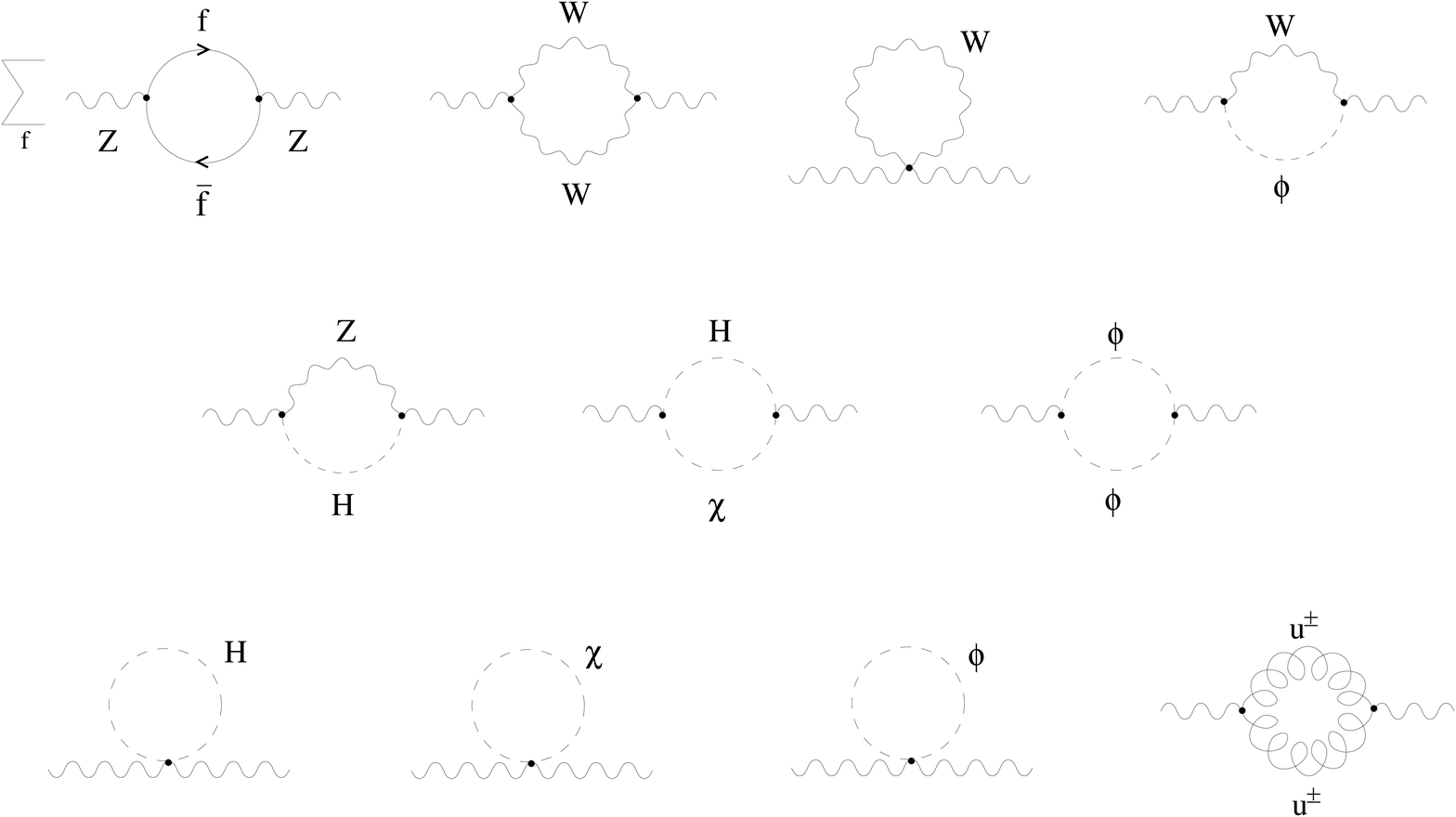}}}
%\put(1.8,-0.25){\footnotesize {\bf Figure \ref{zfd}:} Z boson self-energy.}
\end{picture}
\end{center}
\caption{Z boson self-energy.}
\label{zfd}
\begin{center}
\setlength{\unitlength}{1in}
\begin{picture}(6,4)
\put(0.6,+0.4){\mbox{\epsfxsize=4.5in\epsffile{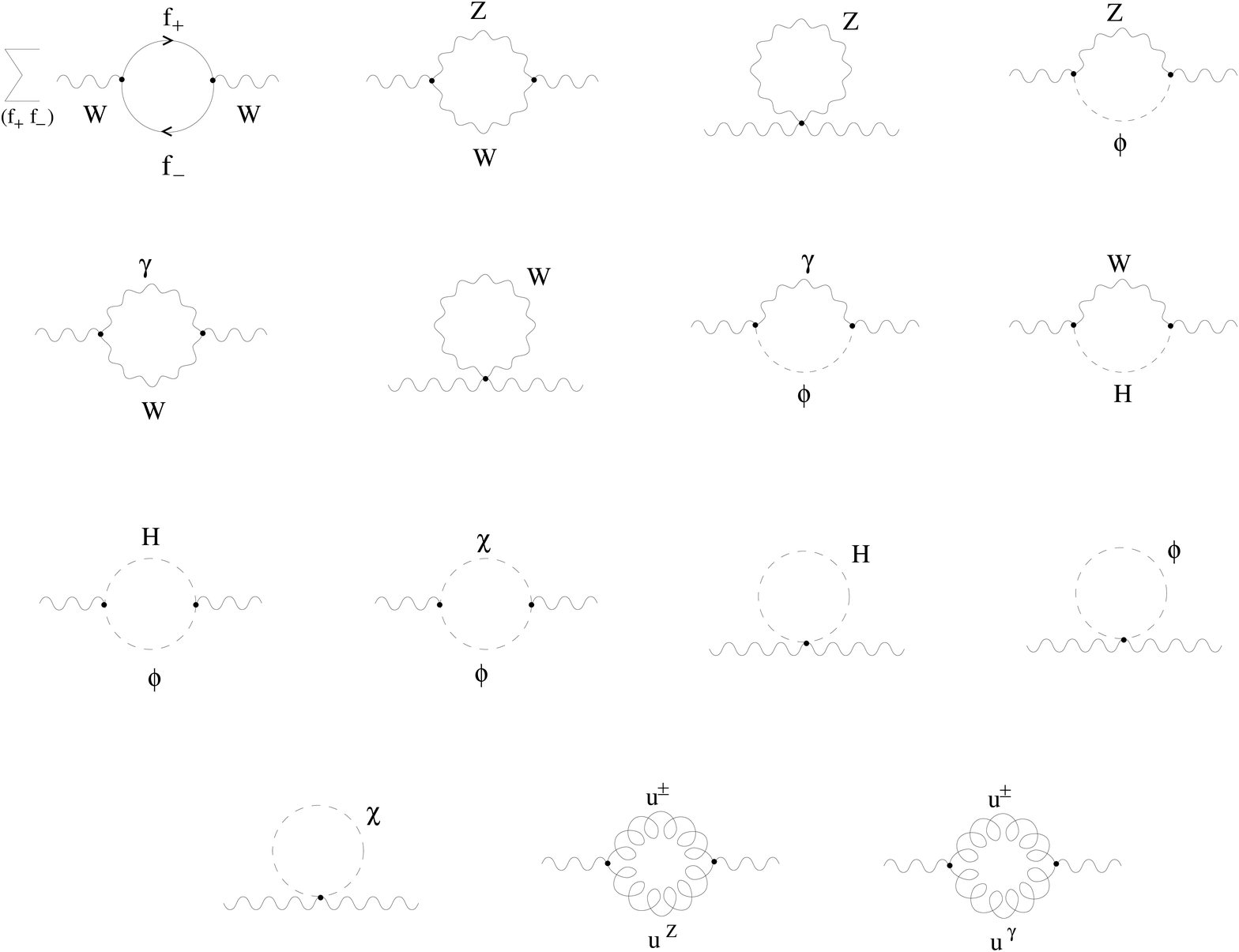}}}
%\put(1.8,-0.25){\footnotesize {\bf Figure \ref{wfd}:} W boson self-energy.}
\end{picture}
\end{center}
\caption{W boson self-energy.}
\label{wfd}
\end{figure}

Z-propagator corrections ${\hat \Pi}_{Z}$  are related to the real part of the
renormalized Z self-energy ${\hat \Sigma}_{Z}$ via
\begin{eqnarray}
{\hat \Pi}_{Z}(M_{Z}^{2}) & = & \frac{\partial \; {\cal R}e 
\:{\hat \Sigma}_{Z}}{\partial
p^{2}}(M_{Z}^{2}),
\end{eqnarray}
where $p$ is the 4-momentum of the Z boson.
${\hat \Sigma}_{Z}$ includes, besides the unrenormalized Z self-energy
$\Sigma_{Z}$, through the renormalization constant $\delta
Z_{2}^{Z}$ (see Eq. \ref{rconstants}), also all other unrenormalized gauge 
boson self-energies
$\Sigma_{W}, \Sigma_{\gamma}$ and $\Sigma_{\gamma Z}$. The diagrams
contributing to these self-energies are in Figs. \ref{pfd} - \ref{wfd}.

The photon self-energy (Fig. \ref{pfd}) and the photon - Z mixing energy 
(Fig. \ref{pzfd})
are not modified by the NHL's (the sum of the fermion loops runs over all 
fermions except neutrinos) and therefore we will use the SM
analytical formulae of Refs.~\cite{key6,cernlib} given in 
Eqs. \ref{agama}, \ref{asedem}. 

\begin{figure}[t]
\begin{center}
\setlength{\unitlength}{1in}
\begin{picture}(6,3)
\put(.25,+0.325){\mbox{\epsfxsize=5.5in\epsffile{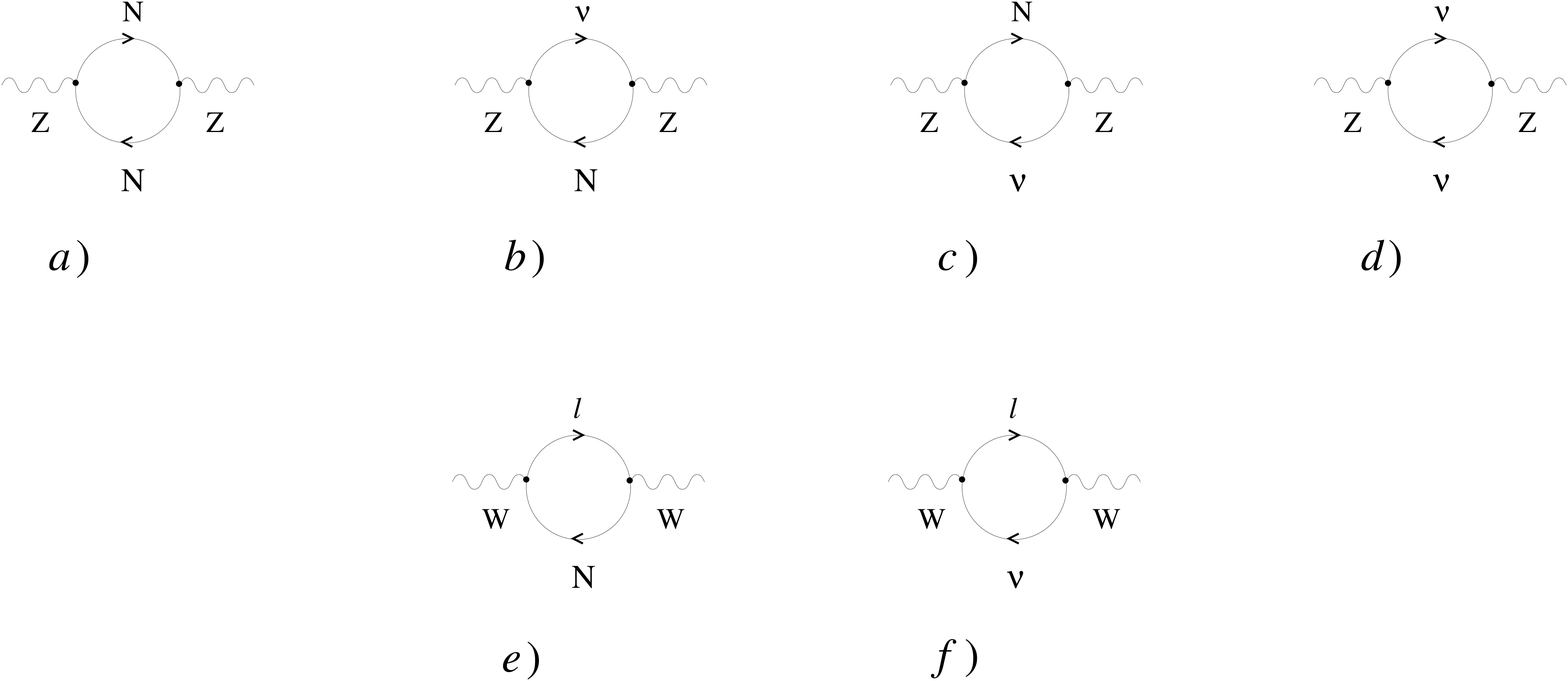}}}
%\put(.1,0){\footnotesize {\bf Figure \ref{obliquefd}:} Non-SM loops with
%NHL's and massless neutrinos.}
\end{picture}
\end{center}
\caption{Non-SM loops with NHL's and massless neutrinos.}
\label{obliquefd}
\end{figure}

The Z self-energy (Fig. \ref{zfd}) and the W self-energy (Fig. \ref{wfd}),
are modified
in our model as NHL's enter the fermion loops. The non-SM graphs from
Figs. \ref{zfd}, \ref{wfd} are
shown explicitly in Fig. \ref{obliquefd}. They include the graphs with 
massless neutrinos
(no NHL's), since these differ from the SM in the mixing factors.
We will calculate these graphs and the resulting amplitudes will replace the
SM neutrino contribution in Eq. \ref{azet}.

$\Sigma_{Z}$ is associated with the transverse ($g^{\mu\nu}$) part of the
Z self-energy tensor $\Sigma_{Z}^{\mu\nu}$:
\begin{eqnarray}
\Sigma_{Z}^{\mu\nu} & = & g^{\mu\nu}\Sigma_{Z} + p^{\mu}p^{\nu}
\tilde{\Sigma}_{Z}.
\end{eqnarray}
The longitudinal part $\tilde{\Sigma}_{Z}$ does not contribute 
to S-matrix elements
\cite{jeger} and we will not consider it here.
The non-SM part of the Z (unrenormalized) self-energy tensor 
$\Sigma_{Z}^{\mu\nu}$ is the sum of four
terms corresponding to Figs. $\ref{obliquefd} \;a - d$ :
\begin{eqnarray}
\Sigma_{Z}^{\mu\nu} & = & \Sigma_{Z}^{\mu\nu}(M_{N},M_{N}) +
\Sigma_{Z}^{\mu\nu}(M_{N},0) + \Sigma_{Z}^{\mu\nu}(0,M_{N}) + 
\Sigma_{Z}^{\mu\nu}(0,0).
\end{eqnarray}
We evaluate the contribution of one of these terms,
$\Sigma_{Z}^{\mu\nu}(M_{N},M_{N})$ in detail below as an example.
Divergent integrals are regularized and evaluated in $n$ dimensions using the 
technique of dimensional regularization due to 't Hooft and Veltman
\cite{thooftdim} (see Appendix \ref{Decko}).

Following the Feynman rules of Appendix \ref{Cecko}, the contribution of Fig. 
$\ref{obliquefd}a$ to the Z self-energy tensor is
\begin{eqnarray}
\label{SZHH}
-i\; \Sigma_{Z}^{\mu\nu}(M_{N},M_{N}) & = & - \sum_{a,b=4,5,6}\int \frac{d^{n}q}
{(2\pi)^{n}}
Tr \Big\{\frac{ie}{4s_{w}c_{w}}\big(K_{H}^{\dagger}K_{H}\big)_{ab}\gamma_{\mu}
\big(1-\gamma_{5}\big) \;\;\;\;\; \Big. \nonumber \\
& \times &
\Big. \frac{i}{\not q - \not p - M_{N}}  
\frac{ie}{4s_{w}c_{w}}\big(K_{H}^{\dagger}K_{H}\big)_{ba}\gamma_{\nu}
\big(1-\gamma_{5}\big)\frac{i}{\not q - M_{N}} \Big\}  \nonumber \\
\nonumber \\                
& = & - \frac{e^{2}}{16s_{W}^{2}c_{W}^{2}}
\sum_{a,b}\big(K_{H}^{\dagger}K_{H}\big)_{ab}
\big(K_{H}^{\dagger}K_{H}\big)_{ba} \int \frac{d^{n}q}{(2\pi)^{n}} \nonumber \\
& \times &
\frac{Tr \big\{\gamma_{\mu}\big(1-\gamma_{5}\big)
\big(\not q - \not p + M_{N}\big)
\gamma_{\nu}\big(1-\gamma_{5}\big)\big(\not q + M_{N}\big)\big\}}
{\big[(q-p)^{2} - M_{N}^{2}\big]\big[q^{2} - M_{N}^{2}\big]}.  \nonumber \\
\end{eqnarray}
In the above we sum over NHL's of all three families ($a,b = 4,5,6$).
Using the relations and theorems of Appendix \ref{gamas} we now evaluate 
the trace:
\begin{eqnarray}
Tr \; \big\{...\big\} & \equiv  &
Tr \; \big\{\gamma_{\mu}\big(1-\gamma_{5}\big)\big(\not q - \not p + M_{N}\big)
\gamma_{\nu}\big(1-\gamma_{5}\big)\big(\not q + M_{N}\big)\big\} \nonumber \\
& = &
2 \; Tr \; \big\{\gamma_{\mu}\big(1-\gamma_{5}\big)\big(\not q - \not p\big)
\gamma_{\nu}
\big(\not q + M_{N}\big)\big\} \nonumber \\
& = &
2 \; Tr \; \big\{\gamma_{\mu}\big(\not q - \not p\big)\gamma_{\nu}\not q\big\}
- 2 \; Tr \; \big\{\gamma_{\mu}\gamma_{5}\big(\not q - \not p\big)\gamma_{\nu}
\not q \big\}.
\end{eqnarray}
The trace with $\gamma_{5}$ does not contribute:
\begin{eqnarray}
Tr \; \big\{\gamma_{\mu}\gamma_{5}\big(\not q - \not p\big)\gamma_{\nu}
\not q \big\} 
%& = &
%- Tr \; \big\{\gamma_{\mu}\big(\not q - \not p\big)\gamma_{\nu}
%\not q \gamma_{5} \big\} \\
& = & 4 i \; \epsilon_{\mu\alpha\nu\beta} \; (q - p)^{\alpha}q^{\beta}
\; = \; 0,
\end{eqnarray}
using $\epsilon_{\mu\alpha\nu\beta}\;q^{\alpha}q^{\beta}\; = \; 0$ and
$\int_{q}q^{\alpha}...\; = \; p^{\alpha}...\;\;\;$.
The original trace is thus given by
\begin{eqnarray}
Tr \; \big\{... \big\} & = &
2 \; Tr \; \big\{\gamma_{\mu}\big(\not q - \not p\big)\gamma_{\nu}\not q \big\}
\nonumber \\
& = &
8 \big[(q-p)_{\mu}q_{\nu} - g_{\mu\nu}(q-p) q + q_{\mu}(q-p)_{\nu} \big].
\end{eqnarray}
We plug this result back in Eq. \ref{SZHH}:
\begin{eqnarray}
-i\; \Sigma_{Z}^{\mu\nu}(M_{N},M_{N}) & = &- \frac{e^{2}}{2s_{W}^{2}c_{W}^{2}}
k_{HH} \nonumber \\
& \times & \int \frac{d^{n}q}{(2\pi)^{n}}
\frac{(q-p)_{\mu}q_{\nu} - g_{\mu\nu}(q-p) q + q_{\mu}(q-p)_{\nu}}
{\big[(q-p)^{2} - M_{N}^{2}\big]\big[q^{2} - M_{N}^{2}\big]} \nonumber \\
& = &
- \frac{e^{2}}{2s_{W}^{2}c_{W}^{2}}k_{HH}
\frac{i {\pi}^{2}}{(2 \pi)^{n}} \Big\{ -p_{\mu} B_{\nu}(p;M_{N},M_{N})
 \nonumber \\
& - &
p_{\nu}B_{\mu}(p;M_{N},M_{N}) + g_{\mu\nu}p_{\alpha}B^{\alpha}(p;M_{N},M_{N}) 
\nonumber \\
& +  & 2 B_{\mu\nu}(p;M_{N},M_{N})
- g_{\mu\nu}g_{\alpha\beta}B^{\alpha\beta}(p;M_{N},M_{N})\Big\},
\end{eqnarray}
where $k_{HH} \equiv
\sum \big(K_{H}^{\dagger}K_{H}\big)_{ab}\big(K_{H}^{\dagger}K_{H}\big)_{ba}$
and functions $B_{\mu}, B_{\mu\nu}$ are 't Hooft scalar \linebreak
$n$-dimensional
integrals defined in Appendix \ref{scalari}. Using (see Eq. \ref{abfunc})
\begin{eqnarray}
B_{\mu} & = & - p_{\mu} B_{1}, \;\;\;\;\;\;\;\;\;\;
B_{\mu\nu} \;\; = \;\; p_{\mu}p_{\nu}B_{21} - g_{\mu\nu}B_{22},
\end{eqnarray}
the $n$-dimensional space-time relation (see Eq. \ref{nalgebra})
\begin{eqnarray}
g^{\mu \nu} g_{\mu \nu} & = & n,
\end{eqnarray}
and recollecting that only the transverse (terms with $g_{\mu\nu}$) part 
contributes to \linebreak S-matrix elements, we get
\begin{eqnarray}
-i\; \Sigma_{Z}^{\mu\nu}(M_{N},M_{N}) & = &- \frac{e^{2}}{2s_{W}^{2}c_{W}^{2}}
k_{HH}
\frac{i {\pi}^{2}}{(2 \pi)^{n}}g_{\mu\nu}\Big[ -p^{2}B_{1} - p^{2}B_{21}
-2B_{22} \Big. \nonumber  \\ 
& + &
 \Big. n B_{22}\Big].   
\end{eqnarray}
With the help of formulae from Appendix \ref{scalari}, 
we arrive at the following result for the self-energy $\Sigma_{Z}(M_{N},M_{N})$:
\begin{eqnarray}
-i\; \Sigma_{Z}(M_{N},M_{N}) & = &- \frac{e^{2}}{2s_{W}^{2}c_{W}^{2}}
k_{HH}
\frac{i {\pi}^{2}}{(2
\pi)^{n}}\frac{1}{3}\Big\{\Big(- 3 M_{N}^{2} 
+ p^{2}\Big)\Delta \Big. \nonumber \\ 
& + &
\Big. \Big[2 A_{0}^{fin}(M_{N})  + 2 M_{N}^{2} 
-\frac{p^{2}}{3} + \Big(p^{2} - M_{N}^{2} 
\Big) B_{0}^{fin}(p;M_{N},M_{N})\Big]\Big\} \nonumber \\
& = &
- \frac{e^{2}}{2s_{W}^{2}c_{W}^{2}}k_{HH}
\frac{i {\pi}^{2}}{(2
\pi)^{n}}\frac{1}{3}\Big\{\Big(- 3 M_{N}^{2}
+ p^{2}\Big)\Delta \Big. \nonumber \\
& + &
\Big. \Big[2M_{N}^{2} \ln M_{N}^{2} - \frac{p^{2}}{3} + 
\Big(F(p;M_{N},M_{N}) - \ln M_{N}^{2}\Big) \nonumber \\
& \times & \Big(p^{2} - M_{N}^{2}\Big)\Big]\Big\},
\end{eqnarray} 
where the function $F$ is related to the function $B_{0}$ by Eq. \ref{baf}
and $A_{0}^{fin}(m) = - m^{2}(- \ln m^{2} + 1)$.
The divergence is displayed as a pole at $n = 4$ (see
Appendix \ref{dimreg}):
\begin{eqnarray}
\Delta & = & \frac{2}{4-n} - \gamma - \ln \pi \; = \; \frac{2}{\epsilon} -
\gamma - \ln \pi.
\end{eqnarray}
In $n$ dimensions, $\alpha$ becomes a dimensional quantity and we should do the
the following replacement:
\begin{eqnarray}
\alpha = \frac{e^{2}}{4\pi} & \rightarrow & \alpha \mu^{\epsilon} \;\; =
\;\; \alpha \Big(1 + \frac{\epsilon}{2}\ln \mu^{2} 
+ O(\epsilon^{2}) + ... \Big),
\end{eqnarray}
where $\mu$ is an arbitrary mass. Together with the expansion
\begin{eqnarray}
\frac{1}{(2\pi)^{n}} & = & \frac{1}{(2\pi)^{4-\epsilon}} \;\; = \;\;
\frac{1}{(2\pi)^{4}}(1 + \epsilon \ln 2 \pi + O(\epsilon^{2}) + ... ),
\end{eqnarray}
this yields
\begin{eqnarray}
\frac{1}{(2\pi)^{n}} \alpha\; \Delta & = & \frac{1}{(2\pi)^{4}} \alpha\; 
\Big( \frac{2}{\epsilon} -\gamma + \ln 4\pi + \ln \mu^{2}\Big) \;\; = \;\;
\frac{1}{(2\pi)^{4}} \alpha \; \Delta_{\mu} \nonumber \\
& = & 
 \frac{1}{(2\pi)^{4}} \alpha \; (\Delta_{m} + \ln m^{2}), \;\;\;\Delta_{m}
\;=\; \frac{2}{\epsilon} -\gamma + \ln 4\pi + \ln \frac{\mu^{2}}{m^{2}}.\;\;\;
\end{eqnarray}
The self-energy thus becomes
\begin{eqnarray}
\Sigma_{Z}(M_{N},M_{N}) & = &
 \frac{\alpha}{8\pi}\frac{1}{s_{W}^{2}c_{W}^{2}}
k_{HH}\frac{1}{3}\Big\{\Big(-3M_{N}^{2} + p^{2}\Big)
\Big(\Delta_{M_{N}} + \ln M_{N}^{2}\Big) \nonumber \\
& + &
\Big. \Big[2M_{N}^{2} \ln M_{N}^{2} - \frac{p^{2}}{3} +
\Big(F(p;M_{N},M_{N}) - \ln M_{N}^{2}\Big)\Big(p^{2} - M_{N}^{2}
\Big)\Big]\Big\} \nonumber \\
& = &
 \frac{\alpha}{8\pi}\frac{1}{s_{W}^{2}c_{W}^{2}}
k_{HH}\Big\{\Delta_{M_{N}}\Big(\frac{p^{2}}{3} - M_{N}^{2}\Big)
-\frac{p^{2}}{9} \nonumber \\
& + & \frac{1}{3} F(p;M_{N},M_{N})\Big(p^{2} - M_{N}^{2}
\Big)\Big\}. 
\end{eqnarray}
For the other three contributions we get, following the same steps,
\begin{eqnarray}
\Sigma_{Z}(M_{N},0) & = &
 \frac{\alpha}{8\pi}\frac{1}{s_{W}^{2}c_{W}^{2}}
k_{HL}\Big\{\Delta_{M_{N}}\Big(\frac{p^{2}}{3} - \frac{M_{N}^{2}}{2}\Big)
+ \frac{2}{9}p^{2} - \frac{M_{N}^{2}}{6} \Big. \nonumber \\ 
& + &
\Big. F(p;M_{N},0)\Big(\frac{p^{2}}{3}
- \frac{M_{N}^{2}}{6} - \frac{M_{N}^{4}}{6p^{2}}\Big)\Big\}, \nonumber \\
\Sigma_{Z}(0,M_{N}) & = & \Sigma_{Z}(M_{N},0), \nonumber \\
\Sigma_{Z}(0,0) & = &
 \frac{\alpha}{8\pi}\frac{1}{s_{W}^{2}c_{W}^{2}}k_{LL}
\frac{p^{2}}{3}\Big(\Delta_{m} + F(p;m,m) - \frac{1}{3} \Big), 
\end{eqnarray}
%where $m \rightarrow 0$,
%$F(p;m,m) \; = \; 1 - \ln (-p^{2}/m^{2} - i \epsilon )$ and therefore
%$\Delta_{m} + F(p;m,m)$ is independent of $m$.
where $m^{2} \ll p^{2}$, otherwise $m$ can be arbitrary since 
$F(p;m,m) \; = \; 1 -
\ln (-p^{2}/m^{2} \linebreak -~i \epsilon )$ 
and therefore $\Delta_{m} + F(p;m,m)$ is independent of $m$.
The mixing factors $k_{HH}, k_{HL}$ and $k_{LL}$ can be cast into a more
convenient form by converting $K_{L}$ matrices into $K_{H}$ matrices with the
help of Eq. \ref{KLKH}:
\begin{eqnarray}
\label{khhmixi}
k_{HH} & = & 
\sum_{a,b} \big(K_{H}^{\dagger}K_{H}\big)_{ab}
\big(K_{H}^{\dagger}K_{H}\big)_{ba} 
\;\; = \;\;
\sum_{a,b,l,j} \big(K_{H}^{\dagger}\big)_{al}\big(K_{H}\big)_{lb} 
\big(K_{H}^{\dagger}\big)_{bj}\big(K_{H}\big)_{ja} \nonumber  \\
& = &
 \sum_{a,b,l,j} \big(K_{H}^{*}\big)_{la}\big(K_{H}\big)_{ja}
\big(K_{H}\big)_{lb}\big(K_{H}^{*}\big)_{jb} 
\;\; = \;\;
ee_{mix}^{2} + |e\mu_{mix}|^{2} + |e\tau_{mix}|^{2} \nonumber \\
& &
+ |e\mu_{mix}|^{2} + \mu\mu_{mix}^{2} + |\mu\tau_{mix}|^{2} 
+ |e\tau_{mix}|^{2} + |\mu\tau_{mix}|^{2} + \tau\tau_{mix}^{2},  \\ 
\nonumber \\
k_{HL} & = &
\sum_{a,i}
\big(K_{H}^{\dagger}K_{L}\big)_{ai}\big(K_{L}^{\dagger}K_{H}\big)_{ia} 
\;\;=\;\; \sum_{a,i,j,k} \big(K_{H}^{\dagger}\big)_{aj} \big(K_{L}\big)_{ji} 
\big(K_{L}^{\dagger}\big)_{ik}\big(K_{H}\big)_{ka} \nonumber \\
& = &
\sum_{a,j,k} \big(K_{H}^{\dagger}\big)_{aj} \delta_{jk} \big(K_{H}\big)_{ka}
 - \sum_{a,b,j,k} \big(K_{H}^{\dagger}\big)_{aj}\big(K_{H}\big)_{jb}
\big(K_{H}^{\dagger}\big)_{bk}\big(K_{H}\big)_{ka} \nonumber \\
& = &
\sum_{a,k} \big(K_{H}^{\dagger}\big)_{ak}\big(K_{H}\big)_{ka} -
\sum_{a,b} \big(K_{H}^{\dagger}K_{H}\big)_{ab}
\big(K_{H}^{\dagger}K_{H}\big)_{ba}
\nonumber \\
& = &
ee_{mix} +  \mu\mu_{mix} + \tau\tau_{mix} - k_{HH}, \\
 \nonumber \\
k_{LL} & = &
\sum_{i,j}
\big(K_{L}^{\dagger}K_{L}\big)_{ji}\big(K_{L}^{\dagger}K_{L}\big)_{ij}
\;\;=\;\; \sum_{i,j,k,l} \big(K_{L}^{\dagger}\big)_{jk}\big(K_{L}\big)_{ki}
\big(K_{L}^{\dagger}\big)_{il}\big(K_{L}\big)_{lj} \nonumber \\
%& = &
%\sum_{j,k,l} \big(K_{L}^{\dagger}\big)_{jk} \delta_{kl} \big(K_{L}\big)_{lj}
%- \sum_{j,k,l,a} \big(K_{L}^{\dagger}\big)_{jk} \big(K_{H}\big)_{ka}
%\big(K_{H}^{\dagger}\big)_{al} \big(K_{L}\big)_{lj} \nonumber \\
%& = & 
%\sum_{j,l} \big(K_{L}\big)_{lj}\big(K_{L}^{\dagger}\big)_{jl}
%- \sum_{j,k,l,a} \big(K_{L}\big)_{lj}\big(K_{L}^{\dagger}\big)_{jk}
%\big(K_{H}\big)_{ka}\big(K_{H}^{\dagger}\big)_{al}
%\nonumber \\
%& = &
%\sum_{l} \delta_{ll} - \sum_{l,b} \big(K_{H}\big)_{lb}
%\big(K_{H}^{\dagger}\big)_{bl} - \sum_{k,l,a} \delta_{lk} \big(K_{H}\big)_{ka}
%\big(K_{H}^{\dagger}\big)_{al} \nonumber  \\
%& + & 
%\sum_{k,l,a,c} \big(K_{H}\big)_{lc} \big(K_{H}^{\dagger}\big)_{ck}
%\big(K_{H}\big)_{ka} \big(K_{H}^{\dagger}\big)_{al}  \nonumber \\
& = & ... \; = \;
3 - 2 \big(ee_{mix} +  \mu\mu_{mix} + \tau\tau_{mix}\big) + k_{HH}.
\end{eqnarray}                                 
The total Z self-energy in our model is obtained by cutting out the neutrino
contribution from the total Z self-energy in the SM (the first line
of Eq. \ref{azet}) and replacing it with the sum $ \Sigma_{Z}(M_{N},M_{N}) +
2 \Sigma_{Z}(M_{N},0)  + \Sigma_{Z}(0,0)$.

%*************** W boson selfenergy ********************************
The W self-energy calculation goes along the same lines yielding
\begin{eqnarray}    
\Sigma_{W}(M_{N},m_{l}) & = &\frac{\alpha}{12\pi s_{W}^{2}}\Big\{
\sum_{l=e,\mu,\tau} ll_{mix} \Big[\frac{\Delta^{M_{N}}}{2}\Big(p^{2}-\frac{5}{2}
M_{N}^{2}-
\frac{m_{l}^{2}}{2}\Big)
\Big. \Big. \Big. \nonumber  \\
& + & \Big. \frac{\Delta^{m_{l}}}{2}\Big(p^{2} 
 - \frac{5}{2}m_{l}^{2} -  \frac{M_{N}^{2}}{2}\Big) \nonumber  \\
& + &
\Big(p^{2}
- \frac{M_{N}^{2}+m_{l}^{2}}{2}-\frac{(M_{N}^{2}-m_{l}^{2})^{2}}{2p^{2}}\Big)
F(p;M_{N},m_{l}) \nonumber \\
& + & 
\Big. \Big.
 \Big(p^{2}-\frac{M_{N}^{2}+m_{l}^{2}}{2}\Big)\Big(1-\frac{M_{N}^{2}+m_{l}^{2}}
{M_{N}^{2}-m_{l}^{2}}\ln\frac{M_{N}}{m_{l}}\Big)-\frac{p^{2}}{3}\Big]\Big\},
  \nonumber  \\
\Sigma_{W}(0,m_{l}) & = & \frac{\alpha}{12\pi s_{W}^{2}}\Big\{
\sum_{l=e,\mu,\tau} (1 - ll_{mix})\Big[\Big(p^{2}-\frac{3}{2}m_{l}^{2}\Big)
\Delta^{m_{l}}  \Big. \Big.
\nonumber \\
& + & 
\Big. \Big. \Big(p^{2}-\frac{m_{l}^{2}}{2}-
\frac{m_{l}^{4}}{2p^{2}}\Big)F(p;0,m_{l})
+\frac{2}{3}p^{2}-\frac{m_{l}^{2}}{2}\Big]
\Big\},
\end{eqnarray}
for the diagrams of Figs. $\ref{obliquefd}$e,f respectively. 
The total W self-energy in
our model is obtained by cutting out the lepton
contribution from the total W self-energy in the SM (the first two
lines
of Eq. \ref{adablju}) 
and replacing it with the sum $\Sigma_{W}(M_{N},m_{l}) + \Sigma_{W}(0,m_{l})$.
The self-energies are then renormalized using (see Eq. \ref{rselfe})
\begin{eqnarray}
\hat{\Sigma}_{Z}\left(p^{2}\right) & = & \Sigma_{Z}\left(p^{2}\right)
\;-\;\delta M_{Z}^{2}\;+\;\delta
Z_{2}^{Z}\;\left(p^{2}-M_{Z}^{2}\right),  \nonumber  \\
\hat{\Sigma}_{W}\left(p^{2}\right) & = & \Sigma_{W}\left(p^{2}\right)
\;-\;\delta M_{W}^{2}\;+\;\delta Z_{2}^{W}\;\left(p^{2}-M_{W}^{2}\right),
\nonumber
\end{eqnarray}
with renormalization constants given by Eq. \ref{rconstants}.
Note the form of the equations above is the same as in the SM.

In order to better see the dependence of ${\hat \Pi}_{Z}$ on $M_{N}$ for 
$M_{N} \gg M_{W}$ (this is the limit we are ultimately interested in, see
Sec. \ref{appelc}.), we split ${\hat \Pi}_{Z}$ as
\begin{eqnarray}
{\hat \Pi}_{Z} & = & {\hat \Pi}_{Z}^{SM} + {\hat \Pi}_{Z}^{NHL},
\end{eqnarray}
where ${\hat \Pi}_{Z}^{SM}$ is the SM limit of ${\hat \Pi}_{Z}$
and ${\hat \Pi}_{Z}^{NHL}$ are corrections due to NHL's.
Expanding the $F$ functions in powers of $M_{N}^{2}$, we obtain in the
limit of $M_{N} \gg M_{W}$
\begin{eqnarray}
\label{aprox1}
{\hat \Pi}_{Z}^{NHL} & = &
\frac{\alpha}{\pi}\Big\{\frac{c_{W}^{2}-s_{W}^{2}}{16
s_{W}^{4}}\frac{M_{N}^{2}}{M_{W}^{2}}k_{HH} + O(\ln M_{N}^{2}/M_{W}^{2}) +
...\Big\}.
\end{eqnarray}
Although this formula looks very simple, one should be aware of one important
fact. The leading term is suppressed by the mixing parameter. Indeed, the
$k_{HH}$ mixing is quadratic in $\tau\tau_{mix}$, while some of the $O(\ln
M_{N}^{2}/M_{W}^{2})$ terms are only linear. As a result, a few of them are
comparable in size to the leading term in $M_{N}^{2}$ expansion up to $\sim$
1 TeV NHL mass. We illustrate this point in Table \ref{t1}. Here we show 
numerical
predictions for the (exact) oblique parameter ${\hat \Pi}_{Z}$ and compare them
with the approximate parameter ${\hat \Pi}_{Zappx} = {\hat \Pi}_{Z}^{SM} +
{\hat \Pi}_{Zappx}^{NHL}$ where
${\hat \Pi}_{Zappx}^{NHL}$  is the first term in Eq. \ref{aprox1}
\footnote{The dependence of ${\hat \Pi}_{Z}^{SM}$ on the NHL mass has its 
origin in a
different value of the input parameter $M_{W}$  (as calculated from $G_{\mu}$,
see Sec. \ref{secimp})
for different NHL masses. Thus the formula for ${\hat
\Pi}_{Z}^{SM}$ comes from the SM, but the choice of $M_{W}$ comes from our
model, not the SM.}.
The contribution of higher order terms from Eq. \ref{aprox1} is given by the
difference $d = {\hat \Pi}_{Z} - {\hat \Pi}_{Zappx}$.
Input numbers used are $M_{Z} = 91.1884$ GeV, $M_{H} = 200$~GeV, $m_{t} = 176$
GeV, $\tau\tau_{mix} = 0.033$ and $ee_{mix} = \mu\mu_{mix} = 0$.
% All numbers in the table with the exception of masses
%are to be multiplied by $10^{-2}$.

\begin{table}[htb]
\begin{center}
\begin{tabular}{|l|r|r|r|r|r|} \hline
%\multicolumn{5}{|c|}{Table 5.1.  Comparison of ${\hat \Pi}_{Zappx}$ with
%${\hat \Pi}_{Z}$ } \\ \hline
$M_{N}$                       & 0.5 TeV  &  1 TeV  &  3 TeV &  5 TeV &  \\ 
  \hline
${\hat \Pi}_{Z}^{SM}$         & - 4.299  & - 4.297  & - 4.265 & - 4.199 &
                                                         $\times 10^{-2}$ \\
${\hat \Pi}_{Z}$              & - 4.313  & - 4.298  & - 4.054 & - 3.526 &
                                                          $\times 10^{-2}$ \\
${\hat \Pi}_{Zappx} = {\hat \Pi}_{Z}^{SM} + {\hat \Pi}_{Zappx}^{NHL}$
                              & - 4.292  & - 4.270  & - 4.013 & - 3.479 &
                                                          $\times 10^{-2}$ \\ 
${\hat \Pi}_{Zappx}^{NHL}$    &  0.007  &  0.027  &  0.252 &  0.720 & 
                                                         $\times 10^{-2}$ \\
$d = {\hat \Pi}_{Z} - {\hat \Pi}_{Zappx}$
                              & - 0.021  & - 0.028  & - 0.041 & - 0.047 &
                                                          $\times 10^{-2}$ \\
\hline
\end{tabular}
\end{center}
\caption{Comparison of ${\hat \Pi}_{Zappx}$ with ${\hat \Pi}_{Z}$}
\label{t1}
\end{table}
   
For $M_{N} = 0.5$ TeV, ${\hat \Pi}_{Zappx}^{NHL}$ contributes $0.007
\times 10^{-2}$ of the total difference between ${\hat \Pi}_{Z}$ and 
${\hat \Pi}_{Z}^{SM}$ (${\hat \Pi}_{Z}$ - ${\hat \Pi}_{Z}^{SM}$ = d + ${\hat
\Pi}_{Zappx}^{NHL}$). The higher order terms contribute more 
(with opposite sign), $d = - 0.021 \times
10^{-2}$.  At $1$ TeV, ${\hat \Pi}_{Zappx}^{NHL} = 0.027 \times 10^{-2}$
is comparable with $|d| = 0.028 \times 10^{-2}$ and at $3$ TeV it already 
dominates.

Overall, ${\hat \Pi}_{Zappx}$ differs from ${\hat \Pi}_{Z}$ by
approximately $1 \%$ in the considered range of NHL masses.

\section{Vertex factor $\delta{\hat \Gamma}_{ll}$}
\label{secver}

\begin{figure}
\begin{center}
\setlength{\unitlength}{1in}
\begin{picture}(6,3)
\put(.38,+0.325){\mbox{\epsfxsize=5.0in\epsffile{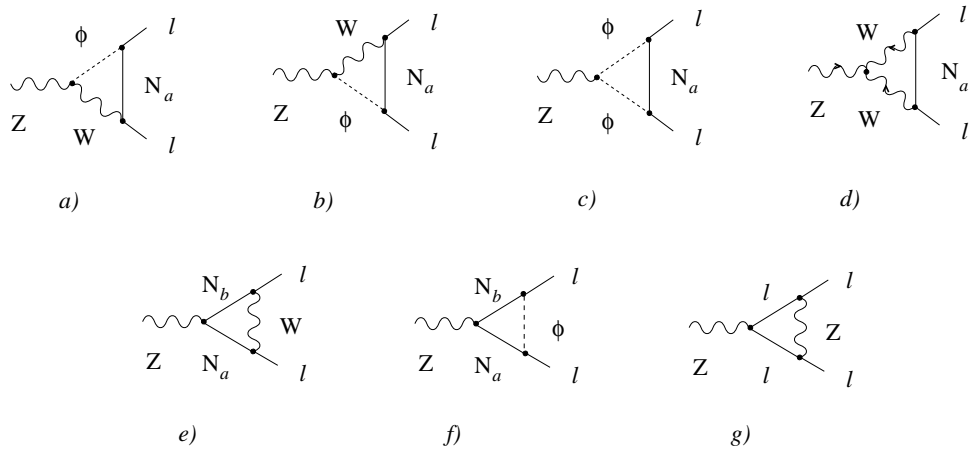}}}
%\put(1.65,0){\footnotesize {\bf Figure \ref{vertexfd}:} Irreducible vertex
% corrections.}
\end{picture}
\end{center}
\caption{Irreducible vertex corrections.}
\label{vertexfd}
\end{figure}

\begin{figure}
\begin{center}
\setlength{\unitlength}{1in}
\begin{picture}(6,3)
\put(.38,+0.325){\mbox{\epsfxsize=5.0in\epsffile{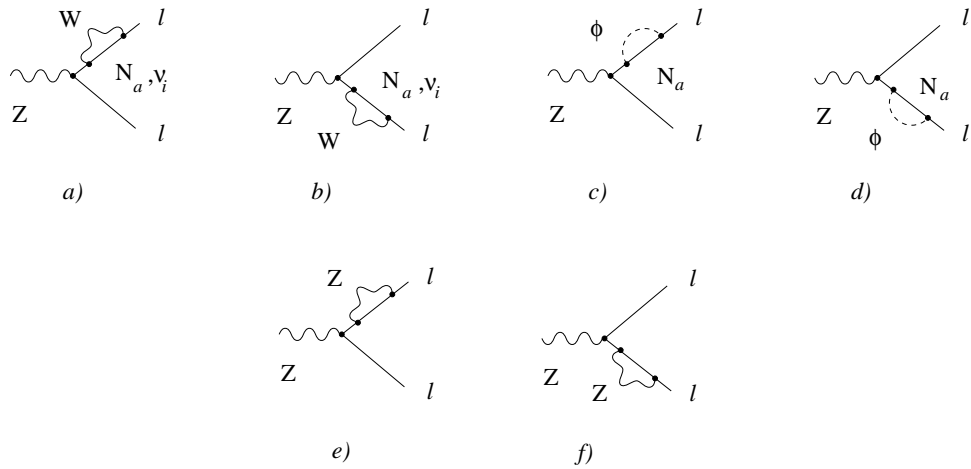}}}
%\put(1.65,0){\footnotesize {\bf Figure \ref{selffd}:} Lepton self-energies.}
\end{picture}
\end{center}
\caption{Lepton self-energies.}
\label{selffd}
\end{figure}

The vertex factor
\begin{eqnarray}
\delta{\hat \Gamma}_{ll} & = & \frac{2}{3} \alpha M_{Z} \Big\{
v_{l}\Big[ {\cal R}e\: {\hat F}_{V}(M_{Z}^{2}) 
- {\hat \Pi}_{\gamma Z}(M_{Z}^{2})\Big] +
a_{l} {\cal R}e\: {\hat F}_{A}(M_{Z}^{2}) \Big\},
\end{eqnarray}
is the source of the lepton universality breaking in $\Gamma_{ll}$. It
includes, besides irreducible
vertex corrections, also lepton wave function renormalization
and the
renormalized mixing energy ${\hat \Sigma}_{\gamma Z}$. The irreducible vertices
(Fig. \ref{vertexfd}) and  lepton wave function corrections (Fig. \ref{selffd})
are absorbed in the
renormalized formfactors ${\hat F}_{V}, {\hat F}_{A}$ while the mixing energy 
(Fig. \ref{pzfd}) comes in as
\begin{eqnarray}
{\hat \Pi}_{\gamma Z}(p^{2}) & = & \frac{{\cal R}e\; 
{\hat \Sigma}_{\gamma Z}(p^{2})}{p^{2}}. 
\end{eqnarray}
As in the case of ${\hat \Sigma}_{Z}$, ${\hat \Sigma}_{\gamma Z}$ (given in
Appendix \ref{recons}) includes besides $\Sigma_{\gamma Z}$ also $\Sigma_{W}$
and
$\Sigma_{Z}$ (through the renormalization constants $\delta Z_{1}^{\gamma Z},
\delta Z_{2}^{\gamma Z}$). This is the source of the $M_{N}$ dependence of 
${\hat
\Sigma}_{\gamma Z}$, in spite of $\Sigma_{\gamma Z}$ being a purely SM quantity.

With $\Sigma_{W}$, $\Sigma_{Z}$ calculated in Sec. \ref{seczprop} we get in 
the limit of
$M_{N} \gg M_{W}$ in the leading order an expression similar to Eq.
\ref{aprox1}:
\begin{eqnarray}
\label{aprox3}
{\hat \Pi}_{\gamma Z}(M_{Z}^{2}) & = & {\hat \Pi}_{\gamma Z}(M_{Z}^{2})^{SM}
- \frac{\alpha}{\pi}\Big\{\frac{c_{W}}{16
s_{W}^{3}} \frac{M_{N}^{2}}{M_{W}^{2}}k_{HH} + O\Big(\ln
\frac{M_{N}^{2}}{M_{W}^{2}}\Big) +
...\Big\}.
\end{eqnarray}

\subsection{Irreducible vertices}
\label{irreduciblev}

\begin{figure}
\begin{center}
\setlength{\unitlength}{1in}
\begin{picture}(6,2)
\put(1.5,+0.325){\mbox{\epsfxsize=2.5in\epsffile{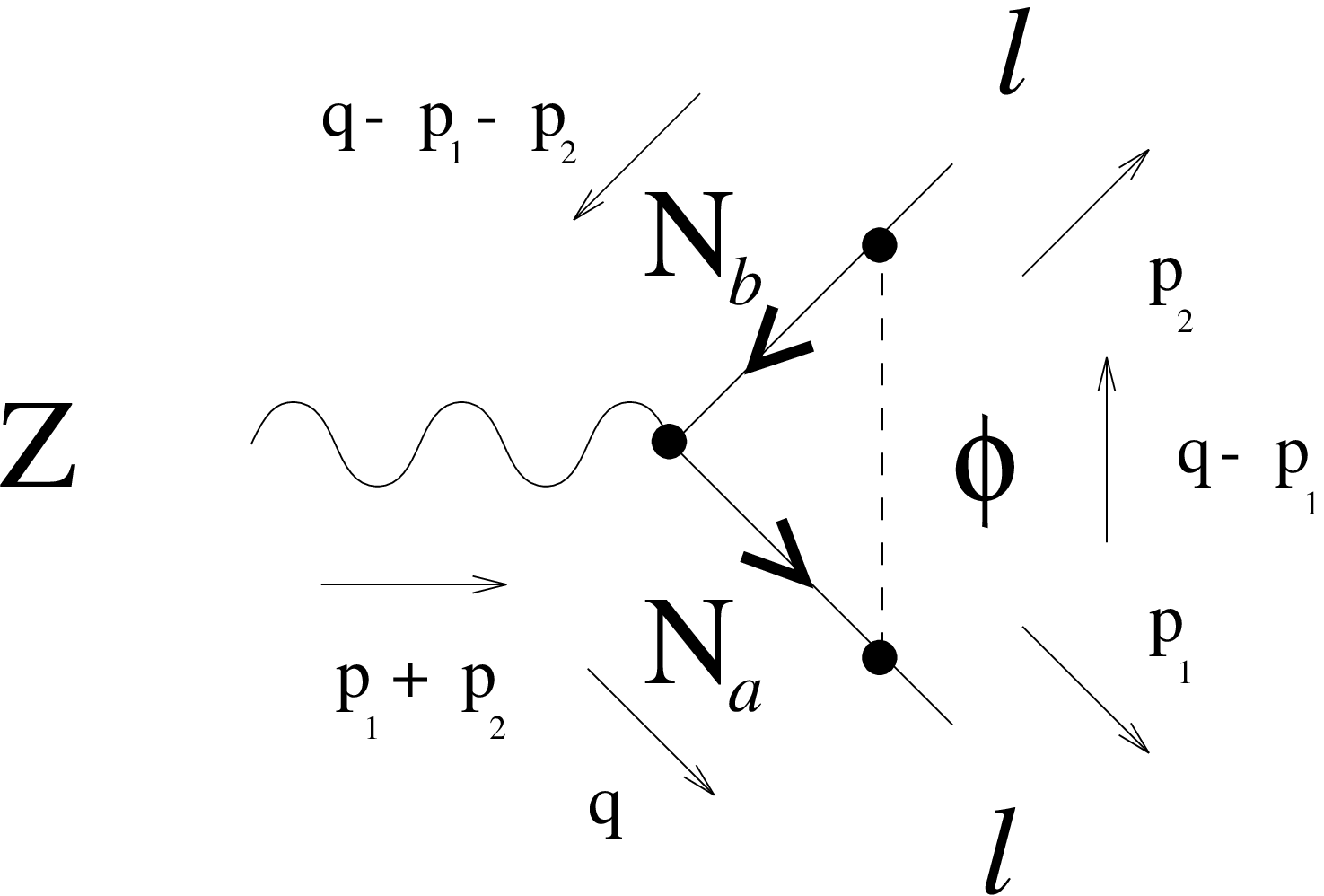}}}
%\put(1.65,0){\footnotesize {\bf Figure \ref{convenfd}:} Momenta and charge flow
%convention.}
\end{picture}
\end{center}
\caption{Momenta and charge flow convention.}
\label{convenfd}
\end{figure}

Here we examine irreducible vertices.
As an example we calculate the contribution to the unrenormalized form factors
$F_{V}, F_{A}$ of the 
diagram of Fig. \ref{vertexfd}f, which we redraw in Fig. \ref{convenfd} 
to show our convention of
momenta flow ($p_{1}, p_{2}, p=p_{1}+p_{2}, q$) and charge flow 
(arrows on internal lines).

\pagebreak
Following the Feynman rules of Appendix \ref{Cecko},
the vertex $V^{\mu}_{NN\phi}$
is given by (terms $m_{l}^{2}/M_{W}^{2}$ neglected)
\begin{eqnarray}
V^{\mu}_{NN\phi} & = & \sum_{a,b} \int \frac{d^{n}q}{(2\pi)^{n}}
\frac{i}{(q-p_{1})^{2}-M_{W}^{2}} \frac{+ig_{2}}{\sqrt{2}M_{W}} \big(K_{H}
\big)_{la}M_{N}\frac{1+\gamma_{5}}{2} \frac{i}{\not q - M_{N}} \nonumber \\
& \times &
\frac{+ie}{4s_{W}c_{W}} \big(K_{H}^{\dagger}K_{H}\big)_{ab}
\gamma_{\mu}\big(1-\gamma_{5}\big)\frac{i}{\not q - \not p_{1} -\not p_{2}
- M_{N}}\frac{+ig_{2}}{\sqrt{2}M_{W}} \big(K_{H}^{\dagger}\big)_{bl}
M_{N} \nonumber \\
& \times &
\frac{1-\gamma_{5}}{2}. 
\end{eqnarray}
Introducing a shorthand $l_{2} \equiv \sum \big(K_{H}\big)_{la}
\big(K_{H}^{\dagger}K_{H}\big)_{ab}\big(K_{H}^{\dagger}\big)_{bl}$,
collecting numerical factors at the front and merging 
$1 \pm \gamma_{5}$ factors we get 
%(${\cal X} = \frac{M_{N}^{2}}{M_{W}^{2}}$) 
\begin{eqnarray}
\label{shorthand}
V^{\mu}_{NN\phi} & = & - \frac{e^{3}}{32
s_{W}^{3}c_{W}} \frac{M_{N}^{2}}{M_{W}^{2}} l_{2}
\nonumber \\
& \times &
 \int \frac{d^{n}q}{(2\pi)^{n}}
\frac{4M_{N}^{2} \gamma^{\mu} (1-\gamma_{5})}
{(q^{2}-M_{N}^{2})
\big[(q-p_{1})^{2}-M_{W}^{2}\big]
\big[(q-p_{1}-p_{2})^{2}-M_{N}^{2}\big]}.\;\;\;
\end{eqnarray}
Now we can identify the integral as a finite $C_{0}$ function (see Eq.
\ref{cnula}): 
\begin{eqnarray}
V^{\mu}_{NN\phi} & = &
- \frac{e^{3}}{8 s_{W}^{3}c_{W}}\frac{M_{N}^{4}}{M_{W}^{2}}l_{2}
\frac{-i \pi^{2}}{(4\pi)^{2}\pi^{2}}
C_{0}(M_{N},M_{W},M_{N}) \gamma^{\mu} (1-\gamma_{5}) \nonumber \\
& = &
+ \:ie \gamma^{\mu} (1-\gamma_{5})\frac{\alpha}{4\pi} l_{2}
\frac{M_{W}^{2}}{8 s_{W}^{3}c_{W}} \frac{M_{N}^{4}}{M_{W}^{4}}
C_{0}(M_{N},M_{W},M_{N}) \nonumber \\
& = &
+ \:ie \gamma^{\mu} (1-\gamma_{5})\frac{\alpha}{4\pi}l_{2}{\cal M}_{NN\phi}.
\end{eqnarray}
The total contribution of the irreducible vertex diagrams and the definition
of the unrenormalized form factors $F_{V}, F_{A}$ is given by
\begin{eqnarray}
V^{\mu} & = & +ie\gamma^{\mu}F_{V} -  ie\gamma^{\mu}\gamma_{5}F_{A} \nonumber \\
& = &
+ie\gamma^{\mu}\frac{\alpha}{4\pi}
\Big\{ l_{1}{\cal M}_{\phi WN} + l_{2}{\cal M}_{NN \phi} +
l_{1}{\cal M}_{\phi \phi N} - (1-l_{1}){\cal M}_{WW\nu} \Big. \nonumber  \\
& + &
\Big. l_{1}{\cal M}_{WWN} + l_{3}{\cal M}_{N\nu W} +
l_{3}{\cal M}_{\nu NW}
+ l_{4}{\cal M}_{\nu \nu W} + l_{2}{\cal M}_{NNW} - \big(c_{L}^{3} +
c_{R}^{3}\big) {\cal M}_{llZ} \Big\}           \nonumber \\
& - &
ie\gamma^{\mu}\gamma_{5}\frac{\alpha}{4\pi}
\Big\{ l_{1}{\cal M}_{\phi WN} + l_{2}{\cal M}_{NN \phi} +
l_{1}{\cal M}_{\phi \phi N} - (1-l_{1}){\cal M}_{WW\nu} + l_{1}{\cal M}_{WWN}
\Big. \nonumber  \\
& + &
\Big. l_{3}{\cal M}_{N\nu W} +
l_{3}{\cal M}_{\nu NW}
+ l_{4}{\cal M}_{\nu \nu W} + l_{2}{\cal M}_{NNW} - \big(c_{L}^{3} -
c_{R}^{3}\big) {\cal M}_{llZ} \Big\},         
\end{eqnarray}
where
\begin{eqnarray}
c_{L} & = & - \frac{1}{2} + s_{W}^{2} ,\;\;\;\;\;\;\;c_{R}\;=\;s_{W}^{2},
\nonumber  \\
{\cal M}_{llZ} & = &
+ \frac{1}{2 s_{W}^{3} c_{W}^{3}} \Big[2M_{Z}^{2}\big(C_{23}(m_{l},M_{Z},m_{l}) 
+ C_{11}(m_{l},M_{Z},m_{l})\big) + 2 \Big. \nonumber  \\
& - &
\Big. 4 C_{24}^{fin}(m_{l},M_{Z},m_{l}) - \Delta_{\mu} \Big] ,
\end{eqnarray}
$\Delta_{\mu}$ is given in Eq. \ref{deltas}
and ${\cal M}_{\phi WN}$, ... were defined before (see Eq. \ref{amplitudes}). 
The mixing factors are obtained from the flavour-violating ones (see 
Eq. \ref{fvmixings}) by setting $l = l^{'}$:
\begin{eqnarray}
\label{elka}
l_{1} & = & \big(K_{H}\big)_{la}\big(K_{H}^{\dagger}\big)_{al} 
\;=\; ll_{mix},      \nonumber         \\
l_{2} & = & \big(K_{H}\big)_{la}\big(K_{H}^{\dagger}K_{H}\big)_{ab}
            \big(K_{H}^{\dagger}\big)_{bl} 
\;=\; |le_{mix}|^{2} + |l\mu_{mix}|^{2} + |l\tau_{mix}|^{2}, \nonumber \\
l_{3} & = & \big(K_{L}\big)_{li}\big(K_{L}^{\dagger}K_{H}\big)_{ia}
            \big(K_{H}^{\dagger}\big)_{al}
 \;=\; l_{1} - l_{2}, \nonumber \\
l_{4} & = & \big(K_{L}\big)_{li}\big(K_{L}^{\dagger}K_{L}\big)_{ij}
            \big(K_{L}^{\dagger}\big)_{jl} 
\;=\; 1 - 2 l_{1} + l_{2}.      
\end{eqnarray}
Renormalized form factors ${\hat F}_{V}, {\hat F}_{A}$ are defined (using 
Eq. \ref{rvertexa}) as
\begin{eqnarray}
{\hat V}^{\mu} & = & +i e \gamma^{\mu} {\hat F}_{V} - i e \gamma^{\mu}
\gamma_{5} {\hat F}_{A} \nonumber \\
& = & + i e \gamma^{\mu} \{F_{V} + v_{l}(\delta Z_{1}^{Z} - \delta Z_{2}^{Z})
- (\delta Z_{1}^{\gamma Z} - \delta Z_{2}^{\gamma Z}) + (v_{l}\:\delta Z_{V}^{l}
+ a_{l}\:\delta Z_{A}^{l})\} \nonumber \\
& - &
i e \gamma^{\mu} \gamma_{5} \{F_{A} + a_{l}(\delta Z_{1}^{Z} - \delta
Z_{2}^{Z}) + (v_{l}\:\delta Z_{A}^{l}
+ a_{l}\:\delta Z_{V}^{l})\}.  
\end{eqnarray}
The check of the cancellation of divergences  in ${\hat V}^{\mu}$ shows that 
the divergence of Fig.~\ref{vertexfd}c is cancelled by those of
Figs. \ref{selffd}c,d and
the divergence of Fig. \ref{vertexfd}g by those of Figs. \ref{selffd}e,f. 
The sum of the remaining
divergences of Figs. \ref{vertexfd}d,e is cancelled by the sum of the 
divergences of
Figs. \ref{selffd}a,b plus those associated with the counterterms 
$\delta Z_{1}^{Z} - \delta Z_{2}^{Z}$ and $\delta Z_{1}^{\gamma Z} - \delta
Z_{2}^{\gamma Z}$ (these come from the bosonic loops of the photon-Z mixing
energy).

\subsection{Lepton self-energies}
\label{seclep}

\begin{figure}[hbtp]
\begin{center}
\setlength{\unitlength}{1in}
\begin{picture}(6,1.3)
\put(.1,+0.325){\mbox{\epsfxsize=3.0in\epsffile{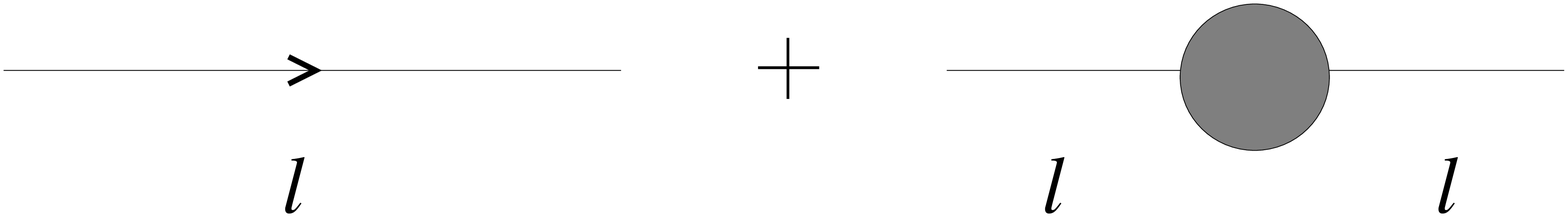}}}
%\put(1.65,0){\footnotesize {\bf Figure 5.1:} QED corrections.}
\put(3.4,0.55){$ \equiv \frac{i}{\not p} + \frac{i}{\not p} \; i \; \Sigma^{l}
\; \frac{i}{\not p} \;\; = \;\; \frac{i}{\not p} - \frac{i}{\not p} \;
 \Sigma^{l} \; \frac{1}{\not p},$}
\end{picture}
\end{center}
\caption{The definition of the lepton self-energy $\Sigma^{l}$}
\label{lepself}
\end{figure}

We define lepton self-energies $\Sigma^{l}$
at the one-loop level as shown in Fig. \ref{lepself}
, i.e., we call the blob $i\:\Sigma^{l}$, rather than $- i\:\Sigma^{l}$
(Hollik's convention).
$\Sigma^{l}$ has the form
\begin{eqnarray}
\label{hasthe}
\Sigma^{l}(p) & = & {\not p} \Sigma_{V}^{l}(p^{2}) + {\not p} \gamma_{5}
\Sigma_{A}^{l}(p^{2}) + m_{l} \Sigma_{S}^{l}(p^{2})  \nonumber \\
& = &
{\not p} \frac{1-\gamma_{5}}{2} \Sigma_{L}^{l}(p^{2}) + {\not p} 
\frac{1+\gamma_{5}}{2}
\Sigma_{R}^{l}(p^{2}) + m_{l} \Sigma_{S}^{l}(p^{2}),
\end{eqnarray}
where $\Sigma_{V,A,S,L,R}^{l}(p^{2})$ are vector, axial vector, scalar,
left-handed and right-handed parts respectively.
Renormalized charged lepton self-energies ${\hat \Sigma^{l}}$ (see Eq.
\ref{rselfe})
do not contribute to the renormalized vector and axial
vector formfactors
${\hat F}_{V}, {\hat F}_{A}$. Indeed, the graphs
of Fig. \ref{selffd} give zero after the on-shell renormalization conditions
(see Eq.~$\ref{RC1}$) are applied. However, the unrenormalized lepton
self-energies $\Sigma^{l}$ do contribute to the formfactors 
${\hat F}_{V}, {\hat F}_{A}$ through the
renormalization constants $\delta Z_{V}^{l}, \delta Z_{A}^{l}$.
The relation between these renormalization constants and $\Sigma^{l}$ can be
found in Eq.~\ref{rconstants}.

The charged lepton self-energy is the sum of four parts
corresponding to loops with $W-N, W-\nu, \phi-N$ and $Z-l$ respectively 
(see Fig. \ref{selffd}): 
%***********lepton selfenergy ***********************
\begin{eqnarray}
\Sigma^{l}(p) & = & \Sigma_{WN}^{l} + \Sigma_{W\nu}^{l} + \Sigma_{\phi N}^{l} +
\Sigma_{Zl}^{l},
\end{eqnarray}
where 
\begin{eqnarray}
\label{leptonself1}
\Sigma_{WN}^{l} & = & - \frac{\alpha}{32\pi s_{W}^{2}}l_{1}
[1- 2 \Delta_{\mu} + 2 \ln M_{W}^{2}
+ 2 f({\cal X})] \not p (1-\gamma_{5}), \nonumber  \\
\Sigma_{W\nu}^{l} & = & - \frac{\alpha}{32\pi s_{W}^{2}}(1-l_{1})
[1 - 2 \Delta_{\mu} + 2 \ln M_{W}^{2}] \not p (1-\gamma_{5}), \nonumber \\
\Sigma_{\phi N}^{l} & = &  + \frac{\alpha}{32\pi s_{W}^{2}}l_{1} {\cal X}
[ \Delta_{\mu} + \frac{1}{2} - \ln M_{W}^{2}  - f({\cal X})]
\not p (1-\gamma_{5}), \nonumber \\
\Sigma_{Zl}^{l} & = & - \frac{\alpha}{16\pi s_{W}^{2}c_{W}^{2}}c_{L}^{2}
[ 1 - 2 \Delta_{\mu} +
2 \ln M_{Z}^{2} ] \not p (1-\gamma_{5}) \nonumber
\\
& - &
\frac{\alpha}{16\pi s_{W}^{2}c_{W}^{2}}c_{R}^{2}
[ 1 - 2 \Delta_{\mu} +
2 \ln M_{Z}^{2} ] \not p (1+\gamma_{5}), \nonumber \\
%\end{eqnarray}
%where (see also Eq. \ref{calex})
%\begin{eqnarray}
f({\cal X}) & = & \frac{{\cal X}^{2}}{({\cal X}-1)^{2}}\ln {\cal
X} + \frac{{\cal X}}{1-{\cal X}}, \;\;\;\;\;\;\; {\cal
X} \; = \; \frac{M_{N}^{2}}{M_{W}^{2}}. 
\end{eqnarray}
%$\Sigma_{V,A,S,L,R}^{l}(p^{2})$ parts of
These expressions are evaluated at 
$p^{2} =
m_{l}^{2}$ as required by $\delta Z_{V}^{l}, \delta Z_{A}^{l}$ (see Eq.
\ref{rconstants}), and terms $\frac{m_{l}^{2}}{M_{W}^{2}}$ were neglected.
Note the $\Sigma_{Zl}^{l}$ part is a pure SM result.
For completeness, we also give $\Sigma_{\gamma l}^{l}$:
\begin{eqnarray}
\label{leptonself2}
\Sigma_{\gamma l}^{l} & = & \frac{\alpha}{4\pi}\big[\Delta_{\mu} - 1 
+ 2 B_{0}^{fin}(p;m_{\lambda},m_{l}) + 2 B_{1}^{fin}(p;m_{\lambda},m_{l})\big]
\not p \nonumber \\
& - & \frac{\alpha}{\pi}\big[\Delta_{\mu} - \frac{1}{2} +
B_{0}^{fin}(p;m_{\lambda},m_{l})\big] m_{l}.
\end{eqnarray}
Here $m_{\lambda}$ is the regularized photon mass and 
$B_{0}^{fin}, B_{1}^{fin}$ (evaluated at $p^{2} = m_{l}^{2}$) 
are given in Eq. \ref{bphoton}.
$\Sigma_{\gamma l}^{l}$ is, in this case, part of the QED subset treated
independently of genuine electroweak corrections (see Sec. \ref{secqed}).
However, in Chapter 7, this photonic
correction will be included as a part of the total lepton self-energy and
the counterterm $\delta Z_{L}^{l}$ together with the genuine electroweak
corrections.

\subsection{Form factors ${\hat F}_{V}, {\hat F}_{A}$ in the limit of large NHL
mass}
\label{secform}

Studying the behaviour of ${\hat F}_{V}, {\hat F}_{A}$ in the limit of 
$M_{N} \gg M_{W}$, we observe in agreement with Sec. \ref{arit} the three 
sources of
quadratic nondecoupling: ${\cal M}_{N N \phi}, {\cal M}_{\phi \phi N}$ 
and ${\cal M}_{\phi N}$ (which contains $\Sigma^{l}_{\phi N}$).
The quadratic nondecoupling of these amplitudes is also expected from 
dimensional analysis considerations. For illustration, 
for ${\cal M}_{N N \phi}$ we get from Eq. \ref{shorthand}
\begin{eqnarray}
{\cal M}_{N N \phi} & \sim & M_{N}^{4} \int \frac{d^{n}q}{(q^{2}-M_{N}^{2})
\big[(q-p_{1})^{2}-M_{W}^{2}\big]
\big[(q-p_{1}-p_{2})^{2}-M_{N}^{2}\big]}. 
\end{eqnarray}
Setting n=4 and neglecting all masses and momenta except $M_{N}$, we obtain
\begin{eqnarray}
{\cal M}_{N N \phi} & \sim & M_{N}^{4} \int \frac{d^{4}q}{(q^{2}-M_{N}^{2})
q^{2} (q^{2}-M_{N}^{2})}.
\end{eqnarray}
The integral has to be of the form $(M_{N})^{k}$; the power counting yields
$k = -2$, so
\begin{eqnarray}
{\cal M}_{N N \phi} & \sim & M_{N}^{4} M_{N}^{-2} \; = \; M_{N}^{2}.
\end{eqnarray}
The exact result is (see Eqs. \ref{amplitudes}, \ref{aproxc})
\begin{eqnarray}
{\cal M}_{N N \phi} & = & +
\frac{1}{8s_{W}^{3}c_{W}}\frac{M_{N}^{2}}{M_{W}^{2}} + ... \; .
\end{eqnarray}

As mentioned in Sec. \ref{fvzb1}, the contribution of ${\cal M}_{\phi \phi N}$
cancels
with that of ${\cal M}_{\phi N}$, 
leaving ${\cal M}_{N N \phi}$ as the only 
amplitude with the nondecoupling behaviour. We will now shed more light on this
curious cancellation.
 The way to go is to replace in the diagrams corresponding to these amplitudes
the Z boson with the photon and to use 
a Ward-identity \cite{key6}, which 
relates the vertex formfactors $F_{V,A}^{\gamma}$ evaluated at 
$(p_{1}+p_{2})^{2}=0$ 
to electron self-energies
represented by the counterterms $\delta Z_{V,A}$:
\begin{eqnarray}
\label{wardi}
F_{V,A}^{\gamma}(0) + \delta Z_{V,A} & = & \frac{1}{4s_{W}c_{W}}
\frac{\Sigma_{\gamma Z}(0)}{M_{Z}^{2}},
\end{eqnarray}
where $\Sigma_{\gamma Z}(0)$ is the term originating in the bosonic loops of
the $\gamma$-Z mixing.
At zero $M_{N}$ the graphs with unphysical Higgs $\phi$ are negligible,
however, with $M_{N}$ rising the two graphs dominate the left-hand side of Eq.
\ref{wardi}: the 
vertex ${\cal M}_{\phi \phi N}^{\gamma}$ and the self-energy ${\cal M}_{\phi
N}^{\gamma}$.
Since the right-hand side of Eq. \ref{wardi} is not
affected by the NHL's, it remains constant and (very) small with respect to
${\cal M}_{\phi \phi N}^{\gamma}$ or ${\cal M}_{\phi N}^{\gamma}$ 
at $M_{N} = O$(TeV). 
Hence the only way to meet the
above formula is to have ${\cal M}_{\phi N}^{\gamma} = - {\cal M}_{\phi \phi
N}^{\gamma}$
in the limit of large $M_{N}$.

If we now return from the photon to the Z boson, it suffices to check 
how the Feynman rules for the vertices change. It turns out that
\begin{eqnarray}
{\cal M}_{\phi N} \; \equiv \; {\cal M}_{\phi N}^{Z}  
                     & = & c {\cal M}_{\phi N}^{\gamma},  \\
{\cal M}_{\phi \phi N} \; \equiv \; {\cal M}_{\phi \phi N}^{Z}
                     & = & c {\cal M}_{\phi \phi N}^{\gamma},
\end{eqnarray}
with the same constant c (a simple function of Weinberg angle) in both eqs.,
so we conclude that
\begin{eqnarray}
     {\cal M}_{\phi N} & = & - {\cal M}_{\phi \phi N}.
\end{eqnarray}
Note that for the photon case the form factors $F_{V,A}^{\gamma}$ are
evaluated at $(p_{1}+p_{2})^{2}=0$, while for the Z case $F_{V,A}$ are
calculated
at $(p_{1}+p_{2})^{2} = M_{Z}^{2}$; this is acceptable, 
since in the limit of a large $M_{N}$ the Z mass can be neglected.

This cancellation implies that the formfactors are dominated by the single
amplitude ${\cal M}_{N N \phi}$:
\begin{eqnarray}
\label{aprox4}
{\hat F}_{V} \; = \; {\hat F}_{A} & = & \frac{\alpha}{4\pi} l_{2} 
\frac{1}{8s_{W}^{3}c_{W}}{\cal X} + ... \; .
\end{eqnarray}
As in the case of the oblique parameter ${\hat \Pi}_{Z}$, we have to carefully
examine the reliability of this approximation.
We note again the leading term is suppressed by the mixing parameter 
$l_{2} = \tau\tau_{mix}^{2}$
\footnote{Assuming $ee_{mix} = \mu\mu_{mix} = 0$ and $Z \rightarrow \tau\tau$
mode}. On the other hand, some of the higher order terms
(found in many
of the irreducible vertex diagrams, not just ${\cal M}_{N N \phi}$) are
only linear in $\tau\tau_{mix}$, which implies they can be larger than expected
on the basis of an $M_{N}$ expansion.   In Table \ref{t2} we show numerical 
predictions
for the (exact) formfactor ${\hat F}_{V}$ and compare them
with the approximate parameter ${\hat F}_{Vappx} = {\hat F}_{V}^{SM} +
{\hat F}_{Vappx}^{NHL}$ where ${\hat F}_{Vappx}^{NHL}$ is the leading
term in Eq. \ref{aprox4}.
The dependence of ${\hat F}_{V}^{SM}$ on the NHL mass has its origin in a
different value of the input parameter $M_{W}$  (as calculated from $G_{\mu}$,
see Sec. \ref{secimp}) for different NHL masses.
Input numbers used are $M_{Z} = 91.1884$~GeV, $M_{H} = 200$ GeV, $m_{t} = 176$
GeV, $\tau\tau_{mix} = 0.033$ and $ee_{mix} = \mu\mu_{mix} = 0$.
%All numbers in the table with the exception of masses
%are to be multiplied by $10^{-3}$.

\begin{table}[htb]
%\vspace{1.75in}
\begin{center}
\begin{tabular}{|l|r|r|r|r|r|} \hline
%\multicolumn{5}{|c|}{Table 5.2.  Comparison of ${\hat F}_{Vappx}$ with
%${\hat F}_{V}$ } \\ \hline
$M_{N}$                       & 0.5 TeV  &  1 TeV  &  3 TeV &  5 TeV & \\
  \hline
${\hat F}_{V}^{SM}$         & 1.938  & 1.939  & 1.949 & 1.971 & 
                                                      $\times 10^{-3}$ \\
${\hat F}_{V}$              & 2.056  & 2.247  & 3.525 & 5.903 & 
                                                      $\times 10^{-3}$ \\
${\hat F}_{Vappx} = {\hat F}_{V}^{SM} + {\hat F}_{Vappx}^{NHL}$
                              & 1.971  & 2.071  & 3.150 & 5.345 &
                                                      $\times 10^{-3}$ \\
${\hat F}_{Vappx}^{NHL}$    &  0.033  &  0.132 & 1.201 &  3.374 &
                                                      $\times 10^{-3}$ \\
$d = {\hat F}_{V} - {\hat F}_{Vappx}$
                              & 0.085   & 0.176  & 0.375 & 0.558 &
                                                      $\times 10^{-3}$ \\
\hline
\end{tabular}
\end{center}
\caption{Comparison of ${\hat F}_{Vappx}$ with ${\hat F}_{V}$ }
\label{t2}
\end{table}

For $M_{N} = 0.5$ TeV, ${\hat F}_{Vappx}^{NHL}$ contributes $0.033
\times~10^{-3}$ of the total difference between ${\hat F}_{V}^{SM}$ and
${\hat F}_{V}$. It is less than the contribution of the higher order terms, 
$d = 0.085 \times
10^{-3}$.  However, at $3$ TeV the roles are switched, with ${\hat
F}_{Vappx}^{NHL}$ dominating as expected.

Overall, ${\hat F}_{Vappx}$ differs from ${\hat F}_{V}$ by approximately
$4 \%$ at $0.5$ TeV and by $9.5 \%$ at $5$ TeV.  This is worse than
the $1 \%$ found for the ${\hat \Pi}_{Z}$ parameter and it signals that 
the terms linear
in mixing are more important in this case. Indeed, the second largest
non-SM contribution to ${\hat F}_{V}$ comes from the graph of Fig. 
\ref{vertexfd}d 
(associated with amplitude ${\cal M}_{WWN}$). 
It is given by (see Eqs. \ref{aproxm}, \ref{aproxc})
\begin{eqnarray}
-\frac{\alpha}{4 \pi}l_{1}\frac{3c_{W}}{4s_{W}^{3}}\Big(\frac{5}{6}  - \ln
M_{N}^{2}\Big),
\end{eqnarray}
and even at $5$ TeV it is as large as 
$2.02 \times 10^{-3}$ compared to ${\hat F}_{Vappx}^{NHL} = 3.374 \times
10^{-3}$. Other terms are also relatively large, partly cancelling the ${\cal
M}_{WWN}$ effect to produce the comparatively small difference 
of $9.5 \%$.
                                
\section{Imprecise $M_{W}$, precise $G_{\mu}$}
\label{secimp}

Our on-shell renormalization scheme takes $\alpha, M_{Z}$ and $M_{W}$ as input
parameters (see Sec. \ref{onshell1}). However, the direct measurement 
of $M_{W}$ 
\cite{mw1},
\begin{eqnarray}
M_{W} & = & 80.410 \pm 0.180 \; {\rm GeV},
\end{eqnarray}
as opposed to that of $M_{Z}$ \cite{mt1},
\begin{eqnarray}
M_{Z} & = & 91.1884 \pm 0.0022 \; {\rm GeV},
\end{eqnarray}
is not yet precise enough for its use as an input parameter. To appreciate
this point, let us examine the sensitivity of $\Gamma_{ll}$ to
$M_{W}$. It mainly comes from the tree-level formula (loops are
suppressed by factors of $\alpha$)
\begin{eqnarray}
\Gamma_{0} & = & \frac{\alpha}{3} M_{Z}(v_{l}^{2} + a_{l}^{2})
\;=\;\frac{\alpha}{3} M_{Z}\frac{1-4s_{W}^{2}+8s_{W}^{4}}{8s_{W}^{2}c_{W}^{2}},
\end{eqnarray}
where $c_{W}^{2} = M_{W}^{2}/M_{Z}^{2}$.
It turns out that $\sigma_{M_{W}} = 0.18$ GeV induces $\sigma_{\Gamma_{0}}
\doteq \sigma_{\Gamma_{ll}}$ $\doteq~1$~MeV. 
This is very large compared to the experimental value
(see Eq. \ref{gamaexp})
\begin{eqnarray}
\Gamma_{ll}^{exp} & = & 83.93 \pm 0.14 \; {\rm MeV}. \nonumber
\end{eqnarray}
As a consequence $M_{W}$, even though convenient as an input parameter
for one-loop calculations, is usually replaced by the more precisely
measured muon decay constant $G_{\mu}$ \cite{pdb} :
\begin{eqnarray}
G_{\mu} & = & 1.16637 \; (2)\times 10^{-5} \; {\rm GeV}^{-2}.
\end{eqnarray}
The replacement is done in such a way that while
$M_{W}$ is still kept in our formulae for one-loop self-energies and vertices,
its actual value is no longer taken from the direct measurement, but is rather
calculated from $G_{\mu}$.
To calculate $M_{W}$ from $G_{\mu}$, 
one first computes the $\mu$-decay rate to one-loop level in the Fermi model 
(this defines
$G_{\mu}$) and then equates it to the one-loop calculation of the same quantity
in the SM \cite{key6}. The result is a formula relating $M_{W}$ to $G_{\mu}$:
\begin{eqnarray}
\label{MGSM}
%M_{W}^{2} s_{W}^{2} & = & \frac{\pi\alpha}{\surd 2 G_{\mu} (1-\Delta r)}
%(1-\frac{1}{2}ee_{mix}-\frac{1}{2}\mu\mu_{mix})
M_{W}^{2} s_{W}^{2} & = & \frac{\pi\alpha}{\sqrt 2 G_{\mu} (1-\Delta r^{SM})},
\end{eqnarray}
with $\Delta r^{SM}$ (the notation $\Delta r$ is reserved for our model)
representing loop effects in $\mu$-decay,
\begin{eqnarray}
\Delta r^{SM} & = & \Delta r^{SM}(\alpha,M_{W},M_{Z},M_{H},m_{t}) \nonumber \\
& = &
\frac{{\cal R}e \:{\hat \Sigma}_{W}^{SM}(0)}{M_{W}^{2}} + 
\frac{\alpha}{4\pi s_{W}^{2}}
\Big(6 + \frac{7-4s_{W}^{2}}{2s_{W}^{2}} \ln c_{W}^{2}\Big) \nonumber \\
& = & \frac{{\cal R}e \:{\hat \Sigma}_{W}^{SM}(0)}{M_{W}^{2}} +
\delta_{V}^{SM}.
\end{eqnarray}
$M_{W}$ is found from Eq. \ref{MGSM} by iterative procedure.

The reason we discuss this in some detail is that our model modifies
this relation between $G_{\mu}$ and $M_{W}$ both at tree and one-loop level
and thus NHL's, 
besides contributing directly to $\Gamma_{ll}$ through
Z-decay loops, contribute also indirectly via $M_{W}$.
This complication in the calculation of $\Gamma_{ll}$ is something we can
actually benefit from since in $M_{W}$ we obtained another observable sensitive
to NHL's.
In our model, Eq.~\ref{MGSM} is modified as
\begin{eqnarray}
\label{MG}
M_{W}^{2} s_{W}^{2} & = & \frac{\pi\alpha}{\sqrt{2} G_{\mu} (1-\Delta r)}
(1-\frac{1}{2}ee_{mix}-\frac{1}{2}\mu\mu_{mix}),
\end{eqnarray}
where $1-\frac{1}{2}ee_{mix}-\frac{1}{2}\mu\mu_{mix}$ represents the tree-level
modification 
\footnote{$\Gamma_{\mu}$ is modified by
$1-ee_{mix}-\mu\mu_{mix}$ and we take the square root of $\Gamma_{\mu}$ to
get Eq. \ref{MG}.}.
The loop quantity $\Delta r$ is calculated from the diagrams depicted in Figs.
\ref{boxmuon} - \ref{vermuon} and the diagrams with the corrected W 
propagator. 
We devote the next chapter to the detailed 
calculation of the $\Delta r$ for the arbitrary values of the mixings
$ll_{mix}$. Here we make a specific choice of mixings which will help 
us to avoid most of the $\mu$-decay loop 
diagrams: we put
$ee_{mix} = \mu\mu_{mix} = 0$. This choice also implies $e\mu_{mix} = 
e\tau_{mix} =
\mu\tau_{mix} = 0$ and leaves $\tau\tau_{mix}$ as the only nonzero mixing
parameter. Of the non-SM $\mu$-decay loops, the vertex corrections,
neutrino self-energy corrections and the boxes are all proportional to either
$ee_{mix}$ or $\mu\mu_{mix}$ and therefore only ${\hat \Sigma}_{W}$
(see Sec. \ref{seczprop}), the $W$
propagator correction remains to modify $\Delta r^{SM}$:
\begin{eqnarray}
\label{delrv}
\Delta r & = & \frac{{\cal R}e \:{\hat \Sigma}_{W}(0)}{M_{W}^{2}} + \delta_{V}
\; = \; \frac{{\cal R}e \:{\hat \Sigma}_{W}(0)}{M_{W}^{2}} + \delta_{V}^{SM}.
\end{eqnarray}
In the limit of $M_{N} \gg M_{W}$ we obtain an expression similar to Eq.
\ref{aprox1},
\begin{eqnarray}
\label{aprox2}
\Delta r & = & \Delta r^{SM} -
\frac{\alpha}{\pi}\Big\{\frac{c_{W}^{2}}{16
s_{W}^{4}} k_{HH} \frac{M_{N}^{2}}{M_{W}^{2}} + O(\ln M_{N}^{2}/M_{W}^{2}) +
...\Big\}.
\end{eqnarray}
The neglect of terms beyond the $O(M_{N}^{2}/M_{W}^{2})$ results in an error
less than $2.5 \%$ for $M_{N} < 5$ TeV.

\section{Violation of the decoupling theorem}
\label{appelc}

The reader may wonder whether NHL loop effects can have any noticeable effect at
all. After all, these loops are suppressed by factors such as $ll_{mix}$ or even
$ll_{mix}^{2}$ (see for example Eq. \ref{aprox2}, where $k_{HH} =
\tau\tau_{mix}^{2}$) compared to their SM 
counterparts.
Although this reasoning is correct, it is not complete. In spite of the 
smallness of the
mixing parameters, the NHL loop effects can actually be larger than SM ones.
The reason is the (possibly) large NHL mass and a violation of the decoupling
theorem.

The decoupling theorem was established and proven by Appelquist and Carazzone
in Ref. \cite{ac}. It describes how the heavy particles of a renormalizable
theory $A$ enter into the low-energy theory $B$ \cite{dono}:
{\it  All effects of
the heavy particle in the low-energy theory B appear either as a 
renormalization of the coupling constants
or else are suppressed by powers of the heavy particle mass.}
For instance, heavy $W$ and $Z$ bosons of the SM (theory $A$) decouple from the
low-energy QED (theory~$B$).

\begin{figure}
\begin{center}
\setlength{\unitlength}{1in}
\begin{picture}(6,1.5)
\put(.5,+0.225){\mbox{\epsfxsize=4.5in\epsffile{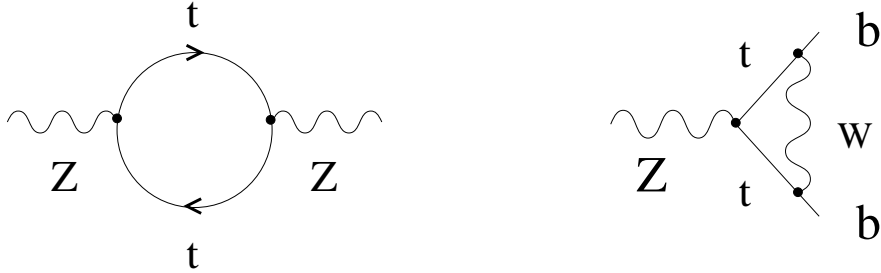}}}
%\put(1.65,0){\footnotesize {\bf Figure \ref{toblique}:} Diagrams with 
%the top quark nondecoupling.}
\end{picture}
\end{center}
\caption{Diagrams with the top quark nondecoupling.}
\label{toblique}
\end{figure}

There are, however, cases in spontaneously broken theories when the
decoupling theorem is violated.
A well-known example can be found in the SM itself with respect to
top quark
behaviour. There are two diagrams (Fig. \ref{toblique}) which exhibit a 
quadratic dependence
on the top quark mass and therefore do not vanish as $m_{t} 
\rightarrow \infty$. 

The
nondecoupling effects are easily visible in this case - the diagrams of Fig.
\ref{toblique}
led to indirect bounds on the top quark mass from LEP I observables. The
bounds, $m_{t} = 170 \pm 10 {}^{+17}_{-19} \;
{\rm GeV}\:$ \cite{mt1}, are actually competitive 
with the value obtained so far from the 
direct observation at CDF at Fermilab, $m_{t} = 176 \pm 8 \pm 10 \;
{\rm GeV}\:$ \cite{mt2}.

%As noted above, the decoupling violation is linked to the non-renormalizability
%of the
%low-energy theory \cite{dono}: after integrating-out the top quark, the
%$b$ quark is missing its partner in the doublet, hence the $SU(2)_{L}$ symmetry
% is
%broken and the theory cannot be renormalized. A similar thing happens in our
%model. NHL's are part of a lepton doublet via their contribution (albeit small)
%to the left-handed weak eigenstate $\nu_{L} = K_{L}\nu_{i} + K_{H}N_{a}$ and,
%after integrating them out, the remaining theory is not renormalizable.
%Indeed, looking back at our results, we can see the quadratic nondecoupling in
%Eqs. \ref{aprox1}, \ref{aprox3}, \ref{aprox2}.
The NHL's also exhibit nondecoupling as can be seen in
Eqs. \ref{aprox1}, \ref{aprox3},
\ref{aprox2}, where the dominant terms are quadratically dependent on $M_{N}$.
We note however, that the nondecoupling of NHL's is the consequence of our 
treatment of the mixings as the parameters independent of $M_{N}$. If we 
replaced
$\tau\tau_{mix}$ with $\tau\tau_{mix} \sim \frac{D^{2}}{M_{N} ^{2}}$ ($K_{H}
\sim \frac{D}{M_{N}}$), the dominant terms would change from $\sim
\tau\tau_{mix}^{2} M_{N}^{2}$ (see Eqs. \ref{aprox1}, \ref{aprox3},
\ref{aprox2}) to
\begin{eqnarray}
\tau\tau_{mix}^{2} M_{N}^{2} & \rightarrow & \frac{D^{4}}{M_{N} ^{4}} 
M_{N}^{2} \; = \; \frac{D^{4}}{M_{N} ^{2}},
\end{eqnarray}
and the decoupling would be recovered.

The analogy with the top quark can be useful in another aspect. We can make a
naive estimate of how large an NHL mass $M_{N}$ should be to produce an effect
comparable to that of the top quark. For example, oblique corrections rise 
with the top quark mass as ${\hat \Pi}_{Z}^{SM} \sim m_{t}^{2}/M_{W}^{2}$,
\footnote{According to Okun \cite{okun}, this point is subtle:
a large positive contribution from the top quark is cancelled by a large
negative contribution from all other virtual particles. As a result, genuine
electroweak corrections are negligible and the bulk of $ \sim 2.5$ MeV loop
corrections is associated with the running coupling constant $\alpha$.}
while in our model, the dependence on $M_{N}$ (besides the dependence on
$m_{t}$) is
\begin{eqnarray}
{\hat \Pi}_{Z} & \sim & k_{HH} \frac{M_{N}^{2}}{M_{W}^{2}} \; \doteq\;
\tau\tau_{mix}^{2} \frac{M_{N}^{2}}{M_{W}^{2}},
\end{eqnarray}
yielding
\begin{eqnarray}
m_{t}^{2} & \sim & \tau\tau_{mix}^{2}M_{N}^{2}.
\end{eqnarray}
As a result, for $\tau\tau_{mix} = 0.033$, $M_{N} \doteq 30\; m_{t} \;\doteq\; 5
$ TeV is required to produce 'the top size' effect. We note this is not the
necessary condition for observing the NHL given the current mixings. As our
numerical results indicate, even NHL's with the mass $M_{N} \sim 3$ TeV  could
lead to non-SM effects in $\Gamma_{ll}$. In the same sense
we do not have to change the top mass by $175$ GeV to see conflict with the
data.

\section{Heavy NHL's and perturbation theory breakdown}
\label{breakdown}

With one-loop corrections rising as $M_{N}^{2}/M_{W}^{2}$, we have to know
at what value of $M_{N}$ these corrections are comparable
in size to the tree-level contribution. At this point the theory becomes
strongly interacting and the perturbative treatment fails. 
Alternatively, we can study the transition to the
strongly interacting regime through the tree-level width of the NHL. We
investigate this latter approach here 
and show the analogy to the the strongly interacting Higgs.

\begin{figure}
\begin{center}
\setlength{\unitlength}{1in}
\begin{picture}(6,1.75)
\put(.3,+0.325){\mbox{\epsfxsize=5.0in\epsffile{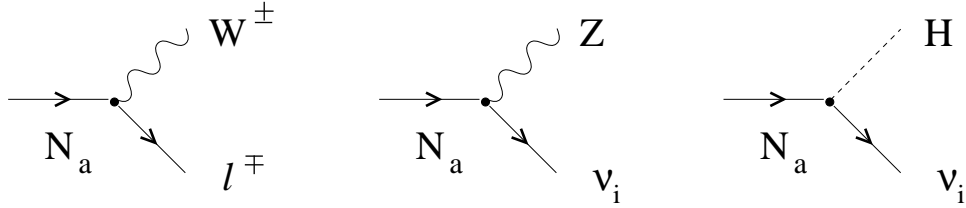}}}
%\put(1.65,0){\footnotesize {\bf Figure \ref{nhldecay}:} Decay modes of an NHL}
\end{picture}
\end{center}
\caption{Decay modes of an NHL.}
\label{nhldecay}
\end{figure}

There are three decay modes of an NHL,
$N_{a} \rightarrow W^{\pm} + l^{\mp}, N_{a} \rightarrow Z + \nu_{i}$ and
$N_{a} \rightarrow H + \nu_{i}$, open for
$M_{N} > M_{W}, M_{Z}, M_{H}$. They are represented by the diagrams of 
Fig. \ref{nhldecay}.
In the limit $M_{N}  \gg  M_{W}, M_{Z}, M_{H}$,
the partial decay widths are given by \nopagebreak 
\footnote{The partial width for 
$N_{a} \rightarrow W^{\pm} + l^{\mp}$ is the same as in the see-saw model
of Ref. \cite{pilaftsis2}, while the widths for
$N_{a} \rightarrow Z + \nu_{i}$ and
$N_{a} \rightarrow H + \nu_{i}$ are half of their see-saw counterparts.}
%\begin{eqnarray}
%\sum_{l=e,\mu,\tau}\Gamma\;(N_{a} \rightarrow W^{\pm} + l^{\mp}) & = & 
%\frac{\alpha}{16 s_{W}^{2}}
%\sum_{l}|\left(K_{H}\right)_{la}|^{2}\frac{M_{N}^{3}}{M_{W}^{2}}
%{\Big(1-\frac{M_{W}^{2}}{M_{N}^{2}}\Big)}^{2}
%\Big(1+\frac{2M_{W}^{2}}{M_{N}^{2}}\Big), \nonumber \\
%\sum_{i=1,2,3}\Gamma\;(N_{a} \rightarrow H + \nu_{i}) & = &
%\frac{\alpha}{32 s_{W}^{2}}\sum_{i} |\left(K_{L}^{\dagger}K_{H}\right)_{ia}|^{2}
%\frac{M_{N}^{3}}{M_{W}^{2}}{\Big(1-\frac{M_{H}^{2}}{M_{N}^{2}}\Big)}^{2},
%\nonumber \\
%\sum_{i=1,2,3}\Gamma\;(N_{a} \rightarrow Z + \nu_{i}) & = &
%\frac{\alpha}{32 s_{W}^{2}}\sum_{i} |\left(K_{L}^{\dagger}K_{H}\right)_{ia}|^{2}
%\nonumber \\
%& \times &
%\frac{M_{N}^{3}}{M_{W}^{2}}{\Big(1-\frac{M_{Z}^{2}}{M_{N}^{2}}\Big)}^{2}
%\Big(1+\frac{2M_{Z}^{2}}{M_{N}^{2}}\Big).   
%\end{eqnarray}
\begin{eqnarray}
\sum_{l=e,\mu,\tau}\Gamma\;(N_{a} \rightarrow W^{\pm} + l^{\mp}) & = &
\frac{\alpha}{16 s_{W}^{2}}
a_{mix}\frac{M_{N}^{3}}{M_{W}^{2}}, \nonumber \\
\sum_{i=1,2,3}\Gamma\;(N_{a} \rightarrow H + \nu_{i}) & = &
\frac{\alpha}{32 s_{W}^{2}}
a_{mix}
\frac{M_{N}^{3}}{M_{W}^{2}}, \nonumber \\
\sum_{i=1,2,3}\Gamma\;(N_{a} \rightarrow Z + \nu_{i}) & = &
\frac{\alpha}{32 s_{W}^{2}}a_{mix}
\frac{M_{N}^{3}}{M_{W}^{2}},
\end{eqnarray}
where the following relations were used:
\begin{eqnarray}
\sum_{i} |\left(K_{L}^{\dagger}K_{H}\right)_{ia}|^{2} & \doteq &
\sum_{l} |\left(K_{H}\right)_{la}|^{2} \; = \; a_{mix}.
\end{eqnarray}
The total tree-level width of the NHL is 
the sum of the partial widths:
\begin{eqnarray}
\Gamma_{N} & = & \frac{3\alpha}{16 s_{W}^{2}}
a_{mix}\frac{M_{N}^{3}}{M_{W}^{2}}.
\end{eqnarray}
For a comparison, the tree-level width of a very heavy Higgs is given by
\begin{eqnarray}
\Gamma_{H} & = & \frac{3\alpha}{32 s_{W}^{2}}
\frac{M_{H}^{3}}{M_{W}^{2}}.
\end{eqnarray}
Both $\Gamma_{N}$ and $\Gamma_{H}$ rise swiftly with the particle mass and
at some critical point they become larger than the masses $M_{N}, M_{H}$ 
themselves - a clear
indication of a perturbative breakdown. The tree-level formulae are no longer
appropriate. We can get a safe estimate on the
critical mass by demanding that
\begin{eqnarray}
\Gamma_{N,H} & \leq & \frac{1}{2}M_{N,H}.
\end{eqnarray}
With $a_{mix} \leq ee_{mix} + \mu\mu_{mix} + \tau\tau_{mix} \doteq
\tau\tau_{mix} = 0.033$, we obtain
\begin{eqnarray}
M_{N}^{crit} & \sim & 4 \; {\rm TeV},
\end{eqnarray}
and for the Higgs we get the well-known bound of $M_{H}^{crit} \sim  1$ TeV
\footnote{A similar value is obtained from considerations of perturbative
unitarity violation.}.
Note that $M_{N}^{crit}$ is not to be interpreted as the upper bound on the NHL
mass. Its sole purpose is to make us aware of the limitations of the 
perturbative treatment.

\section{Results}
\label{seresu}

In this section we present our numerical results. 

Our FORTRAN program is written as a modification of the routines from the
CERN electroweak library \cite{cernlib} used for the SM predictions of
LEP I parameters. This applies namely to the oblique parameter, ${\hat
\Pi}_{Z}$, with most of the contributing diagrams (see Fig. \ref{pfd}
 - \ref{wfd})
being SM. The non-SM contributions were implemented as specified in
Sec. \ref{seczprop}. The vertex factor $\delta{\hat \Gamma}_{ll}$ was 
derived independently 
of the CERN electroweak library, including the SM diagrams.

\begin{figure}
\begin{center}
\setlength{\unitlength}{1in}
\begin{picture}(6,6.5)
\put(.1,+0.4){\mbox{\epsfxsize=5.0in\epsffile{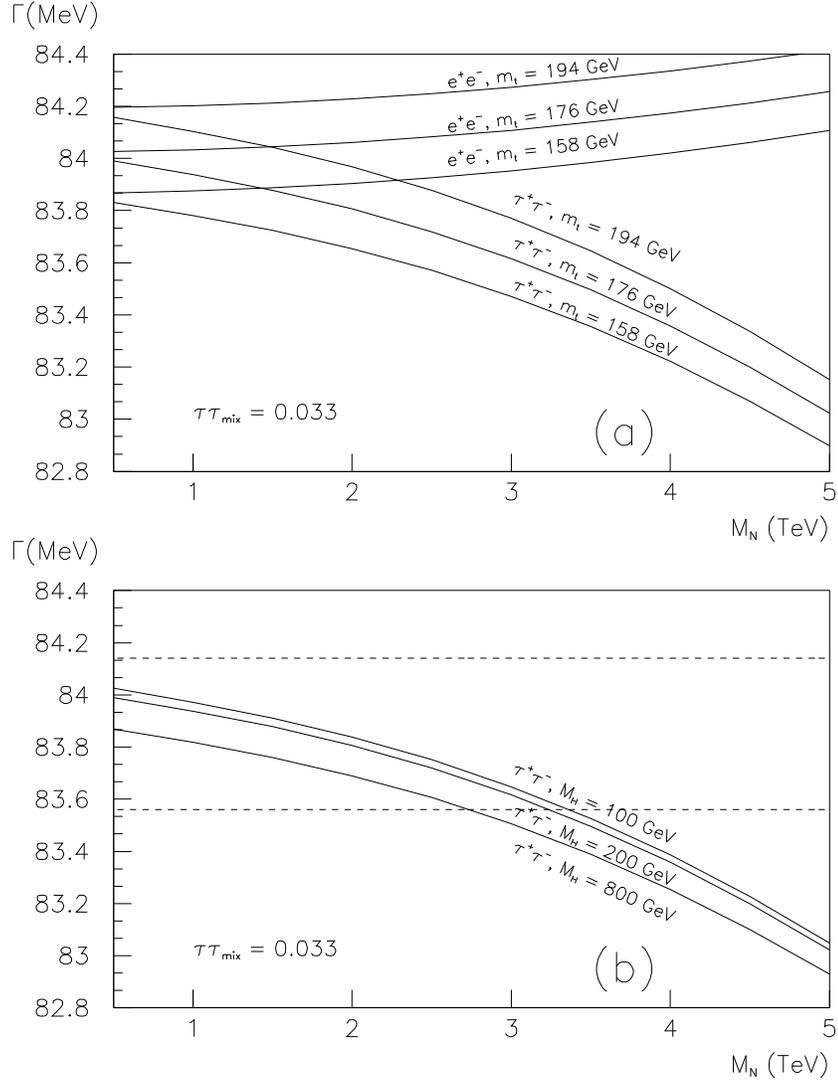}}}
\end{picture}
\end{center}
\caption{Z leptonic width as a function of
$M_{N}$ in the interval $0.5$ TeV $\leq M_{N} \leq 5$~TeV
for (a) fixed mixing parameter $\tau\tau_{mix}$, fixed Higgs mass and different
 values of $m_{t}$; 
both $Z \rightarrow \tau\tau$ and $Z \rightarrow ee$ modes shown,
(b) fixed mixing parameter $\tau\tau_{mix}$, fixed $m_{t}$ and different 
values of the Higgs mass $M_{H}$; only $Z \rightarrow \tau\tau$ mode shown.
The dashed lines
represent $1 \sigma$ band about the current experimental value
$\Gamma_{\tau\tau}^{exp} =
83.85 \pm 0.29$~MeV.}
\label{numw1}
\end{figure}

\begin{figure}
\begin{center}
\setlength{\unitlength}{1in}
\begin{picture}(6,6.5)
\put(.1,+0.4){\mbox{\epsfxsize=5.0in\epsffile{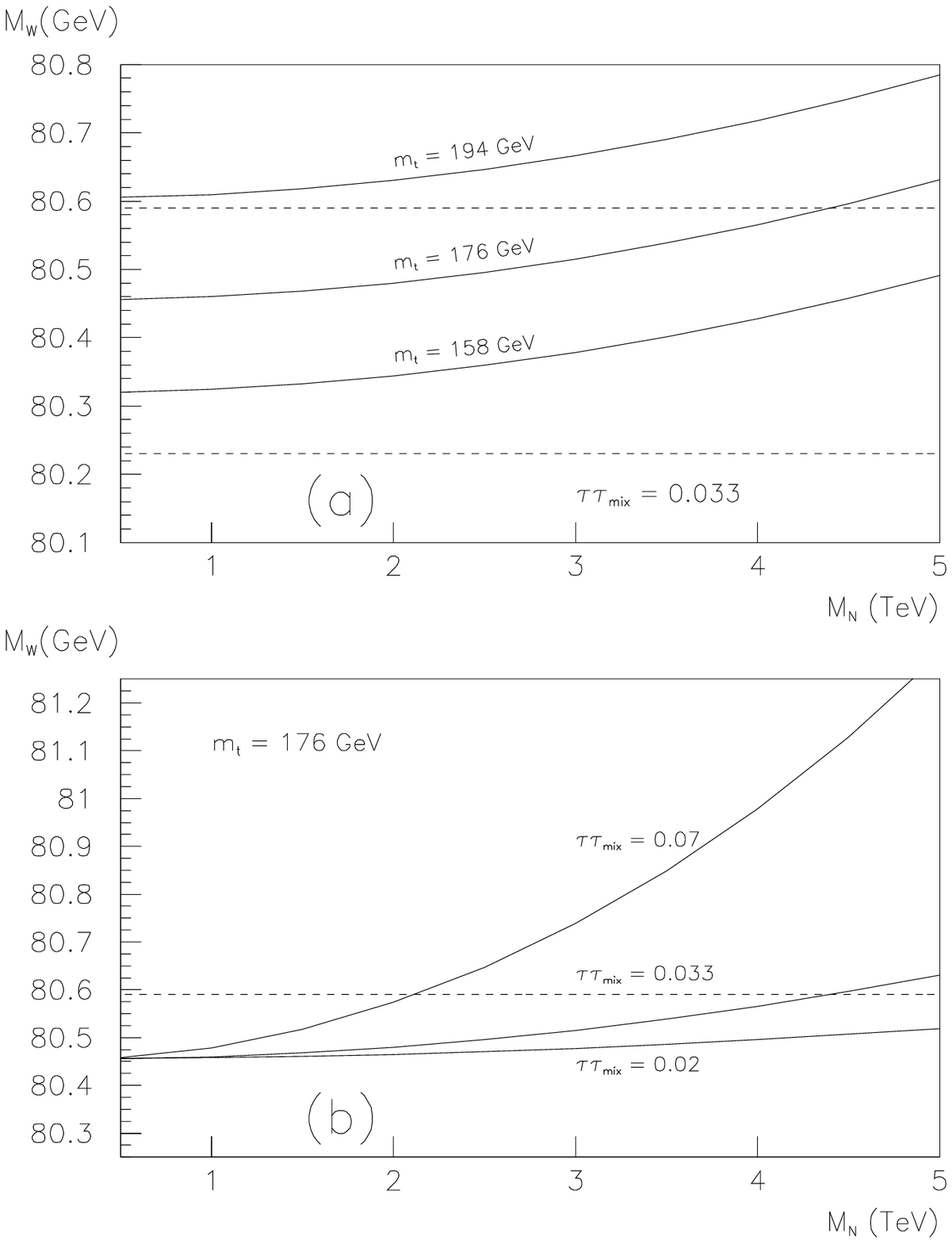}}}
\end{picture}
\end{center}
\caption{W mass as a function of $M_{N}$
for (a) fixed mixing parameter ($\tau\tau_{mix}=0.033$) and different values of 
$m_{t}$, (b) fixed top quark mass
($m_{t}=176$ GeV) and different values of the mixing parameter. The dashed
lines represent $1 \sigma$ band about the current experimental 
value $M_{W} =
80.410 \pm 0.180$ GeV.}
\label{nummw}
\end{figure}

As a 'standard set' of input parameters, we used
$M_{Z} = 91.1884$ GeV, 
$\alpha ^{-1} = 137.036$,
$A \equiv \frac{\pi \alpha}{\sqrt{2} G_{\mu}} = 37.281 \: {\rm GeV}$.
Also part of this set are $M_{H} =  200$ GeV and $m_{t} = 176$ GeV. Below we 
only make it explicit when values different from the standard set are used,
e.g., $M_{H}, m_{t}$ in Figs. \ref{numw1}, \ref{nummw}a. 
To keep the discussion simple (without losing any qualitative features), we
have reduced the free parameter space by
assuming degenerate masses for the three NHL's.
In this chapter, we have also imposed restrictions on the mixing parameters.
We assume
that $ee_{mix}$ and $\mu \mu_{mix}$ are very small relative to $\tau
\tau_{mix}$. The model and NHL mass independent limits quoted in Eq.
\ref{limits1} are
more stringent for $e$ and $\mu$ than for $\tau$. In addition, our assumption
is also partially
\footnote{Partially, because small $e\mu_{mix}$ does not necessarily imply
small $ee_{mix}$ and $\mu \mu_{mix}$.}
supported by the smallness of $e\mu_{mix}$ (see Eq. \ref{limits3}), 
as determined from
$\mu \rightarrow e \gamma$, in combination with the inequality Eq. \ref{ineq}.
This
neglect of $ee_{mix}$ and $\mu \mu_{mix}$ proves useful practically in that
many of the muon decay loops (boxes and vertex corrections, but not $W$ oblique
correction) are eliminated as a result.
The general case of arbitrary $ee_{mix}$, $\mu \mu_{mix}$ is considered in the
next chapter.
We present results mainly for the NHL mass range $0.5 \: {\rm TeV} \leq M_{N}
 \leq 5$
TeV, as motivated
by the non-decoupling arguments given in Sec. \ref{appelc}.

Given the relative complexity of the formulae involved, we looked for
alternative ways of checking them, other than the special care
taken in their derivation. 

First, a logical step to take is to run our program
with all mixing parameters zero, to see if it reduces to the SM, as predicted
by the CERN electroweak library. For the standard set of input parameters we
get precise agreement, see Table~\ref{t3}.

\begin{table}[bthp]
%\vspace{1.75in}
\begin{center}
\begin{tabular}{|l|c|c|c|} \hline
%\multicolumn{4}{|c|}{Table 5.3. SM limit of our model } \\ \hline
$   $     & ${\hat \Pi}_{Z}^{SM}$  &  $\delta{\hat \Gamma}_{ll}^{SM}$ & 
$\Gamma_{ll}^{SM}$ [MeV] \\
  \hline
CERN library      & -4.29769 $\times 10^{-2}$ & -1.16976 $\times
10^{-3}$  & 
                84.0297   \\
Our program &  -4.29769 $\times 10^{-2}$  &  -1.16976 $\times
10^{-3}$ & 
                84.0297  \\
\hline
\end{tabular}
\end{center}
\caption{SM limit of our model}
\label{t3}
\end{table}

%Results agree exactly for the ${\hat \Pi}_{Z}^{SM}$, and quite well for
%$\delta{\hat \Gamma}_{ll}^{SM}$ and $\Gamma_{ll}^{SM}$. 
%The small difference
%for the latter is understandable, since $\delta{\hat \Gamma}_{ll}^{SM}$
%was derived independently of the CERN library.
% and therefore subject to
%different round-off errors.

Second, we checked that infinities cancelled out in all renormalized
quantities. 
%This implies that those multiplicative factors in finite parts 
%which are
%shared between finite and infinite parts, are sensible. For example, in the
%first line of Eq. \ref{amplitudes}, it is the factor
%$\frac{-\frac{1}{2}+s_{W}^{2}}{4 s_{W}^{3}c_{W}}$.

Third, throughout this chapter we tried to separate the dominating
contributions in the limit of large NHL mass and represent them by simple
formulae. Here, we collect these approximations and see, if we can understand
the behaviour of partial leptonic widths on their basis.
Thus, collecting Eqs. \ref{oneloop}, \ref{aprox1}, \ref{aprox3} and
\ref{aprox4}, we get
\begin{eqnarray}
\label{aprox5}
\Gamma_{ll} & = & \frac{\Gamma_{0}
 + \delta{\hat \Gamma_{ll}}}{1+{\hat \Pi}_{Z}
(M_{Z}^{2})}(1+\delta_{QED}) \; \doteq \; \Gamma_{ll}^{appx} \nonumber \\
 & = & 
\frac{\Gamma_{0} + \delta{\hat \Gamma_{Z}}^{SM} + \frac{2}{3}\alpha M_{Z}
\big(v_{l}{\hat F}_{Vappx}^{NHL} + a_{l}{\hat F}_{Vappx}^{NHL} - v_{l}
{\hat \Pi}_{\gamma Zappx}^{NHL}\big)}
{1 + {\hat \Pi}_{Z}^{SM} + {\hat \Pi}_{Zappx}^{NHL}}
\nonumber \\
& \times & (1+ \delta_{QED}), 
\end{eqnarray}
where
\begin{eqnarray}
{\hat \Pi}_{Zappx}^{NHL} & = &
\frac{\alpha}{\pi}\frac{c_{W}^{2}-s_{W}^{2}}{16
s_{W}^{4}}\frac{M_{N}^{2}}{M_{W}^{2}}k_{HH}, \nonumber \\
{\hat \Pi}_{\gamma Zappx}^{NHL} & = &
- \frac{\alpha}{\pi}\frac{c_{W}}{16
s_{W}^{3}} \frac{M_{N}^{2}}{M_{W}^{2}}k_{HH}, \nonumber \\
{\hat F}_{Vappx}^{NHL} & = &
\frac{\alpha}{4\pi} 
\frac{1}{8s_{W}^{3}c_{W}}\frac{M_{N}^{2}}{M_{W}^{2}} l_{2} . 
\end{eqnarray}
The third, fourth and the fifth term in the numerator in $\Gamma_{ll}^{appx}$
are all increasingly
negative with rising $M_{N}$, therefore they make the partial width smaller.
The denominator rises with $M_{N}$ and also makes the partial width smaller.
On the other hand, the tree-level result, $\Gamma_{0}$, whose indirect
dependence on NHL's via $M_{W}$ was discussed in Sec. \ref{secimp}, rises with 
$M_{N}$ (see Table \ref{t4}).  As a result, depending on the particular values
of the mixing parameters $l_{2}$ and $k_{HH}$, the partial width can either
 increase with NHL mass or decrease. 
An example of the latter case can be found in Table \ref{t4} which shows 
(for the standard set of input parameters, $\tau\tau_{mix}
= 0.033$ and the above assumptions on $ee_{mix}, \mu\mu_{mix}$) the $Z
\rightarrow \tau\tau$ mode partial width $\Gamma_{\tau\tau}$ for different 
NHL masses. Also shown
are $\Gamma_{0}$ and $\Gamma_{\tau\tau}^{appx}$, the prediction of Eq.
\ref{aprox5}.

\begin{table}[bthp]
%\vspace{1.75in}
\begin{center}
\begin{tabular}{|l|r|r|r|r|} \hline
%\multicolumn{5}{|c|}{Table 5.4.  $\Gamma_{0}$ and $\Gamma_{\tau\tau}$ as a
%function of $M_{N}$ } \\ \hline
$M_{N}$                       & 0.5 TeV  &  1 TeV  &  3 TeV &  5 TeV  \\
  \hline
$\Gamma_{0}$ [MeV]         & 81.43  & 81.45  & 81.82 & 82.60  \\
$\Gamma_{\tau\tau}$ [MeV]  &  83.99  &  83.94 & 83.62 & 83.02  \\
$\Gamma_{\tau\tau}^{appx}$ [MeV]
                              & 84.00  & 83.97  & 83.70 & 83.18  \\
\hline
\end{tabular}
\end{center}
\caption{$\Gamma_{0}$ and $\Gamma_{\tau\tau}$ as a function of $M_{N}$ }
\label{t4}
\end{table}

Once we can predict partial leptonic widths, we can take advantage of that
and study the universality breaking parameter defined as 
\cite{bernabeu2}
\begin{eqnarray}
U_{br} & = &
\displaystyle \left| \frac{\Gamma_{\tau\tau} -
\Gamma_{ee}}
{\Gamma_{\tau\tau} + \Gamma_{ee}}
              \right| \; = \;  \left|
\frac{\delta{\hat \Gamma_{\tau\tau}} -
\delta{\hat \Gamma_{ee}}}
{2 \Gamma_{0} + \delta{\hat \Gamma_{\tau\tau}} + \delta{\hat \Gamma_{ee}}}
              \right|.
\end{eqnarray}
The new feature here is that the universality breaking parameter depends only 
on the vertex factor $\delta{\hat \Gamma_{ll}}$; the indirect influence of 
$M_{W}$ via $\Gamma_{0}$ is suppressed and the direct oblique corrections 
cancell out in the ratio.
Thus $U_{br}$ represents an independent and complementary quantity to the
partial leptonic widths for the study of the NHL's. The experimental limit
on the universality breaking parameter can be derived from the limits on
partial leptonic widths as follows. From the limits on partial leptonic widths
 (see Eq. \ref{pwidths}) we get
\begin{eqnarray}
\frac{\Gamma_{\tau\tau}}{\Gamma_{ee}} & = & 0.9991 \pm 0.0040.
\end{eqnarray}
This ratio implies
\begin{eqnarray}
U_{br}^{exp} & \doteq & \displaystyle \left| \frac{\Gamma_{\tau\tau} -
\Gamma_{ee}}
{\Gamma_{ee} + \Gamma_{ee}}
              \right| \; = \;
 \left| \frac{\Gamma_{\tau\tau}}{
2 \Gamma_{ee}} - \frac{1}{2}          \right| \; = \; 0.00045 \pm 0.00200,
\end{eqnarray}
leading to the upper limit (at $2 \sigma$ level)
\begin{eqnarray}
U_{br}^{exp} & < & 0.00445.
\end{eqnarray}

The third experimental parameter sensitive to NHL's is the W mass, $M_{W}$. 
Under the
assumptions made above it is only sensitive to $W$ oblique corrections and thus
complements the former two quantities.

\begin{figure}
\begin{center}
\setlength{\unitlength}{1in}
\begin{picture}(6,6.5)
\put(.1,+0.4){\mbox{\epsfxsize=5.0in\epsffile{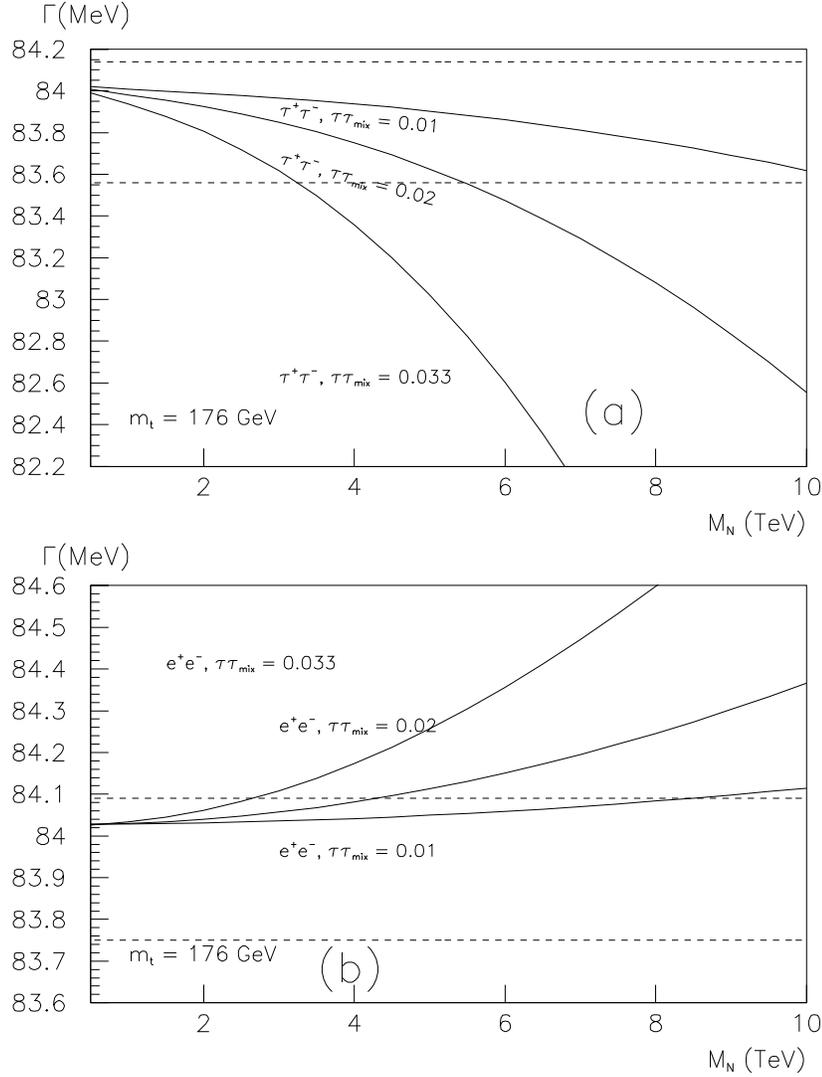}}}
\end{picture}
\end{center}
\caption{Z leptonic width as a function of
$M_{N}$ in the interval $0.5$ TeV $\leq M_{N} \leq 5$~TeV for fixed
$m_{t}$, fixed Higgs mass and different values of the mixing parameter,
(a) $Z \rightarrow \tau\tau$ mode,
(b) $Z \rightarrow ee$ mode.
The dashed lines
represent $1 \sigma$ band about the current experimental value 
(a)$\Gamma_{\tau\tau}^{exp} = 83.85 \pm 0.29$ MeV,
(b)$\Gamma_{ee}^{exp} = 83.92 \pm 0.17$~MeV.}
\label{numw2}
\end{figure}

\begin{figure}
\begin{center}
\setlength{\unitlength}{1in}
\begin{picture}(6,3)
\put(.1,+0.325){\mbox{\epsfxsize=5.0in\epsffile{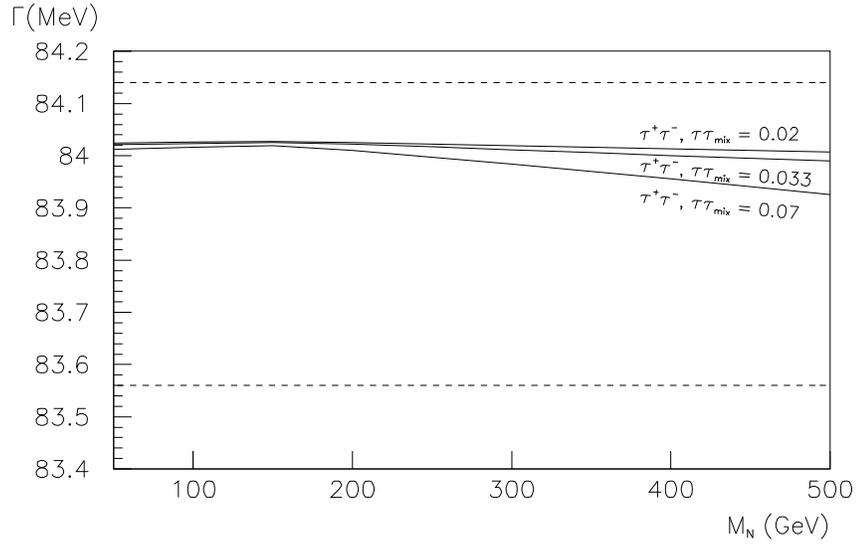}}}
\end{picture}
\end{center}
\caption{$Z \rightarrow \tau\tau$ width as a function of
$M_{N}$ in the interval $50$ GeV $\leq M_{N} \leq 500$~GeV
for a fixed top mass $m_{t} = 176$ GeV and different values of the mixing
parameter $\tau\tau_{mix}$. The dashed lines
represent the $1 \sigma$ band about the current experimental value
$\Gamma_{\tau\tau}^{exp} = 83.85 \pm 0.29$ MeV.}
\label{numw3}
\end{figure}

The $Z$ leptonic widths are given as a function of NHL mass in Figs.
\ref{numw1} - \ref{numw3}. We mostly show $\Gamma_{\tau\tau}$, as
the most NHL-sensitive mode under the stated assumptions on mixing
parameters.  $\Gamma_{ee} = \Gamma_{\mu\mu}$ is also plotted in Figs.
\ref{numw1}a, \ref{numw2}b . The dashed lines represent the $1\sigma$
variation about the current experimental results for the individual $Z$ 
leptonic widths $\Gamma_{\tau \tau}^{exp}$ or $\Gamma_{ee}^{exp}$, see
Eq. \ref{pwidths}.

In Fig. \ref{numw1}a, the widths $\Gamma_{ee}, \Gamma_{\tau\tau}$ are
shown for three values of the top
quark mass, $m_{t} = 158, 176$ and $194$ GeV. 
The mixing was fixed at $\tau\tau_{mix} =
0.033$.  
The striking difference
between the two modes can be seen to arise from the competition between
the $\Gamma_{0}$ on one side and the $\delta{\hat \Gamma_{Z}}$ and ${\hat
\Pi}_{Z}$ on the other side (see the discussion above). For the $\tau\tau$
mode, the mixing parameter $l_{2} = k_{HH} = \tau\tau_{mix}^{2}$, while for
the $ee$ mode, $k_{HH} = \tau\tau_{mix}^{2}$ and $l_{2} = 0$ (see Eq.
\ref{elka}).
Zero parameter $l_{2}$ eliminates the third and the fourth terms in the
numerator in the second line of Eq.
\ref{aprox5} and the rising $\Gamma_{0}$ dominates as a result.
In Fig. \ref{numw1}b, 
we again set $\tau\tau_{mix} = 0.033$
and show the $Z$ width to $\tau^{+}\tau^{-}$
for three values of the Higgs mass, $M_{H} = 100, 200, 800$ GeV.  
The dependence on the top quark mass and the weak dependence on the Higgs mass
got transferred from the SM as expected.

In Fig. \ref{numw2} we vary the mixings 
from $\tau\tau_{mix} = 0.02$ to $\tau\tau_{mix} = 0.07$. Fig.~\ref{numw2}a
shows $\Gamma_{\tau\tau}$, Fig. \ref{numw2}b  shows $\Gamma_{ee}$. 
The quadratic dependence on the mixing parameter, anticipated from Eq.
\ref{aprox5}, is confirmed.

Fig. \ref{numw3} is the only figure where we look at a 'lower energy' 
scale. It is a
continuation of Fig. \ref{numw2}a to the lower NHL mass range. As
expected, the partial width is well within the $1 \sigma$ band and the NHL
effects are negligible.

\begin{figure}
\begin{center}
\setlength{\unitlength}{1in}
\begin{picture}(6,7)
\put(.1,+0.325){\mbox{\epsfxsize=5.0in\epsffile{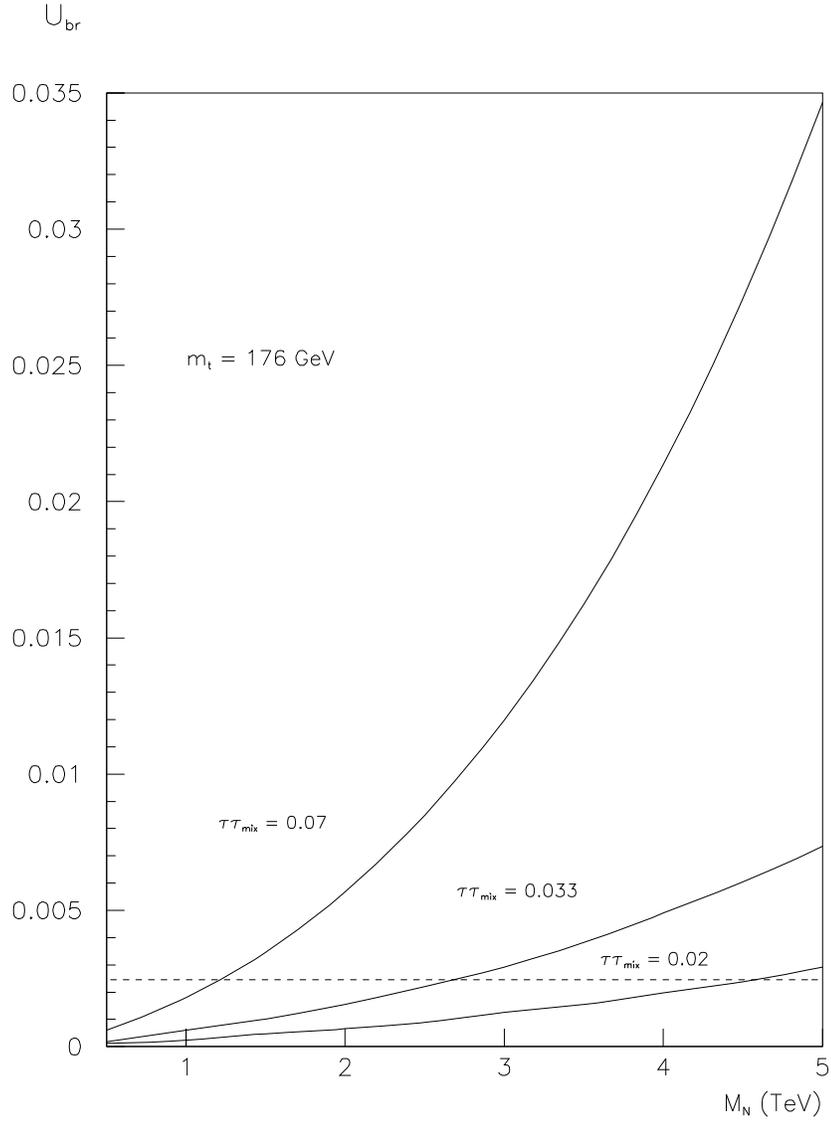}}}
\end{picture}
\end{center}
\caption{Universality breaking parameter
$U_{br}$ as a function of $M_{N}$ for fixed top quark mass ($m_{t}=176$ GeV)
and different values of the
mixing parameter. The dashed line represents $1 \sigma$ experimental limit ($<
0.00245$).}
\label{numubr}
\end{figure}

The universality breaking factor $U_{br}$ is plotted in Fig. \ref{numubr}
as a function
of $M_{N}$, with the mixing parameter varied about $\tau\tau_{mix} =
0.033$.
The $1\sigma$ experimental limit on $U_{br}$ is indicated as the dashed line.

Finally, we present the NHL mass dependence of the $W$ mass in Figs.
\ref{nummw}a,b.
The top quark mass is varied about $m_{t} = 176$ GeV in Fig. \ref{nummw}a, 
while the mixing is held 
constant at
$\tau\tau_{mix} = 0.033$. In Fig. \ref{nummw}b, the mixing is varied about
$\tau\tau_{mix} = 0.033$ for a fixed top quark mass.

%\begin{figure}
%\begin{center}
%\setlength{\unitlength}{1in}
%\begin{picture}(6,7)
%\put(.1,+0.325){\mbox{\epsfxsize=5.0in\epsffile{fviol.eps}}}
%\end{picture}
%\end{center}
%\caption{The branching ratio $Z \rightarrow l_{1}^{\pm}l_{2}^{\mp}$ 
%as a function
%of $M_{N}$
%for (a) $\delta = -1$, (b) $\delta = +1$ 
%.}
%\label{fviolfig}
%\end{figure}

\section{Discussion and Conclusions}\label{conc}

Our primary consideration here has been the inclusion of neutral heavy
leptons in the calculation of the flavour-conserving $Z$ decays to
charged leptons at one-loop level. The dependence of the $Z$ leptonic
widths on the NHL mass, $M_N$, and on the mixing parameter $\tau \tau_{mix}$
 was given in Figs. \ref{numw1} - \ref{numw3}.
We see for the experimentally
allowed upper limit of $\tau \tau_{mix} = 0.033$, and assuming 
$ee_{mix} = \mu\mu_{mix} = 0, m_{t} = 176$~GeV, 
the $Z$ decay width to $\tau$ leptons becomes
sensitive to NHL masses of about $4.3$~TeV at the $2\sigma$ level.
Curiously, this is an upper limit,
\begin{equation}
\label{upper1}
M_{N}  \leq  4.3 \; {\rm TeV},
\end{equation}
rather than the lower one. The cause is the nondecoupling of the NHL's.

Apart from this comparison of each
leptonic width prediction with experiment we can also exploit the lepton flavour
universality violation which takes place in the model. The universality
breaking ratio, $U_{br}$ (see Fig. \ref{numubr}), leads to a yet better
upper limit, 
\begin{equation}
\label{upper2}
M_{N}  \leq  3.8 \; {\rm TeV},
\end{equation}
at the $2\sigma$ level.  
The importance of $U_{br}$ is underlined by the fact that 
it is sensitive only to the vertex parameter $\delta{\hat \Gamma_{ll}}$,
unlike $\Gamma_{ll}$, which besides the vertex parameter also depends on 
the Z oblique corrections and indirectly, via the $W$ boson mass, on
the W oblique corrections. Thus the universality breaking complements
the Z leptonic partial widths as far as sensitivity to NHL's is concerned.

The $W$ boson mass also exhibits some sensitivity to NHL parameters
arising from the mixing factor modifications and the presence of one-loop
diagrams containing NHL's, as described in Sec. \ref{secimp}.
From Figs. \ref{nummw}a,b we see that the sensitivity of the $W$ mass, 
currently measured as
$M_{W} = 80.410 \pm 0.180$ GeV \cite{mw1}, to NHL's
depends to a large degree on the top mass. The experimental error on $M_W$
might be expected
to come down to about $0.05$ GeV once LEP II measures $W$ pair
production \cite{mw2}.

%Current data from LEP I on $Z$ leptonic widths and the present Collider
%Detector at Fermilab (CDF) and
%${\rm D}\emptyset$ collaboration
%measurements of $M_W$ are sensitive to NHL masses greater than about
%2.5 -- 3.5 TeV. With the accumulation of about 60
%$({\rm pb})^{-1}$ at LEP I
%in 1994 and the prospect of the very precise $W$ mass measurement at
%LEP II, these sensitivities will certainly be improved considerably.

In conclusion to this chapter, the Z boson flavour-conserving leptonic widths 
along with the lepton
universality breaking parameter and the W mass represent some of the best 
quantities  sensitive to the NHL mass. This applies especially to NHL's that 
are far too heavy to be produced directly at present colliders.  The only way to
probe their mass in this case is via their loop contributions.
Much of the previous effort on NHL studies has been so far concentrated on 
flavour-violating processes,
either at very low energies ($\mu, \tau$ decays, see Eq. \ref{fvdecay3}),
or at $Z$-peak energy ($Z$ decays, see Eq. \ref{fvz3} and Sec. \ref{fvzb1}). 
We feel there are
at least two reasons which give the processes studied in this thesis a
distinct advantage over the flavour-violating decays.
                                
First, we were able to actually reduce the allowed region in the 
mixings - NHL mass
parameter space here (see Eqs. \ref{upper1}, \ref{upper2}) using the 
{\it current} experimental data. 
The only flavour-violating process that competes with the limits of Eqs.
\ref{upper1}, \linebreak \ref{upper2} is the decay $\mu \rightarrow e e e$, 
which sets an upper
limit on NHL mass (see Eqs. \ref{mueee4}, \ref{mnlimit4}, assuming
$ee_{mix} = 0.0071, |e\mu_{mix}| = 0.00024$) at $2 \sigma$ level
\begin{equation}
M_{N}  \leq  3 \;{\rm TeV}.
\end{equation}
The flavour-violating decay rates for $\tau$ and $Z$ are below the current 
experimental
sensitivity (see Secs. \ref{numeres}, \ref{fvple}). 
%This is caused to a large degree by the quartic dependence
%of the dominant contribution to these rates on the small mixing parameters
%(see, e.g., Eqs.~\ref{quartica4}, \ref{quarticb4})
%\footnote{, as opposed to the
%quadratic dependence of the oblique and vertex corrections to $\Gamma_{ll},
%U_{br}$ and $M_{W}$.

Second, the inequality Eq. \ref{ineq} can further suppress the 
flavour-violating processes against the flavour-conserving ones 
via the 'conspiracy of the phases' in the sum of complex
terms making up the flavour-violating parameters $ll^{'}_{mix}$. In this case, 
flavour-conserving processes are the only way to probe very heavy NHL's.

\newpage

\chapter{Muon decay and W mass}
\label{chap7}

Here we generalize our analysis from the previous chapter by considering the
case of arbitrary mixings
$ee_{mix}, \mu\mu_{mix}$ and $\tau\tau_{mix}$. In this case non-SM
box, vertex and
self-energy diagrams contributing to the muon decay (see Figs. \ref{boxmuon}
- \ref{vermuon}) may become important for the calculation of $M_{W}$.
This is in contrast to the previous chapter
where, as a result of the assumption $ee_{mix} = \mu\mu_{mix} = 0$,
only
oblique corrections (corrections to the W propagator) had to be considered (see
Sec. \ref{secimp}).
Still, we assume in this chapter $e\mu_{mix} = 0$ and $\mu\tau_{mix} = 0$.
The last parameter, $\mu\tau_{mix}$ does not contribute to the muon decay.
The neglect of $e\mu_{mix}$ is supported by experiment (see Eq.
\ref{limits3}) and the neglect of $\mu\tau_{mix}$ is motivated
by our intention to keep the discussion as straightforward as possible.
                                
\section{Box diagrams}
\label{secbox}

\begin{figure}[hbtp]
\begin{center}
\setlength{\unitlength}{1in}
\begin{picture}(6,3)
\put(.3,+0.325){\mbox{\epsfxsize=5.0in\epsffile{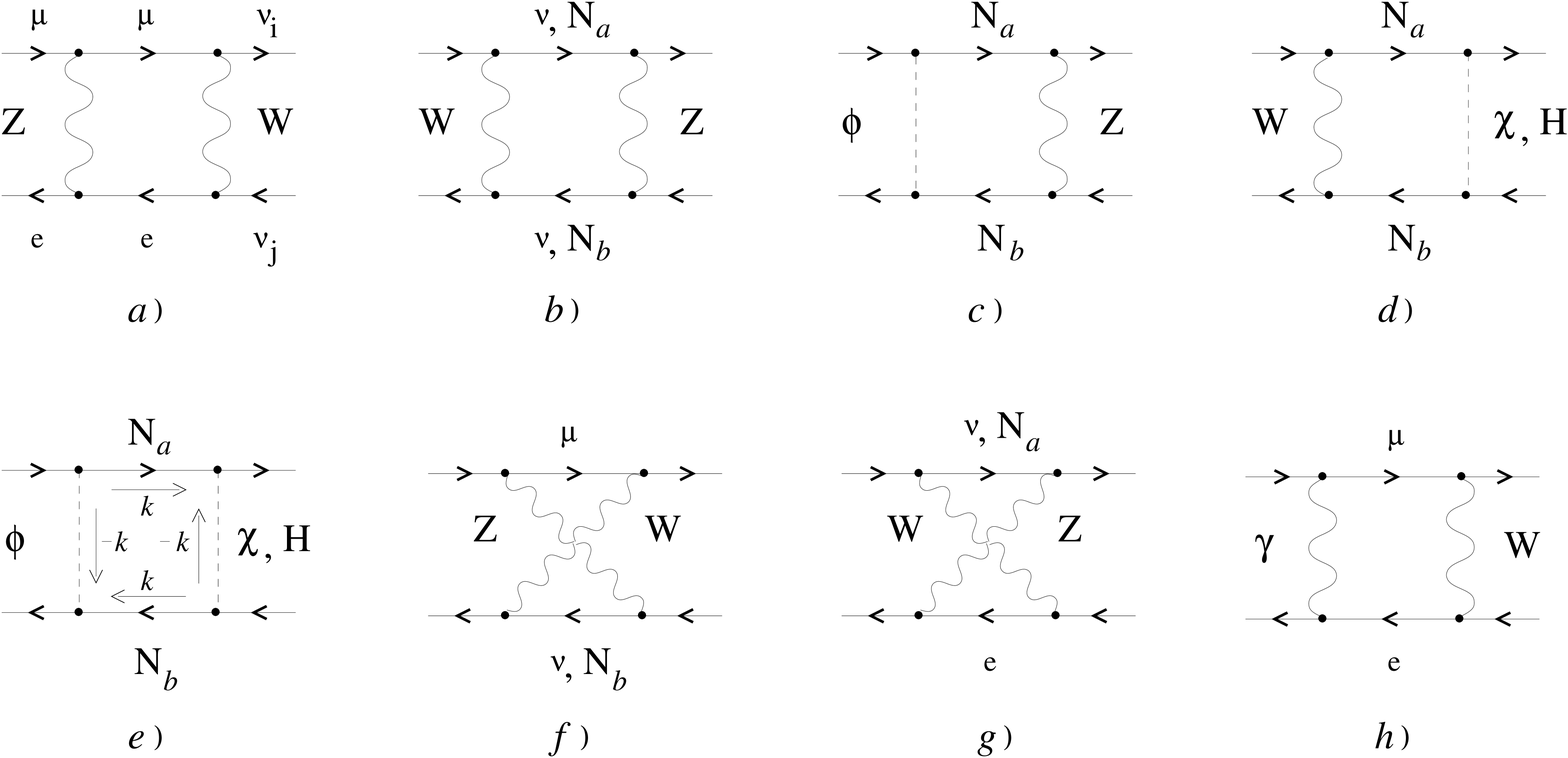}}}
\end{picture}
\end{center}
\caption{Box diagrams for muon decay}
\label{boxmuon}
\end{figure}

The set of box diagrams contributing to the muon decay is depicted in Fig.
\ref{boxmuon}.

As an example, we will calculate the diagrams of Fig. \ref{boxmuon}e.
These two were chosen since, as we will show, they exhibit quadratic
nondecoupling.

The amplitude for the diagram with the Higgs boson $H$ is given by
(we sum over NHL's $N_{a}, N_{b}$ with $M_{N_{a}} = M_{N_{b}} = M_{N}$ and
we neglect external momenta in the internal propagators)
\begin{eqnarray}
{\cal M}_{\phi N H N} & = & \sum_{a,b} \int \frac{d k^{4}}{(2\pi)^{4}}
\overline{u_{\nu_{\mu}}}\frac{-i
g_{2}}{2}\frac{M_{N}}{M_{W}}\big(K_{L}^{\dagger}K_{H}\big)_{ia}
\frac{1+\gamma_{5}}{2}\frac{i}{\not k - M_{N}}\frac{ig_{2}}{2\sqrt{2}}
\frac{M_{N}}{M_{W}}\big(K_{H}^{\dagger}\big)_{a\mu} \nonumber \\
& \times &
(1-\gamma_{5})u_{\mu}\;\;\overline{v_{e}}\frac{ig_{2}}{2\sqrt{2}}
\frac{M_{N}}{M_{W}}\big(K_{H}\big)_{eb}(1+\gamma_{5})
\frac{i}{\not k - M_{N}}\frac{-i
g_{2}}{2}\frac{M_{N}}{M_{W}}\big(K_{H}^{\dagger}K_{L}\big)_{bj} \nonumber \\
& \times  &
\frac{1-\gamma_{5}}{2}v_{\nu_{e}}\frac{i}{k^{2}-M_{W}^{2}}
\frac{i}{k^{2}-M_{H}^{2}} \nonumber \\
\nonumber \\
& = & 
\frac{g_{2}^{4}}{128}\frac{M_{N}^{4}}{M_{W}^{4}}\sum_{a,b} 
\big(K_{L}^{\dagger}K_{H}\big)_{ia}\big(K_{H}^{\dagger}\big)_{a\mu}
\big(K_{H}\big)_{eb}\big(K_{H}^{\dagger}K_{L}\big)_{bj}
\int \frac{d k^{4}}{(2\pi)^{4}} \nonumber \\
& \times   &
\frac{\big[\overline{u_{\nu_{\mu}}}(1+\gamma_{5})(\not k + M_{N})(1-\gamma_{5})
u_{\mu}\big]\big[\overline{v_{e}}(1+\gamma_{5})(\not k + M_{N})(1-\gamma_{5})
v_{\nu_{e}}\big]}
{\big(k^{2}-M_{N}^{2}\big)^{2} \big(k^{2}-M_{W}^{2}\big)
\big(k^{2}-M_{H}^{2}\big)}   \nonumber \\
\nonumber \\
& = &
\frac{g_{2}^{4}}{32}\frac{M_{N}^{4}}{M_{W}^{4}}k_{mix}\int \frac{d
k^{4}}{(2\pi)^{4}}
\frac{\big[\overline{u_{\nu_{\mu}}}(1+\gamma_{5})\not k
u_{\mu}\big]\big[\overline{v_{e}}(1+\gamma_{5}) \not k 
v_{\nu_{e}}\big]}
{\big(k^{2}-M_{N}^{2}\big)^{2} \big(k^{2}-M_{W}^{2}\big)
\big(k^{2}-M_{H}^{2}\big)},  
\end{eqnarray}
where 
\begin{eqnarray}
k_{mix} \equiv 
\big(K_{L}^{\dagger}K_{H}\big)_{ia}\big(K_{H}^{\dagger}\big)_{a\mu}
\big(K_{H}\big)_{eb}\big(K_{H}^{\dagger}K_{L}\big)_{bj} 
& = & {\big(K_{L}^{\dagger}\big)_{i\mu}
\big(K_{L}\big)_{ej}}ee_{mix}\mu\mu_{mix}.\;\;\;
\end{eqnarray}
Using $\int k^{\epsilon}k^{\nu} ... = \frac{1}{4} 
g^{\epsilon \nu} \int k^{2} ...\;\;$
we get
\begin{eqnarray}
{\cal M}_{\phi N H N} & = & 
\frac{g_{2}^{4}}{32}\frac{M_{N}^{4}}{M_{W}^{4}}k_{mix}
\big[\overline{u_{\nu_{\mu}}}(1+\gamma_{5})\gamma_{\epsilon}
u_{\mu}\big]\big[\overline{v_{e}}(1+\gamma_{5}) \gamma_{\nu}
v_{\nu_{e}}\big]
\nonumber \\
& \times &
\frac{1}{4}g^{\epsilon \nu}\frac{i}{(4 \pi)^{2}}{\cal I}_{2}
(M_{H}),
\end{eqnarray}
where ${\cal I}_{2}(m)$ is the integral
\begin{eqnarray}
{\cal I}_{2}(m) & = & \frac{{(4\pi)}^{2}}{i}\int \frac{d^{4}k}{{(2\pi)}^{4}}
\frac{k^{2}}{(k^{2}-M_{N}^{2})^{2}(k^{2}-M_{W}^{2})(k^{2}-m^{2})}.
\end{eqnarray}
In the next step we use the tree-level amplitude
\begin{eqnarray}
{\cal M}_{tree} & = & - \frac{i g_{2}^{2}}{8 M_{W}^{2}}
\big[\overline{u_{\nu_{\mu}}}
(1+\gamma_{5})\gamma_{\alpha}u_{\mu}\big]
\big[\overline{v_{e}}(1+\gamma_{5}) \gamma^{\alpha} v_{\nu_{e}}\big]
{\big(K_{L}^{\dagger}\big)_{i\mu} \big(K_{L}\big)_{ej}},
\end{eqnarray}
and $g_{2} = e/s_{W}$ to obtain
%\begin{eqnarray}
%{\cal M}_{\phi N H N} & = &
%\frac{g_{2}^{2}e^{2}}{128 s_{W}^{2}}\frac{i}{(4 \pi)^{2}}k_{mix}
%\frac{8 M_{W}^{2}}{-i
%g_{2}^{2}}\frac{{\cal M}_{tree}}{\big(K_{L}^{\dagger}\big)_{i\mu}
%\big(K_{L}\big)_{ej}}
%\frac{M_{N}^{4}}{M_{W}^{4}}{\cal I}_{2}(M_{H}) \\
%& = &
%- \frac{\alpha}{4 \pi}\frac{1}{16s_{W}^{2}}k_{mix} \frac{M_{N}^{4}}{M_{W}^{2}}
%{\cal I}_{2}(M_{H})\frac{{\cal M}_{tree}}{\big(K_{L}^{\dagger}\big)_{i\mu}
%\big(K_{L}\big)_{ej}} \nonumber
%\end{eqnarray}
%Looking at the mixing factor $k_{mix}$, the (by now) routine steps along with
%the assumptions stated in the beginning of this chapter yield
%\begin{eqnarray}
%\end{eqnarray}
%so the amplitude is
\begin{eqnarray}
{\cal M}_{\phi N H N} & = &
- \frac{\alpha}{4\pi}ee_{mix}\mu\mu_{mix}\frac{1}{16s_{W}^{2}}
\frac{M_{N}^{4}}{M_{W}^{2}}I_{2}(M_{H}){\cal M}_{tree}.
\end{eqnarray}
The integral ${\cal I}_{2}(M_{H})$ is calculated in 
Appendix \ref{Decko}. The result is
\begin{eqnarray}
{\cal I}_{2}(M_{H}) & = &
\frac{1}{M_{H}^{2}-M_{W}^{2}}
\biggl\{ \frac{1}{1-\frac{M_{W}^{2}}{M_{N}^{2}}} +
\frac{\frac{M_{W}^{4}}{M_{N}^{4}} \ln \frac{M_{W}^{2}}{M_{N}^{2}}}
{{(1-\frac{M_{W}^{2}}{M_{N}^{2}})}^{2}}
-\frac{1}{1-\frac{M_{H}^{2}}{M_{N}^{2}}} \nonumber \\
& - &
\frac{\frac{M_{H}^{4}}{M_{N}^{4}} \ln \frac{M_{H}^{2}}{M_{N}^{2}}}
{{(1-\frac{M_{H}^{2}}{M_{N}^{2}})}^{2}} \biggr\}. 
\end{eqnarray}
The amplitude ${\cal M}_{\phi N \chi N}$ is obtained from ${\cal M}_{\phi N H
N}$ by replacing $M_{H}$ with $M_{Z}$.

%%%%%%%%%%%%%%%%%%%%% Boxes %%%%%%%%%%%%%%%%%%%%%%%%%%%%%%%%%%%%%%%%%%%%%%%%%
The total contribution of the box diagrams (Figs. \ref{boxmuon} a-g) is
\begin{eqnarray}
{\cal M}_{box} & = & {\cal M}_{ZeW\mu} +{\cal M}_{W \nu Z\nu} +
{\cal M}_{W\nu Z N} +
{\cal M}_{W N Z\nu}
+ {\cal M}_{W N Z N}  \nonumber \\
& + &
{\cal M}_{\phi N Z N} + {\cal M}_{W N H N} + {\cal M}_{W N \chi N} + 
{\cal M}_{\phi N H N} + {\cal M}_{\phi N \chi N}  \nonumber \\
& + &
{\cal M}_{Z \nu W \mu} + {\cal M}_{Z  N W \mu} + {\cal M}_{W e Z\nu}
+ {\cal M}_{W e Z N} \nonumber   \\
\nonumber \\
& = & {\cal M}_{tree}\frac{\alpha}{4 \pi}\Biggl\{
\frac{-1}{4s_{W}^{2}c_{W}^{2}}M_{W}^{2}\biggl[
4{(-\frac{1}{2}+s_{W}^{2})}^{2}{\cal I}_{0}
 + {\cal I}_{0}(1-{\mu \mu}_{mix}) (1-ee_{mix})
 \Biggr. \biggr.   \nonumber \\
& + & 
{\cal I}_{1}(M_{Z})(1-{\mu \mu}_{mix})ee_{mix}
+{\cal I}_{1}(M_{Z}){\mu \mu}_{mix}(1-ee_{mix})
+{\cal I}_{2}(M_{Z}) ee_{mix} \nonumber \biggr. \\
& \times & 
\biggl. {\mu \mu}_{mix}  \biggr]
+ \frac{1}{4 s_{W}^{2}}M_{N}^{4}\biggl[
\frac{1}{c_{W}^{2}} {\cal I}_{3}(M_{Z}) + {\cal I}_{3}(M_{H}) +
{\cal I}_{3}(M_{Z}) - \frac{1}{4 M_{W}^{2}} {\cal I}_{2}(M_{H})\biggr.
\nonumber \\
& - & 
\frac{1}{4 M_{W}^{2}}{\cal I}_{2}(M_{Z})\biggr]ee_{mix}
{\mu\mu}_{mix} +\frac{2(-\frac{1}{2}+s_{W}^{2})}{s_{W}^{2} c_{W}^{2}}M_{W}^{2}
 \biggl[ {\cal I}_{0}(1-ee_{mix}) \nonumber  \\
& + & 
{\cal I}_{1}(M_{Z})ee_{mix} +
{\cal I}_{0}(1-{\mu \mu}_{mix}) 
+ {\cal I}_{1}(M_{Z}){\mu \mu}_{mix} \biggr] \Biggr\},  
\end{eqnarray}
where the integrals ${\cal I}_{0}, {\cal I}_{1}(m), {\cal I}_{3}(m)$ are
\begin{eqnarray}
{\cal I}_{0} & = & \frac{{(4\pi)}^{2}}{i}\int \frac{d^{4}k}{{(2\pi)}^{4}}
\frac{1}{k^{2}(k^{2}-M_{W}^{2})(k^{2}-M_{Z}^{2})} \; = \;
\frac{1}{M_{Z}^{2}-M_{W}^{2}}\ln
\frac{M_{W}^{2}}{M_{Z}^{2}}, \\
& & \nonumber \\
{\cal I}_{1}(m) & = & \frac{{(4\pi)}^{2}}{i}\int \frac{d^{4}k}{{(2\pi)}^{4}}
\frac{1}{(k^{2}-M_{N}^{2})(k^{2}-M_{W}^{2})(k^{2}-m^{2})} \nonumber \\
& & \nonumber \\
& = &
\frac{1}{m^{2}-M_{W}^{2}}
\biggl\{\ln\frac{M_{W}^{2}}{m^{2}} +
\frac{M_{N}^{2}}{M_{W}^{2}-M_{N}^{2}} \ln \frac{M_{W}^{2}}{M_{N}^{2}} -
\frac{M_{N}^{2}}{m^{2}-M_{N}^{2}} \ln
\frac{m^{2}}{M_{N}^{2}}\biggr\},\;\;\;\;\;
\\
& & \nonumber \\
{\cal I}_{3}(m) & = & \frac{{(4\pi)}^{2}}{i}\int \frac{d^{4}k}{{(2\pi)}^{4}}
\frac{1}{(k^{2}-M_{N}^{2})^{2}(k^{2}-M_{W}^{2})(k^{2}-m^{2})}
 \nonumber \\
& & \nonumber \\
& = &
\frac{1}{m^{2}-M_{W}^{2}}
\biggl\{ \frac{1}{M_{N}^{2}-M_{W}^{2}} +
\frac{M_{W}^{2} \ln \frac{M_{W}^{2}}{M_{N}^{2}}}{{(M_{N}^{2}-M_{W}^{2})}^{2}}
%\biggr. \nonumber \\
%& & \hspace*{2.in} \biggl.
-\frac{1}{M_{N}^{2}-m^{2}} \nonumber \\
& - &
\frac{m^{2} \ln \frac{m^{2}}{M_{N}^{2}}}{{(M_{N}^{2}-m^{2})}^{2}}
\biggr\}. 
\end{eqnarray}
For the calculation of these integrals see Appendix \ref{Decko}.
In the limit of $M_{N} \gg M_{W}$, the leading contribution to 
${\cal I}_{2}(m)$ is very simple,
\begin{eqnarray}
{\cal I}_{2}^{appx}(m) & = & - \frac{1}{M_{N}^{2}},
\end{eqnarray}
implying the $M_{N}^{2}$ dependence of the ${\cal M}_{\phi N H N}$ and
${\cal M}_{\phi N \chi N}$,
\begin{eqnarray}
\label{pnhn}
{\cal M}_{\phi N H N}^{appx} & = & {\cal M}_{tree} \frac{\alpha}{4 \pi}
\frac{1}{16s_{W}^{2}}\frac{M_{N}^{2}}{M_{W}^{2}} ee_{mix} \mu\mu_{mix}.
\end{eqnarray}
The leading contribution to ${\cal I}_{3}(m)$ goes as $\sim 1/M_{N}^{4}$  and
to ${\cal I}_{1}(m)$ as $\sim 1/M_{N}^{2}$, therefore no other box depends 
quadratically on the NHL mass.

The last box diagram, Fig. \ref{boxmuon} h can be written as \cite{key6}:
\begin{eqnarray}
{\cal M}_{\gamma e W \mu} & = & {\cal M}_{tree} \frac{\alpha}{4 \pi} 
\Big(\ln \frac{M_{W}}{m_{e}} + \ln \frac{M_{W}}{m_{\mu}} - 2 \ln
\frac{m_{e}}{\lambda} - 2 \ln \frac{m_{\mu}}{\lambda} + \frac{9}{2} \Big)
+ ... \;\;\;
\end{eqnarray}
The ellipses denote additional terms discussed in  Ref. \cite{key6}.

\section{Neutrino self-energy and its renormalization}
\label{secneus}

\begin{figure}[hbtp]
\begin{center}
\setlength{\unitlength}{1in}
\begin{picture}(6,2)
\put(.4,+0.325){\mbox{\epsfxsize=5.0in\epsffile{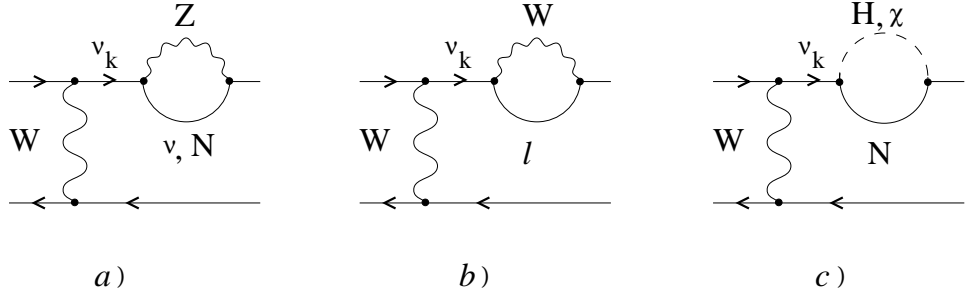}}}
\end{picture}
\end{center}
\caption{Neutrino self-energy diagrams for muon decay}
\label{selfmuon}
\end{figure}

One half of the neutrino self-energy diagrams contributing to muon decay
is shown in Fig. \ref{selfmuon}. 
The corresponding self-energy is denoted as $\Sigma^{\nu_{\mu}}$. 
The other half consists
of the same loops sitting on the bottom neutrino leg with the corresponding
self-energy $\Sigma^{\nu_{e}}$.
In all these diagrams, we sum over the internal massless neutrinos 
$\nu_{k}, k=1,2,3$. In principle,
the graphs with $\nu_{k}$ replaced by $N_{a}$ are also present, however, they
are suppressed by the large mass $M_{N}$.

%%%%%%%%%%%%%%%%%%% Neutrino selfenergy %%%%%%%%%%%%%%%%%%%%%%%%%%%%%%%%%%
 Without derivation, we present results for the unrenormalized
 neutrino self-energy $\Sigma^{\nu_{l}}$ ($l = e, \mu$). It has the form
\begin{equation}
\Sigma^{\nu_{l}}  =  \frac{1}{2}\Sigma_{L}^{\nu_{l}}
\not p (1-\gamma_{5}),
\end{equation}
where $\Sigma_{L}^{\nu_{l}}$ receives the following contributions from the
diagrams of Fig. \ref{selfmuon}:
\begin{eqnarray}
\label{neuself}
\Sigma_{L}^{\nu_{l}} & = &
\Sigma_{L}^{H}(p)+\Sigma_{L}^{\chi}(p)+\Sigma_{L}^{Z,N}(p)
+\Sigma_{L}^{Z,\nu}(p)+
        \Sigma_{L}^{W}(p) \nonumber \\
& = &
\frac{\alpha}{2\pi}(1-ll_{mix})\Biggl\{
\frac{1}{8s_{W}^{2}}
ll_{mix}
\frac{M_{N}^{2}}{M_{W}^{2}}
\biggl[ \frac{1}{2}\Delta_{\mu}+
B_{0}^{fin}(p;M_{H},M_{N})+B_{1}^{fin}(p;M_{H},M_{N})\biggr]\nonumber \\
& + &
\frac{1}{8s_{W}^{2}}ll_{mix} \frac{M_{N}^{2}}{M_{W}^{2}}
\biggl[ \frac{1}{2}\Delta_{\mu}+
B_{0}^{fin}(p;M_{Z},M_{N})+B_{1}^{fin}(p;M_{Z},M_{N})\biggr]\nonumber \\
& + &
\frac{1}{4s_{W}^{2}c_{W}^{2}}ll_{mix}
\biggl[ \frac{1}{2}\Delta_{\mu}-\frac{1}{2}+
B_{0}^{fin}(p;M_{Z},M_{N})+B_{1}^{fin}(p;M_{Z},M_{N})\biggr]\nonumber \\
& + &
\frac{1}{4s_{W}^{2}c_{W}^{2}}(1-ll_{mix})
\biggl[ \frac{1}{2}\Delta_{\mu}-\frac{1}{2}+
B_{0}^{fin}(p;M_{Z},0)+B_{1}^{fin}(p;M_{Z},0)\biggr] \nonumber \\
& + &
\frac{1}{2s_{W}^{2}}
\biggl[ \frac{1}{2}\Delta_{\mu}-\frac{1}{2}+
B_{0}^{fin}(p;M_{W},m_{l}\rightarrow 0)+B_{1}^{fin}(p;M_{W},m_{l}\rightarrow 0)
\biggr]\Biggr\}.
\end{eqnarray}
Here $s = p^{2} = 0 \;\; \ll \;\; M_{H}^{2},M_{Z}^{2},M_{W}^{2},M_{N}^{2}$, 
therefore
Eqs. \ref{males} apply for $B_{0}^{fin}, B_{1}^{fin}$.

The amplitude for the diagrams of Fig. \ref{selfmuon} in terms of
$\Sigma_{L}^{\nu_{l}}$ can be shown to be equal to
\begin{equation}
{\cal M}_{self} = - {\cal M}_{tree}\frac{\Sigma_{L}^{\nu_{l}}}{2},
\end{equation}
where the one half comes from our dealing with the external wave
function rather than the neutrino propagator.

\subsection{Renormalization of the neutrino self-energies} 
Let us now investigate the question of the renormalization of the diagrams of
Fig.~\ref{selfmuon}. It is the only case in this work when the counterterms
are modified from their SM form \footnote{So far we have used
the SM form of the counterterms, see Appendix \ref{recons}. The actual value 
of the
counterterms was, of course, different from SM.}. The problem is how to
renormalize a part of a theory where interaction eigenstates are different 
from mass eigenstates. Curiously, this also happens in the SM quark sector. The
difference is that in the SM the problem is circumvented by arguing the 
off-diagonal quark mixings are too small (with the required accuracy in mind) 
to have any effect in the loops and the
renormalization procedure is effectively simplified to that of mass eigenstates
being also the flavour eigenstates.  In our model, we cannot neglect the 
off-diagonal
mixings ($ll_{mix}$), since they (in combination with TeV NHL masses)
lead to the dominant terms in the predicted signals
\footnote{Just a few days before the submission of this thesis an interesting
paper appeared on the preprint bulletin board \cite{kniehl}, where the
arguments of this paragraph are also made. The authors then continue to develop
the first formal framework ever for the renormalization of theories with
nonnegligible mixings between mass and interaction eigenstates. Their treatment
is more general than ours below.}.

\begin{figure}[hbtp]
\begin{center}
\setlength{\unitlength}{1in}
\begin{picture}(6,2)
\put(1.4,+0.025){\mbox{\epsfxsize=2.8in\epsffile{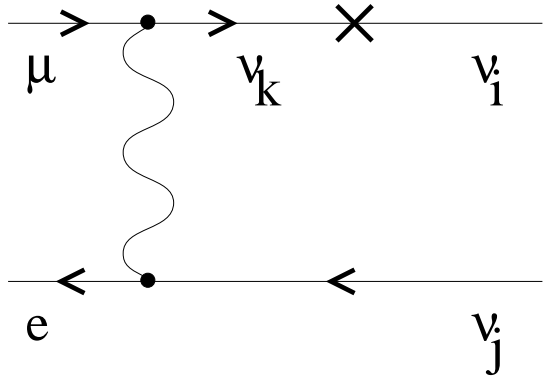}}}
\end{picture}
\end{center}
\caption{Counterterm diagram for neutrino self-energy in muon decay}
\label{countl}
\end{figure}

To derive the required counterterm, we start with the counterterm Lagrangian,
which has the same form in both SM and our model:
\begin{eqnarray}
\label{ourcounter}
i \;\delta Z_{L}^{e} \; \overline{\nu_{e}}  {\not \partial} \nu_{e}
+ i\; \delta Z_{L}^{\mu} \overline{\nu_{\mu}}  {\not \partial} \nu_{\mu}
+ i \;\delta Z_{L}^{\tau} \overline{\nu_{\tau}}  \not {\partial} \nu_{\tau},
\end{eqnarray}
where $\delta Z_{L}^{l}$ is the sum of $\delta Z_{V}^{l}$ and $\delta
Z_{A}^{l}$ renormalization constants which are given in Eq. \ref{rconstants}.
Weak eigenstates $\nu_{l}$ can be expressed in our model in terms 
of mass eigenstates $\nu_{i}, N_{a}$ (see Eq. \ref{alter})
\begin{eqnarray}
\nu_{l} & = & \sum_{i}\big(K_{L}\big)_{li} \nu_{i} + \sum_{a}
\big(K_{H}\big)_{la}N_{a} \; = \; \sum_{n_{\alpha}=\nu_{1},...,N_{6}}
K_{ln_{\alpha}}n_{\alpha}, \nonumber \\
\overline{\nu_{l}} & = & \overline{n_{\alpha}}K^{\dagger}_{n_{\alpha} l}.
\end{eqnarray}
This gives us for the product $\overline{\nu_{l}}\nu_{l}$
\begin{eqnarray}
\overline{\nu_{l}}\nu_{l} & = & \overline{n}K^{\dagger}K n \; = \;
\sum_{k,i=1,2,3}\overline{\nu_{i}}\big(K_{L}^{\dagger}\big)_{il}
\big(K_{L}\big)_{lk}\nu_{k} + ... (\overline{\nu_{i}}N, \overline{N}\nu_{k},
\overline{N}N),
\end{eqnarray}
and Eq. \ref{ourcounter} thus contributes the following terms
as the massless neutrino counterterm Lagrangians:
\begin{equation}
\sum_{k,i=1,2,3}\Big\{\delta
Z_{L}^{e}\big(K_{L}^{\dagger}\big)_{ie}\big(K_{L}\big)_{ek}
+ \delta Z_{L}^{\mu}\big(K_{L}^{\dagger}\big)_{i\mu}\big(K_{L}\big)_{\mu k}
+ \delta Z_{L}^{\tau}\big(K_{L}^{\dagger}\big)_{i\tau}\big(K_{L}\big)_{\tau k}
\Big\}\overline{\nu_{i}} {\not \partial} \nu_{k}.
\end{equation}
In our case, however, we sum over internal $\nu_{k}$ but not over external
$\nu_{i}$.
The graphic representation of the relevant counterterm (embedded in muon decay)
is in Fig. \ref{countl}.
The amplitude for this diagram is
%Including the factor $\big(K_{L}^{\dagger}\big)_{i\mu}$, the new counterterm
%Lagrangian for the massless neutrinos is  given as
\begin{equation}
{\cal M}_{C} = - \frac{1}{2}{\cal M}_{tree}^{SM}\sum_{l=e,\mu,\tau}
\sum_{k=1,2,3} \delta Z_{L}^{l}
\big(K_{L}^{\dagger}\big)_{il}\big(K_{L}\big)_{lk}
\big(K_{L}^{\dagger}\big)_{k\mu}.
\end{equation}
Again, the factor $\frac{1}{2}$
comes from our dealing with the external wave function
rather than the internal propagator and the mixing factor 
$\big(K_{L}^{\dagger}\big)_{k\mu}$ originates at the $\mu W \nu_{k}$ vertex. 
The amplitude ${\cal M}_{C}$ can be further simplified,
\begin{eqnarray}
{\cal M}_{C} & = & - \frac{1}{2}{\cal M}_{tree}^{SM} \sum_{l=e,\mu,\tau}
\delta Z_{L}^{l}\big(K_{L}^{\dagger}\big)_{il}
\sum_{k=1,2,3}\big(K_{L}\big)_{lk}\big(K_{L}^{\dagger}\big)_{k\mu} \nonumber \\
& = & - \frac{1}{2}
M_{tree}^{SM} \sum_{l=e,\mu,\tau}\delta Z_{L}^{l}\big(K_{L}^{\dagger}\big)_{il}
\big(\delta_{l\mu} - l\mu_{mix}\big) 
\; = \; -  \frac{1}{2}
M_{tree}^{SM} \delta Z_{L}^{\mu} \big(1 - \mu \mu_{mix} \big)
\big(K_{L}^{\dagger}\big)_{i\mu} \nonumber \\
& = & - \frac{1}{2}
\delta Z_{L}^{\mu} \big(1 - \mu \mu_{mix} \big) {\cal M}_{tree}.
\end{eqnarray}
The factor $\big(K_{L}^{\dagger}\big)_{i\mu}$ was absorbed by $M_{tree} =
M_{tree}^{SM} \big(K_{L}^{\dagger}\big)_{i\mu}$.

Now we can write down the final expressions for the renormalized amplitude 
${\hat {\cal M}}_{self}$  and the renormalized neutrino self-energy:  
\begin{eqnarray}
\label{modren}
{\hat {\cal M}}_{self} & = & {\cal M}_{self} + {\cal M}_{C} \; = \;
- \frac{\Sigma_{L}^{\nu_{l}}}{2} {\cal M}_{tree} - \frac{\delta Z_{L}^{l}}{2}
\big(1 - ll_{mix}\big) {\cal M}_{tree}, \\
\label{modren1}
{\hat \Sigma_{L}}^{\nu_{l}} & = & \Sigma_{L}^{\nu_{l}} + \delta Z_{L}^{l}
 \big(1 - ll_{mix}\big).
\end{eqnarray}
The constant $\delta Z_{L}^{l}$ is found from 
%(see Eq. \ref{recons} for
%relations between $\delta Z_{L}^{l}, \delta Z_{R}^{l}$ and $\delta Z_{V}^{l},
%\delta Z_{A}^{l}$; and Eq. \ref{rconstants})
\begin{eqnarray}
\delta Z_{L}^{l} & = & - \Sigma_{L}^{l} (m_{l}^{2}) - m_{l}^{2}
\big[\Sigma_{L}^{l'}(m_{l}^{2}) + \Sigma_{R}^{l'}(m_{l}^{2}) +
 2 \Sigma_{S}^{l'}(m_{l}^{2}) \big], \nonumber \\
\Sigma^{l'}_{L,R,S}(m_{l}^{2}) & = &
\frac{\partial \Sigma_{L,R,S}^{l}}{\partial p^{2}}(m_{l}^{2}),
\end{eqnarray}
where $\Sigma_{L}, \Sigma_{R}$ and $\Sigma_{S}$ are respectively the 
left-handed, right-handed and the scalar part of the lepton self-energy
given in Eqs. \ref{hasthe} - \ref{leptonself2} 
\footnote{Now we include the photonic loop $\Sigma_{\gamma l}^{l}$ in the 
total lepton self-energy,
see Eq. \ref{leptonself2} and the following comments.}.
The only graph with significant contribution to the term with
derivatives, is the photonic loop (Eq. \ref{leptonself2}). The other diagrams
lead to derivatives of $B_{0}^{fin}, B_{1}^{fin}$ functions which, when
multiplied by $m_{l}^{2}$, are of the order $m_{l}^{2}/M_{W}^{2}$ and therefore
negligible. This can be easily seen from Eqs. \ref{males}, \ref{bphoton}, 
wherein also
derivatives of the $B_{0}^{fin}, B_{1}^{fin}$ functions corresponding to
the photonic loop are given.
The final answer is
\begin{eqnarray}
\label{finans}
\delta Z_{L}^{l} & = & - \Sigma_{L}^{l} (m_{l}^{2}) + \frac{\alpha}{2 \pi}
\Big( 2 \ln \frac{m_{l}}{\lambda} - 1 \Big).
\end{eqnarray}

To prove the cancellation of the infinities, we note that the infinite part of
$\delta Z_{L}^{l}$ is given by
\begin{eqnarray}
\delta Z_{L}^{l,\infty} & = & - \frac{\alpha}{4 \pi}\frac{1}{s_{W}^{2}}
 \Big\{\frac{1}{2} + \frac{1}{4c_{W}^{2}} + \frac{{\cal X}}{4}ll_{mix}\Big\}
\Delta_{\mu},
\end{eqnarray}
and the infinite part of the neutrino self-energy by
\begin{eqnarray}
\Sigma_{L}^{\nu_{l},\infty} & = & \frac{\alpha}{4 \pi}\frac{1}{s_{W}^{2}}
\Big\{\frac{{\cal X}}{4}ll_{mix}(1-ll_{mix}) + \frac{1}{2}(1-ll_{mix})
+ \frac{1}{4c_{W}^{2}}ll_{mix}(1-ll_{mix}) \Big. \nonumber \\
& + &
\Big. \frac{1}{4c_{W}^{2}}(1-ll_{mix})^{2} \Big\} \Delta_{\mu}.
\end{eqnarray}
From the formulae above it can be easily seen that infinities cancel out in
Eq. \ref{modren1}.

\subsection{Limit $M_{N} \gg M_{W}, M_{Z}, M_{H}$}

We will investigate the large $M_{N}$ behaviour of the renormalized neutrino
self-energy ${\hat \Sigma}_{L}^{\nu_{l}}$.
For large $M_{N}$ we get from Eq. \ref{males}
\begin{eqnarray}
B_{0}(p;M_{H,Z,W},M_{N}) & = &  1 - 2 \ln M_{N}, \nonumber \\
B_{1}(p;M_{H,Z,W},M_{N}) & = &  - 0.25 + \ln M_{N}.
\end{eqnarray}
This implies the quadratic nondecoupling for
$\Sigma_{L}^{H}(p)$ and $\Sigma_{L}^{\chi}(p)$, see Eq. \ref{neuself}:
\begin{eqnarray}
\Sigma_{L}^{H}(p) + \Sigma_{L}^{\chi}(p) & = & \frac{\alpha}{2\pi}
\frac{1}{4s_{W}^{2}} ll_{mix}(1-ll_{mix}) \frac{M_{N}^{2}}{M_{W}^{2}}
\biggl[ \frac{1}{2}\Delta_{\mu} + 0.75 - \ln M_{N} \biggr], \;\;\;
\end{eqnarray}
and for the $\Sigma_{L}^{\phi N}$, a left-handed part of $\Sigma_{\phi N}^{l}$
(see Eq. \ref{leptonself1}), 
which contributes to the ${\hat \Sigma}_{L}^{\nu_{l}}$ via the
counterterm $\delta Z_{L}^{l}$, see Eq. \ref{finans}:
\begin{eqnarray}
\Sigma_{L}^{\phi N} & = &  + \frac{\alpha}{16\pi s_{W}^{2}}ll_{mix} 
\frac{M_{N}^{2}}{M_{W}^{2}}
\biggl[ \Delta_{\mu} + \frac{3}{2}  - 2 \ln M_{N} \biggr].
\end{eqnarray}
From here we can see the $\Sigma_{L}^{\phi N}$ not only cancels out
infinities in the $\Sigma_{L}^{H}(p)$ and $\Sigma_{L}^{\chi}(p)$, but,
in the limit investigated, it also cancels out the finite parts.
As a result, there is no quadratic nondecoupling in the renormalized
neutrino self-energy.

\section{Vertex diagrams}

\begin{figure}[hbtp]
\begin{center}
\setlength{\unitlength}{1in}
\begin{picture}(6,3)
\put(0.93,+0.325){\mbox{\epsfxsize=4.0in\epsffile{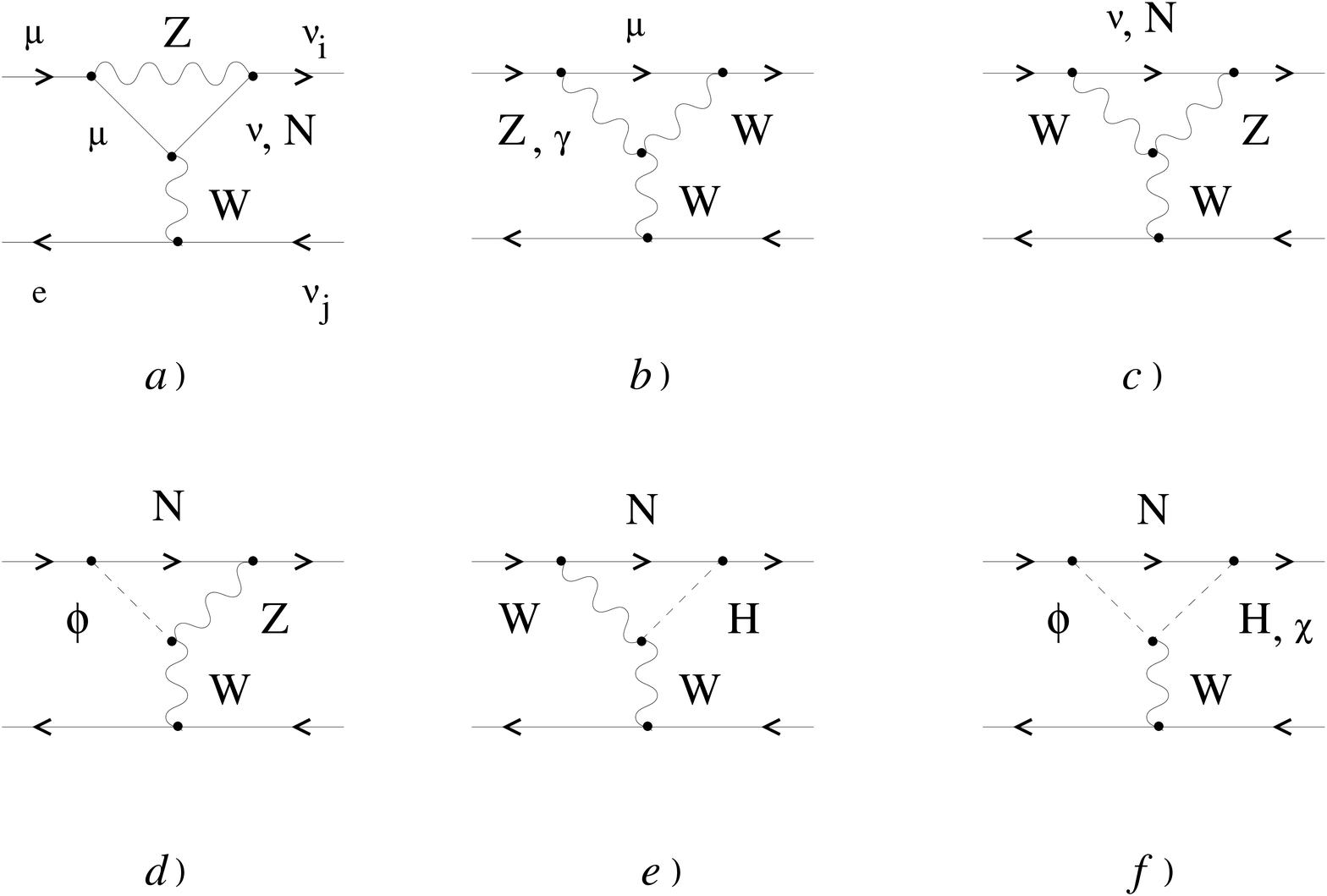}}}
\end{picture}
\end{center}
\caption{Vertex diagrams for muon decay}
\label{vermuon}
\end{figure}

Diagrams modifying the $W\mu\nu_{i}$ vertex are depicted in Fig.
\ref{vermuon}. Another set, one that modifies $We\nu_{j}$
vertex, is not shown.

The sum over the depicted set of diagrams gives the muon
vertex amplitude ${\cal M}_{vertex}^{\mu}$:
%%%%%%%%%%%%%%%%%%%%%%%% Vertex  graphs %%%%%%%%%%%%%%%%%%%%%%%%%%%%%%%%%%%%%
\begin{eqnarray}
{\cal M}_{vertex}^{\mu} & = & {\cal M}_{\mu \nu Z} + {\cal M}_{\mu N Z} +
{\cal M}_{Z W \mu} + {\cal M}_{\gamma W \mu} + {\cal M}_{WZ\nu}\nonumber   \\
& + &
{\cal M}_{WZN} + {\cal M}_{\phi ZN} + {\cal M}_{WHN}
+ {\cal M}_{\phi HN} + {\cal M}_{\phi \chi N}.
\end{eqnarray}
The computation of the diagrams yields
\begin{eqnarray}
{\cal M}_{vertex}^{\mu} & = & {\cal M}_{tree} \frac{\alpha}{4 \pi} \Biggl\{
\frac{2s_{W}^{2}-1}{4s_{W}^{2}c_{W}^{2}}\Big(\Delta_{M_{Z}}-\frac{1}{2}\Big)
(1-{\mu \mu}_{mix}) \Biggr.\nonumber \\
& + & 
\frac{2s_{W}^{2}-1}{4s_{W}^{2}c_{W}^{2}}\Big(\Delta_{M_{Z}}-\frac{1}{2}-
\frac{M_{N}^{2}}{M_{Z}^{2}-M_{N}^{2}}\ln \frac{M_{Z}^{2}}{M_{N}^{2}}\Big)
{\mu \mu}_{mix}\nonumber     \\
& + & 
\frac{\frac{1}{2}-s_{W}^{2}}{s_{W}^{2}}\Big(3 \Delta_{M_{W}} + \frac{5}{2} +
\frac{3}{s_{W}^{2}}\ln c_{W}^{2}\Big) 
+ 3\Big(\Delta_{M_{W}} + \frac{5}{6}\Big)\nonumber \\
& + & 
\frac{3}{2s_{W}^{2}}\Big(\Delta_{M_{W}} + \frac{5}{6} + \frac{1}{s_{W}^{2}}
\ln c_{W}^{2}\Big)(1-{\mu \mu}_{mix})\nonumber  \\
& + & 
\frac{3}{2s_{W}^{2}}\biggl[ \Delta_{M_{W}} + \frac{5}{6} +
\frac{1}{s_{W}^{2}}\ln c_{W}^{2} + \frac{M_{N}^{2}}{M_{Z}^{2}-M_{W}^{2}}
v(M_{Z})\biggr]{\mu \mu}_{mix} \nonumber \\
& + & 
\frac{1}{2c_{W}^{2}} \frac{-M_{N}^{2}}{M_{Z}^{2}-M_{W}^{2}}v(M_{Z})
{\mu \mu}_{mix}\nonumber  \\
& + & 
\frac{1}{2s_{W}^{2}} \frac{-M_{N}^{2}}{M_{H}^{2}-M_{W}^{2}}v(M_{H})
{\mu \mu}_{mix}\nonumber  \\
& + & 
\frac{1}{8s_{W}^{2}}\frac{M_{N}^{2}}{M_{W}^{2}}\biggl[
\Delta_{M_{W}} + \frac{3}{2} -\frac{M_{H}^{2}}{M_{W}^{2}-M_{H}^{2}}\ln \frac
{M_{W}^{2}}{M_{H}^{2}} + \frac{M_{N}^{2}}{M_{H}^{2}-M_{W}^{2}}v(M_{H})
\biggr]{\mu \mu}_{mix}\nonumber  \\
& + & 
\frac{1}{8s_{W}^{2}}\frac{M_{N}^{2}}{M_{W}^{2}}\biggl[
\Delta_{M_{W}} + \frac{3}{2} -\frac{M_{Z}^{2}}{M_{W}^{2}-M_{Z}^{2}}\ln \frac
{M_{W}^{2}}{M_{Z}^{2}} + \frac{M_{N}^{2}}{M_{Z}^{2}-M_{W}^{2}}v(M_{Z})
\biggr]
\nonumber \\
& \times & {\mu \mu}_{mix}  \Biggr\},  
\end{eqnarray}
where 
\begin{eqnarray}
v(m) & = & \ln \frac{M_{W}^{2}}{m^{2}} +
\frac{M_{N}^{2}}{M_{W}^{2}-M_{N}^{2}}\ln \frac{M_{W}^{2}}{M_{N}^{2}}
- \frac{M_{N}^{2}}{m^{2}-M_{N}^{2}}\ln \frac{m^{2}}{M_{N}^{2}}. 
\end{eqnarray}
The vertices are renormalized as (see Eq. \ref{rvertexa})
\begin{eqnarray}
{\hat \Lambda}^{\mu} & = & \Lambda^{\mu} + \delta Z_{1}^{W} - \delta Z_{2}^{W}
+ \delta Z_{L}^{\mu},
\end{eqnarray}
where
\begin{eqnarray}
\Lambda^{\mu} & = & {\cal M}_{vertex}^{\mu} / {\cal M}_{tree},  \\
\delta Z_{1}^{W} - \delta Z_{2}^{W} & = & - \frac{\alpha}{2 \pi s_{W}^{2}}
\Delta_{M_{W}}.
\end{eqnarray}
Looking for the dominant graphs in the limit $M_{N} \gg M_{W},
M_{Z}, M_{H}$, we note
the leading contribution to the function $v(m)$ is
\begin{eqnarray}
v(m)^{appx} & = & \frac{1}{M_{N}^{2}}\Big(m^{2}\ln \frac{m^{2}}{M_{N}^{2}} -
M_{W}^{2} \ln \frac{M_{W}^{2}}{M_{N}^{2}}\Big),
\end{eqnarray}
which implies the graphs of Fig. \ref{vermuon}f have quadratic nondecoupling
\footnote{Another interesting point the authors of Ref. \cite{kniehl} make
(see also footnote 2 on page 122) is that the diagrams for the muon
decay ($\pi$ decay in Ref. \cite{kniehl}), Figs. \ref{boxmuon}e,
\ref{selfmuon}c, \ref{vermuon}f can be singled out as dominant in an elegant 
way by using the Goldstone boson equivalence theorem.}.
However, as in the case of the neutrino self-energy, both infinite and 
finite terms of these graphs are cancelled by the $\Sigma_{L}^{\phi N}$ term 
in the counterterm $\delta Z_{L}^{l}$. Therefore 
there are no $M_{N}^{2}$ dependent terms in the 
renormalized vertex diagrams either.

\section{Results}

The muon decay loops modify the $\Delta r$ quantity in the implicit relation
between $M_{W}$ and $G_{\mu}$, see Eq. \ref{MG}:
\begin{eqnarray}
\label{mwg}
M_{W}^{2} s_{W}^{2} & = & \frac{\pi\alpha}{\sqrt{2} G_{\mu} (1-\Delta r)}
\Big(1-\frac{1}{2}ee_{mix}-\frac{1}{2}\mu\mu_{mix}\Big),
\end{eqnarray}
and $\Delta r$ can be written as (see Eq. \ref{delrv})
\begin{eqnarray}
\Delta r & = & \frac{{\cal R}e \:{\hat \Sigma}_{W}(0)}{M_{W}^{2}} + \delta_{V}.
\end{eqnarray}
Here the parameter $\delta_{V}$ is the sum of the loop diagrams calculated in
this section:
\begin{eqnarray}
\delta_{V} & = & \frac{{\cal M}_{\gamma e W \mu}}{{\cal M}_{tree}}
+ {\hat \Lambda}^{\mu} + {\hat \Lambda}^{e} - 
\frac{1}{2} {\hat \Sigma}^{\nu_{e}}
- \frac{1}{2} {\hat \Sigma}^{\nu_{\mu}} + \frac{{\cal M}_{box}}{{\cal
M}_{tree}}.
\end{eqnarray}

Numerical results are shown in Table \ref{muonloops}. As input data we used
masses from the standard set, mixings $ee_{mix} = 0.0071$, $\mu\mu_{mix} = 
0.0014$ and $\tau\tau_{mix} = 0$. We suppressed the $\tau\tau_{mix}$, which at
its maximal value currently allowed (0.033) would make corrections 
to ${\hat \Sigma}_{W}(0)/M_{W}^{2}$ much larger than the
corrections  to $\delta_{V}$
(these only depend on $ee_{mix},\mu\mu_{mix}$) and we would like to show the
case when also the latter are important.
In the first three lines of the table we show how much the self-energy,
vertex and box diagrams contribute to $\delta_{V}$ (line 4) for
NHL masses $M_{N}$ of up to 30 TeV.
Also shown (lines 5,6) are ${\hat \Sigma}_{W}(0)/M_{W}^{2}$ and $\Delta r$.
Ultimately we are interested in NHL effects in the observable $M_{W}$ (line 7). 

\begin{table}[htb]
\begin{center}
\begin{tabular}{|l|r|r|r|r|r|r|} \hline
                  & SM  &  0.5 TeV  &  5 TeV &  15 TeV & 30 TeV &            \\
  \hline
${\hat \Sigma}^{\nu_{e}}+ {\hat \Sigma}^{\nu_{\mu}}$  
      & - 4.995  & - 4.972  & - 4.982 & - 4.988 & - 4.992 & $\times 10^{-2}$ \\
${\hat \Lambda}^{\mu}$
      & - 1.441  & - 1.442 &  - 1.444 & - 1.444 & -1.445  & $\times 10^{-2}$ \\
${\cal M}_{box}/{\cal M}_{tree}$
      &   4.273  &   4.300  &   4.315 &   4.457 &  4.950  & $\times 10^{-3}$ \\
$\delta_{V}$
      &   6.670  &   6.539  &   6.525 &   6.652 &  7.133  & $\times 10^{-3}$ \\
${\hat \Sigma}_{W}(0)/M_{W}^{2}$
      &   2.396  &   2.346  &   2.301 &   1.872 &  0.329  & $\times 10^{-2}$ \\
$\Delta r$
      &   3.063  &   3.000  &   2.954 &   2.537 &  1.043  & $\times 10^{-2}$ \\
$M_{W}$           
      &  80.459  &  80.537  &  80.545 &  80.612 & 80.846  & $\times 1$       \\ 
\hline
\end{tabular}
\end{center}
\caption{Contribution of the muon decay loops to $\delta_{V}$ and $\Delta_{r}$}
\label{muonloops}
\end{table}                             

The results confirm expectations from the previous sections. There is no
nondecoupling for self-energies and vertices and there is a quadratic
dependence on $M_{N}$ for the boxes. The boxes are becoming important
at very high masses. Still, they are small compared to the change in
${\hat \Sigma}_{W}(0)/M_{W}^{2}$. This is due to the fact the dominant
boxes depend on the product $ee_{mix}\mu\mu_{mix}$ (see Eq. \ref{pnhn}),
while the correction to the $W$ propagator is proportional to 
$k_{HH} = ee_{mix}^{2} + \mu\mu_{mix}^{2}$ (see Eq. \ref{khhmixi}), 
which is allowed to be larger given the current bounds on the mixings.

The $W$ mass jumps from $M_{W}^{SM} = 80.459$ GeV to $M_{W} = 80.537$ GeV
at $M_{N} = 0.5$ TeV mainly as a result of the tree-level correction
$\Big(1-\frac{1}{2}ee_{mix}-\frac{1}{2}\mu\mu_{mix}\Big)$, see Eq.~\ref{mwg}.
After that it rises very slowly until the $M_{N}$ dependent
amplitudes become dominant above $5$ TeV.

In conclusion,
the numerical analysis of Chapter 6 turns out to be basically valid even
after the restriction $ee_{mix} = \mu\mu_{mix} = 0$ is relaxed. It can be
improved by the inclusion of the tree-level correction
$\Big(1-\frac{1}{2}ee_{mix}-\frac{1}{2}\mu\mu_{mix}\Big)$,
while the largest loop corrections, 
the box diagrams of Fig. \ref{boxmuon}e, are only marginally important.

\newpage

\chapter{Conclusions}

Two theoretical schemes were discussed here as possible solutions of
the problem of the small neutrino masses:   
the see-saw mechanism of Yanagida and Gell-Mann, Ramond and Slansky  
\cite{guts,seesaw} (Sec. \ref{see-saw32}); and the superstring-inspired 
low-energy model of neutrino
masses suggested in Refs. \cite{vallemo,wolfe} (Chapter 3). 
Both these solutions introduce NHL's into
the theory as a necessary ingredient.
Our main consideration has been the phenomenology of NHL's in the 
superstring-inspired low-energy model. The
qualitative features of our analysis are applicable also in the context of
see-saw models with enhanced mixings. 

The superstring-inspired low-energy model (Chapter 3) is a simple extension of
the SM. It enriches only the neutral fermion spectrum of the
SM, leaving the gauge and Higgs sectors intact.
We found that among the new parameters of the model, seven are especially
important: 'flavour-conserving' mixing parameters $ee_{mix}, \mu\mu_{mix},
\tau\tau_{mix}$; 'flavour-violating'
mixing parameters $e\mu_{mix}, e\tau_{mix}, \mu\tau_{mix}$ and the mass scale 
$M_{N}$ of NHL's (we assumed all three NHL's have mass $M_{N}$). 
The bounds on the six mixing
parameters (Sec. \ref{review3}) are largely independent of the
mass $M_{N}$. On the other hand bounds on $M_{N}$ always depend on the mixings.

The mass $M_{N}$, if larger than $M_{Z}$, can presently 
only be probed in radiative
corrections (loops). A traditional approach was mostly limited to hypothetical
lepton flavour-violating processes 
such as $\mu \rightarrow e \gamma;\: \mu, \tau \rightarrow e
e^{+} e^{-}; \: Z \rightarrow e^{\pm}\mu^{\mp}$ etc
\cite{Ng1,Ng2,bernabeu1,ggjv,Ilakovac,Jarlskog,Valle2,Korner,pilaftsis1}. 
We reviewed constraints from these processes in Chapter 5.

Besides flavour-violating processes NHL's
could also induce (via radiative corrections) deviations from the SM
in currently observed processes. We calculated radiative corrections due
to NHL's to three such observables: leptonic widths of the Z boson
$\Gamma_{ll}$,
lepton universality breaking parameter $U_{br}$ and the mass of the W boson 
$M_{W}$
(Chapters 6 and 7). We found that these observables form three complementary 
quantities as far as sensitivity to NHL masses and mixings is concerned.  
$\Gamma_{ll}$ depends on three kinds of radiative corrections: the vertex
corrections (Sec. \ref{secver}), the Z oblique corrections and the W oblique
corrections (Sec. \ref{seczprop}); the universality breaking parameter depends
only on the vertex corrections; the W mass $M_{W}$ depends to a large degree
only on the W oblique corrections.

The effect of the NHL mass $M_{N}$ in radiative corrections is, on the one
hand, 
suppressed by the small mixings; on the other hand it is enhanced due to the
violation of the Appelquist-Carazzone theorem (Sec. \ref{appelc}). These
competing tendencies are reflected by the typical behaviour of the dominant
terms in radiative corrections due to NHL's (see Eqs. \ref{aprox1},
\ref{aprox4}),
\begin{eqnarray}
\label{domterm}
\sim  k_{HH} \frac{M_{N}^{2}}{M_{W}^{2}} & \sim & (\tau \tau_{mix})^{2}
\frac{M_{N}^{2}}{M_{W}^{2}}. 
\end{eqnarray}
To make up for the small mixings, only NHL's with masses in the TeV range
can lead to significant deviations from the SM.

Assuming $\tau \tau_{mix} = 0.033$, 
$ee_{mix} = \mu\mu_{mix} = 0$ and $m_{t} = 176$ GeV, we derived the following
limit on the NHL mass at the $2\sigma$ level from the $Z$ decay width to 
$\tau$~leptons:
\begin{equation}
\label{upp1}
M_{N}  \leq  4.3 \; {\rm TeV}.
\end{equation}
The universality breaking ratio, $U_{br}$ , leads to a yet better upper limit, 
\begin{equation}
\label{upp2}
M_{N}  \leq  3.8 \; {\rm TeV},
\end{equation}
at the $2\sigma$ level.
We can use Eq. \ref{domterm} to display the approximate dependence of the above
limits on $\tau \tau_{mix}$:
\begin{eqnarray}
M_{N} & < & 4.3 \times \frac{0.033}{\tau \tau_{mix}} \;{\rm TeV}
\end{eqnarray}
from Eq. \ref{upp1} and
\begin{eqnarray}
M_{N} & < & 3.8 \times \frac{0.033}{\tau \tau_{mix}} \;{\rm TeV}
\end{eqnarray}
from Eq. \ref{upp2}. 
Note the limits of Eqs. \ref{upp1}, \ref{upp2} are comparable to the limit
\begin{eqnarray}
M_{N} & < & 4 \;{\rm TeV},
\end{eqnarray}
derived from the considerations of perturbation theory breakdown in Sec.
\ref{breakdown}.

We also found some sensitivity of the W mass $M_{W}$ to 
the NHL mass and
mixings, which depends to a large degree on the top quark mass (Figs.
\ref{nummw}a,b).

In Chapter 7 we generalized our analysis of Chapter 6
by relaxing the restriction $ee_{mix}=\mu\mu_{mix}=0$.
We found that while the numerical results of Chapter 6 remain basically valid, 
they can be improved by the inclusion of the tree-level correction to the muon
decay,
$\Big(1-\frac{1}{2}ee_{mix}-\frac{1}{2}\mu\mu_{mix}\Big)$.
                                    
%We found that only two box diagrams exhibit quadratic
%nondecoupling and could have some importance in parts of the parameter space
%(when $\tau\tau_{mix} \rightarrow 0$). Otherwise Eqs. \ref{upp1}, \ref{upp2}
%remain valid for arbitrary mixings.

As already noted in Chapter 6, we feel there are
at least two reasons which give the (flavour-conserving) processes studied in 
this thesis a
distinct advantage over the flavour-violating ones.
First, the limits on $M_{N}$ which we derived are only matched by those
from $\mu \rightarrow e e e$. 
The flavour-violating decay rates for
$\tau$ and $Z$ are below the current
experimental sensitivity (see Sec. \ref{numeres} and \ref{fvple}).
Moreover, the $\mu \rightarrow e e e$ decay
depends only on $ee_{mix}, e\mu_{mix}$, the two of the six mixing
parameters (see Eq.~\ref{mnlimit4}), and may be unobservable if $ee_{mix}$ 
or $e\mu_{mix}$ are very small.
Second, the inequality Eq. \ref{ineq} can further suppress the 
flavour-violating processes
against the flavour-conserving ones
via the 'conspiracy of the phases' in the sum of complex
terms making up the flavour-violating parameters. 

For these two reasons, first signatures of neutral heavy leptons could come
from flavour-conserving observables. At this time, LEP has stopped its runs at
the Z-peak energy 
and is running at $130 - 140$ GeV. It will eventually be producing W
pairs which will allow the mass $M_{W}$ to be measured with a precision of 
$0.05$ GeV \cite{mw2}
(currently $M_{W} = 80.410 \pm 0.180$ \cite{mw1}). 
Combined with more precise measurements of the top quark mass we might be in
a position to place even more stringent limits on NHL masses and mixings from
our prediction of $M_{W}$ (see Figs. \ref{nummw}a,b).

The observation of neutral heavy leptons
is essential for our understanding of the small neutrino masses. It would
provide us with significant hints on grand unified theories and possibly 
superstring theories.

\newpage

\appendix

\chapter{Dirac algebra and trace theorems}
\label{Acko}

In this Appendix we define the Dirac gamma matrices
and collect their properties. We also show spin sums, some Fierz identities
and the proof of the identity  $\;\overline{\nu_{L}^{c}}\nu_{R}^{c}
 = \overline{\nu_{L}}\nu_{R}$. 

\section{Gamma matrices}
\label{gamas}

The so-called Dirac representation of the gamma matrices is given by
\begin{eqnarray}
\gamma^{0}  =
\left( \begin{array}{rr}
                       {\bf I} & 0  \\
                       0 & -{\bf I}
               \end{array} \right) ,\;\;\;
\gamma^{i}  =
\left( \begin{array}{rr}
                       0 & \sigma^{i}  \\
                   - \sigma^{i} &  0
               \end{array} \right) ,\;\;\;
\gamma_{5}\;\;\equiv\;\;i \gamma^{0}\gamma^{1}\gamma^{2}\gamma^{3}\;\; = \;\;
\left( \begin{array}{rr}
                       0 & {\bf I}  \\
                       {\bf I} & 0
               \end{array} \right),
\end{eqnarray}
where ${\bf I}$ is $2 \times 2$ identity matrix and the Pauli matrices 
$\sigma^{i}$ are defined as
\begin{eqnarray}
\sigma^{1}  =
\left( \begin{array}{rr}
                       0 & 1  \\
                       1 & 0
               \end{array} \right) ,\;
\sigma^{2}  =
\left( \begin{array}{rr}
                       0 & -i  \\
                       i & 0
               \end{array} \right) ,\;
\sigma^{3}  =
\left( \begin{array}{rr}
                       1 & 0  \\
                       0 & -1
               \end{array} \right) .
\end{eqnarray}
Some properties of the gamma matrices follow:
\begin{eqnarray}
\gamma_{0} & = & \gamma^{0},\;\;\;\;\;\gamma_{i}\;\;=\;\;-\gamma^{i},
\nonumber  \\
{\gamma_{0}}^{\dagger} & = & \gamma_{0}, \nonumber   \\
\gamma_{0} \gamma_{0}  & = &   {\bf 1}, \nonumber   \\
\gamma_{0} {\gamma_{\mu}}^{\dagger} \gamma_{0}  & = & \gamma_{\mu}, 
\nonumber   \\
\gamma_{0}^{T} & = & \gamma_{0}, \nonumber  \\
\gamma_{5}^{T} & = & \gamma_{5}, \nonumber   \\
\gamma_{5}^{2} & = & {\bf 1},     \nonumber       \\
\nu_{1}^{T} \: \Gamma \: \nu_{2} & = &  - {\nu_{2}}^{T} \: \Gamma^{T} \:\nu_{1},
\end{eqnarray}
where $\nu_{1}, \nu_{2}$ are spinors, ${\bf 1}$ is $4 \times 4$ unit matrix,
and $ \Gamma$ represents product of gamma matrices.

Trace theorems:
\begin{eqnarray}
Tr({\bf 1}) & = & 4, \nonumber \\
Tr(\gamma^{\mu}\gamma^{\nu}) & = & 4 g^{\mu\nu}, \nonumber   \\
Tr(\gamma^{\mu}\gamma^{\nu}\gamma^{\lambda}\gamma^{\sigma}) & = & 4\left(
g^{\mu\nu}g^{\lambda\sigma} - g^{\mu\lambda}g^{\nu\sigma} + 
g^{\mu\sigma}g^{\nu\lambda}\right),  \nonumber  \\
Tr(\gamma_{5}) & = & 0, \nonumber  \\
Tr(\gamma_{5}\gamma^{\mu}\gamma^{\nu}) & = & 0, \nonumber  \\
Tr(\gamma_{5}\gamma^{\mu}\gamma^{\nu}\gamma^{\lambda}\gamma^{\sigma}) & = & 
4 i \epsilon^{\mu\nu\lambda\sigma}, \nonumber  \\
Tr(odd\;number\;of\;gamma\;matrices) & = & 0.
\end{eqnarray}

Product rules:
\begin{eqnarray}
\gamma^{\mu}\gamma^{\nu} + \gamma^{\nu}\gamma^{\mu} & = & 2 g^{\mu\nu}
, \nonumber   \\
\gamma_{\mu}\gamma^{\mu} & = & 4, \nonumber \\
\gamma_{\mu}\gamma^{\nu}\gamma^{\mu} & = & - 2 \gamma^{\nu}, \nonumber \\
\gamma_{\mu}\gamma^{\nu}\gamma^{\lambda}\gamma^{\mu} & = & 4
g^{\nu\lambda}, \nonumber \\
\gamma_{\mu}\gamma^{\nu}\gamma^{\lambda}\gamma^{\sigma}\gamma^{\mu} & = & 
- 2 \gamma^{\sigma}\gamma^{\lambda}\gamma^{\nu}, \nonumber  \\
\gamma_{5}\gamma^{\mu} + \gamma^{\mu}\gamma_{5} & = & 0 \nonumber \\
g^{\mu\nu}g_{\mu\nu} & = & 4.
\end{eqnarray}

An explicit representation of the charge conjugation matrix C is
\begin{eqnarray}
C = i \gamma^{2} \gamma^{0} & = &
\left( \begin{array}{rrrr}
                       0 & 0  & 0 & -1  \\
                       0 & 0  & 1 &  0  \\
                       0 & -1 & 0 &  0  \\
                       1 & 0  & 0 &  0
               \end{array} \right),\;\;\;\;\;
C^{T}C^{T}\;\; = \;\; - {\bf 1}.
\end{eqnarray}

\section{Spin sums}
\label{spins}

\begin{eqnarray}
\sum_{s}u_{\alpha}(p,s)\overline{u_{\beta}}(p,s) & = & {({\not p} +
m)}_{\alpha\beta}\;, \nonumber \\
\sum_{s}v_{\alpha}(p,s)\overline{v_{\beta}}(p,s) & = & {({\not p} -
m)}_{\alpha\beta}\;, \nonumber \\
\sum_{\lambda}\epsilon_{\mu}(p,\lambda)\epsilon_{\nu}^{*}(p,\lambda) & = &
 - g_{\mu\nu} + \frac{p_{\mu}p_{\nu}}{M_{V}^{2}}\;,  
\end{eqnarray}
where $u_{\alpha}(p,s), v_{\alpha}(p,s)$ are spinors with momentum $p$, 
spin $s$ and mass $m$; $\epsilon_{\mu}(p,\lambda)$ is a polarization vector
of a weak boson $V = W, Z$ with momentum $p$, spin $\lambda$ and mass $M_{V}$.

\section{Fierz identities used for the calculation of the boxes}

\begin{eqnarray}
\big[\bar u \gamma_{\alpha} (1-\gamma_{5}) \gamma_{\epsilon} \gamma_{\gamma}
u \big] \times \big[{\overline v_{e}} \gamma^{\alpha} (1-\gamma_{5})
\gamma^{\epsilon} \gamma^{\gamma} v_{\nu_{e}} \big]
& = &
16 \big[\bar u \gamma_{\mu} (1-\gamma_{5}) u \big] \nonumber \\
& \times &
\big[{\overline
v_{e}} \gamma^{\mu} (1-\gamma_{5}) v_{\nu_{e}} \big], \\
\big[\bar u \gamma_{\alpha} (1-\gamma_{5}) \gamma_{\epsilon} \gamma_{\gamma}
u \big] \times \big[{\overline v_{e}} \gamma^{\gamma} (1-\gamma_{5})
\gamma^{\epsilon} \gamma_{\alpha} v_{\nu_{e}} \big]
& = &
4 \big[\bar u \gamma_{\mu} (1-\gamma_{5}) u \big] \nonumber \\
& \times &
\big[{\overline
v_{e}} \gamma^{\mu} (1-\gamma_{5}) v_{\nu_{e}} \big].
\end{eqnarray}

\section{Proof of the identity $\;\;\;\overline{\nu_{L}^{c}}\nu_{R}^{c}
 = \overline{\nu_{L}}\nu_{R}$} 
\label{proof}

Using the definition of the charge conjugate field,
\begin{eqnarray}
\nu^{c} & = & C\gamma_{0}\nu^{*}, \;\;\;(C= i \gamma^{2} \gamma^{0})
 \nonumber \\
\overline{\nu^{c}} & = & \nu^{T} C ,
\end{eqnarray}
we get
\begin{eqnarray}
\overline{\nu_{L}^{c}}\nu_{R}^{c} & = & \left[\frac{1-\gamma_{5}}{2}
\nu^{c}\right]^{\dagger} \gamma_{0} \frac{1+\gamma_{5}}{2} \nu^{c}
 = {\nu^{c}}^{\dagger}{\frac{1-\gamma_{5}}{2}}^{\dagger} \gamma_{0}
\frac{1+\gamma_{5}}{2}\nu^{c} \nonumber \\
& = & {\nu^{c}}^{\dagger} \gamma_{0}\frac{1+\gamma_{5}}{2} 
\frac{1+\gamma_{5}}{2}
 \nu^{c} = \overline{\nu^{c}}\frac{1+\gamma_{5}}{2}\nu^{c} \nonumber \\
& = & \nu^{T} C \frac{1+\gamma_{5}}{2} C \gamma_{0}\nu^{*}
  =  - {\nu^{*}}^{T} \left[C \frac{1+\gamma_{5}}{2} C
\gamma_{0}\right]^{T}\nu  \nonumber \\
& = & - \nu^{\dagger}\left[\gamma_{0} C^{T} \frac{1+\gamma_{5}}{2} C^{T}
\right]\nu = - \overline{\nu} C^{T} \frac{1+\gamma_{5}}{2} C^{T}\nu
\nonumber \\
& = & - \overline{\nu} C^{T}C^{T} \frac{1+\gamma_{5}}{2} \nu
  =  \overline{\nu} \frac{1+\gamma_{5}}{2} \nu \nonumber \\
& = & \overline{\nu_{L}}\nu_{R}.  
\end{eqnarray}

\newpage

\chapter{Couplings of $\nu^{'}$ and $N$ to Higgs}
\label{Becko}

We begin with some useful properties of the rotation matrix $G$ (see Eq.
\ref{matrixg}). From $GG^{\dagger} = G^{\dagger}G = 1$ we have
\begin{eqnarray}
\label{relationsg}
U_{1}U_{1}^{\dagger} + U_{2}U_{2}^{\dagger} = 1, &\;\;\;\;     \;\;\;\;&
U_{1}^{\dagger}U_{1} + U_{3}^{\dagger}U_{3} = 1,  \nonumber    \\
U_{3}U_{3}^{\dagger} + U_{4}U_{4}^{\dagger} = 1, &\;\;\;\;     \;\;\;\;&
U_{2}^{\dagger}U_{2} + U_{4}^{\dagger}U_{4} = 1,  \nonumber    \\
U_{1}U_{3}^{\dagger} + U_{2}U_{4}^{\dagger} = 0, &\;\;\;\;     \;\;\;\;&     
U_{1}^{\dagger}U_{2} + U_{3}^{\dagger}U_{4} = 0,  \nonumber    \\
U_{3}U_{1}^{\dagger} + U_{4}U_{2}^{\dagger} = 0, &\;\;\;\;     \;\;\;\;&     
U_{2}^{\dagger}U_{1} + U_{4}^{\dagger}U_{3} = 0.
\end{eqnarray}
Further, from (see Sec. \ref{diagon3})
\begin{eqnarray}
U_{1}D + U_{2}M & = & 0,      \nonumber \\
U_{3}D + U_{4}M & = & M^{'},
\end{eqnarray}
we get
\begin{eqnarray}
U_{2}^{\dagger}U_{1}D + U_{2}^{\dagger}U_{2}M & = & 0,      \nonumber \\
U_{4}^{\dagger}U_{3}D + U_{4}^{\dagger}U_{4}M & = & U_{4}^{\dagger}M^{'}.
\end{eqnarray}
Adding these two equations and using Eq. \ref{relationsg} we find the following
relation between $M$ and $M{'}$:
\begin{equation}
\label{mmu}
M = U_{4}^{\dagger}M^{'}.
\end{equation}

To derive the Lagrangian describing the couplings of $\nu^{'}$ and $N$ to $H$,
${\cal L}_{H}$, we rewrite Eq. \ref{yukawa1} as
\begin{eqnarray}
{\cal L} & = & - \frac{g_{2}}{\sqrt{2} M_{W}}
\left(\overline{\nu_{L}}\:\overline{l_{L}}\right)D \tilde{\Phi}n_{R} + h.c.\;,
\end{eqnarray}
where $D$ is a $3 \times 3$ matrix in family space and, (see Eqs.
\ref{higgsplus}, \ref{higgsminus}),
\begin{eqnarray}
\tilde{\Phi} = i \tau_{2} \Phi^{*} = \left( \begin{array}{c}
                                  {\phi^{0}}^{*} \\
                                   -\phi^{-}
             \end{array} \right)
= \left( \begin{array}{c}
                       \frac{1}{\sqrt{2}}(v + H - i \chi) \\
                         - \phi^{-}
               \end{array} \right).
\end{eqnarray}
Selecting the $H$ part we get
\begin{eqnarray}
{\cal L}_{H} & = & - \frac{g_{2}}{2 M_{W}}\overline{\nu_{L}} D n_{R}H
 + h.c.\;.
\end{eqnarray}
In the next step we add and subtract a term:
\begin{eqnarray}
{\cal L}_{H} & = &  - \frac{g_{2}}{2 M_{W}}\overline{\nu_{L}} D n_{R}H
 + h.c. \nonumber \\
& - & \frac{g_{2}}{2 M_{W}}\overline{S_{L}} M n_{R}H
 + h.c. \nonumber  \\
& + & \frac{g_{2}}{2 M_{W}}\overline{S_{L}} M n_{R}H
 + h.c.\;.
\end{eqnarray}
The first two lines of this relation can be compared with Eq. \ref{fa3}.
We can now use the results of Sec. \ref{diagon3}, which give
\begin{eqnarray}
{\cal L}_{H} & = & - \frac{g_{2}}{2 M_{W}}
\overline{S_{L}^{'}}M^{'}n_{R}H + h.c. \nonumber \\
& + & \frac{g_{2}}{2 M_{W}} \left(\overline{\nu_{L}^{'}}U_{2} +
\overline{S_{L}^{'}}U_{4}\right) M n_{R} H + h.c. \nonumber \\
& = & - \frac{g_{2}}{2 M_{W}}\overline{S_{L}^{''}}T M^{'} Z^{\dagger}
n_{R}^{''}H + h.c. \nonumber \\
& + & \frac{g_{2}}{2 M_{W}}\left(\overline{\nu_{L}^{'}}U_{2}M Z^{\dagger} +
\overline{S_{L}^{''}}T U_{4} M Z^{\dagger}\right) n_{R}^{''}H + h.c.\;.
\end{eqnarray}
Now we use $M = U_{4}^{\dagger} M^{'}$ (see Eq. \ref{mmu})
and $M^{''} = T M^{'} Z^{\dagger}$ (see Eq. \ref{mprimed}):
\begin{eqnarray}
{\cal L}_{H}
& = & - \frac{g_{2}}{2 M_{W}}\overline{S_{L}^{''}}T M^{'} Z^{\dagger}
n_{R}^{''}H + h.c. \nonumber \\
& + & \frac{g_{2}}{2 M_{W}}\left(\overline{\nu_{L}^{'}}U_{2}U_{4}^{\dagger}
M^{'} Z^{\dagger} +
\overline{S_{L}^{''}}T U_{4} U_{4}^{\dagger} M^{'} Z^{\dagger}\right)
 n_{R}^{''}H + h.c.
\nonumber \\
& = & - \frac{g_{2}}{2 M_{W}}\overline{S_{L}^{''}}M^{''}
n_{R}^{''}H + h.c. \nonumber \\
& + & \frac{g_{2}}{2 M_{W}}\left(\overline{\nu_{L}^{'}}U_{2}U_{4}^{\dagger}
T^{\dagger} T M^{'} Z^{\dagger} +
\overline{S_{L}^{''}}T U_{4} U_{4}^{\dagger} T^{\dagger} T
M^{'} Z^{\dagger}\right)
 n_{R}^{''}H + h.c. \nonumber \\
& = & - \frac{g_{2}}{2 M_{W}}\overline{S_{L}^{''}}M^{''}
n_{R}^{''}H + h.c. \nonumber \\
& + & \frac{g_{2}}{2 M_{W}}\left(\overline{\nu_{L}^{'}}U_{2}U_{4}^{\dagger}
T^{\dagger} M^{''} +
\overline{S_{L}^{''}}T U_{4} U_{4}^{\dagger} T^{\dagger}
M^{''}\right)
 n_{R}^{''}H + h.c.\;. 
\end{eqnarray}
Using $U_{2}U_{4}^{\dagger} = - U_{1}U_{3}^{\dagger}$ and
$U_{4}U_{4}^{\dagger} = 1 - U_{3}U_{3}^{\dagger}$, see Eq. \ref{relationsg},
\begin{eqnarray}
{\cal L}_{H} & = & - \frac{g_{2}}{2 M_{W}}\overline{S_{L}^{''}}M^{''}
n_{R}^{''}H + h.c. \nonumber \\
& + & \frac{g_{2}}{2 M_{W}}\big[\overline{\nu_{L}^{'}}(-U_{1}U_{3}^{\dagger})
T^{\dagger} M^{''} +
\overline{S_{L}^{''}}T (1 - U_{3} U_{3}^{\dagger}) T^{\dagger}
M^{''}\big]
 n_{R}^{''}H + h.c.\; ,\;\;\; 
\end{eqnarray}
and putting
\begin{eqnarray}
K_{L} & = &  U_{1}^{\dagger},\;\;\;\;\; K_{H}=U_{3}^{\dagger}T^{\dagger},
\end{eqnarray}
we get
\begin{eqnarray}
{\cal L}_{H} & = & - \frac{g_{2}}{2 M_{W}}\overline{S_{L}^{''}}M^{''}
n_{R}^{''}H + h.c. \nonumber \\
& + & \frac{g_{2}}{2 M_{W}}\overline{\nu_{L}^{'}}(-K_{L}^{\dagger}K_{H})
M^{''} n_{R}^{''}H + h.c. \nonumber \\
& + & \frac{g_{2}}{2 M_{W}} \overline{S_{L}^{''}} (1 - K_{H}^{\dagger} K_{H})
M^{''} n_{R}^{''}H + h.c. \nonumber \\
& = & - \frac{g_{2}}{2 M_{W}} \overline{S_{L}^{''}}(K_{H}^{\dagger} K_{H})
M^{''} n_{R}^{''}H + h.c. \nonumber \\
& - & \frac{g_{2}}{2 M_{W}} \overline{\nu_{L}^{'}}(K_{L}^{\dagger}K_{H})
M^{''} n_{R}^{''}H + h.c. \nonumber \\
& = & - \frac{g_{2}}{2 M_{W}} \overline{N_{L}}(K_{H}^{\dagger} K_{H})
M^{''} N_{R}H + h.c. \nonumber \\
& - & \frac{g_{2}}{2 M_{W}} \overline{\nu_{L}^{'}}(K_{L}^{\dagger}K_{H})
M^{''} N_{R}H + h.c. \nonumber \\
& = & - \frac{g_{2}}{2 M_{W}}\overline{N} (K_{H}^{\dagger} K_{H}) M_{N}
N H  \nonumber \\
& - & \frac{g_{2}}{2 M_{W}} \overline{\nu^{'}}(K_{L}^{\dagger}K_{H}) M_{N}
\frac{1+\gamma_{5}}{2}N H  \nonumber \\
& - & \frac{g_{2}}{2 M_{W}} \overline{N}(K_{H}^{\dagger}K_{L}) M_{N}
\frac{1-\gamma_{5}}{2}\nu^{'} H.  
\end{eqnarray}
The couplings of $\chi, \phi^{+}$ and $\phi^{-}$ are found by analogy.

\newpage

\chapter{Feynman rules}
\label{Cecko}

We list here the Feynman rules needed for the 
computation
of the non-SM diagrams contributing to the processes studied in this thesis.
They are given in the 't~Hooft-Feynman gauge (see Sec. \ref{quant1}).
The rules for the vertices correspond to the interaction Lagrangians of Sec.
\ref{inter3}. The SM case is obtained in the limit 
\begin{equation}
K_{H} \rightarrow 0, \;\;\;\;\; K_{L} \rightarrow 1.
\end{equation}
%The rules for the propagators are purely SM. They come from the kinetic energy
%parts of the SM Lagrangian.
In vertices, where applicable, the arrows indicate in addition to the flow 
of the charge also the flow of momenta. 
%In propagators, the arrows indicate the flow 
%of the the particle's momentum $q$.

\begin{figure}
\begin{center}
\setlength{\unitlength}{1in}
\begin{picture}(6,8)
\put(0.3,+0.3){\mbox{\epsfxsize=1.2in\epsffile{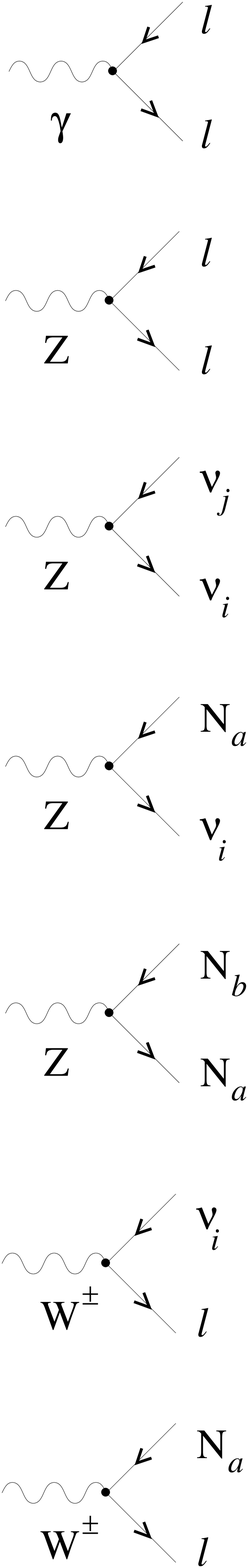}}}
\put(2.6,7.72){$+ \;i e \gamma_{\mu}$}
\put(2.6,6.58){$+ \;\frac{ie}{s_{W}c_{W}}\gamma_{\mu}
\left[\left(-\frac{1}{2}+s_{W}^{2}\right)
\frac{(1-\gamma_{5})}{2}+s_{W}^{2}\frac{(1+\gamma_{5})}{2}\right]$}
\put(2.6,5.47){$+ \;\frac{ie}{4s_{W}c_{W}}
\left(K_{L}^{\dagger}K_{L}\right)_{ij}
\gamma_{\mu}(1-\gamma_{5})$}
\put(2.6,4.27){$+ \;\frac{ie}{4s_{W}c_{W}}\left(K_{L}^{\dagger}K_{H}
\right)_{ia}\gamma_{\mu}(1-\gamma_{5})$}
\put(2.6,3.03){$+ \;\frac{ie}{4s_{W}c_{W}}\left(K_{H}^{\dagger}K_{H}
\right)_{ab}\gamma_{\mu}(1-\gamma_{5})$}
\put(2.6,1.78){$+ \;\frac{ig_{2}}{2\sqrt{2}}(K_{L})_{li}
\gamma_{\mu}\left(1-\gamma_{5}\right)$}
\put(2.6,0.65){$+ \;\frac{ig_{2}}{2\sqrt{2}}(K_{H})_{la}
\gamma_{\mu}\left(1-\gamma_{5}\right)$}
\end{picture}
\end{center}
\end{figure}
\begin{figure}
\begin{center}
\setlength{\unitlength}{1in}
\begin{picture}(6,7.5)
\put(0.3,+0.7){\mbox{\epsfxsize=1.2in\epsffile{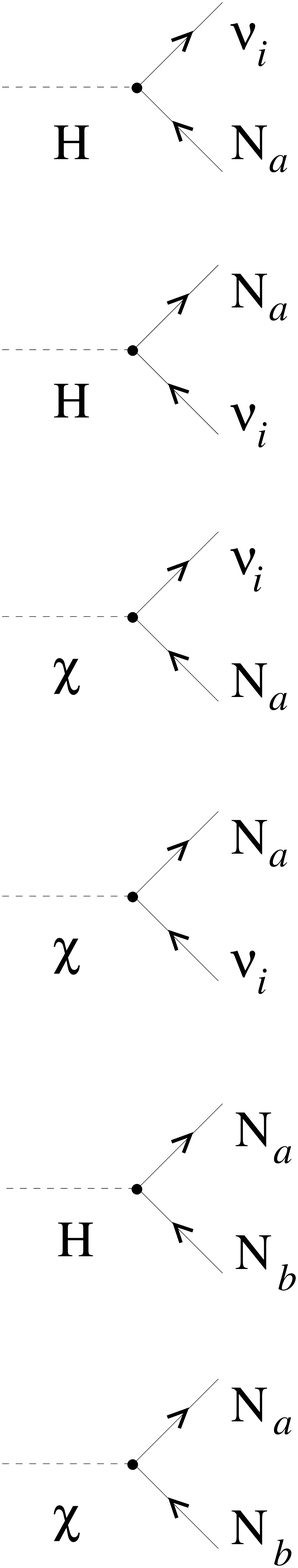}}}
\put(2.6,6.79){$- \;\frac{ig_{2}}{2M_{W}}M_{N}\left(K_{L}^{\dagger}
  K_{H}\right)_{ia}\frac{\left(1+\gamma_{5}\right)}{2}$}
\put(2.6,5.72){$- \;\frac{ig_{2}}{2M_{W}}M_{N}\left(K_{H}^{\dagger}
  K_{L}\right)_{ai}\frac{\left(1-\gamma_{5}\right)}{2}$}
\put(2.6,4.60){$- \;\frac{g_{2}}{2M_{W}}M_{N}\left(K_{L}^{\dagger}
  K_{H}\right)_{ia}\frac{\left(1+\gamma_{5}\right)}{2}$}
\put(2.6,3.44){$+ \;\frac{g_{2}}{2M_{W}}M_{N}\left(K_{H}^{\dagger}
  K_{L}\right)_{ai}\frac{\left(1-\gamma_{5}\right)}{2}$}
\put(2.6,2.24){$- \;\frac{ig_{2}}{2M_{W}}M_{N}\left(K_{H}^{\dagger}
  K_{H}\right)_{ab}$}
\put(2.6,1.11){$- \;\frac{g_{2}}{2M_{W}}M_{N}\left(K_{H}^{\dagger}K_{H}
  \right)_{ab}\gamma_{5}$}
\end{picture}
\end{center}
\end{figure}
\begin{figure}
\begin{center}
\setlength{\unitlength}{1in}
\begin{picture}(6,7.5)
\put(0.3,+0.7){\mbox{\epsfxsize=1.5in\epsffile{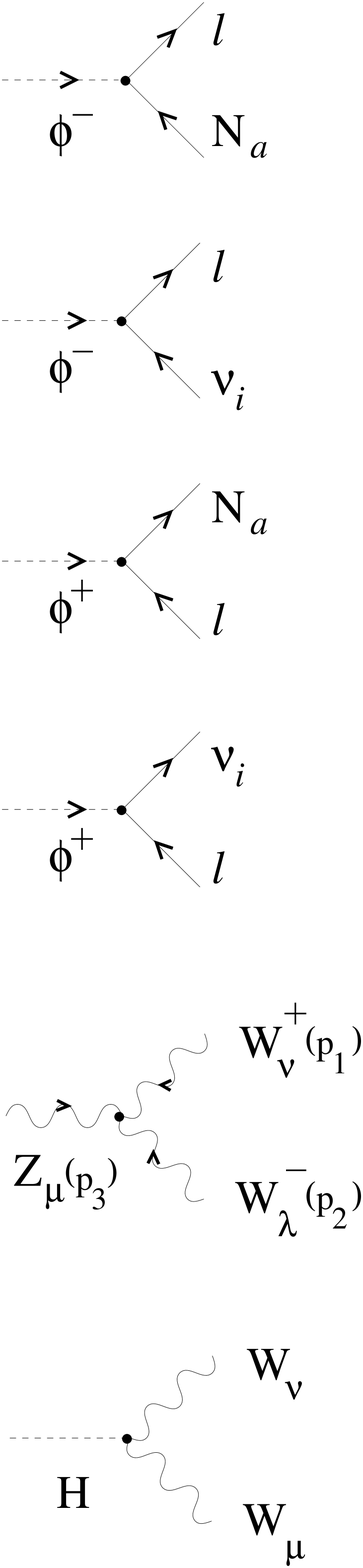}}}
\put(2.6,6.88){$+ \;\frac{ig_{2}}{\sqrt{2}M_{W}}\left(K_{H}\right)_{la}
   \left[
   M_{N}\frac{(1+\gamma_{5})}{2} - m_{l}\frac{(1-\gamma_{5})}{2}\right]$}
\put(2.6,5.88){$- \;\frac{ig_{2}m_{l}}{\sqrt{2}M_{W}}\left(K_{L}\right)_{li}
   \frac{(1-\gamma_{5})}{2}$}
\put(2.6,4.86){$+ \;\frac{ig_{2}}{\sqrt{2}M_{W}}\left(K_{H}^{\dagger}
   \right)_{al}
   \left[
   M_{N}\frac{(1-\gamma_{5})}{2} - m_{l}\frac{(1+\gamma_{5})}{2}\right]$}
\put(2.6,3.82){$- \;\frac{ig_{2}m_{l}}{\sqrt{2}M_{W}}\left(K_{L}^{\dagger}
   \right)_{il}
   \frac{(1+\gamma_{5})}{2}$}
\put(2.6,2.53){$- \;\frac{iec_{W}}{s_{W}}\left[g_{\nu\lambda}
   (p_{1}-p_{2})_{\mu}
  +g_{\lambda\mu}(p_{2}-p_{3})_{\nu}\right.$}
\put(2.6,2.25){$+ \;\left.g_{\mu\nu}(p_{3}-p_{1})_{\lambda}\right]$}
\put(2.6,1.17){$+ \;ig_{2}M_{W}g^{\mu\nu}$}
\end{picture}
\end{center}
\end{figure}
\begin{figure}
\begin{center}
\setlength{\unitlength}{1in}
\begin{picture}(6,7.5)
\put(0.3,+0.7){\mbox{\epsfxsize=1.3in\epsffile{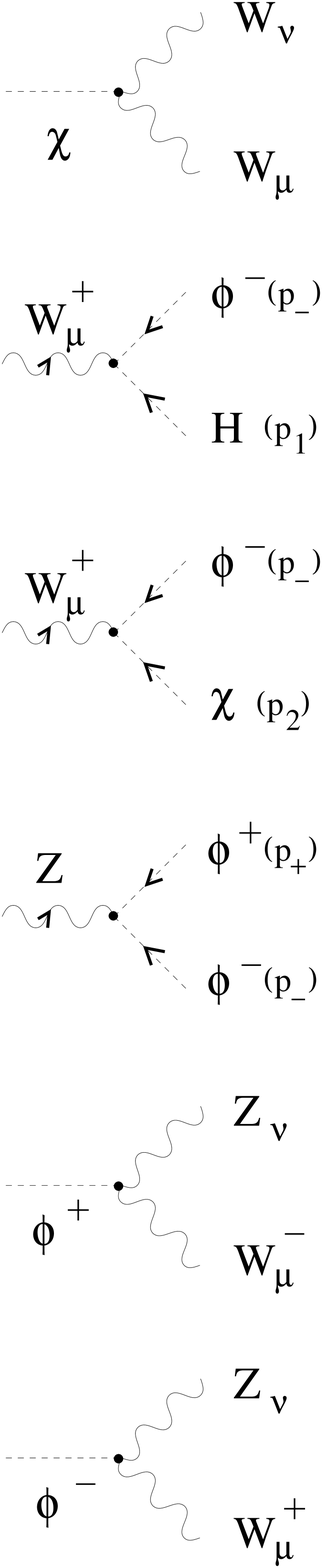}}}
\put(2.6,6.67){$0$}  
\put(2.6,5.58){$+\frac{ig_{2}}{2}\left(p_{1}-p_{-}\right)_{\mu}$}
\put(2.6,4.47){$+\frac{g_{2}}{2}\left(p_{-}-p_{2}\right)_{\mu}$}
\put(2.6,3.32){$-\frac{ig_{2}}{2}\frac{1-2s_{W}^{2}}{c_{W}}
\left(p_{-}-p_{+}\right)_{\mu}$}
\put(2.6,2.19){$-ig_{2}M_{W}\frac{s_{W}^{2}}{c_{W}}g^{\mu\nu}$}
\put(2.6,1.07){$-ig_{2}M_{W}\frac{s_{W}^{2}}{c_{W}}g^{\mu\nu}$}
\end{picture}
\end{center}
\end{figure}
\begin{figure}
\begin{center}
\setlength{\unitlength}{1in}
\begin{picture}(6,7.5)
\put(0.3,+1.5){\mbox{\epsfxsize=1.5in\epsffile{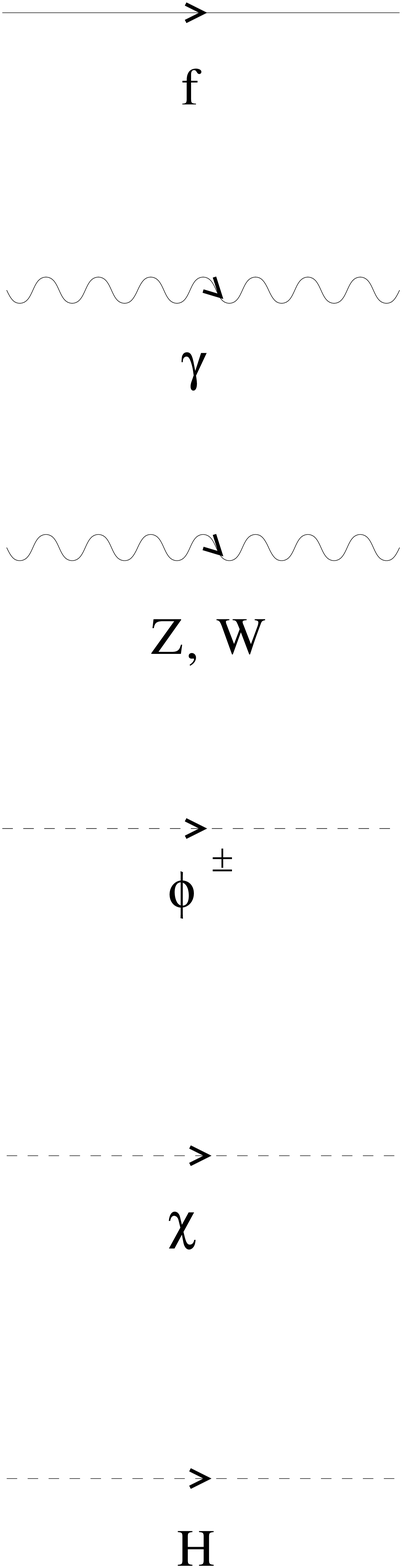}}}
\put(2.6,7.25){$+\frac{i}{\not q - m}$}
\put(2.6,6.21){$-\frac{ig^{\alpha\beta}}{q^{2}}$}
\put(2.6,5.25){$-\frac{ig^{\alpha\beta}}{q^{2}-M_{V}^{2}}$}
\put(2.6,4.18){$+\frac{i}{q^{2}-M_{W}^{2}}$}
\put(2.6,3){$+\frac{i}{q^{2}-M_{Z}^{2}}$}
\put(2.6,1.8){$+\frac{i}{q^{2}-M_{H}^{2}}$}
\end{picture}
\end{center}
\end{figure}

\newpage

\chapter{Dimensional regularization and some useful integrals}
\label{Decko}

\section{Dimensional regularization}
\label{dimreg}

Before one-loop amplitudes with divergent momentum integrals can be
renormalized, they have to be regularized. Regularization defines integrals,
parametrizes their divergences, and separates their finite parts.
Dimensional regularization \cite{thooftdim} defines integrals by analytically
continuing them from 4-dimensional to $n$-dimensional space-time. The
computation of integrals in $n$ dimensions typically yields (see Eq.
\ref{lambda})
\begin{eqnarray}
%\label{lambda}
\Lambda_{V}(0) & = & -\frac{\alpha_{137}}{4\pi}\left(
\frac{2}{\epsilon} + finite \;\;constants\right), \;\;\;\;\epsilon = 4 - n,
\nonumber
\end{eqnarray}
that is, the divergence is parametrized as a simple pole at $n = 4$. In our
calculations, we used the following momentum integrals in $n$ dimensions:

\begin{eqnarray}
\label{momint}
I_{0}(l) & = & \int \frac{d^{n}k}{(2\pi)^{n}} \frac{1}{\big(k^{2} + 2k\cdot s +
t\big)^{l}} \; = \; \frac{i (- \pi)^{n/2}}{(2\pi)^{n}}
\frac{\Gamma(l- n/2)}{\Gamma(l)} \frac{1}{(t-s^{2})^{(l- n/2)}}\nonumber  \\
 & \equiv &
N(l) \frac{1}{(t-s^{2})^{(l- n/2)}},   \\
I_{\mu}(l) & = & \int \frac{d^{n}k}{(2\pi)^{n}} \frac{k_{\mu}}{\big(k^{2}
 + 2k\cdot s + t\big)^{l}} \; = \; - s_{\mu} I_{0}(l),  \\
I_{\mu\nu}(l) & = & \int \frac{d^{n}k}{(2\pi)^{n}}
\frac{k_{\mu}k_{\nu}}{\big(k^{2}
 + 2k\cdot s + t\big)^{l}} \; = \;  I_{0}(l) \Big[ s_{\mu}s_{\nu} +
\frac{1}{2}g_{\mu\nu}(t-s^{2}) \nonumber \\
& \times & \frac{1}{l - n/2 -1}\Big].
\end{eqnarray}
Specifically, for $l = 2,3$ we have
\begin{eqnarray}
\label{endva}
N(2) & = & \frac{i (- \pi)^{\frac{n}{2}}}{(2\pi)^{n}} \frac{\Gamma(2-
\frac{n}{2})}{\Gamma(2)} \; = \; \frac{i (- 1)^{\frac{n}{2}}}{(4\pi)^{2}}
\Big(1+\frac{\epsilon}{2}\ln 4\pi\Big) \Big(\frac{2}{\epsilon} - \gamma\Big),
  \\
N(3) & = & \frac{i (- \pi)^{\frac{n}{2}}}{(2\pi)^{n}} \frac{\Gamma(3-
\frac{n}{2})}{\Gamma(3)} \; = \;  \frac{i (- 1)^{\frac{n}{2}}}{2(4\pi)^{2}}
\Big(1+\frac{\epsilon}{2}\ln 4\pi\Big) \Big(1- \frac{\epsilon}{2}\gamma\Big),
\end{eqnarray}
where we used
\begin{eqnarray}
\Gamma\Big(2-\frac{n}{2}\Big) = \Gamma\Big(2- \frac{4 - \epsilon}{2}\Big) =
\Gamma\Big(\frac{\epsilon}{2}\Big) = \frac{2}{\epsilon} - \gamma, \nonumber \\
\Gamma\Big(3-\frac{n}{2}\Big) = \Gamma\Big(1+ \frac{\epsilon}{2}\Big) =
\frac{\epsilon}{2}\Gamma\Big(\frac{\epsilon}{2}\Big) = 1-
\frac{\epsilon}{2}\gamma,
\end{eqnarray}
where $\gamma \doteq 0.5772$ is Euler-Mascheroni constant.
In $n = 4 - \epsilon$ dimensions, $\alpha$ becomes a dimensional quantity:
\begin{eqnarray}
\label{alfamu}
\alpha = \frac{e^{2}}{4\pi} \rightarrow \alpha \mu^{\epsilon} = \alpha
\Big(1 + \frac{\epsilon}{2} \ln \mu^{2} + ... \Big),
\end{eqnarray}
where $\mu$ is an arbitrary mass scale. The combination of Eq. \ref{endva}
and Eq. \ref{alfamu} yields
\begin{eqnarray}
\label{alfamu1}
\alpha \mu^{\epsilon} \Big(1+\frac{\epsilon}{2}\ln 4\pi\Big)
\Big(\frac{2}{\epsilon} - \gamma\Big) & = & \alpha \Big(\frac{2}{\epsilon} -
\gamma + \ln 4\pi + \ln \mu^{2} \Big) \;\;=\;\; \alpha \Delta_{\mu}.
\end{eqnarray}
It is this $\Delta_{\mu}$ rather than $\frac{2}{\epsilon}$, which is usually
thought of as the parameterization of the divergence since factors $\gamma, \ln
4\pi$ and $\ln \mu^{2}$ are always present along with $\frac{2}{\epsilon}$ and
they all together cancel out in renormalized quantities. We use also other
variants of the $\Delta$ symbol:
\begin{eqnarray}
\label{deltas}
\Delta & = & \frac{2}{4-n}-\gamma-\ln \pi \;=\;\frac{2}{\epsilon}
-\gamma-\ln \pi, \nonumber  \\
\Delta_{\mu} & = & \frac{2}{\epsilon}-\gamma + \ln 4\pi + \ln \mu^{2},
\nonumber  \\
\Delta_{m} & = & \frac{2}{\epsilon}-\gamma + \ln 4\pi - \ln
\frac{m^{2}}{\mu^{2}}. 
\end{eqnarray}

To cast momentum integrals into the form of Eq. \ref{momint}, the following
Feynman parameterization is used:
\begin{eqnarray}
\frac{1}{a_{0}a_{1}a_{2}...a_{n}} &  = & \Gamma (n+1) \int_{0}^{1} dx_{1}
\int_{0}^{x_{1}} dx_{2} ... \int_{0}^{x_{n-1}} dx_{n} \nonumber \\
& \times &
\frac{1}{[a_{0}+(a_{1}-a_{0})x_{1}+...(a_{n}-a_{n-1})x_{n}]^{n+1}}.
\end{eqnarray}
Higher powers in $a_{i}$ are obtained by differentiation with respect to this
parameter. In this work, these specific expressions were used:
\begin{eqnarray}
\frac{1}{ab} & = & \int_{0}^{1} dx \frac{1}{[a+(b-a)x]^{2}}, \\
\frac{1}{abc} & = & 2 \int_{0}^{1}dx \int_{0}^{x} dy
\frac{1}{[a+(b-a)x+(c-b)y]^{3}}, \\
\label{pabc}
\frac{1}{abc} & = & 2 \int_{0}^{1}dx \int_{0}^{1} dy \frac{y}{\big[axy +
by(1-x)
+ c(1-y)\big]^{3}},                \\
\frac{1}{abc^{2}} & = & 6 \int_{0}^{1}dx \int_{0}^{x} dy
\frac{y}{[a+(b-a)x+(c-b)y]^{4}}.
\end{eqnarray}

In $n$ dimensions, the algebra of the Dirac matrices described in Appendix A 
is generalized as follows:
\begin{eqnarray}
\label{nalgebra}
\gamma^{\mu}\gamma^{\nu} + \gamma^{\nu}\gamma^{\mu} & = & 2 g^{\mu \nu},
\nonumber \\
g^{\mu \nu} g_{\mu \nu} & = & n, \nonumber \\
\gamma_{\mu}\gamma^{\mu} & = & n, \nonumber \\
Tr(\gamma^{\mu} \gamma^{\nu}) & = & n g^{\mu \nu}, \nonumber \\
Tr(\gamma^{\mu}\gamma^{\nu}\gamma^{\lambda}\gamma^{\sigma}) & = & n \left(
g^{\mu\nu}g^{\lambda\sigma} - g^{\mu\lambda}g^{\nu\sigma} +
g^{\mu\sigma}g^{\nu\lambda}\right), \nonumber  \\
\gamma_{\mu}\gamma^{\nu}\gamma^{\mu} & = & (2 - n) \gamma^{\nu}, \nonumber \\
\gamma_{\mu}\gamma^{\nu}\gamma^{\lambda}\gamma^{\mu} & = & 4
g^{\nu\lambda} + (n - 4) \gamma^{\nu}\gamma^{\lambda}, \nonumber \\
\gamma_{\mu}\gamma^{\nu}\gamma^{\lambda}\gamma^{\sigma}\gamma^{\mu} & = &
- 2 \gamma^{\sigma}\gamma^{\lambda}\gamma^{\nu} 
- (n -4) \gamma^{\nu}\gamma^{\lambda}\gamma^{\sigma}, \nonumber  \\
\gamma_{5}\gamma^{\mu} + \gamma^{\mu}\gamma_{5} & = & 0. 
\end{eqnarray}

\section{The computation of ${\cal I}_{0}, {\cal I}_{1}(m), {\cal I}_{2}(m)$ and
${\cal I}_{3}(m)$ integrals}

We start with two useful integrals one often encounters during the computation
of momentum integrals:
\begin{eqnarray}
\int_{0}^{1} dx \ln (Cx + D) & = & \ln (C + D) + \frac{D}{C} \ln \frac{C +
D}{D} - 1, \nonumber \\  \nonumber \\
\int_{0}^{1} dx \big[x \ln (Ex + F)\big] & = &  \frac{(E+F)^{2}}{2E^{2}}
\ln (E+F)
- \frac{F^{2}}{2E^{2}} \ln F - \frac{1}{2E^{2}} \frac{(E+F)^{2}}{2} \nonumber \\
& + & \frac{1}{4}\frac{F^{2}}{E^{2}} - \frac{F(E+F)}{E^{2}}\ln (E+F)
+ \frac{FE}{E^{2}} \nonumber \\
& + & \frac{F^{2}\ln F}{E^{2}}. 
\end{eqnarray}

To compute ${\cal I}_{0}, {\cal I}_{1}(m), {\cal I}_{2}(m)$ and
${\cal I}_{3}(m)$  we note
these integrals (defined in \linebreak Sec.~\ref{secbox}) are related via
\begin{equation}
{\cal I}_{2}(M_{Z}) = {\cal I}_{1} + M_{N}^{2} \;\; {\cal I}_{3}(M_{Z}),
\end{equation}
and the
integral ${\cal I}_{0}$ is obtained as a special case of ${\cal I}_{1}(m)$
for $M_{N} \rightarrow 0$.
To calculate the ${\cal I}_{1}(m)$ integral, we use the 
parameterization of Eq. \ref{pabc}:
\begin{eqnarray}
\frac{1}{abc} & = & 2 \int_{0}^{1}dx \int_{0}^{1}dy \frac{y}{\big[axy + by(1-x)
+ c(1-y)\big]^{3}}.  \nonumber
\end{eqnarray}
The term in the denominator is
\begin{eqnarray}
axy + by(1-x) + c(1-y) & = & (k^{2} - M_{W}^{2})xy + (k^{2} - m^{2})(y-xy)
\nonumber \\
& + & (k^{2} - M_{N}^{2})(1-y) \nonumber \\
& = &
k^{2} - M_{W}^{2}xy + m^{2}xy - m^{2}y \nonumber \\
& - &
M_{N}^{2}(1-y).
\end{eqnarray}
Using the momentum integral Eq. \ref{momint}, we can write for ${\cal
I}_{1}(m)$
\begin{eqnarray}
\frac{i}{(4\pi)^{2}}{\cal I}_{1}(m) & = &
\int \frac{d^{4}k}{{(2\pi)}^{4}}
\frac{1}{(k^{2}-M_{N}^{2})(k^{2}-M_{W}^{2})(k^{2}-m^{2})} \nonumber \\
& = & 2 \int_{0}^{1}dx \int_{0}^{1}dy
\int \frac{d^{4}k}{(2\pi)^{4}} \nonumber \\
& \times &
\frac{y}{\big[k^{2} - M_{W}^{2}xy + m^{2}xy - m^{2}y - M_{N}^{2}(1-y)
\big]^{3}} \nonumber  \\
& = &
\frac{i}{(4\pi)^{2}}\int_{0}^{1}dx \int_{0}^{1}dy \nonumber  \\
& \times &
\frac{y}{\big(m^{2} -
M_{W}^{2}\big)yx - m^{2}y - M_{N}^{2}(1-y)}.
\end{eqnarray}
First we integrate over the $x$ parameter:
\begin{eqnarray}
{\cal I}_{1}(m) & = & \int_{0}^{1}dy\:y \int_{0}^{1}dx 
\frac{-1}{\big(M_{W}^{2} -
m^{2}\big)yx + m^{2}y + M_{N}^{2}(1-y)} \nonumber  \\
 & = & \int_{0}^{1}dy\:y \int_{0}^{1}dx
\frac{-1}{Ax+B}  \; = \; - \int_{0}^{1}dy\:y 
\frac{1}{A}\big[\ln (A+B) - \ln B\big] \nonumber  \\
 & = &
\int_{0}^{1}dy\:y 
\frac{1}{\big(m^{2} - M_{W}^{2}\big)y}\big[\ln (M_{W}^{2}y + M_{N}^{2}
\; - \;  M_{N}^{2}y) \big. \nonumber  \\
& - & \big. \ln (m^{2}y + M_{N}^{2} - M_{N}^{2}y)\big],
\end{eqnarray}
and then the integration over $y$:
\begin{eqnarray}
{\cal I}_{1}(m)  & = &
\int_{0}^{1}dy \frac{1}{\big(m^{2} - M_{W}^{2}\big)}\big[\ln (M_{W}^{2}y +
M_{N}^{2} -
M_{N}^{2}y) - \ln (m^{2}y + M_{N}^{2} - M_{N}^{2}y)\big] \nonumber  \\
& = &
\frac{1}{\big(m^{2} - M_{W}^{2}\big)}\Big[\int_{0}^{1}dy \ln (Cy + D)
- \int_{0}^{1}dy \ln (Ey + F)\Big] \nonumber \\
& = &
\frac{1}{\big(m^{2} - M_{W}^{2}\big)}\Big[\ln (C + D)+ \frac{D}{C}\ln
\frac{C+D}{D} - 1 - \ln (E + F) \Big. \nonumber \\
& - & \Big. \frac{F}{E}\ln \frac{E+F}{F} + 1 \Big]
\nonumber \\
& = &
\frac{1}{\big(m^{2} - M_{W}^{2}\big)}
\Big[\ln \frac{M_{W}^{2}}{m^{2}} + \frac{M_{N}^{2}}{M_{W}^{2} - M_{N}^{2}}
\ln \frac{M_{W}^{2}}{M_{N}^{2}} - \frac{M_{N}^{2}}{m^{2} - M_{N}^{2}}
\ln \frac{m^{2}}{M_{N}^{2}} \Big].
\end{eqnarray}
   The ${\cal I}_{3}(m)$ integral is found using similar steps:
\begin{eqnarray}
{\cal I}_{3}(m) & = & \frac{{(4\pi)}^{2}}{i}\int \frac{d^{4}k}{{(2\pi)}^{4}}
\frac{1}{(k^{2}-M_{N}^{2})^{2}(k^{2}-M_{W}^{2})(k^{2}-m^{2})}
 \nonumber \\
& = &
6 \int \frac{d^{4}k}{{(2\pi)}^{4}} \int_{0}^{1}dx \int_{0}^{x}dy
\frac{y}{\big[k^{2}-M_{W}^{2}+(M_{W}^{2}-m^{2})x+(m^{2}-M_{N}^{2})y\big]^{4}}
\nonumber \\
& = &
\int_{0}^{1}dx \int_{0}^{x}dy
\frac{y}{\big[(M_{W}^{2}-m^{2})x -M_{W}^{2} +(m^{2}-M_{N}^{2})y\big]^{2}}
 \nonumber \\
& = &
\frac{1}{m^{2}-M_{W}^{2}}
\biggl\{ \frac{1}{M_{N}^{2}-M_{W}^{2}} +
\frac{M_{W}^{2} \ln \frac{M_{W}^{2}}{M_{N}^{2}}}{{(M_{N}^{2}-M_{W}^{2})}^{2}}
-\frac{1}{M_{N}^{2}-m^{2}} \biggr. \nonumber \\
& - & \biggl.
\frac{m^{2} \ln \frac{m^{2}}{M_{N}^{2}}}{{(M_{N}^{2}-m^{2})}^{2}} \biggr\}. 
\end{eqnarray}

\newpage

\chapter{Renormalization constants, unrenormalized self-energies and
't Hooft scalar integrals} 
\label{Ecko}

\section{Renormalization constants}
\label{recons}

It is convenient to define the following linear combinations of the
renormalization constants ($i=1,2$):
\begin{eqnarray}
\delta Z_{i}^{\gamma} & = & s_{W}^{2}\;\delta Z_{i}^{W}\;+\;c_{W}^{2}\;\delta
Z_{i}^{B}, \nonumber \\
\delta Z_{i}^{Z}      & = & c_{W}^{2}\;\delta Z_{i}^{W}\;+\;s_{W}^{2}\;\delta
Z_{i}^{B}, \nonumber \\
\delta Z_{i}^{\gamma Z} & = & \frac{c_{W}s_{W}}{c_{W}^{2}-s_{W}^{2}}
\left(\delta Z_{i}^{Z}\;-\;\delta
Z_{i}^{\gamma}\right), \nonumber   \\
\delta Z_{V}^{f} & = & (\delta Z_{L}^{f} + \delta Z_{R}^{f})/2\;,\;\;\;\;\;\;\;
\delta Z_{A}^{f}\; = \; (\delta Z_{L}^{f} - \delta Z_{R}^{f})/2.   
\end{eqnarray}
The renormalized self-energies are obtained from unrenormalized ones by adding
appropriate counterterms:
\begin{eqnarray}
\label{rselfe}
\hat{\Sigma}^{\gamma}\left(p^{2}\right) & = & \Sigma^{\gamma}\left(p^{2}\right)
\;+\;\delta Z_{2}^{\gamma}\;p^{2}, \nonumber \\
\hat{\Sigma}^{Z}\left(p^{2}\right) & = & \Sigma^{Z}\left(p^{2}\right)
\;-\;\delta M_{Z}^{2}\;+\;\delta
Z_{2}^{Z}\;\left(p^{2}-M_{Z}^{2}\right), \nonumber \\
\hat{\Sigma}^{W}\left(p^{2}\right) & = & \Sigma^{W}\left(p^{2}\right)
\;-\;\delta M_{W}^{2}\;+\;\delta Z_{2}^{W}\;\left(p^{2}-M_{W}^{2}\right),
\nonumber \\
\hat{\Sigma}^{\gamma Z}\left(p^{2}\right) & = & \Sigma^{\gamma Z}
\left(p^{2}\right)
\;-\;\delta Z_{2}^{\gamma Z}p^{2}\;+\;\left(\delta Z_{1}^{\gamma Z}\;-
\;\delta Z_{2}^{\gamma Z}\right)M_{Z}^{2}, \nonumber \\
\hat{\Sigma}^{f}\left(p\right) & = & {\not p}\left(\Sigma_{V}^{f}(p^{2})
 + \delta
Z_{V}^{f}\right)\;+\;{\not p}\gamma_{5}\left(\Sigma_{A}^{f}(p^{2}) - \delta
Z_{A}^{f}\right) \nonumber \\
& + & m_{f}\left(\Sigma_{S}^{f}(p^{2}) - \delta Z_{V}^{f}
- \frac{\delta m_{f}}{m_{f}}\right) ,
\end{eqnarray}
where $\Sigma_{V}^{f}, \Sigma_{A}^{f}$ and $\Sigma_{S}^{f}$ are vector, axial
vector and scalar part of the fermion self-energy, respectively (see Eq.
\ref{hasthe}).

The renormalized electromagnetic, weak neutral and charged current vertices are
given by
\begin{eqnarray}
\label{rvertexa}
\hat{\Gamma}^{\gamma ff} & = & \Gamma^{\gamma ff} \;+\; ie_{137}\gamma_{\mu}
\left(\delta Z_{1}^{\gamma} - \delta Z_{2}^{\gamma} + \delta Z_{V}^{f} -
\delta Z_{A}^{f}\gamma_{5}\right) \nonumber \\
& - & ie_{137}\gamma_{\mu}(v_{f}-a_{f}\gamma_{5})\left(\delta Z_{1}^{\gamma Z}
- \delta Z_{2}^{\gamma Z}\right), \nonumber \\
\hat{\Gamma}^{Zff} & = & \Gamma^{Zff} \;+\; ie_{137}\gamma_{\mu}
(v_{f}-a_{f}\gamma_{5})\left(\delta Z_{1}^{Z}
- \delta Z_{2}^{Z}\right)\;-\;ie_{137}\gamma_{\mu}\left(\delta Z_{1}^{\gamma Z}
- \delta Z_{2}^{\gamma Z}\right) \nonumber \\
& & + ie_{137}\gamma_{\mu}\left(v_{f}\delta Z_{V}^{f} + a_{f}\delta
Z_{A}^{f}\right)\;-\;ie_{137}\gamma_{\mu}\gamma_{5}\left(v_{f}\delta Z_{A}^{f}
+ a_{f}\delta Z_{V}^{f}\right), \nonumber \\
\hat{\Gamma}^{Wl\nu} & = & \Gamma^{Wl\nu} \;+\; i\frac{e_{137}}{2\sqrt{2}s_{W}}
\gamma_{\mu}(1-\gamma_{5})\left(1+\delta Z_{1}^{W}-\delta Z_{2}^{W}+\delta
Z_{L}^{f}\right). 
\end{eqnarray}
The renormalization constants are obtained from the OS renormalization
conditions Eqs. \ref{RC1}-\ref{RC2} (we show only constants needed for our
calculations):
\begin{eqnarray}
\label{rconstants}
\delta M_{W}^{2} & = & Re\;\Sigma^{W}(M_{W}^{2}), \nonumber \\
\delta M_{Z}^{2} & = & Re\;\Sigma^{Z}(M_{Z}^{2}), \nonumber \\
\delta Z_{2}^{\gamma} & = & - \frac{\partial \Sigma^{\gamma}}{\partial
p^{2}}(0), \nonumber \\
\delta Z_{1}^{\gamma} & = & - \frac{\partial \Sigma^{\gamma}}{\partial
p^{2}}(0)\;-\;\frac{s_{W}}{c_{W}}\frac{\Sigma^{\gamma Z}(0)}{M_{Z}^{2}},
\nonumber  \\
\delta Z_{2}^{Z} & = & - \frac{\partial \Sigma^{\gamma}}{\partial
p^{2}}(0)\;-\;2\frac{c_{W}^{2}-s_{W}^{2}}{s_{W}c_{W}}\frac{\Sigma^{\gamma
Z}(0)}{M_{Z}^{2}}\;+\;\frac{c_{W}^{2}-s_{W}^{2}}{s_{W}^{2}}
\left(\frac{\delta M_{Z}^{2}}{M_{Z}^{2}}-\frac{\delta M_{W}^{2}}{M_{W}^{2}}
\right), \nonumber \\
\delta Z_{1}^{Z} & = & - \frac{\partial \Sigma^{\gamma}}{\partial
p^{2}}(0)\;-\;\frac{3c_{W}^{2}-2s_{W}^{2}}{s_{W}c_{W}}\frac{\Sigma^{\gamma
Z}(0)}{M_{Z}^{2}}\;+\;\frac{c_{W}^{2}-s_{W}^{2}}{s_{W}^{2}}
\left(\frac{\delta M_{Z}^{2}}{M_{Z}^{2}}-\frac{\delta M_{W}^{2}}{M_{W}^{2}}
\right), \nonumber \\
\delta Z_{2}^{W} & = & - \frac{\partial \Sigma^{\gamma}}{\partial
p^{2}}(0)\;-\;2\frac{c_{W}}{s_{W}}\frac{\Sigma^{\gamma
Z}(0)}{M_{Z}^{2}}\;+\;\frac{c_{W}^{2}}{s_{W}^{2}}\left(\frac{\delta
M_{Z}^{2}}{M_{Z}^{2}}-\frac{\delta M_{W}^{2}}{M_{W}^{2}}
\right), \nonumber  \\
\delta Z_{V}^{f} & = & - \Sigma_{V}^{f}(m_{f}^{2}) - m_{f}^{2} 
\big[2 \Sigma^{f'}_{V}(m_{f}^{2})
+ 2 \Sigma^{f'}_{S}(m_{f}^{2})\big],\;\;\;\;\;\Sigma^{f'}_{V,S}(m_{f}^{2})\;=\;
\frac{\partial \Sigma_{V,S}^{f}}{\partial p^{2}}(m_{f}^{2}),  \nonumber  \\
\delta Z_{A}^{f} & = & + \Sigma_{A}^{f}(m_{f}^{2}).
\end{eqnarray}

\section{Unrenormalized self-energies in the SM}

Below we present complete SM gauge boson self-energies corresponding to Figs.
\ref{pfd} - \ref{wfd}. They were calculated in Ref. \cite{key6}. For the
definition of the function $F$ see Sec. \ref{scalari}; for the definition of
the $\Delta$ factors see Eq. \ref{deltas}; $s = p^{2}$, where $p$ is the
4-momentum of the gauge boson; $w=M_{W}^{2}, z=M_{Z}^{2}, h=M_{H}^{2}$. 
\begin{eqnarray}
\label{agama}
\Sigma^{\gamma}(s) & = & \frac{\alpha}{4\pi}\Big\{\frac{4}{3}\sum_{f}
Q_{f}^{2}\Big[ s\Delta_{f} + (s+2m_{f}^{2})F(p;m_{f},m_{f}) -
\frac{s}{3}\Big]\Big. \nonumber \\
& - &
\Big. 3 s \Delta_{W} - (3s+4w)F(p;M_{W},M_{W})\Big\},  \\
\nonumber \\
%%%%%%%%%%%%    gamma - Z    %%%%%%%%%%%%%%%%%%%%%%%%%%%%%%%%%%%%%%%%%
\label{asedem}
\Sigma^{\gamma Z}(s) & = & \frac{\alpha}{4\pi}\Big\{-\frac{4}{3}\sum_{f}
Q_{f} v_{f}\Big[ s\Delta_{f} + (s+2m_{f}^{2})F(p;m_{f},m_{f}) -
\frac{s}{3}\Big]\Big. \nonumber \\
& + &
\frac{1}{c_{W}s_{W}}\Big[\Big(3 c_{W}^{2} + \frac{1}{6}\Big)s+2w \Big]
\Delta_{W} \nonumber \\
& + &
\Big.\frac{1}{c_{W}s_{W}}\Big[\Big(3 c_{W}^{2} + \frac{1}{6}\Big)s+
\Big(4c_{W}^{2}+\frac{4}{3}\Big)w\Big] F(p;M_{W},M_{W}) +
\frac{s}{9c_{W}s_{W}}\Big\},  \\
\nonumber \\
%%%%%%%%%%%%    Z    %%%%%%%%%%%%%%%%%%%%%%%%%%%%%%%%%%%%%%%%%%%%%%%%%%
\label{azet}
\Sigma^{Z}(s) & = & \frac{\alpha}{4\pi}\Big\{\frac{4}{3}\sum_{l=e,\mu,\tau}
2a_{l}^{2}s\Big(\Delta_{l} + \frac{5}{3} - \ln\Big(-\frac{s}{m_{l}^{2}}
-i\epsilon\Big)\Big)\Big. \nonumber \\
& + &
\frac{4}{3}\sum_{f\neq\nu}\Big[(v_{f}^{2}+a_{f}^{2})\Big(s\Delta_{f}+
(s+2m_{f}^{2})F(p;m_{f},m_{f})-\frac{s}{3}\Big)\Big. \nonumber \\
& - &
\Big.\frac{3}{8c_{W}^{2}s_{W}^{2}}m_{f}^{2}
(\Delta_{f}+F(p;m_{f},m_{f}))\Big]
\nonumber \\
& + &
\Big[\Big(3-\frac{19}{6s_{W}^{2}}+\frac{1}{6c_{W}^{2}}\Big)s+
\Big(4+\frac{1}{c_{W}^{2}}-\frac{1}{s_{W}^{2}}\Big)M_{Z}^{2}\Big]
\Delta_{W}  \nonumber \\
& + &
\Big[\left(-c_{W}^{4}(40s+80w) + (c_{W}^{2}-s_{W}^{2})^{2}(8w+s) + 12w\right)
F(p;M_{W},M_{W}) \Big. \nonumber \\
& + &
\Big(10z-2h+s+\frac{(h-z)^{2}}{s}\Big)F(p;M_{H},M_{Z})- 2h\ln\frac{h}{w} -
2z\ln{z}{w}  \nonumber \\
& + &
(10z-2h+s)\Big(1-\frac{h+z}{h-z}\ln\frac{M_{H}}{M_{Z}} - \ln
\frac{M_{H}M_{Z}}{w} \Big) \nonumber \\
& + &
 \Big.\Big.\frac{2}{3}s\Big(1 + (c_{W}^{2}-s_{W}^{2})^{2} -
4c_{W}^{2}\Big)\Big]\frac{1}{12c_{W}^{2}s_{W}^{2}}\Big\},  \\
\nonumber \\
%%%%%%%%%%%%%%%%%%     W      %%%%%%%%%%%%%%%%%%%%%%%%%%%%%%%%%%%%%%%%%%%%%%%
\label{adablju}
\Sigma^{W}(s) & = & \frac{\alpha}{4\pi} \frac{1}{s_{W}^{2}}\Big\{\frac{1}{3}
\sum_{l=e,\mu,\tau}\Big[\Big(s-\frac{3}{2}m_{l}^{2}\Big)\Delta_{l}\Big.\Big. 
\nonumber \\
& + &
\Big.\Big(s-\frac{m_{l}^{2}}{2} - \frac{m_{l}^{4}}{2s}\Big)F(p;0,m_{l}) +
\frac{2}{3}s - \frac{m_{l}^{2}}{2}\Big]  \nonumber \\
& + &
\sum_{q-doublets} \frac{1}{3} \Big[ \frac{\Delta_{+}}{2}\Big(s -
\frac{5}{2}m_{+}^{2} + \frac{m_{-}^{2}}{2}\Big) + \frac{\Delta_{-}}{2}
\Big(s -
\frac{5}{2}m_{-}^{2} + \frac{m_{+}^{2}}{2}\Big)\Big. \nonumber \\
& + &
\Big(s- \frac{m_{+}^{2}+m_{-}^{2}}{2} -
\frac{(m_{+}^{2}-m_{-}^{2})^{2}}{2s}\Big) F(p;m_{+},m_{-}) \nonumber \\
& + &
\Big. \Big(s - \frac{m_{+}^{2}+m_{-}^{2}}{2}\Big)\Big(1 -
\frac{m_{+}^{2}+m_{-}^{2}}{m_{+}^{2}-m_{-}^{2}} \ln{m_{+}}{m_{-}}\Big)
-\frac{s}{3}\Big] \nonumber \\
& - &
\Big[ \frac{19}{2} s + 3w\Big(1 - \frac{s_{W}^{2}}{c_{W}^{2}}\Big)\Big]
\frac{\Delta_{W}}{3} \nonumber \\
& + &
\Big[ s_{W}^{4} z -\frac{c_{W}^{2}}{3}\Big(7z + 7w +10s -
2\frac{(z-w)^{2}}{s}\Big)\Big. \nonumber \\
& - &
\Big.\frac{1}{6}\Big( w + z -\frac{s}{2} 
 - \frac{(z-w)^{2}}{2s}\Big)\Big]
F(p;M_{Z},M_{W}) \nonumber \\
& + &
\frac{s_{W}^{2}}{3}\Big(-4w-10s + \frac{2w^{2}}{s}\Big) F(p;0,M_{W})
\nonumber \\
& + &
\frac{1}{6}\Big(5w -h +\frac{s}{2} + \frac{(h-w)^{2}}{2s} \Big)
F(p;M_{H},M_{W}) \nonumber \\
& + &
\Big[\frac{c_{W}^{2}}{3}\Big(7z+7w+10s-4(z-w)\Big) - s_{W}^{4}z +
\frac{1}{6}\Big(2w-\frac{s}{2}\Big)\Big]\frac{3z}{z-w} \ln\frac{z}{w}
\nonumber \\
& - &
\Big(\frac{2}{3}w+\frac{s}{12}\Big)\frac{h}{h-w}\ln\frac{h}{w} -
\frac{c_{W}^{2}}{3}\Big(7z +7w+\frac{32}{3}s\Big)+s_{W}^{4}z \nonumber \\
& + &
\Big.\frac{1}{6}\Big(\frac{5}{3}s+4w-z-h\Big) - 
\frac{s_{W}^{2}}{3}\Big(4w +
\frac{32}{3}s\Big)\Big\}. 
\end{eqnarray}

\section{'t Hooft scalar integrals}
\label{scalari}

Here we define various $C, B, A$ and $F$ functions and reduce them to scalar
integrals $C_{0}$, $B_{0}$ and $A_{0}$. For the calculation of $C_{0}$ and 
$B_{0}$ we refer the reader to the original work of 't Hooft and Veltman,
Ref. \cite{thooft}.

%***********C functions definition ******************
The $C_{0}$ function is defined as (with finite parts indicated by the 
superscript):
\begin{eqnarray}
\label{cnula}
C_{0}(m_{1},m_{2},m_{3}) & \equiv & C_{0}(p_{1},p_{2};m_{1},m_{2},m_{3})
\; \equiv \; C_{0}^{fin}(m_{1},m_{2},m_{3}) \nonumber   \\
& = &  -\int \frac{d^{n}q}{i\pi^{2}}
\frac{1}{D}, 
\end{eqnarray}
where 
\begin{eqnarray}
D & = & (q^{2}-m_{1}^{2}+i\epsilon)\;\big[(q-p_{1})^{2}-m_{2}^{2}
+i\epsilon\big]\; 
\big[(q-p_{1}-p_{2})^{2}-m_{3}^{2}+i\epsilon\big].  
\end{eqnarray}
%************ C24,C23,C11 reductions ****************
The functions  $C_{ij}$ are defined by:
\begin{eqnarray}
C_{\mu} & = & -\int \frac{d^{n}q}{i\pi^{2}}\frac{q_{\mu}}{D} = -p_{1\mu}C_{11}
-
p_{2\mu}C_{12}, \nonumber \\        \nonumber \\
C_{\mu\nu} & = & -\int \frac{d^{n}q}{i\pi^{2}} \frac{q_{\mu}q_{\nu}}{D}
\nonumber \\
& = & p_{1\mu}p_{1\nu}C_{21} + p_{2\mu}p_{2\nu}C_{22} + (p_{1\mu}p_{2\nu} +
p_{1\nu}p_{2\mu})C_{23} - g_{\mu\nu}C_{24}. 
\end{eqnarray}
The functions  $C_{11}, C_{24}, C_{23}$ are reduced (in the limit
$p_{1}^{2} = p_{2}^{2} = m_{l}^{2} \ll  (p_{1}+p_{2})^{2}=M_{Z}^{2}$,
applicable for our considerations of the leptonic decays of the Z boson) to:
\begin{eqnarray}
C_{11}(m_{1},m_{2},m_{3}) & = & C_{11}^{fin}(m_{1},m_{2},m_{3}) =
 -\frac{1}{M_{Z}^{2}}[f_{2}C_{0}(m_{1},m_{2},m_{3})  \nonumber \\
& - &
B_{0}^{fin}(p_{1}+p_{2};m_{1},m_{3}) + B_{0}^{fin}(p_{1};m_{1},m_{2})],
 \nonumber \\   \nonumber \\
C_{24}(m_{1},m_{2},m_{3}) & = & \frac{1}{4}\Delta +
 C_{24}^{fin}(m_{1},m_{2},m_{3}), \nonumber  \\
C_{24}^{fin}(m_{1},m_{2},m_{3}) & = & [m_{1}^{2}C_{0}(m_{1},m_{2},m_{3}) +
f_{1}C_{11}(m_{1},m_{2},m_{3})  \nonumber \\
& + &
B_{1}^{fin}(p_{1}+p_{2};m_{1},m_{3})](-\frac{1}{2}) + \frac{1}{4},
\nonumber \\    \nonumber \\
C_{23}(m_{1},m_{2},m_{3}) & = &  C_{23}^{fin}(m_{1},m_{2},m_{3}) =
 -\frac{1}{M_{Z}^{2}}
[B_{1}^{fin}(p_{1}+p_{2};m_{1},m_{3}) 
\nonumber \\
& + & 
B_{0}^{fin}(p_{2};m_{2},m_{3}) + f_{1}C_{11}(m_{1},m_{2},m_{3})] \nonumber \\
& + & C_{24}^{fin}(m_{1},m_{2},m_{3})
 \frac{2}{M_{Z}^{2}},                                 
\end{eqnarray}
where
\begin{eqnarray}
f_{2} & = & M_{Z}^{2}+m_{2}^{2}-m_{3}^{2}, \nonumber \\
f_{1} & = & m_{1}^{2}-m_{2}^{2}. 
\end{eqnarray}
%*************B functions definition *****************
The functions $B_{0}, B_{1}$ are defined as:
\begin{eqnarray}
\label{abfunc}
B_{0}(p;m_{1},m_{2}) & = & \int \frac{d^{n}q}{i\pi^{2}}
\frac{1}{(q^{2}-m_{1}^{2}+i\epsilon)\big[(q-p)^{2}-m_{2}^{2}+i\epsilon\big]}
\nonumber  
\\
& = & \Delta + B_{0}^{fin}(p;m_{1},m_{2}),
 \nonumber \\
B_{0}^{fin}(p;m_{1},m_{2}) & = & -\int_{0}^{1} dx\; 
\ln \big[p^{2}x^{2} + m_{1}^{2} - (p^{2}+m_{1}^{2}-m_{2}^{2})x\big],
 \nonumber \\   \nonumber \\
B_{\mu}(p;m_{1},m_{2}) & = & \int \frac{d^{n}q}{i\pi^{2}}
\frac{q_{\mu}}{(q^{2}-m_{1}^{2}+i\epsilon)\big[(q-p)^{2}-m_{2}^{2}+i\epsilon
\big]}
\;\;=\;\;-p_{\mu}B_{1}, \nonumber \\
B_{1}(p;m_{1},m_{2}) & = & -\frac{1}{2}\Delta
+ B_{1}^{fin}(p;m_{1},m_{2}), \nonumber \\
B_{1}^{fin}(p;m_{1},m_{2}) & = & \int_{0}^{1} dx \; \ln \big[p^{2}x^{2} + 
m_{1}^{2}-(p^{2}+m_{1}^{2}-m_{2}^{2})x\big] x,   
\nonumber \\ \nonumber \\
B_{\mu\nu}(p;m_{1},m_{2}) & = & \int \frac{d^{n}q}{i\pi^{2}}
\frac{q_{\mu}q_{\nu}}{(q^{2}-m_{1}^{2}+i\epsilon)\big[(q-p)^{2}-m_{2}^{2}
+i\epsilon\big]}
\nonumber \\
& = & p_{\mu}p_{\nu}B_{21} - g_{\mu\nu}B_{22}.   
\end{eqnarray}
%*************B functions reduction *****************
The functions $B_{1}, B_{21}$ and $B_{22}$ can be reduced to
\begin{eqnarray}
B_{1} & = & \frac{1}{2p^{2}}\big\{ -A_{0}(m_{1})+A_{0}(m_{2}) -
(p^{2} + m_{1}^{2} - m_{2}^{2})B_{0}\big\}, \nonumber \\ 
B_{21} & = & \frac{1}{3p^{2}}\big\{-A_{0}(m_{2})-2(p^{2}+m_{1}^{2}-
m_{2}^{2})B_{1}
-m_{1}^{2}B_{0}-1/2(m_{1}^{2}+m_{2}^{2}-p^{2}/3)\big\}, \nonumber \\
B_{22} & = & \frac{1}{6}\big\{+A_{0}(m_{2})-(p^{2}+m_{1}^{2}-m_{2}^{2})B_{1}
-2m_{1}^{2}B_{0}-(m_{1}^{2}+m_{2}^{2}-p^{2}/3)\big\}. 
\end{eqnarray}
%**************A function definition*********************
The functions $A$ are defined as 
\begin{eqnarray}
A_{0}(m) & = & - \int \frac{d^{n}q}{i\pi^{2}}\frac{1}{(q-p)^{2}-m^{2}}  
\;\;\; = \;\;\; - \int \frac{d^{n}q}{i\pi^{2}}\frac{1}{q^{2}-m^{2}} \nonumber 
    \\
& = & - m^{2}(\Delta - \ln m^{2} + 1),  \nonumber \\  \nonumber \\
A_{\mu}(p,m) & = & \int \frac{d^{n}q}{i\pi^{2}}\frac{q_{\mu}}{(q-p)^{2}-m^{2}}
\;\;\; = \;\;\; - p_{\mu}A_{0}(m).
\end{eqnarray}
%*************F function definition *********************
Relations between F and  B functions:
\begin{eqnarray}
\label{baf}
F(p;m_{1},m_{2}) & = & -1 + \frac{m_{1}^{2}+m_{2}^{2}}
{m_{1}^{2}-m_{2}^{2}}\;\ln \frac{m_{1}}{m_{2}} + \ln m_{1} + \ln m_{2}+
 B_{0}(p;m_{1},m_{2}),  \nonumber \\
F(p;0,m) & = & -1 + \ln m^{2} + B_{0}(p;0,m),  \nonumber \\
B_{1}(p;m_{1},m_{2}) & = & \frac{m_{2}^{2}-m_{1}^{2}}{2}
\frac{F(p;m_{1},m_{2})}{p^{2}} -
\frac{1}{2}B_{0}(p;m_{1},m_{2}). 
\end{eqnarray}
%%%%%%%%%%%%%B and F for small s %%%%%%%%%%%%%%%%%%%%%%%%%%%%%%%%%%%%%%%%
For $s = p^{2}$ small with respect to $m_{1}^{2}, m_{2}^{2}, m^{2}$,  we have
\begin{eqnarray}
\label{males}
F(p;m_{1},m_{2}) & = & \frac{s}{{(m_{1}^{2}-m_{2}^{2})}^{2}}\bigg[
\frac{m_{1}^{2}+m_{2}^{2}}{2} - \frac{m_{1}^{2}m_{2}^{2}}{m_{1}^{2}-m_{2}^{2}}
\ln \frac{m_{1}^{2}}{m_{2}^{2}}\biggr],   \nonumber \\
B_{0}(p;m_{1},m_{2}) & = & 1 - \frac{m_{1}^{2}+m_{2}^{2}}
{m_{1}^{2}-m_{2}^{2}}\ln \frac{m_{1}}{m_{2}} -\ln m_{1} -\ln m_{2} +O(s),
\nonumber \\   
B_{0}(p;0,m) & = & 1 -2\ln m +O(s),  \nonumber \\
B_{1}(p;m_{1},m_{2}) & = & \frac{1}{2}\frac{1}{m_{2}^{2}-m_{1}^{2}}\bigg[
\frac{m_{1}^{2}+m_{2}^{2}}{2} - \frac{m_{1}^{2}m_{2}^{2}}{m_{1}^{2}-m_{2}^{2}}
\ln \frac{m_{1}^{2}}{m_{2}^{2}}\biggr] -\frac{1}{2}B_{0}(p;m_{1},m_{2}),
\nonumber \\
B_{1}(p;0,m) & = & -\frac{1}{4} + \ln m + O(s),  \nonumber \\
B_{1}(p;m,0) & = & -\frac{3}{4} + \ln m + O(s). 
\end{eqnarray}
%%%%%%%%%%%%%% B0,B1(lambda) %%%%%%%%%%%%%%%%%%%%%%%%%%%%%%%%%%%%%%%%%%%%%%
Finally, in photon loops we encounter functions with regularized photon mass
\linebreak $m_{\lambda}~\rightarrow~0$:
\begin{eqnarray}
\label{bphoton}
\displaystyle \left. B_{0}(p;m_{\lambda},m_{l}) \right|_{p^{2}=m_{l}^{2}} 
& = & 2-2\ln m_{l}, \nonumber  \\
\displaystyle \left. B_{1}(p;m_{\lambda},m_{l}) \right|_{p^{2}=m_{l}^{2}}
& = & -\frac{1}{2} + \ln m_{l}, \nonumber \\
\displaystyle \left. B_{1}(p;m_{l},m_{\lambda}) \right|_{p^{2}=m_{l}^{2}}
& = & -\frac{3}{2} + \ln m_{l}, \nonumber \\
\displaystyle \left. 
\frac{\partial B_{0}}{\partial p^{2}}(p;m_{\lambda},m_{l})
\right|_{p^{2}=m_{l}^{2}}
& = & 
\frac{1}{m_{l}^{2}}\Big(-1 - \ln \frac{m_{\lambda}}{m_{l}}\Big), \nonumber \\
\displaystyle \left.
\frac{\partial B_{1}}{\partial p^{2}}(p;m_{\lambda},m_{l})
\right|_{p^{2}=m_{l}^{2}} 
& = &
- \frac{1}{2 m_{l}^{2}}. 
\end{eqnarray}

\newpage

%%%%%%%%%%%%%%%%%%%%%%%%%%REFERENCES%%%%%%%%%%%%%%%%%%%%%%%%%%%
%\begin{references}

%\end{references}

\end{document}